%% file: main.tex
\documentclass[a4paper, 12pt,oneside]{book} 
\pdfoutput=1 
\usepackage[a4paper,top=3.5cm,left=3cm,right=3cm,bottom=2.5cm]{geometry}
\usepackage{times}
\usepackage[utf8]{inputenc}
\usepackage[T1]{fontenc}

\usepackage{fancyhdr}
\usepackage{graphicx}
\graphicspath{{\subfix{./images/}}}
\usepackage{caption}
\usepackage{subcaption}
\usepackage{setspace}

\usepackage{epigraph}
\usepackage{bm}
\usepackage{dsfont}
\usepackage{mathrsfs}
\usepackage[retainorgcmds]{IEEEtrantools}
\usepackage{amsmath}
\usepackage{amssymb}
\usepackage{amsfonts}
\usepackage{nicefrac}
\usepackage{physics}
\usepackage[amssymb]{SIunits}
\usepackage{afterpage}
\usepackage{color}
\usepackage{lipsum}
\usepackage{stackengine}
\usepackage{wrapfig} 

\stackMath
\usepackage{wasysym} 
\usepackage{skak} 
\usepackage{pdfpages} 

\usepackage[dvipsnames]{xcolor}
\usepackage{tikz}
\usetikzlibrary{decorations.pathmorphing}
\usetikzlibrary{arrows}
\usetikzlibrary{decorations.markings}

\usepackage{tikz-3dplot}
\usepackage{pgfplots}
\tdplotsetmaincoords{60}{115} 
\pgfplotsset{compat=newest}   
    
\usepackage[colorlinks=true,linkcolor=blue,urlcolor=blue,citecolor=blue,pdfusetitle]{hyperref}

\usepackage{mmacells}

\usepackage[framemethod=TikZ]{mdframed}

\newenvironment{boxedCode}
{
	\begin{mdframed}[  roundcorner=10pt,backgroundcolor=RoyalBlue!20 ,linecolor=RoyalBlue!70!black,linewidth=1.5pt]
	}
	{
	\end{mdframed}

}

\DeclareUnicodeCharacter{2217}{*}
\DeclareUnicodeCharacter{0301}{*************************************}

\usepackage{subfiles}

\usepackage{tikz}
\usepackage[mode=build]{standalone} 

\usepackage[labelfont=bf]{caption}

\newcommand\blankpage{%
    \null
    \thispagestyle{empty}%
    \addtocounter{page}{-1}%
    \newpage}

\usepackage[style=phys, biblabel=brackets, doi=false,isbn=false, url=false, eprint=false,backend=bibtex]{biblatex}

\newcommand{\avg}[2]{\mathbb{E}_{#2}\left[ #1 \right]}
\DeclareMathOperator{\Var}{Var}

\newcommand{\ppx}[2]{ \frac{\partial #1}{\partial #2}}

\newcommand{\expo}[1]{\exp\left( #1 \right)}
\DeclareMathOperator{\EX}{\mathbb{E}}
\newcommand{\choi}[1]{\text{vec}{\left( #1 \right)}} 
\newcommand{\expec}[1]{\langle {#1} \rangle}          
\newcommand{\trp}[1]{\text{Tr}(#1)}

\DeclareMathOperator*{\argmax}{arg\,max}

\title{Bayesian estimation for collisional thermometry and time-optimal holonomic quantum computation}
\author{Gabriel Oliveira Alves}
\date{09 February 2023}
\begin{document}

\include{titlepage_portuguese}

\includepdf[pages=-]{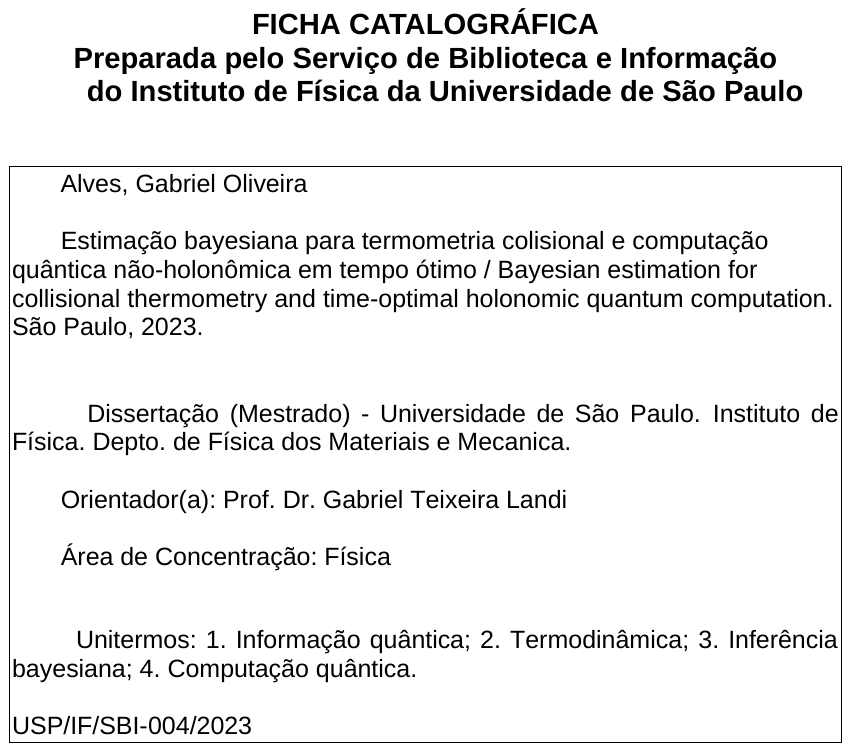}

\include{titlepage}

\blankpage

\frontmatter

\newpage

\thispagestyle{plain}
\vspace*{\fill}
\begin{center}
  \Large \textit{Ao meu avô, Neleu Alves, in memoriam.}
\end{center}
\vspace*{\fill}

\chapter*{\centering \normalfont\emph{Acknowledgements}}

Dada a natureza deste trabalho, acredito que talvez seja apropriado começar direcionando meus agradecimentos aos meus importantes professores e mestres. Em particular, gostaria de começar agradecendo ao meu orientador Gabriel Landi, que é o tipo de professor e pesquisador que ensina -- e inspira -- através do exemplo. Seu carinho pela pesquisa e pelo ensino é contagiante. Mais importante ainda, agradeço por você sempre ter se esforçado para nos proporcionar o melhor ambiente possível no grupo, tanto em aspectos humanos quanto científicos. Tenho certeza de que não havia lugar melhor para eu começar minha carreira, sou grato por tudo. 

Estendo também meus agradecimentos ao meu  orientador no período do IFSP e queridíssimo amigo Marcio Matsumoto. O privilégio que eu tive de receber tanto apoio em um momento de formação tão crucial, seja como cientista ou como pessoa, certamente foi fundamental para todas as minhas empreitadas posteriores. Não só isso, o respeito e humildade com que você trata todas as pessoas nos ensina a sempre a fazer o mesmo. É reconfortante ter a mentoria de pessoas assim. Não menos importante foi também o apoio do professor José Otávio, que reiteradamente demonstrou carinho tremendo pelos alunos, encarando o ensino e a sala de aula com muita seriedade e profissionalismo. 

Em seguida, não posso deixar de mencionar o apoio perene de toda a minha família. Começo pelos meus pais, que são um exemplo de resiliência e de bondade. Aqueles que, além de sempre terem me apoiado e me dado o luxo de crescer e aprender nas melhores condições possíveis, constantemente se preocuparam em mostrar as coisas que fazem um  bom ser humano. Igualmente importantes foram meus avós. Agradeço meus avós paternos Neleu e Maria Lucia por terem cuidado de mim com muito amor e com ensinamentos igualmente fundamentais. Agradeço também meus avôs maternos Valdete e Selma, pelo carinho e pelo exemplo de força. Sempre presentes e também muito importantes  foram também minhas tias e tios: Mariana, Rosana, Thiago e Diego.    
Incluo também neste grupo minha querida prima Vanessa e o John.

Agora, incluo meus importantes amigos nestes agradecimentos. Agradeço todos os amigos do grupo, em especial o grande Naim, pela   agradável e constante companhia. Agradeço aos meus amigos Yassin e Lefundes, que são de inteligência singular. Agradeço também o brilhante Roberto Ceccato. 

Ademais, não posso deixar de citar meus muito importante e muito talentosos companheiros da época de IFSP, -- especialmente pela paciência por terem me aturado por uma década: Gustavo Barranova, Melissa, Matheus, Nat, Toninho e Tairan. Incluo também meus bons amigos Rodolfo, Pedro e Minoru pelos ótimos momentos e por todos os projetos nos quais participamos.  

De maneira similar, estendo também meus agradecimentos aos meus queridíssimos (e geniais!) amigos do curso de Ciências Moleculares: Luiz, Samir e Shen! Agradeço também meus veteranos muito gentis e igualmente capazes: Magno e Ramon. Obrigado por sempre cuidarem de mim. 

Agradeço aos meninos: André, Ian, Manuel e Matheus pelo apoio incondicional, pela companhia duradoura e por serem incríveis. Infelizmente as páginas dessa dissertação não são o suficiente para tecer os elogios apropriados (dos quais vocês inclusive já estão cientes). Vocês são meu orgulho. 

Agradeço a Coordenação de Aperfeiçoamento de Pessoal de Nível
Superior (CAPES) e a Fundação de Amparo à Pesquisa do Estado de São Paulo (Fapesp), processo nº 2020/16050-0, pelo financiamento. Também agradeço todos os funcionários do IFUSP por manter este ambiente funcionando bem. Em especial, agradeço a Sandra e o pessoal da CPG por terem sido sempre muito solícitos e prestativos.

Finally, I would also like to thank prof. Erik Sj\"oqvist, who kindly accepted to supervise me during my exchange studies in Uppsala. Thanks to him I could make the most of my time there. Likewise, I extend my thanks to prof. Carlos Moysés Graça Araujo. Without his support this exchange wouldn't even be possible in the first place. 
I am also thankful to all the friends I made during my stay in Uppsala: Paula, Nader, Umer, Sorana, Arsalan, Felix, Artur, Shila, Ingrid, Mostafa, Marco, Mo, Bilal, Ali, Umair and Zeeshan. Finally, I also acknowledge the staff in the Uppsala university, the staff in University of São Paulo and the financial support given by the European Commission through the Swedish Council for Higher Education in the framework of Erasmus+ KA107.

\newgeometry{top=0cm}

\thispagestyle{plain}
\vspace*{\fill}
\epigraph{
\itshape 
``
Deixe-me ir\\
Preciso andar\\
Vou por aí a procurar\\
Rir pra não chorar\\
Deixe-me ir\\
Preciso andar\\
Vou por aí a procurar\\
Sorrir pra não chorar
\newline

Quero assistir ao sol nascer\\
Ver as águas dos rios correr\\
Ouvir os pássaros cantar\\
Eu quero nascer\\
Quero viver
''
}{--- Cartola}
\vspace*{\fill}

\restoregeometry

\chapter*{\centering \normalfont\emph{Abstract}}

\newcommand{\forceindent}{\leavevmode{\parindent=2em\indent}}

\forceindent In this thesis we deal with two different topics.
In the first half we investigate how the Bayesian formalism can be introduced into the problem of quantum thermometry -- a field which exploits the high level of control in coherent devices to offer enhanced precision for
temperature estimation. 
In particular, we investigate concrete estimation strategies, with focus on collisional thermometry, a protocol where a series of ancillae are sent sequentially to probe the system’s temperature. We put forth a complete framework for analyzing collisional thermometry using Bayesian inference. 
The approach is easily implementable and experimentally friendly. Moreover, it is guaranteed to always saturate the Cramér-Rao bound in the long-time limit. 
Subtleties concerning the prior information about the system’s temperature are also discussed and analyzed in terms of a modified Cramér-Rao bound associated with Van Trees and Schützenberger.

Meanwhile, in the last part of the thesis we approach the problem of non-adiabatic holonomic computation. Namely, we investigate the implementation based on $\Lambda$-systems. 
It is known that a three-level system can be used in a $\Lambda$-type configuration in order to construct a universal set of quantum
gates through the use of non-Abelian nonadiabatic geometrical phases. Such construction allows for high-
speed operation times which diminish the effects of decoherence. This might be, however, accompanied by a
breakdown of the validity of the rotating-wave approximation (RWA) due to the comparable timescale between
counter-rotating terms and the pulse length, which greatly affects the dynamics. Here, we investigate the trade-off
between dissipative effects and the RWA validity, obtaining the optimal regime for the operation of the holonomic
quantum gates.

\vspace{1.5cm}

\noindent \textbf{Keywords}: Quantum thermodynamics; Quantum thermometry; Quantum metrology; Open quantum systems; Geometrical phases; Non-adiabatic quantum computing; Holonomic quantum computing.

\chapter*{\centering \normalfont\emph{Resumo}}

\forceindent Realizamos, nesta dissertação, uma investigação a respeito de dois tópicos distintos. Inicialmente lidamos com o formalismo Bayesiano e com o desafio de como introduzi-lo na termometria quântica -- um campo que explora o alto nível de controle em dispositivos coerentes a fim de oferecer precisão aprimorada na inferência de temperaturas. 
Em particular, nós investigamos estratégias de estimação concretas, com foco na termometria colisional, um protocolo onde uma série de ancilas interage sequencialmente com um sistema de interesse, sondando sua temperatura. 
Esta abordagem é facilmente implementável e também propícia experimentalmente. Ademais, ela também assegura que a desigualdade de Cramér-Rao seja saturada no limite assintótico.
Sutilezas a respeito da informação prévia da temperatura do sistema também são discutidas e analisadas em termos de uma desigualdade de Cramér-Rao modificada, associada a Van Trees e Sch\"utzenberger.

Já na parte final deste trabalho, abordamos o problema de computação quântica não-adiabática. Isto é, nós investigamos uma implementação baseada em sistemas $\Lambda$.
Sabe-se que um sistema de três níveis pode ser utilizado na configuração de tipo $\Lambda$ de modo que um conjunto universal de portas lógicas quânticas seja construído. 
Este tipo de implementação permite tempos de operação muito curtos, que amenizam os efeitos da decoerência. Esta escolha pode ser, contudo, acompanhada da invalidação 
da aproximação de onda girante (RWA) devido às escalas de tempo equivalentes entre 
os termos girantes e a duração do pulso; o que afeta a dinâmica do sistema de maneira significativa.
Nosso objetivo consiste em investigar a competição entre os efeitos dissipativos e a validade da RWA, obtendo um regime ótimo de operação para as portas quânticas holonômicas. 

\vspace{1.5cm}

\noindent \textbf{Palavras-chave}: Termodinâmica quântica; Termometria quântica; Metrologia quântica; Sistemas quânticos abertos; Fases geométricas; Computação quântica não-adiabática; Computação quântica holonômica.

\tableofcontents

\listoffigures

\listoftables

%

\mainmatter

\begin{spacing}{1.5}

\chapter{Introduction}
\input{chapters/introduction}


\part{Bayesian estimation for collisional thermometry}

\chapter{Estimation theory}\label{chp:frequentist}
\input{chapters/inference}

\chapter{Bayesian statistics}\label{chp:bayes}
\input{chapters/bayes}

\chapter{Quantum metrology and thermometry}\label{chp:metrology}

\input{chapters/metrology}

\chapter{Bayesian quantum thermometry}\label{chp:results}

\input{chapters/results}


\part{Time-optimal holonomic quantum computation}

\chapter{Geometrical phases in quantum mechanics}\label{chp:geometric}

\input{chapters/geometric}

\chapter{Non-adiabatic holonomic quantum computing}\label{chp:lambda}

\input{chapters/lambda}

\chapter{Time-optimality}\label{chp:results_qc}

\input{chapters/results_qc}

\chapter{Conclusion}
\input{chapters/conclusion}

\appendix

\chapter{Vectorization}\label{appendix:vectorization}
\input{chapters/vectorization}

\chapter{Bayesian inference for correlated ancillae}\label{appendix:correlations}

\input{chapters/correlations}

\chapter{Uniform sampling over the Bloch sphere}\label{appendix:fibonacci}

\input{chapters/fibonacci}

\end{spacing}
\printbibliography

\end{document}

%% file: titlepage_portuguese.tex
\begin{titlepage}
\begin{center}
	{\fontsize{16}{16} \selectfont Universidade de S\~ao Paulo \\}
	\vspace{0.1cm}
	{\fontsize{16}{16} \selectfont Instituto de F\'{i}sica}
    \vspace{3.3cm}

	{\fontsize{22}{22}\selectfont Estimação Bayesiana para termometria colisional e  computação quântica não-holonômica em tempo ótimo\par}
    \vspace{2cm}


    {\fontsize{18}{18}\selectfont Gabriel Oliveira Alves \par}

    \vspace{2cm}

\end{center}

\noindent \leftskip 6cm Orientador: Prof. Dr. Gabriel Teixeira Landi \vspace*{\fill}

\vspace{0.8cm}    


\par
\leftskip 6cm
\noindent {Disserta\c{c}\~{a}o de mestrado apresentada ao Instituto de F\'{i}sica da Universidade de S\~{a}o Paulo, como requisito parcial para a obten\c{c}\~{a}o do t\'{i}tulo de Mestre em Ci\^{e}ncias.}
\par
\leftskip 0cm
\vskip 2cm


\noindent Banca Examinadora: \\
\noindent Prof. Dr. Luis Gregorio Godoy de Vasconcellos Dias da Silva - Universidade de S\~ao Paulo\\
Prof. Dr. Saulo Vicente Moreira - Universidade Federal do ABC\\
Prof. Dr. Fabrício de Souza Luiz - Universidade Estadual de Mato Grosso do Sul\\
\vspace{2.0cm}

\begin{center}
    {S\~ao Paulo \\  2023}
\end{center}
\vfill

\end{titlepage}

%% file: titlepage.tex
\begin{titlepage}
\begin{center}

	{\fontsize{16}{16} \selectfont University of S\~ao Paulo \\}
	\vspace{0.1cm}
	{\fontsize{16}{16} \selectfont Physics Institute}
    \vspace{3.3cm}

	{\fontsize{22}{22}\selectfont Bayesian estimation for collisional thermometry and time-optimal holonomic quantum computation \par}
    \vspace{2cm}

    {\fontsize{18}{18}\selectfont Gabriel Oliveira Alves \par}
    \vspace{2cm}

\end{center}

\noindent \leftskip 6cm Supervisor: Prof. Dr. Gabriel Teixeira Landi


    \vspace{0.8cm}    


\par
\leftskip 6cm
\noindent {Dissertation submitted to the Physics Institute of the University of S\~{a}o Paulo in partial fulfillment of the requirements for the degree of Master of Science.}
\par
\leftskip 0cm
\vskip 2cm


\noindent Examining Committee: \\
\noindent Prof. Dr. Luis Gregorio Godoy de Vasconcellos Dias da Silva - University of S\~ao Paulo\\
Prof. Dr. Saulo Vicente Moreira - Federal University of ABC\\
Prof. Dr. Fabrício de Souza Luiz - State University of Mato Grosso do Sul \\
\vspace{2.8cm}

\begin{center}
    {S\~ao Paulo \\  2023}
\end{center}
\vfill
\end{titlepage}

%% file: chapters/introduction.tex
In this dissertation we engage with two different problems. From  chapters~\ref{chp:frequentist} through \ref{chp:results} we tackle different aspects of (Bayesian) estimation theory and quantum metrology, applying these concepts into collisional quantum thermometry. 
Meanwhile, chapters~\ref{chp:geometric} - \ref{chp:results_qc} constitute the latter half of the dissertation, where we deal  with holonomic quantum computing. Further contextualization and a more complete introduction to each subtopic can be found at the beginning of each of these chapters.

The contents of the first and second parts of this work were conducted under the supervision of prof.~Dr.~Gabriel T. Landi and prof.~Dr.~Erik Sj\"oqvist, respectively. I worked on the latter half mainly during my exchange studies at the University of Uppsala. Any mistakes, inaccuracies or oversights in this work are entirely of my own responsibility. This research has been supported by CAPES, FAPESP (Grant No.2020/16050-0) and the framework of Erasmus + KA107 International Credit Mobility, financed by the European Commission.

\section{Bayesian estimation for collisional quantum \\ thermometry}

Temperature is an ubiquitous notion which is prevalent in the everyday life and, despite finding its roots in thermodynamics, is also of widespread importance in several fields \cite{goldanskiiChemicalReactionsVery1976, disalvoThermoelectricCoolingPower1999, uchidaObservationSpinSeebeck2008, walterSeebeckEffectMagnetic2011, iftikharPrimaryThermometryTriad2016}.
It is thus not surprising that further extending our comprehension of thermodynamics -- especially the formalization of the concept of temperature -- and how it integrates with relevant technologies is constant endeavor in physics. 
In this context, quantum thermodynamics evolved as a natural extension of thermodynamics into the domain of quantum mechanics~\cite{vinjanampathyQuantumThermodynamics2016, gooldRoleQuantumInformation2016}.

In a similar fashion, the precise manipulation of parameters in quantum systems is, at the same time, both a central requirement for cutting-edge quantum technologies 
and also a very hallmark of the ever-growing technological prowess of quantum science. In a sense, the engineering of quantum systems and our conceptual grasp of quantum theory evolve in a symbiotic relationship, where a deeper understanding of fundamental concepts allows us to develop more advanced technology and vice-versa~\cite{leanhardtCoolingBoseEinsteinCondensates2003, hoferQuantumThermalMachine2017, fujiwaraRealtimeNanodiamondThermometry2020}. 
With such motivations in mind, quantum metrology provides a well-established foundation accompanied by many versatile tools~\cite{Paris2009}. 
Many parameters of interest in real applications cannot be directly measured in a laboratory. For that, their value must be \emph{inferred} from other quantities which are directly accessible.
This is where \emph{statistics} and \emph{estimation theory} come into rescue; they provide the proper framework to deal with this type of challenge. 

These lines of investigation converge into what is known as \emph{quantum thermometry}, a field which is concerned with the problem of temperature estimation in quantum systems, taking into account the distinct traits of quantum thermodynamics~\cite{staceQuantumLimitsThermometry2010}. 
Translating the well known notions and ideas from classical theory into quantum thermometry and quantum thermodynamics is a herculean effort due to the very small scales and novel phenomena from quantum physics~\cite{gooldNonequilibriumQuantumLandauer2015, kammerlanderCoherenceMeasurementQuantum2016, micadeiReversingDirectionHeat2019}. 
The challenges range from the reformulation of the laws of the thermodynamics~\cite{landiIrreversibleEntropyProduction2021, silvaQuantumMechanicalWork2021} and the very definition of temperature~\cite{lipka-bartosikWhatTemperatureQuantum2022}, to how we model interactions and how we design experiments~\cite{rivasOpenQuantumSystems2012, batalhaoExperimentalReconstructionWork2014, timpanaroThermodynamicUncertaintyRelations2019, aliQuantumThermodynamicsSingle2020}. 
 
Our objective in this dissertation is to incorporate the Bayesian framework into quantum thermometry, a research venue which, apart from recent works~\cite{rubioBayesianMultiparameterQuantum2020, boeyensUninformedBayesianQuantum2021}, has been hitherto seldom explored. In particular, we employ tools from Bayesian estimation to study a concrete thermometric platform which makes use of collisional models~\cite{ciccarelloCollisionModelsQuantum2017a} in what is known as \emph{collisional quantum thermometry}~\cite{seahCollisionalQuantumThermometry2019}.
In contrast to the paradigm known as \emph{local} quantum thermometry, where one usually has a reasonably accurate estimate for the parameter beforehand, our approach falls into what is known as \emph{global} quantum thermometry~\cite{mokOptimalProbesGlobal2021}, which demonstrates good applicability even when previous knowledge is insufficient and the temperature lies on a broader interval. The Bayesian approach has been regarded recently as very efficient and less restrictive in such scenarios~\cite{mehboudiFundamentalLimitsBayesian2022}.

In chapter~\ref{chp:frequentist} we introduce the central concepts from statistics and classical estimation theory, mainly from the optics of the frequentist approach. Afterwards, in chapter~\ref{chp:bayes}, we extend these investigations to the Bayesian framework. These two chapters lay the groundwork for the contents of the chapter~\ref{chp:metrology}, where we discuss quantum metrology in generality, and later on, quantum thermometry as well. We conclude the first half of this dissertation with our original contributions in chapter~\ref{chp:results}.

\section{Time-optimal holonomic quantum computation}

The advent of quantum mechanics in the past century has brought a plethora of different technologies upon us. Quantum computing, for instance, has seen an auspicious development throughout the years. 
Seminal works such as the ones by Deutsch \cite{Deutsch1992} and Shor \cite{Shor1997} have established a new ground for quantum technologies, showcasing how quantum mechanics can be used to optimize tasks in computation and information processing.  

However, quantum computing naturally requires a suitable platform to be performed on.  Those, in turn, should exhibit several desirable properties, such as scalability \cite{DiVincenzo2000}, robustness against error \cite{Shor1995, Steane1996, Preskill1998}, long decoherence time \cite{Joos1985, Unruh1995} and universality \cite{Barenco1995}. 
The search for such type of implementation is, in itself, an extensive area of research -- one which has seen considerable progress over the last decades, with proposals ranging from trapped ions \cite{Cirac1995} to topological systems \cite{Kitaev2003}.

In this context, holonomic quantum computing emerged at the end of the nineties  as an encouraging alternative \cite{Zanardi1999, Duan2001}. 
This approach is based on the use of geometrical phases for quantum computing due to their inherent robustness against certain types of errors and noise. Even within this field, several different implementations and protocols have been proposed \cite{Zhang2021}. In particular, early proposals were very often restricted to adiabatic implementations. These, however are more susceptible to open quantum system phenomena due to their long operation time. For that reason, non-adiabatic implementations~\cite{xiang-binNonadiabaticConditionalGeometric2001, zhuImplementationUniversalQuantum2002} are regarded as an effective strategy in mitigating these effects~\cite{Sjoqvist2012, Shen2021}.

In this work we examine one of these approaches, focusing our attention on a non-adiabatic scheme for quantum computing based on three-level $\Lambda$-type systems, proposed in \cite{Sjoqvist2012}. 
This framework has seen a couple of generalizations and alternative constructions in the last few years, together with concrete experimental implementations. 
Our main focus here is to study this implementation outside the regime of validity of the rotating wave approximation (RWA). Namely, we show that the breakdown in the RWA and decoherence arise as competing effects in the model, which suggests a time-optimal implementation.

In chapter~\ref{chp:geometric} we review basic concepts about geometrical phases, with special focus on non-Abelian non-adiabatic phases. In chapter~\ref{chp:lambda} we review the proposal for non-adiabatic quantum computing using $\Lambda$-type systems and we introduce some of our results for the non-RWA case. Finally, in chapter~\ref{chp:results_qc} we discuss our main results.

%% file: chapters/inference.tex
Statistics is one of the most ubiquitous branches of mathematics due to its inherent applicability in real-life scenarios and experimental investigations~\cite{Cowan1956, Kaplan1958, Fisher1971, Cox1972, Haslett1989, Lange1997, Hughes1999, Dochain2003, Vapnik2015, Abbott2016, Bernhard2019}. It deals with questions ranging from fundamental and rigorous formulations of the theory \cite{Jaynes2003, Klenke2014, Bickel2015}, in a large intersection with probability theory, to the development of practical statistical methods and algorithmic approaches~\cite{Nelder1972, Dempster1977, Efron1979, Gilks1995, Breiman2001}. Statistics, from its centuries of development, provides a plethora of available tools and theoretical frameworks~\cite{Stigler1999, Tabak2004}. Our objective in this section is to give a brief overview of one of its sub-branches: the estimation theory.  


Just like its parent field, estimation theory is a quite extensive subject and is contained within two major frameworks: the frequentist and the Bayesian approaches. In this section we will deal with the former. As we shall see throughout the next few chapters, there is no single answer or clear-cut formulation to every circumstance we will encounter. For that reason, we discuss relevant definitions and several tools that might be of use in order to construct a versatile toolbox. Ultimately, which approach will be the most useful will depend on the problem at hand and practical constraints. 

Nevertheless, in most of these situations we are interested in drawing useful information from empirical data. This lead us to one of the central problems in estimation theory, which is known as point estimation. In this scenario, one is interested in obtaining a single numerical value which serves as an estimation for a given parameter. Typically, practical formulations are guided either by the estimator which is the most accurate or, at the expense of some precision, an estimator which is reasonably simple to construct. Beyond the point estimation, other approaches for parameter estimation might be based on confidence intervals or posterior distributions. In later chapters we shall study a few of these concepts.

In the current section, we will discuss some essential definitions regarding the point estimation problem with some concrete examples. We also introduce a central result from information and estimation theory, known as the Cramér-Rao bound (CRB) \cite{Rao1992, Cramer1999}. 
In chapter~\ref{chp:results}, these concepts will be used to investigate a quantum thermometer based on the so-called collision models.

\section{Estimation theory}\label{sec:estimation_theory}

Let us start by considering a string of discrete outcomes ${\bf X} = (X_1, ..., X_n)$ obtained from a given probability distribution. The sample space, i.e., the values that an element $X_i$ in ${\bf X}$ can assume, will depend on the problem. For instance, the sample space can binary, where $X_i$ is given by either $0$ or $1$, describing a coin-toss experiment. Or, similarly, $X_i = 1, ..., 6$. describing the outcomes of a dice. We might also have discrete outcomes with a continuous sample space $X_i \in \mathbb{R}$ describing, e.g., the readout of a sensor at specific time intervals \cite{Rice2006}.  

These outcomes, and their underlying probability density function (pdf) $p_\theta(X)$, will usually depend on a parameter $\theta$ (or on a vector of parameters $\boldsymbol{\theta}$). We define an estimator as an arbitrary function of ${\bf X}$ which is used to estimate $\theta$. In other words, we can denote the estimator $\hat{\theta}({\bf X})$ as
\begin{equation}\label{eq:est_def}
\hat{\theta}({\bf X}) = g(X_1, ..., X_n),
\end{equation}
where $g(X_1, ..., X_n)$ is a function of the outcomes. While this definition might seem trivial, the most important insight is that since $X_1, ..., X_n$ are random variables, the estimator $\hat{\theta}({\bf X})$ is \emph{also} a random variable itself. Therefore, it is clear that the estimator will depend on a particular \emph{realization} (or trajectory) of the problem, that is, on the particular streak of results $X_1, ..., X_n$. Therefore, a different sequence of results might yield a different estimate. In this case, how can one judge the quality of an estimator? A certain estimator might perform very well for a particular realization of the problem and very bad for the others, for example.

Another important aspect is that we obviously do not want the estimator in Eq.~\eqref{eq:est_def} to depend on the unknown parameter $\theta$, because this is the very object we are trying to estimate. However, since the parametrization $\theta$ will influence which ${\bf X}$ we might observe and how likely certain realizations are to occur, by consequence, any error measure for the estimator $\hat{\theta}({\bf X})$, as well as its distribution, will implicitly depend on the unknown parameter $\theta$ itself. Hence,

\begin{itemize}
	\item The estimator is a random variable which depends on a particular realization of $X_1, ..., X_n$, but does not depend explicitly on the parametrization $\theta$.
	\item An error measure associated with this estimator will typically depend both on the outcomes $X_1, ..., X_n$ and also on the particular value of $\theta$. 
\end{itemize}
In other words, one can define error measures $\epsilon$ which depend only on ${\bf X}$ and are conditioned solely on the parameter:
\begin{equation}\label{eq:classical_error}
\epsilon_\theta =\epsilon(\hat{\theta}({\bf X})|\theta),
\end{equation}
or some which depend only on the parameter but are conditioned on the data, so in this case we can think of the data as being fixed:
\begin{equation}\label{eq:bayese_error_pure}
\epsilon_{\bf X} = \epsilon(\theta|{\bf X}),
\end{equation}
and finally, others which are conditioned on neither and depend only on the estimator:
\begin{equation}
\epsilon_B = \epsilon(\hat{\theta}({\bf X}); \theta).
\end{equation}
This is one of the main points of divergence between different frameworks in estimation theory. The first case, as illustrated by Eq.~\eqref{eq:classical_error}, is the approach taken by \emph{frequentist} statistics and it is known as \emph{classical} estimation theory. In this scenario only the outcomes are random variables, and the parameter $\theta$ is regarded as \emph{fixed} but unknown. This will be our main framework during this chapter. The latter formulations are Bayesian approaches, which we will discuss in the next sections.

$\opposbishops$ \textbf{\textit{Example:}}  In order to make these definitions less abstract, we will work on a concrete example in order to motivate the first definition. For that, we shall discuss a coin-toss experiment, which will be a recurrent example in this dissertation. For that, we consider a coin which is not necessarily fair. That is, the outcomes are either head or tails, which we denote by $X = 0$ or $X = 1$, and they are obtained with a probability $1-\theta$ or $\theta$, respectively. The probability $p(X|\theta)$ of obtaining an outcome $X$, given $\theta$, is  thus determined by
\begin{equation}\label{eq:bernoulli}
p_\theta(X) := p(X|\theta) = \theta^X(1-\theta)^{(1-X)},
\end{equation}
which we cal the \emph{Bernoulli} distribution. If we perform sequential tosses and assume that they are all independent and identically distributed \footnote{The probability of each outcome is computed from the same probability distribution, in this case from Eq.~\eqref{eq:bernoulli}, and each trial is independent.} (i.i.d), the probability of obtaining $k$ heads and $n-k$ tails on $n$ trials is given by the \emph{binomial} distribution:
\begin{equation}\label{eq:binomial}
p(k, n|\theta) = {n \choose k}\theta^k(1-\theta)^{(n-k)}.
\end{equation}
Note that Eq.~\eqref{eq:binomial} does not depend on the particular order of the outcomes, and only on the total number of heads (and tails). This is due to the particular form of the distribution in Eq.~\eqref{eq:bernoulli} and the fact that trials are i.i.d. This property is also related to the concept of sufficient statistics, which we are going to discuss in later sections. 

\begin{figure}[t!]
	\centering
	\includegraphics[width=0.49\textwidth]{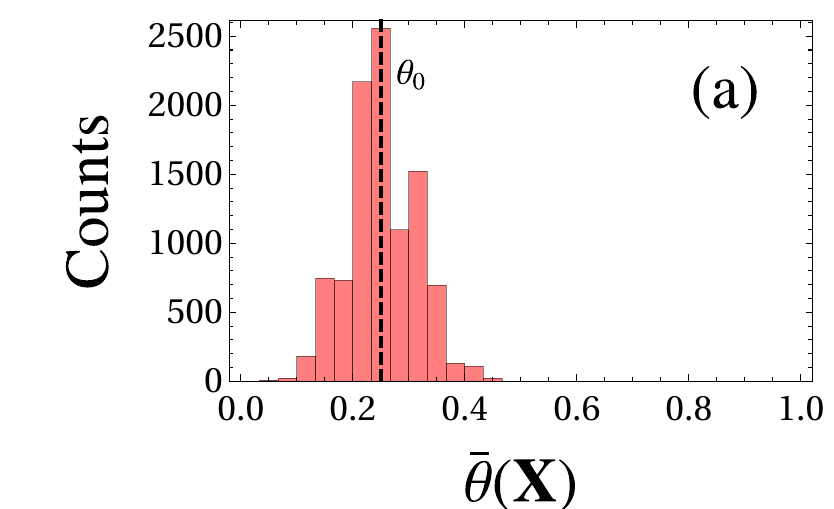}
	\includegraphics[width=0.49\textwidth]{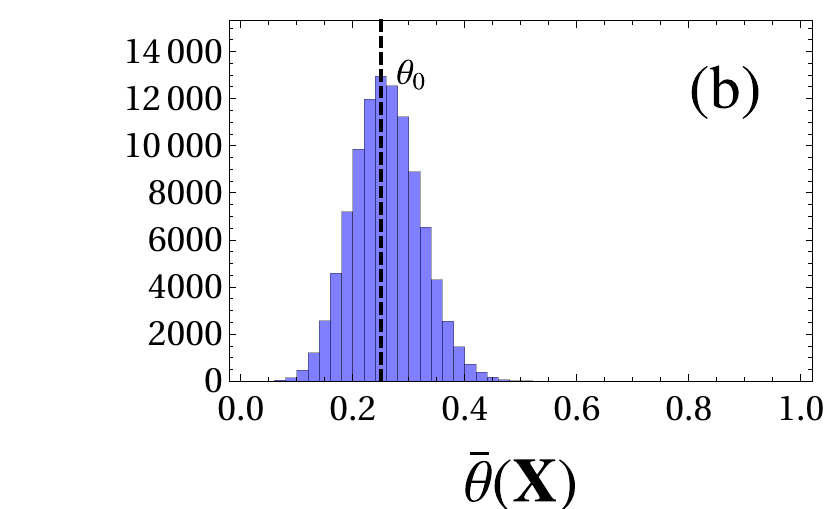}
	\caption[Histogram count for the sample mean estimator.]{Histogram count for the sample mean estimator. Results for $\theta = 0.25$ and (a) $10^4$ realizations and (b) $10^5$ realizations.}
	\label{fig:histogramEstimator}
\end{figure}

A very well-known estimator we can use to guess the true value of $\theta$ in Eqs.~\eqref{eq:bernoulli} and \eqref{eq:binomial} is the \emph{sample mean}, defined as:
\begin{equation}\label{eq:sample_mean}
\bar{\theta}({\bf X}) 
=
\overline{{\bf X}}
:= 
\frac{X_1 + ... + X_n}{n}. 
\end{equation}
As we will see, this estimator is useful due to the fact that, besides its simplicity, it is also optimal in a few senses. It is important to discuss what type of results  Eq.~\eqref{eq:sample_mean} would yield in real experiments. For that, we perform a few numerical simulations with $\theta_0 = 0.25$ being the true value of the parameter. We will consider that a single stochastic realization consists in a trial of $50$ coin-tosses. Thus, in each realization we have a stochastic trajectory ${\bf X}$ which is just a string of $50$ zeroes or ones, representing heads or tails. 
We are interested in what values the estimator~\eqref{eq:sample_mean} might result in. We have claimed before that estimators are random variables themselves, so numerical simulations are a very convenient way of visualizing their distribution. One way to see this is by calculating $\bar{\theta}({\bf X})$ for different stochastic realizations. In Fig.~\ref{fig:histogramEstimator} we plot a histogram for the sample mean $\bar{\theta}({\bf X})$ for a few thousand trajectories. 

We can see that the estimation will depend on a particular realization. We might get an estimate which yields $\bar{\theta} = 0.3$ in one trial, $\bar{\theta} = 0.17$ in another one, and so on. Still, in either case in Fig.~\ref{fig:histogramEstimator} we can see that in this example the estimator is most likely to be found at its true value $\theta_0 = 0.25$. 
That of course makes sense, we can expect to get different proportions of heads and tails by pure chance, and different values of $\theta$ could have resulted in the same sequence outcomes we have observed, so it is not possible to find the exact $\theta_0$ from a finite sample. It would be unnatural to expect to learn \emph{everything} about the model in a single-shot or just a few realizations. Nevertheless, it is true that $\theta_0$ is more likely to have generated the particular values of ${\bf X}$ that we have observed. That is why we actually get a histogram which peaks around $\theta_0$ but which is still bound to assume other values. $\bishoppair$ 


From Fig.~\ref{fig:histogramEstimator} we can see that the estimation might vary a lot depending on the realization. A few outlier trajectories might result in values which are somewhat far away from the true value of the parameter. For that reason, we arrive at two natural questions. 
The first one is whether there is any sensible error figure to quantify how close, on average, the estimator is to $\theta$. 
The second one is whether we can quantify how spread-out the distribution for the estimator is going to be. 
While an estimator might result in the true value of the parameter \emph{on average}, as shown in Fig.~\ref{fig:histogramEstimator}, for example, it still might fluctuate a lot on certain occasions. Meanwhile, other estimators might yield a much tighter distribution. We will now define a very standard error measure, which will give further insight into these two questions.

\section{The mean square error}

One of the standard error measures in statistics is the mean square error (MSE),  defined as:\footnote{In order to make the notation less cluttered,  we express $\hat{\theta}(\cdot)$ simply as $\hat{\theta}$.}
\begin{equation}\label{eq:mse}
\epsilon(\hat{\theta}|\theta) 
= 
\avg{(\hat{\theta}-\theta)^2}{}.
\end{equation}
Here, $\avg{\cdot}{}$ denotes the statistical average over the appropriate sample space.
The pdf $p(X)$ over which the integration is performed is specified in the index as $\avg{\cdot}{p}$, however we omit it whenever the distribution is clear from the context. We can expand the equation above as
\[
\epsilon(\hat{\theta}|\theta) 
=
\avg{\hat{\theta}^2}{}
- 2 \theta  \avg{\hat{\theta}}{}
+ \theta^2.
\]
Here we have used the fact that $\theta$ is deterministic. We can add extra terms on the RHS,
\[
\epsilon(\hat{\theta}|\theta) 
=
\avg{\hat{\theta}^2}{}
{\color{blue}
- \avg{\hat{\theta}}{}^2
+ \avg{\hat{\theta}}{}^2
}
- 2 \theta  \avg{\hat{\theta}}{}
+ \theta^2.
\]
to rewrite the MSE in terms of the famous bias-variance relationship:
\begin{equation}\label{eq:bias_var}
\epsilon(\hat{\theta}|\theta) 
= 
\Var{(\hat{\theta})} + b^2(\hat{\theta}).
\end{equation}
The term $\Var{(\hat{\theta})}$ denotes the \emph{variance} and it is written as:
\begin{equation}\label{eq:variance}
\Var{(\hat{\theta})}
=
\avg{\hat{\theta}^2}{}
- \avg{\hat{\theta}}{}^2.
\end{equation}
Meanwhile, we call the second term the \emph{bias}, and it is defined as:
\begin{equation}\label{eq:bias}
b(\hat{\theta})
=
\avg{\hat{\theta}}{}
-
\theta.
\end{equation}
\begin{figure}[t!]
	\centering
	\includegraphics[width=\textwidth]{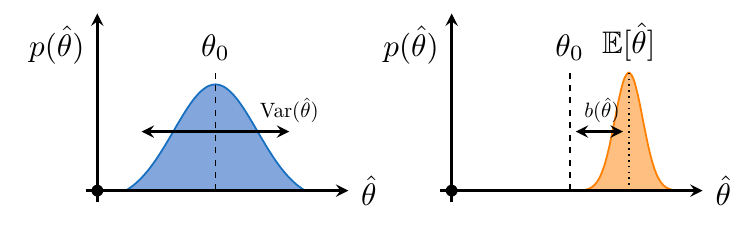}
	\caption[Illustration of the bias and the variance of an estimator.]{Illustration of the bias and the variance of an estimator. On the left we depict the variance. On the right we show the bias. Note how we have a distribution with no bias but large variance, while on the right we see the converse; a distribution with a smaller variance but with the presence of a bias.}
	\label{fig:bias_variance}
\end{figure}
The decomposition from Eq.~\eqref{eq:bias_var} is also known as the bias-variance trade-off, which is very meaningful in fields such as, e.g., machine learning \cite{Shalev-Shwartz2014, Bishop2006}. The terms in Eqs.~\eqref{eq:variance} and \eqref{eq:bias} also have very important interpretations in their own right. 
We illustrate their significance in Fig.~\ref{fig:bias_variance}. We can see that the bias is simply the difference between the expected value of the estimator and the true value of the parameter. Whenever $b(\hat{\theta})=0$ for \emph{all} $\theta$, we say that the estimator is \emph{unbiased}.
Similarly, the variance gives us information about how disperse the distribution for the estimator is. While only a proper probability distribution can give complete information about $\hat{\theta}$, the variance and the bias themselves already give a pretty important picture of the behavior of the estimation.

\begin{figure}[b!]
	\centering
	\includegraphics[width=\textwidth]{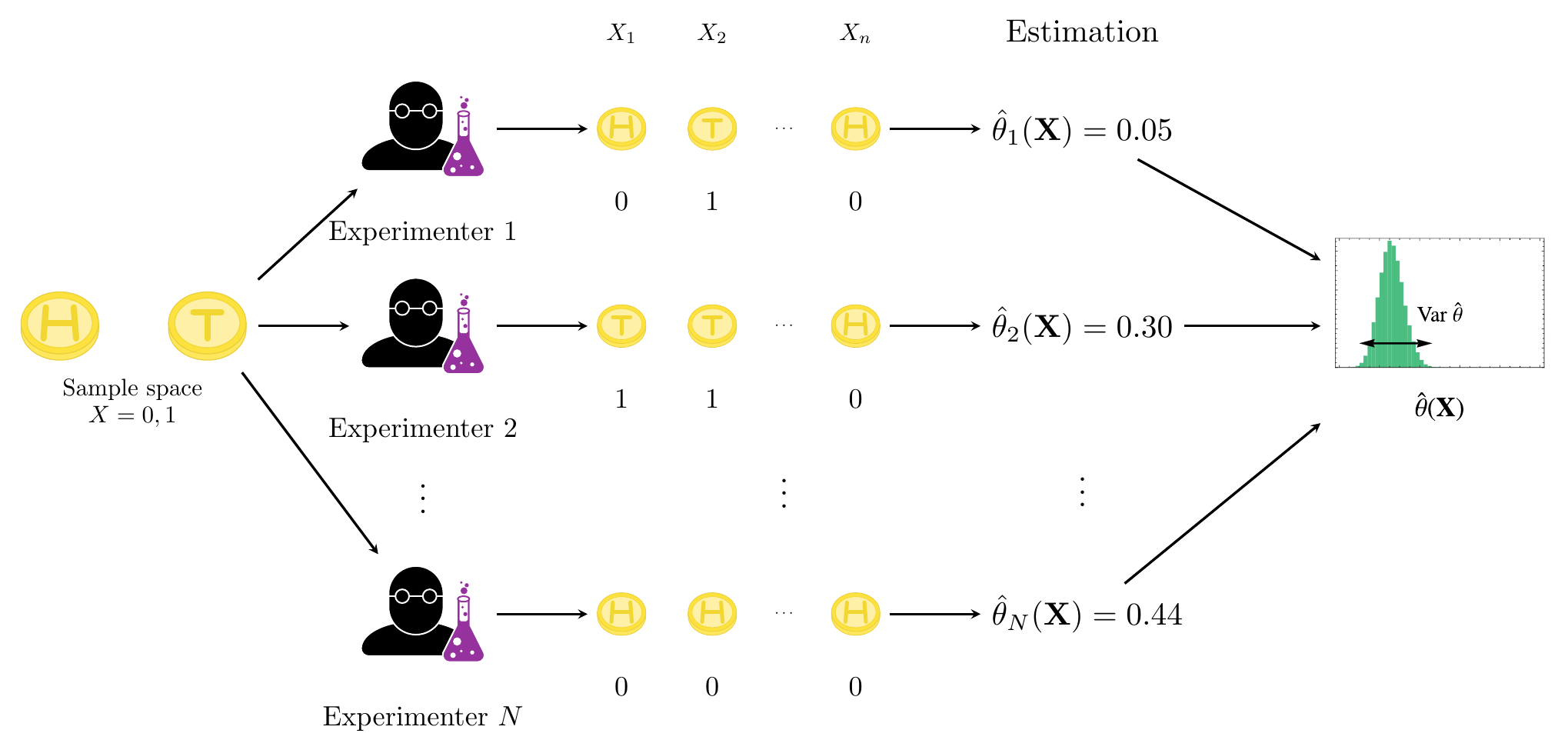}
	\caption[Illustration of a frequentist estimation procedure.]{Illustration of an estimation procedure.}
	\label{fig:frequentist_diagram}
\end{figure}

To make these definitions more intuitive, in Fig.~\ref{fig:frequentist_diagram} we describe a simple illustration of a usual estimation procedure in the frequentist sense. Going back to our example for the coin-toss experiment, we can suppose that $N$ different agents will run a series of $n$ coin tosses "in parallel". What we called a realization before is simply the $n$ coin tosses that \emph{each} of these experimenters perform. Of course, each of them will typically get different results. 
This means that the estimator $\theta_i({\bf X})$ that the $i$-th agent will obtain  will not necessarily agrees with others. If we put all of their results in a histogram, we obtain a plot like the one seen in Fig.~\ref{fig:histogramEstimator}. In this scenario, the variance quantifies how well the different scientists will (dis)agree, quantifying how spread-out this distribution is going to be. The bias, on the other hand, is simply the average of their results, which is hopefully close to the true value of the parameter.  Finally, the MSE will quantify how close to being correct their results are on average.

\section{Minimum variance (unbiased) estimators}\label{sec:MVU}

\begin{wrapfigure}{r}{0.45\textwidth}
	\includegraphics[width=0.45\textwidth]{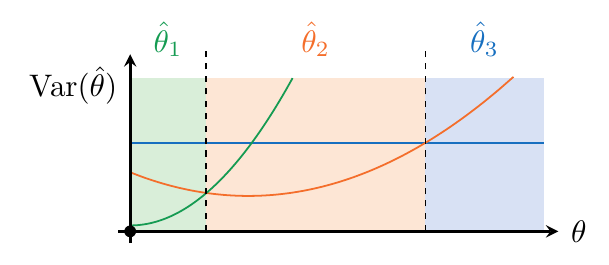}
	\caption[Variance of an estimator as a function of the parameter.]{Variance of an estimator as a function of the parameter. Its variance will depend on $\theta$. This means that there is possibly no single optimal estimator for the whole domain of the parameter. In this illustration there is no MVU estimator. A similar argument holds for the MMSE.}
	\label{fig:minimumvariance}
\end{wrapfigure}
In Sec.~\ref{sec:estimation_theory} we saw a practical example of an estimation protocol, but note that we discussed the sample mean from Eq.~\eqref{eq:sample_mean} without any regard to its optimality or possible alternatives. 
A topic which naturally arises in this context is whether it is possible to minimize the MSE from Eq.~\eqref{eq:mse}, or even more generally, if there exists a procedure to construct other types of estimators. One could be tempted to define a minimum mean square error (MMSE) estimator, which minimizes Eq.~\eqref{eq:mse}. 
However, minimizing the MSE for \emph{all} values of $\theta$ with a single estimator is, in general, not possible~\cite{Kay1993} (p.19). For that reason, there is a couple of standard rules that one might follow in order to choose an appropriate estimator. One of them is to look for \emph{minimum variance unbiased} (MVU) estimators. That is, we restrict ourselves to a class of unbiased estimators only, and choose the one with smallest variance. 
Nonetheless, even this strategy comes with a few caveats. For instance, the MVU estimator still might not exist. See Ref.~\cite{Kay1993} (p.20) or Ref.~\cite{Lehmann1998} (p.57) for examples. In Fig.~\ref{fig:minimumvariance} we depict a scenario where, depending on the value of $\theta$, different estimators are required for optimality. Even worse, there might be situations where \emph{biased} estimators are a better choices; we arrive at a smaller MSE at the expense of an estimation which is, on average, slightly different from the true value. 
This is the case of ridge regression or regularization, for example \cite{Hoerl1970, Ng2004}. Regardless of these details, for now we focus on the MVU case and follow with a concrete example where we can discuss several estimators, their performance and some relevant definitions, commenting on further details about the optimality of estimators along the way. 

$\opposbishops$ \textbf{\textit{Example:}} We discuss a non-trivial example from Ref.~\cite{Devore2016} (pp.252-257, 274) where three very natural estimators arise. 
Consider that we sample a random variable $X$ from the uniform distribution with height $p_U(X) = 1/\theta$ in the interval $[0, \theta]$. Our objective is to estimate $\theta$ from the array ${\bf X} = (X_1, ..., X_N)$, that is, to estimate the largest value that $X$ can assume.
\footnote{The \emph{German tank problem} is a historical problem which is very similar to this example \cite{Clark2021}.}

The first estimator we can try to define is the sample maximum $\hat{\theta}_1 = \max{\bf X}$. We need a few tricks in order to calculate the bias and the variance of this estimator. First, notice that for $n$ samples, the probability that $\hat{\theta}_1 = Y = \max({\bf X})$ is smaller than a certain value $y$ is given by
\begin{equation}
P(Y \leq y) = P(X_1 \leq y) P(X_2 \leq y) ...P(X_n \leq y) = \frac{y^n}{\theta^n}, 
\end{equation}
since all outcomes are independent. We can differentiate the expression above to obtain:
\begin{equation}\label{eq:pdf_maxest}
p_Y(y) =  \frac{n y^{n-1}}{\theta^n}, \quad 0 \leq y \leq \theta
\end{equation}
with $p_Y(y) = 0$ otherwise. The corresponding average reads:
\[
\avg{\hat{\theta}_1}{} =\int_0^\theta y p_Y(y)dy = \frac{n}{n+1}\theta.
\]
We then find that the bias is
\begin{equation}\label{eq:bias_estUni1}
b(\hat{\theta}_1)
=
\avg{\hat{\theta}_1}{} - \theta 
=
-\frac{\theta}{n + 1},
\end{equation}
showing that the estimator $\hat{\theta}_1$ is biased. Fortunately, we can see that the bias goes to zero as $n \rightarrow 0$, so in this case we say that the estimator is \emph{asymptotically unbiased}. 
The minus sign in the equation above also makes sense: since we are taking the maximum of a finite sample, we are guaranteed to underestimate the actual maximum of the full distribution. 
From this point of view, the presence of a negative bias in the estimation is quite reasonable. In the same vein, we can then define a modified estimator
\begin{equation}\label{eq:modified_samplemaximum}
\hat{\theta}_2 = \frac{n+1}{n} \max({\bf X})
\end{equation}
which is unbiased. Following the same steps we can actually show that $\avg{\hat{\theta}_2}{} = \theta$. The extra factor of $(n+1)/n$ correct for this underestimation that we mentioned before by overestimating some guesses, averaging out the bias. We can also calculate the variance of this estimator, finding \cite{Miller2014}:
\begin{equation}\label{eq:var_estUni2}
\Var{(\hat{\theta}_2)}=\frac{\theta^2}{n(n+2)}
\end{equation}

\begin{figure}[b!]
	\centering
	\includegraphics[width=\textwidth]{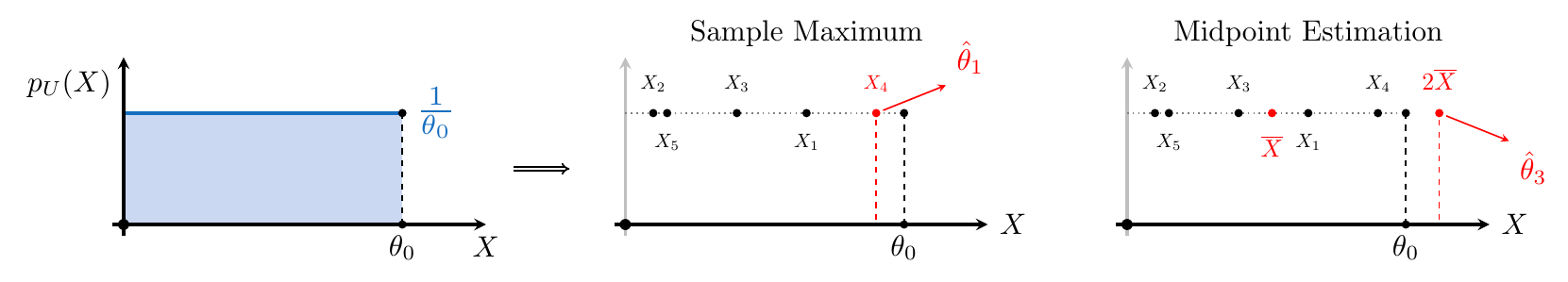}
	\caption[Estimators for the uniform sampling problem.]{Estimators for the uniform sampling problem. We show (left) the uniform distribution with unknown support, (center) the sample maximum estimator $\hat{\theta}_1$ and (right) the midpoint estimator $\hat{\theta}_3$.}
	\label{fig:uniform_problem}
\end{figure}

This, however, is not the only unbiased estimator we can define. Another sensible choice is the estimator
\begin{equation}
\hat{\theta}_3 = 2 \overline{{\bf X}}.
\end{equation}
In this case the midpoint of the uniform distribution is approximated by the sample mean $\overline{{\bf X}}$, and we double it in order to estimate $\theta$. However, we can show that the variance of this estimator is actually worse for $n>1$, because it reads:
\begin{equation}
\Var{(\hat{\theta}_3)}=\frac{\theta^2}{3n}.
\end{equation}
Therefore, we can see that the modified sample maximum estimator given by Eq.~\eqref{eq:modified_samplemaximum} is better than the midpoint estimator $\hat{\theta}_3$. It is actually possible to show that this estimator is the MVU estimator of this distribution due to the fact that ${\bf X}$ is a complete sufficient statistic \cite{Casella2002} (p.342). In fact, we point Ref.~\cite{Galili2016} to the more interested reader for a more extensive discussion on this example and how different techniques can be used to build or improve upon estimators. We briefly go back to this discussion at the end of this chapter.

As a final comment, we check which of the estimators $\hat{\theta}_1$ and $\hat{\theta}_2$ has a smaller MSE. By using Eqs.~\eqref{eq:bias_estUni1} and~\eqref{eq:var_estUni2} we can show, by direct calculation, that
\begin{equation}
\epsilon(\hat{\theta}_1|\theta) = \frac{2}{(n+1)(n+2)}\theta^2. 
\end{equation}
Meanwhile, since the modified sample mean estimator $\hat{\theta}_2$ is unbiased, its MSE is simply given by its variance in Eq.~\eqref{eq:var_estUni2}, that is $\epsilon(\hat{\theta}_2|\theta)=n\theta^2/(n+2)$. Thus, we can see that $\hat{\theta}_2$ is a better estimator even when we consider the MSE. One could be tempted to define a MMSE estimator for this example. However, as we claimed at the beginning of this section, this is not possible in general. 
As far as we are aware, this is a problem where there is no such strategy. To see this, consider the very trivial example of the constant estimator $\hat{\theta}_c = c$, with $c > 0$. In this case it is clear that the MSE is given by $\epsilon(\hat{\theta}_c|\theta) = (c-\theta)^2$. It is clear to see that this estimator is the best one for $c=\theta$ but it is a very bad estimator otherwise, specially because its accuracy does not improve asymptotically. 
While this is a somewhat artificial example, it illustrates how we might come up with some estimators which are very good for some particular values of $\theta$ but which very bad otherwise. That is why it is hard to obtain MMSE estimators in general (in the frequentist sense). 

What we \emph{can} do in this example, is to find an optimal estimator of the form $\hat{\theta}_\alpha = \alpha \max({\bf X}) = \alpha \hat{\theta}_1$. To see this, note that we can simply write the MSE of this estimator as
\begin{equation}
\epsilon(\hat{\theta}_\alpha|\theta)
=
\avg{(\alpha \hat{\theta}_1-\theta)^2}{}
=
\alpha^2\avg{\hat{\theta}_1^2}{}
-2 \alpha \avg{\hat{\theta}_1}{}
+\theta^2
.
\end{equation}
Using the results from Eqs.~\eqref{eq:bias_estUni1} and~\eqref{eq:var_estUni2} once again, we get
\begin{equation}
\epsilon(\hat{\theta}_\alpha|\theta)
=
n \theta^2
\left(
\frac{\alpha^2}{n+2}
+2\frac{\alpha}{n+1}
-\frac{1}{n}
\right).
\end{equation}
Hence, by minimizing the equation above we can see that $\alpha = (n+2)/(n+1)$ minimizes the MSE for this class of estimators. This result is quite interesting. 
Even though the modified sample mean $\hat{\theta}_2$ is the best within the class of unbiased estimators, this new estimator $\hat{\theta}_\alpha$, which is a biased one, still wins when we use the MSE as a metric of error. 
The takeaway message is that looking for biased estimators might be useful in several scenarios, and that unbiasedness does not guarantees that the MSE is minimal. $\bishoppair$

This example was very meaningful because  i) it shows how we can construct an unbiased estimator from a biased one, going from $\hat{\theta}_1$ to $\hat{\theta}_2$, and ii) it also illustrates a concrete scenario where we can directly compare a couple of different estimators in terms of their MSE and, more restrictively, unbiased estimators in terms of their variance. 
While we have introduced the sample maximum and the midpoint estimator in a somewhat arbitrary way in this example, in Sec.~\ref{sec:MLE} we will show how these estimators can actually be obtained. Moreover, note how most of these estimators that we mentioned perform very well in the asymptotic limit. We will give some further thoughts on their asymptotic behavior in Sec.~\ref{sec:est_properties}. However, in the next section we will take a slight detour, discussing a concept form statistics and information theory which will be central to the derivation of the Cramér-Rao bound in Sec.~\ref{sec:CRB}.

\section{The Fisher information}

Suppose we are given a continuous distribution $p(X)$. How can we introduce some notion of "closeness" with respect to another distribution $q(X)$? A prevalent quantity in mathematical statistics which deals with this problem is the Kullback–Leibler (KL) divergence, also known as relative entropy, which is defined as:
\begin{equation}\label{eq:KLdef}
D(p||q) := 
\avg{\ln\frac{p(X)}{q(X)}}{p}
=
\int p(x) \ln \frac{p(x)}{q(x)} dx.
\end{equation} 
In the discrete case we simply substitute the integral by a summation. Strictly speaking, this is not really a measure of distance because it is not symmetric, nor it obeys the triangle inequality. Nevertheless, the KL divergence is still a intuitive way of comparing distributions.
\footnote{As its alternative name indicates, it is more correct to think of the KL divergence as a notion of entropy. In a way, information theory provides a better interpretation of this quantity. If we write the Shannon entropy of $p$ as $H(p):= - \sum_x p(x) \log p(x)$ and its cross-entropy with $q$ as $H(p,q):=- \sum_x p(x) \log q(x)$, we can actually show that the identity $D(p||q) = H(p,q) - H(q)$ holds. Therefore, we might think of $D(p||q)$ as an entropy gain, or how much extra information (bits) should be provided in the context of e.g., optimal encoding, when we use a distribution $q$ instead of a distribution $p$ \cite{Cover2005} (pp.19 - 21). This also makes clear why the relative entropy is not symmetric in general.}

Now, consider that the distribution $p_\theta(x)$ is parametrized by $\theta$, as we had in the previous sections, and twice differentiable. We are interested in calculating the KL divergence for this distribution with itself but for \emph{different} parametrizations. That is, given two parameters $\theta$ and $\theta_0$, where we regard the latter as the "true" one, we can compute the KL divergence as:
\begin{equation}\label{eq:KL-fisher}
D(p_{\theta_0}||p_\theta) 
= 
\int p_{\theta_0}(x) 
\ln
\frac{p_{\theta_0}(x)}{p_\theta(x)} 
dx.
\end{equation}
Since we expect to have $D(p||q)\geqslant 0$, with equality if, and only if, $p = q$, we can conclude that $\theta = \theta_0$ is a minimum point for Eq.~\eqref{eq:KL-fisher}, as any other value of $\theta$ should increase the KL divergence. So, if we regard $p_\theta(x)$ as an estimated distribution, the KL divergence is a measure of how close we are to the true model $p_{\theta_0}(x)$. 
\footnote{A statistical-oriented interpretation of this distribution is to express the KL divergence as the expected value of the likelihood ratio \cite{Etz2018} between $p$ and $q$ \cite{Etz2018b}, as we see in the middle term of Eq.~\eqref{eq:KLdef}.} 
Additionally, this also guarantees that the first derivative of Eq.~\eqref{eq:KL-fisher} evaluates to zero at $\theta = \theta_0$ and that it is locally convex around this minimum \cite{Rioul2018}. These observations allow us to Taylor-expand the KL divergence between the two parametrizations as:
\begin{equation}
D(p_{\theta_0}||p_\theta) 
= 
\frac{(\theta-\theta_0)^2}{2}F_{\theta_0}(X) + O\left((\theta-\theta_0)^3\right).
\end{equation}
The term $F_{\theta_0}(X)$ appearing in the equation above is defined as the \emph{Fisher information}:
\begin{equation}\label{eq:Fisher_def1}
F(\theta_0)
:=
\frac{\partial^2}{\partial \theta^2}D(p_{\theta_0}||p_\theta)\bigg |_{\theta=\theta_0}\geqslant 0.
\end{equation}
We will now follow some algebraic steps in order to make the definition above more operational and its interpretation more apparent. First, we re-express Eq.~\eqref{eq:Fisher_def1} in terms of the expected value at the point $\theta = \theta_0$:
\begin{equation}\label{eq:Fisher_def1b}
F(\theta) 
= 
- \avg{\frac{\partial^2}{\partial \theta^2} \ln p_\theta(X)}{p_\theta}.
\end{equation} 
By using the product and chain rules and explicitly differentiating the expression inside the expected value we get:
\begin{equation}
F_{\theta}(X)
=
- \avg{\frac{1}{p_\theta(X)}\frac{\partial^2 p_\theta}{\partial \theta^2}(X)}{p_\theta}
+ \avg{\left(\frac{1}{p_\theta(X)}\frac{\partial p_\theta}{\partial \theta}(X)\right)^2}{p_\theta}.
\end{equation}
By using the normalization condition of the pdf, we can show that the first term vanishes:
\[
-\avg{\frac{1}{p_\theta(X)}\frac{\partial^2 p_\theta}{\partial \theta^2}(X)}{p_\theta}
=
-\frac{\partial^2}{\partial \theta^2} \int p_{\theta}(x)dx
=
0.
\]
Therefore, the alternative expression for the Fisher information reads:
\begin{equation}\label{eq:Fisher_def2}
F_{\theta}(X)
=
\int
\frac{1}{p_\theta(x)}
\left[ \frac{\partial p_\theta}{\partial \theta}(X) \right]^2
dx
=
\avg{\left( \frac{\partial}{\partial \theta} \ln p_\theta(X) \right)^2}{p_\theta}.
\end{equation}

The Fisher information is a quantity which basically tell us how sensitive a distribution is to changes around $\theta_0$. If $F_\theta$ is large, this means that the relative entropy $D(p_{\theta_0}||p_\theta) $ is very sharp in the neighborhood of $\theta_0$. We will probe a little bit further on this interpretation after we discuss the maximum likelihood mathod in Sec.~\ref{sec:MLE}. Moreover, the usefulness of the FI will become even clearer after we discuss the Cramér-Rao bound in the next section.

\begin{wrapfigure}{r}{0.45\textwidth}
	\includegraphics[width=0.45\textwidth]{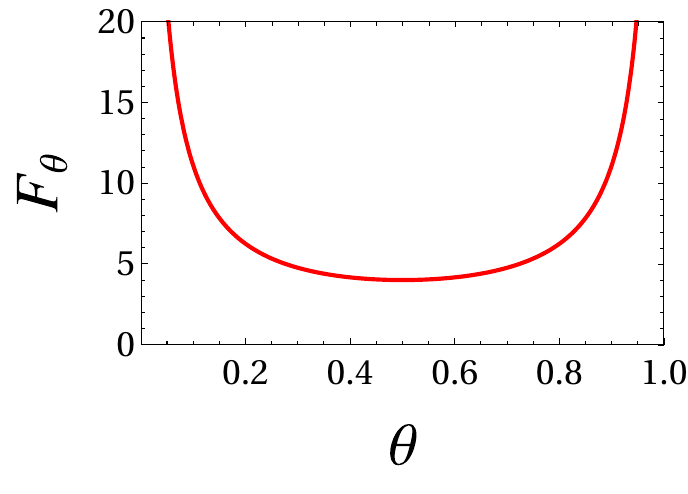}
	\caption[Fisher information of the Bernoulli distribution.]{Fisher information of the Bernoulli distribution.}
	\label{fig:bernoulliFisher}
\end{wrapfigure}

In this derivation we have implicitly assumed some regularity conditions on the probability distribution $p_\theta(x)$. Besides the differentiability condition that we have briefly mentioned, it is also required that the support of the distribution does not depend on the parameter $\theta$ itself. Unfortunately, this include some cases such as the example which we have discussed in the last section for the uniform distribution. In that problem, the support is given by $[0, \theta]$, with an explicit dependence on the parameter $\theta$. For now, we go back to the Bernoulli distribution from Sec.~\ref{sec:estimation_theory}, discussing a concrete example:

$\opposbishops$ \textbf{\textit{Example:}} Consider the Bernoulli distribution from Eq.~\eqref{eq:bernoulli}. We have:
\begin{equation}\label{eq:bernoulliFisher1}
\frac{\partial}{\partial \theta} \ln p_\theta(X)
=
\frac{X}{\theta} - \frac{(1-X)}{(1-\theta)}.
\end{equation}
It holds that $\avg{X}{p_\theta} = 0.p_\theta(X=0) + 1.p_\theta(X=1) = \theta$ and $\avg{X^2}{p_\theta} = 0^2p_\theta(X=0) + 1^2p_\theta(X=1) = \theta$. Thus, we can simply use Eqs.~\eqref{eq:Fisher_def2} and~\eqref{eq:bernoulliFisher1} to obtain the Fisher information for the Bernoulli distribution
\begin{equation}\label{eq:bernoulliFisher}
F_{\theta}(X)
=
\frac{1}{\theta(1-\theta)}.
\end{equation}
Thus, we can see that the FI is minimum for $\theta=1/2$ and that it diverges for $\theta \rightarrow 0$ or $\theta \rightarrow 1$. See the results in Fig.~\ref{fig:bernoulliFisher}.
$\bishoppair$

Notice, both from Eq.~\eqref{eq:Fisher_def2} and from Fig.~\ref{fig:bernoulliFisher} in the example, that the Fisher information in general will depend on the parameter $\theta$. Thus, the information we have about a distribution will in general depend on the very parameter we are trying to estimate. It might therefore be difficult to be aware of the attainable information about a model beforehand. This is one of the difficulties we expect to address in subsequent sections.

\section{The Cramér-Rao bound}\label{sec:CRB}

We saw that the concept of Fisher information introduces a degree of knowledge, or of sensibility upon a parameter $\theta$, that we have about a model or a distribution. Is there a rigorous way of relating this with the framework of estimation theory? The very famous result by Cramér and Rao \cite{Rao1992, Cramer1999}, where one is able to relate the variance of an estimator with the Fisher information of the distribution, answers this question. This is one of the main results of the literature that we are going to review in this work and it will be fundamental for our review on metrology and thermometry in Chapter~\ref{chp:metrology}.

Here we introduce the Cramér-Rao bound for the more general case of \emph{biased} estimator~\cite{Lehmann1998} (p.325). We follow a derivation similar to the one found in Ref.~\cite{Kay1993} (pp. 67-69). First, we require that the regularity condition
\begin{equation}\label{cramerrao_regularity1}
\mathbb{E}_{p_\theta}\left[\ppx{\ln p_\theta(x)}{\theta}\right] = 0.
\end{equation}
holds for all $\theta$. The strategy to derive the CRB is to simply use the Cauchy-Schwartz inequality for integrals:

\begin{equation}
\int f(x)^2 w(x) dx \int g(x)^2 w(x)dx \geqslant \left( \int f(x)g(x)w(x)dx \right)^2.    
\end{equation}

It is intuitive to pick the weight here as the distribution, that is $w(x) = p_\theta(x)$ so every integral becomes an expected value with respect to the $p_\theta(x)$. The other two choices we make are:
\begin{equation}
\begin{split}
f(x) &= \frac{\partial \ln{p_\theta(x)}}{\partial \theta}, \\
g(x) &= \hat{\theta}(x)-\mathbb{E}[\hat{\theta}(x)].
\end{split}
\end{equation}
The first function is also commonly called \emph{score} in the literature, and $g(x)$ are the fluctuations of the estimator around the mean. We can begin by solving the RHS of the inequality. Note that we can write this integral as:
\begin{equation}
\int f(x)g(x)w(x)dx 
=
\int 
\frac{\partial \ln{p_\theta(x)}}{\partial \theta}
\left(\hat{\theta}(x)-\mathbb{E}[\hat{\theta}(x)]\right)
p_\theta(x)
dx. 
\end{equation}
The first term is given by:
\begin{equation}
\int 
\hat{\theta}(x) 
\frac{\partial \ln{p_\theta(x)}}{\partial \theta}
p_\theta(x)
dx    
=
\int 
\hat{\theta}(x) 
\frac{\partial p_\theta(x)}{\partial \theta}
dx    
=
\frac{\partial}{\partial \theta}
\int 
\hat{\theta}(x) 
p_\theta(x)
dx.    
\end{equation}

Note that here we used one of the regularity conditions which enable us to swap the integration and the derivative and also the assumption that the estimator is independent of $\theta$. The expression above is thus simply the derivative of the expected value of the estimator. We can then rearrange the term above in terms of the bias and the real parameter as:

\begin{equation}
\frac{\partial}{\partial \theta}
\int 
\hat{\theta}(x) 
p_\theta(x)
dx    
=
1 + b'(\theta),
\end{equation}
since $b(\theta) = \mathbb{E}[\hat{\theta}(x)]-\theta$. Thus, we can write the inequality as:
\begin{equation}\label{cramerraobias}
\int f(x)^2 w(x) dx \int g(x)^2 w(x)dx \geq (1 + b'(\theta))^2.    
\end{equation}
Now we can show that the integral of the score is simply the Fisher information:
\begin{equation}\label{cramerraofisher}
\int f(x)^2 w(x) dx 
=
\int \left(\ppx{\ln p_\theta(x)}{\theta}\right)^2 p_\theta(x) dx 
=
F(\theta),
\end{equation}
and finally, note that 
\begin{equation}
\int g(x)^2 w(x) dx 
=
\int \left(\hat{\theta}(x)-\mathbb{E}[\hat{\theta}(x)]\right)^2 
p_\theta(x) dx 
=
\Var{\hat{\theta}(x)}.
\end{equation}
However, because of the bias-variance decomposition we have $\epsilon(\hat{\theta}(x)|\theta) = \Var \hat{\theta}(x) + b(\theta)^2$~[Eqs.~\eqref{eq:mse} and~\eqref{eq:bias_var}], and we can write

\begin{equation}\label{cramerraovariance}
\int g(x)^2 w(x) dx =
\epsilon\left(\hat{\theta}(x)|\theta\right) - b(\theta)^2.
\end{equation}

Now we can simply rearrange the equations \eqref{cramerraobias}, \eqref{cramerraofisher} and \eqref{cramerraovariance} in order to arrive at the desired inequality \cite{Lehmann1998}:
\smallskip

\textbf{The biased Cramér-Rao bound}

If a pdf $p_\theta(x)$ obeys the regularity condition \eqref{cramerrao_regularity1} for all $\theta$, then any estimator $\hat{\theta}(x)$ must satisfy:

\begin{equation}
\epsilon(\hat{\theta}(x)|\theta) 
\geqslant
\frac{(1 + b'(\theta))^2}{F(\theta)} + b(\theta)^2.
\end{equation}

If the estimator in unbiased, we obtain the standard and more famous result:
\smallskip

\textbf{The Cramér-Rao bound}

Under the same conditions described above, any unbiased estimator $\hat{\theta}(x)$ must satisfy:

\begin{equation}\label{eq:efficient_estimator}
\Var{\left(\hat{\theta}(X)\right)}
\geqslant
\frac{1}{F(\theta)}.
\end{equation}
This very important result establishes an ultimate limit of precision for \emph{any} possible estimator. It is also clear that the existence of this bound also provides a very practical advantage: if we can find an estimator $\hat{\theta}^*$ which saturates this bound, we know that $\hat{\theta}^*$ will be optimal. This greatly simplifies the very troublesome quest of looking for MVU estimators. Any estimator which saturates Eq.~\eqref{eq:efficient_estimator} is said to be \emph{efficient}. 

A important property of the FI is its additivity: given two pdfs $p_{1;\theta}(X_1)$ and $p_{2;\theta}(X_2)$ with FI $F_1(\theta)$ and $F_2(\theta)$, respectively, the FI of $p_{1;\theta}(X_1)p_{2;\theta}(X_2)$ is given by their sum $F_1(\theta) + F_2(\theta)$. See Refs.~\cite{Zegers2015, Ly2017} for an extensive discussion on the properties of the FI.  From the additivity  of the FI, the CRB can be rewritten as
\begin{equation}\label{eq:efficient_estimatorN}
\Var{\left(\hat{\theta}({\bf X})\right)}
\geqslant
\frac{1}{nF(\theta)}.
\end{equation}
for a sequence ${\bf X} = X_1, ..., X_n$ of i.i.d. outcomes. If the estimator saturates the bound above asymptotically, it is said to be \emph{asymptotically efficient}.

$\opposbishops$ \textbf{\textit{Example:}} We found in the last example that the FI of the Bernoulli distribution is given by $F_{\theta}(X)=1/\theta(1-\theta)$. This means that the CRB for the binomial distribution (with $n=1$ for the Bernoulli distribution) reads $\Var{\left(\hat{\theta}({\bf X})\right)} \geqslant \theta(1-\theta)/n$. 
We would like to check how the variance of a real estimator compares with this bound. The natural candidate is simply the sample mean which we have discussed in Sec.~\ref{sec:estimation_theory}. The variance of $\overline{\theta}({\bf X})$ is given by:
\begin{equation}\label{eq:MSEbinomial}
\Var{\left(\overline{\theta}({\bf X})\right)}
=
\frac{\Var{X_1}+...+\Var{X_n}}{n^2}
=
\frac{\theta(1-\theta)}{n}.
\end{equation}
Thus, we can see that the sample mean actually \emph{saturates} the CRB in this case. It is not only a MVU estimator, but also an efficient estimator. However, the converse is not true in general; the existence of a MVU estimator does not guarantee its efficiency. See Ref.\cite{Romano2017} (p.194) for example. $\bishoppair$

The example above shows that the sample mean is the \emph{best} unbiased estimator one can construct for the Bernoulli and the binomial distributions. This lead us to a very important question: under which conditions is it possible obtain efficient estimators which saturate the CRB? It can be shown that this is only possible for the exponential family \cite{Host-Madsen2000, Schreier2010}. In more general scenarios, one will obtain estimators which, at best, satisfy the weaker condition of asymptotic efficiency. We will learn a method for obtaining such estimators in the next section.

\section{The maximum likelihood estimation}\label{sec:MLE}

R. A. Fisher introduced, during the early decades of the 20th century, a method known as \emph{maximum likelihood estimation} (MLE). See Refs.~\cite{Aldrich1997} and \cite{Stigler2007} for a historical contextualization. This method is a quite general and practical approach, which is specially useful when obtaining a MMSE or a MVU estimator is either not trivial or not possible. Moreover, MLE results in estimators with several useful asymptotic properties. 

The pdf $p(X_1, ..., X_n|\theta)$ for $X_1, ..., X_n$ conditioned of the parameter $\theta$ is also known as the \emph{likelihood}. The strategy of the MLE is very simple. Given the pdf $p({\bf X}|\theta)$ and the outcomes ${\bf X}=X_1 , ..., X_n$, what is the value of $\theta$ which makes the observation of ${\bf X}$ most likely? In other words, we define the maximum likelihood estimator as:
\begin{equation}
\hat{\theta}_\mathrm{MLE}({\bf X}) 
:=
\argmax_{\theta}
p({\bf X}|\theta).
\end{equation}
Since the logarithm is a monotonic function, it is sometimes more convenient, or numerically efficient, to minimize the \emph{log-likelihood} $\ln p(X_1, ..., X_n|\theta)$ instead:
\begin{equation}
\hat{\theta}_\mathrm{MLE}({\bf X}) 
=
\argmax_{\theta}
\ln p({\bf X}|\theta).
\end{equation}
$\opposbishops$ \textbf{\textit{Example:}} Let us to back to the binomial distribution example. In this case we can take the derivative of the log-likelihood and equate it with zero to find:
\[
\frac{\partial \ln  p(n, k|\theta)}{\partial \theta} 
=
\frac{k}{\theta}
+
\frac{n-k}{1-\theta}
=
0,
\]
which yields the sample mean estimator
\begin{equation}\label{eq:MLEbinomial}
\hat{\theta}_\mathrm{MLE}({\bf X}) 
=
\frac{k}{n}
=
\frac{1}{n}
\sum
X_i.
\end{equation}
$\bishoppair$

Similarly, it can be shown that the MLE also yields the sample maximum estimator $\hat{\theta}_\mathrm{MLE}({\bf X}) = \max {\bf X}$ that we have discussed in the example of Sec.~\ref{sec:MVU}. This also shows that the MLE is not necessarily unbiased at finite sample size and that it is not guaranteed to coincide with the MVU. 

We will briefly discuss some of the properties of the maximum-likelihood estimators. While we will not present the proofs, those can be obtained in any standard reference for point estimation theory (see e.g. \cite{Lehmann1998, Casella2002}). 

\begin{itemize}
	\item The MLE is asymptotically distributed as:
	\begin{equation}
	\hat{\theta}_\mathrm{MLE}({\bf X})
	\underset{n \rightarrow \infty}{\sim}
	\mathcal{N}\left(\theta, \frac{1}{nF(\theta)}\right).
	\end{equation}
	Here $\mathcal{N}(\mu, \sigma^2)$ denotes a Gaussian distribution with mean $\mu$ and variance $\sigma^2$.
	\item The result above shows that the MLE is asymptotically unbiased and it is also asymptotically efficient, that is, its asymptotic variance is proportional to the reciprocal of the Fisher information. This implies that the maximum likelihood estimation is, at least asymptotically, an estimator of minimum variance.
	\item If $g(\theta)$ is a one-to-one function of $\theta$, the MLE of $g(\theta)$ is simply $\hat{g}_\mathrm{MLE} = g(\hat{\theta}_\mathrm{MLE})$.
\end{itemize}

These properties show why the MLE can be regarded as an asymptotically optimal estimator and why this method is a good rule-of-thumb when looking for estimators. This method is also numerically friendly. One can simply estimate the MLE numerically by using, e.g., Monte-Carlo methods. A disadvantage of this method which is worth mentioning is that it might be hard deciding beforehand how quickly it converges to a good asymptotic approximation \cite{Kay1993}. Other procedures for obtaining estimators also exist, such as the method of moments.

\section{Properties of estimators}\label{sec:est_properties}

In this section we give a brief summary the most relevant properties of estimators. We will review the properties that we have already encountered and also mention extra ones for completeness. Some of these properties will also motivate discussions in future chapters.

\textbf{Unbiasedness}. The estimator is said to be unbiased when its expected value coincides with the true value of the parameter for \emph{every} $\theta$: 
		\begin{equation}
		\avg{\hat{\theta}}{} = \theta.
		\end{equation}
		The sample mean estimator is an unbiased estimator for the binomial distribution, for example. 
We also mentioned the property of asymptotic unbiasedness before. This a weaker condition, which happens when the estimator is unbiased in the asymptotic limit $n \rightarrow \infty$. The sample maximum estimator that we have encountered in Sec.~\ref{sec:MVU} was biased, but \emph{asymptotically} unbiased.

\begin{wrapfigure}{r}{0.45\textwidth}
		\includegraphics[width=0.5\textwidth]{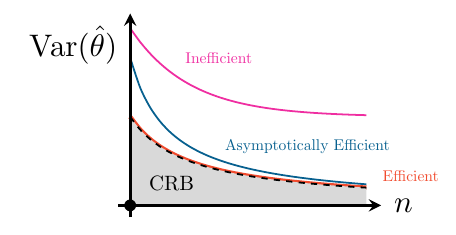}
		\caption[Comparison of estimators with the CRB.]{Comparison of estimators with the CRB.}
		\label{fig:CRB}
\end{wrapfigure}
\textbf{Efficiency}. An estimator is efficient when the equality in the CRB \eqref{eq:efficient_estimator} holds. The sample mean for the binomial distribution is an example of efficient estimator. Analogously, an estimator is asymptotically efficient when the CRB is saturated in the asymptotic limit $n\rightarrow \infty$.  See the chapter 10 of Ref.~\cite{Casella2002} for a discussion on the asymptotic properties of estimators.  
Note that even when the variance of an estimator goes to zero with $n \rightarrow \infty$, its asymptotic efficiency is not guaranteed. One such example is the median estimator for the average of a normal distribution \cite{Miller2014} (p.290). That is, even in the asymptotic limit, some estimators do not attain the CRB.

Finally, we can also define the concept of \emph{relative efficiency}. This is a property of unbiased estimators. If $\Var{\hat{\theta}}^* < \Var{\hat{\theta}}$, then $\hat{\theta}^*$ is relatively more efficient than $\hat{\theta}$. Remember that if this inequality holds for \emph{every} $\hat{\theta}$, then $\hat{\theta}^*$ is the minimum variance unbiased estimator, as we saw in Sec.~\ref{sec:MVU}.

\textbf{Consistency}. This property means that by increasing $n$ the estimator can approach the true value of the true parameter with arbitrary precision (in probability). This can be written as:

\begin{equation}
\lim_{n \to \infty} P(|\hat{\theta} - \theta| < \epsilon) = 1,
\end{equation}
for all $\epsilon > 0$. While it is sometimes possible to explicitly check the consistency of an estimator by direct calculation, it is more usual to check for other properties: if the estimator is asymptotically unbiased and if its variance goes to zero with $n\rightarrow \infty$, then the estimator is consistent \cite{Casella2002} (p.~469). For instance, note that this also means that the MLE is asymptotically consistent.

\textbf{Sufficiency}. An estimator is said to be sufficient if, and only if, the pdf $p({\bf X}|\hat{\theta})$ of $X_1, ..., X_n$ conditioned on $\hat{\theta}$ is independent of $\theta$~\cite{Rice2006} (pp.~305-307). 
This property means that no particular function or combination of the samples $X_1, ..., X_n$ in ${\bf \theta}$ is more likely than others. 
One such example is the binomial distribution from Eq.~\eqref{eq:binomial}: the sample mean $\overline{\bf X} = (X_1+...X_N)/n = k/n$ happens to be a sufficient estimator in this case \cite{Miller2014} (p.~295). Intuitively, this is due to the fact that only the number of heads and tails matter in the estimation, and not the order or  any other complicated functions of the outcomes. 
In analogy, we can also express an estimator, such as the one above, in terms of a function $T({\bf X})$ of the outcomes ${\bf X}$, that is, of a statistic $T$. In the case of the sample mean, the sum $X_1 + ... + X_N$ of the outcomes is a sufficient statistic.  A very helpful result, known as the \emph{Fisher–Neyman factorization theorem},  can be used to verify whether a statistic is sufficient~\cite{Kay1993} (p. 104).

The idea of sufficiency is not only of conceptual interest: it can also aid us in finding a good estimator. Given an unbiased estimator $\hat{\theta}$, a sufficient statistic $T$ and the conditional expectation $\mathbb{E}[X|Y = x]:=\sum_x x \quad p(X = x| Y = y)$, the Rao-Blackwell theorem \cite{Lehmann1998} (p.~47) states that $\hat{\theta}_T = \avg{\hat{\theta}|T}{}$ is a better unbiased estimator of $\theta$ for all $\theta$. `
In other words, this theorem gives a procedure for improving estimators simply by conditioning them on a sufficient statistics \cite{Kay1993} (p.~109). This is of great practical aspect: one can start with a crude estimator and improve on it by using the Rao-Blackwell theorem. Equally importantly, as a corollary this theorem also implies that the MVU estimator is necessarily a function of a sufficient statistic.

This result from Rao and Blackwell leads to the Lehmann–Scheffé theorem, which states that if a statistics possesses a property called completeness, a unbiased estimator $\hat{\theta}_T$ based on $T$ is the \emph{best} and \emph{unique} unbiased estimator of $\theta$. We will not discuss this extra property here, but a statement of the theorem can be found in Ref.~\cite{Casella2002} (p.~369). With the aid of the factorization theorem, this construction can be used to show that the sample maximum is a sufficient statistics for the uniform distribution that we discussed in Sec.~\ref{sec:MVU}, for example. This proves that the unbiased sample maximum is precisely the MVU estimator \cite{Kay1993} (p.~113).

Finally, it is worth stressing that regardless of these conclusions, biased estimators might still be preferable, so one should still be careful when looking for an optimal estimation strategy. See Ref.~\cite{Hardy2003} for an interesting example.

\begin{figure}[t!]
	\centering
	\includegraphics[width=\textwidth]{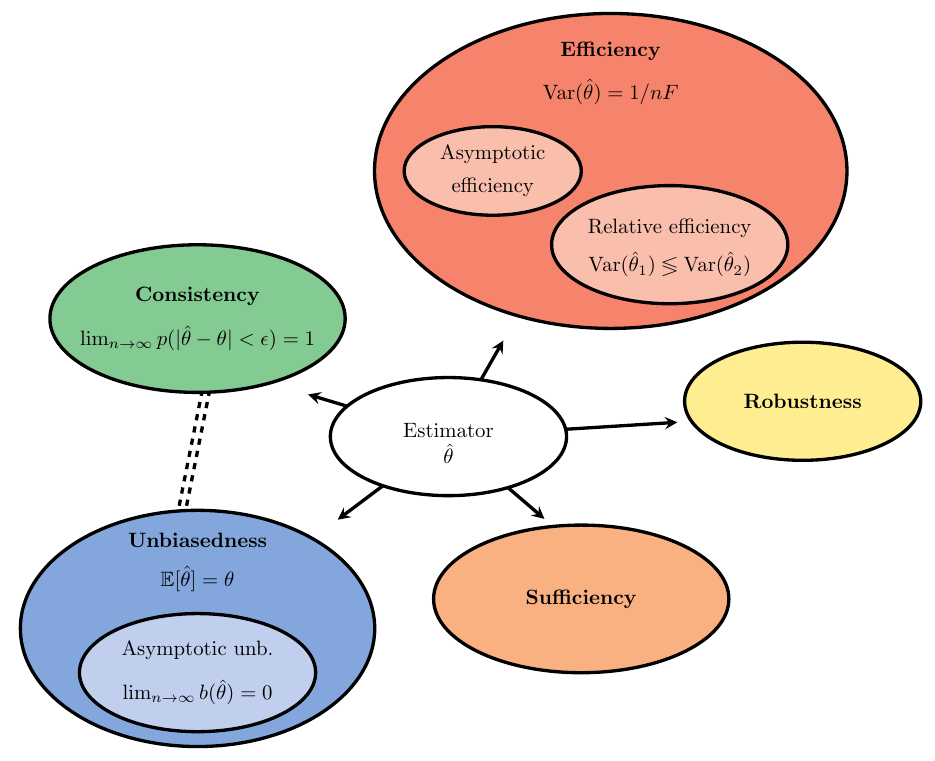}
	\caption[Common definitions for estimators.]{Common definitions for estimators.}
	\label{fig:estimator_properties}
\end{figure}

\textbf{Robustness}. This is another property which we are not going to define mathematically, but we will nevertheless mention it for completeness. This is a more recent concept in relation to the others, and it may very often have different definitions in the literature. Nonetheless, robustness is related to the idea that an estimator should still perform well enough even if underlying assumptions about the models are not entirely correct or precise. One might, for instance, sacrifice optimality in exchange for an estimator which performs reasonably well even for a probability distribution (a model) slightly different from what is ideally expected \cite{Devore2016} (p.~272). A concrete example is given in Ref.~\cite{Casella2002} (p.482).

A visual summary of these properties can also be found in Fig.~\ref{fig:estimator_properties}. While we will not give further attention to the theoretical aspects of some of these properties in this work, it was important mentioning them in order to justify the use of a few estimation strategies that we are going to be focusing from hereafter, such as the maximum likelihood-estimation, and a few Bayesian strategies that we are going to encounter in the next section. The interested reader can find several useful discussions and counter-examples for the theory of point estimation in Ref.~\cite{Romano2017}.

%% file: chapters/bayes.tex
Bayesian statistics constitutes a departure from the frequentist framework that we discussed in the last chapter.
In the frequentist scenario we were mainly concerned with likelihood functions, which are conditioned on the deterministic, but unknown parameter, and also with quantities averaged over different realizations of the problem, such as the MSE. 
Meanwhile, here we focus on probability distributions over the parameters instead. Now, unknown parameters of a distribution are regarded as random variables themselves \cite{Stark2010}, while the relevant distributions are conditioned on the \emph{data} instead. This is a useful approach in experimental scenarios, specially in problems where the notion of repeatability, very natural in frequentism, is less intuitive or difficult to implement. In those situations, the experimenter is probably interested in extracting information of his or her data as much as possible, and not from every possible realization. Similarly, maybe one is interested on how the model and how the estimators would behave for a large set (or interval) of different parameters. 

This paradigm shift introduces a few conceptual and practical differences, as we shall see \cite{Held2014}. Many standard texts give a broad overview of these two fields. In fact, some authors even argue that Bayesian methods have a broader applicability nowadays \cite{Jaynes2003}. We abstain from these discussions: the contents of this chapter are mainly concerned with some of the cornerstones of the Bayesian inference and their practical implications. In particular, in Secs.~\ref{sec:bayes} and~\ref{sec:bayes_est} we show how the Bayesian formulation can be used as an algorithmic procedure to obtain concrete estimators, in consonance with much of the development and the concepts which we have encountered in the last chapter. In the same vein, the Bayesian approach results in a modified version of the Cramér-Rao bound \cite{VanTrees2007} which we discuss in Sec.~\ref{sec:VTSB}. 


\section{Bayes' theorem}\label{sec:bayes}

The joint probability $p(A, B)$ of two events $A$ and $B$ can be decomposed in a symmetric manner with the use of conditional probabilities:
\begin{equation}
p(A, B) = p(A|B)p(B) = p(B|A)p(A).
\end{equation}
A rearrangement leads us to the \emph{Bayes theorem}, 
\begin{equation}\label{eq:bayes_decomp1}
p(A|B) = \frac{p(B|A)p(A)}{p(B)},
\end{equation}
which is the basis of the Bayesian statistical inference. 
See Ref.~\cite{Robert1994} for a very complete introductory discussion and some brief historical remarks. 
A simple change of notation will bring an interpretation of the equation above which is closer to our current context. In the problem of estimation theory we can consider $A$ to be the parameter $\theta$ and $B$ to be the sequence of random variables ${\bf X} = X_1, ..., X_n$. By doing so, we rewrite Eq.~\eqref{eq:bayes_decomp1} as:
\begin{equation}\label{eq:bayestheorem}
p(\theta|{\bf X}) 
= 
\frac{p({\bf X}|\theta)p(\theta)}{p({\bf X})}
=
\frac{p({\bf X}|\theta)p(\theta)}{\int p({\bf X}|\theta)p(\theta)}
=
\frac{1}{\mathcal{N}}p({\bf X}|\theta)p(\theta)
.
\end{equation}
The familiar term $p({\bf X}|\theta)$ is called the \emph{likelihood}, as seen in the the problem of classical estimation. This terms tell us what is the probability of obtaining a particular trajectory (or detection record) ${\bf X}$ if the distribution is parametrized by $\theta$. The term $\mathcal{N}:= p({\bf X}) = \int p({\bf X}|\theta)p(\theta)$ is just the marginalized distribution of ${\bf X}$, and it is essentially a normalization factor in the Bayes theorem.

The novelty of the Bayesian approach comes from the two other terms. The distribution $p(\theta)$ is called the \emph{prior distribution}, and it embeds any previous knowledge we might have about the parameter $\theta$. For instance, we might know that $\theta$ is truncated at a certain value, or that it lies within a particular interval. All this information can be incorporated into the prior, ideally improving the estimates by including any previous information about the model. 

The final term, $p(\theta|{\bf X})$ is called the \emph{posterior distribution}, and it is the most important object in the Bayesian setting. One possible interpretation for this distribution is the following: given the realization $X_1, ..., X_n$, how likely it is that these outcomes were generated by the parameter $\theta$? 
Another very important view is that the posterior distribution represents an update of the prior: given $p(\theta)$, how does the observation of the outcomes $X_1, ..., X_n$ updates our state-of-knowledge about the model? 

\begin{figure}[t!]
		\centering
		\includegraphics[width=\textwidth]{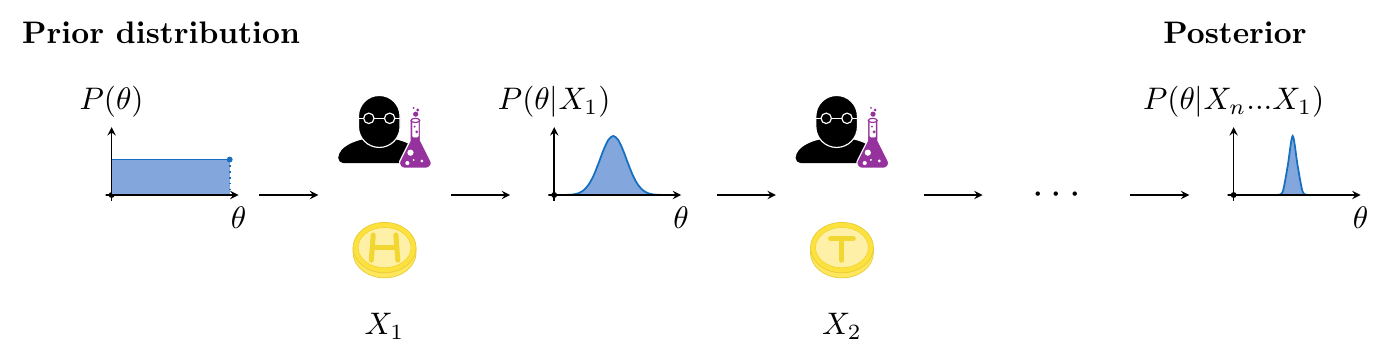}
		\caption[Illustration of the sequential updating scheme.]{Illustration of the sequential updating scheme.}
		\label{fig:bayesdiagram}
\end{figure}

The latter interpretation thus allows for a sequential update scheme which is very natural in, e.g., real experiments or investigations. Suppose we start with a prior distribution $p(\theta)$ and that we obtain an outcome $X_1$. The posterior distribution, given this outcome, is given by
$p(\theta|X_1) \propto p(X_1|\theta) p(\theta)$ up to the normalization factor. Now, suppose that the outcomes are i.i.d. Upon obtaining a second outcome $X_2$, we can use the first posterior $p(\theta|X_1)$ as the prior in Eq.~\eqref{eq:bayestheorem} and the new likelihood as the probability $p(X_2|\theta)$, so the updated distribution becomes $p(\theta|X_2 X_1) \propto p(X_2|\theta)P(\theta|X_1) \propto p(X_2|\theta)p(X_1|\theta)p(\theta)$.  
With each new outcome $X_{n+1}$, one can update the distribution $p(X_n...X_1|\theta)$ to $p(X_{n+1}X_n...X_1|\theta)$.
In this case, after we observe a new outcome, the current posterior will become the prior in the subsequent step, in other words, we can depict this sequential updating scheme as:
\footnote{In the more general case where correlations might be present, the updated distribution would need to be conditioned both on $\theta$ and on the previous outcomes $X_1, ..., X_n$. We would then rewrite Eq.~\eqref{eq:bayes_update} as 
$p(\theta|X_{n+1}X_n...X_1)
\propto
p(X_{n+1}|\theta,X_n...X_1)p(\theta|X_n...X_1)$.
}
\begin{equation}\label{eq:bayes_update}
p(\theta|X_{n+1}X_n...X_1)
\propto
p(X_{n+1}|\theta)p(\theta|X_n...X_1)
\propto
p(X_{n+1}|\theta)
...
p(X_{1}|\theta)
p(\theta)
.
\end{equation}
A pictorial representation of this scheme is Shown in Fig.~\ref{fig:bayesdiagram}. Of course, if one is not interested in studying or visualizing the intermediate distributions, it is possible to update the prior with all the outcomes $X_1, ..., X_n$ at once in a single calculation; this avoids the trouble of normalazing the distribution at every step.~ 
Explicitly calculating the intermediate distribution is not strictly necessary, and both strategies are equivalent. We have included this description motivated by the fact that this is a strategy which is going to be used in later chapters. For now, we discuss in an example how the Bayes theorem can be used in practice.

\framebox{$\opposbishops$ \textbf{\textit{Example:}}} Let us go back to the coin-toss example. In order to emulate ignorance about the parameter $\theta$, we will assume that the prior is given by a flat distribution $p(\theta) = 1$.
The likelihood, as we are aware, is given by the Bernoulli distribution in Eq.~\eqref{eq:bernoulli}. Now, suppose that one obtains an outcome $X_1$. The posterior distribution is given by:
\begin{equation}
p(\theta|X_1)
=
\frac{1}{\mathcal{N}_1}
\theta^{X_1}(1-\theta)^{1 - X_1}.
\end{equation}
Now, consider that we obtain a second outcome given by $X_2$. The posterior in this case becomes:
\begin{equation}\label{eq:bayesbernoulliex1}
p(\theta|X_2 X_1)
=
\frac{1}{\mathcal{N}_2}
\theta^{X_1 + X_2}(1-\theta)^{2 - X_1 - X_2}.
\end{equation}
For concreteness, let us take a realization where $X_1 = 1$ and that $X_2 = 0$. By doing so, the posterior in the equations above becomes:
\begin{equation}
\begin{split}
p(\theta|X_1 = 1) & \propto \theta, \\
p(\theta|X_2 = 0, X_1 = 1) & \propto \theta(1-\theta)
\end{split}
\end{equation}
By normalizing the distributions, we get $\mathcal{N}_1 = \int_0^1 \theta d\theta =1/2$ and $\mathcal{N}_2 = 1/6$, respectively. Thus, the posterior for the particular realization $(X_1, X_2) = (1, 0)$ in the coin toss experiment is given by:
\begin{equation}
\begin{split}
p(\theta|X_1 = 1) & = 2 \theta, \\
p(\theta|X_2 = 0, X_1 = 1) &= 6 \theta(1-\theta).
\end{split}
\end{equation}

\begin{wrapfigure}{r}{7cm}
	\centering
	\includegraphics[width=7cm]{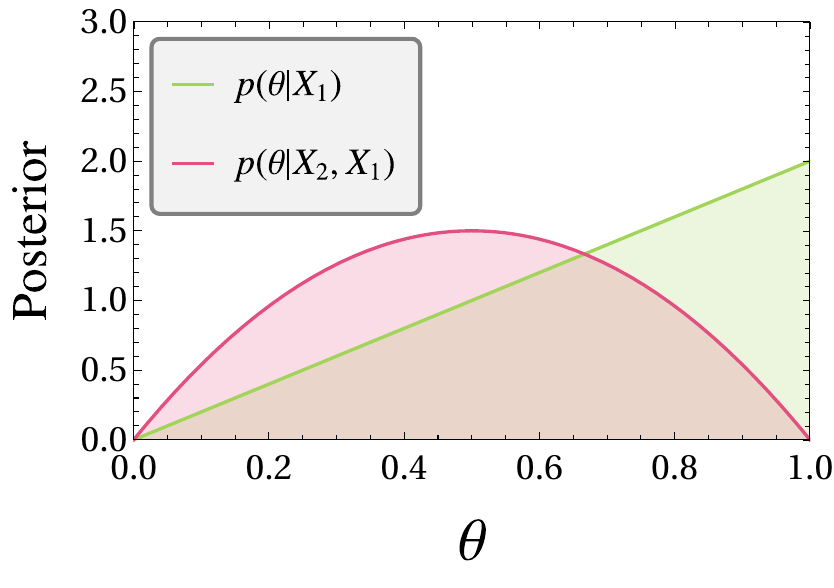}
	\caption[Posterior distribution in the coin toss experiment.]{Posterior distribution in the coin toss experiment. Here we assume a flat prior and the realization $(X_1, X_2) = (1, 0)$.}
	\label{fig:posteriorBernoulli}
\end{wrapfigure}
We show the results in Fig.~\ref{fig:posteriorBernoulli}. The plot for $p(\theta|X_1)$ tell us that as we increase $\theta$, it is increasingly likely that one would obtain $X_1 = 1$. This probability is obviously zero (one) for $\theta = 0$ ($\theta = 1$), corresponding to the scenario where the coin is fully biased. 
Similarly, $p(\theta|X_2 X_1)$ tells us that the greatest chance of obtaining exactly one head and one tail in two trials happens when the coin is perfectly fair, that is, when we have $\theta=1/2$, as seen in the plot. Equivalently, we can say that parametrizing the Bernoulli distribution with $\theta = 1/2$ maximizes our chances of obtaining the observed realization $(X_1, X_2) = (1, 0)$.       

Finally, note that for an arbitrary number of outcomes $X_1, ..., X_n$, one would simply obtain the posterior distribution:
\begin{equation}\label{eq:bayesbernoulliex2}
p(\theta|X_n ... X_1)
\propto
\theta^{\sum_i X_i}(1-\theta)^{n - \sum_i X_i}.
\end{equation}
Normalizing the distribution above is not entirely trivial. We will go back to this example in the next section. $\bishoppair$

The example above has shown how we can obtain the posterior for a desired parameter $\theta$ that we expect to estimate. Additionally, we could explicitly see how the stochastic outcomes result in distributions which are updated with new measurements. However, differently from the point estimation scenario where we had a \emph{single number} as an estimate, here we have a whole range of values that $\theta$ might assume, with an associated probability $p(\theta|{\bf X})$. This is a departure from the classical estimation theory and one of the main differences which we encounter with the  Bayesian procedures. 
Nevertheless, it is still possible to perform point estimation even in the Bayesian case, as we shall discuss in Sec.~\ref{sec:bayes_est}. This means, of course, that it would be erroneous to think that point estimation is restricted to the classical paradigm and that interval estimation is restricted to the Bayesian theory: the Bayesian and the frequentist approaches are, in different ways, capable of both \cite{Lehmann1998}. 

For now, let us take a slight detour in order to discuss a potential obstacle which might have been apparent from the last example. Obtaining the normalization factor in Eq.~\eqref{eq:bayestheorem} is, in general, a very taxing effort. 
This is valid both in the theoretical and numerical scenarios \cite{Gilks1995}. The analytical normalization factor for the posterior from Eq.~\eqref{eq:bayesbernoulliex2} in the example, for instance, requires some work to be found. Even so, a non-standard prior could result in a even more difficult integral, and, in the worst case scenario, one which might not have a solution in terms of elementary functions. 

This complication motivates the definition of \emph{conjugate priors}. We say that a prior $p(\theta)$ is conjugate to a likelihood function $p({\bf X}|\theta)$ whenever the application of the Bayes theorem results in a posterior $p(\theta|{\bf X})$ which belongs to the same family of probability distributions as the prior \cite{Berger1985}. 
This provides the enormous advantage of simplifying the normalization of the posterior: instead of calculating a (possibly highly dimensional) integral for several outcomes at once, or re-normalizing it at every step, as we did in the last example, one can simply find the normalization factor \emph{once} for pertinent family of distributions and update its parameters accordingly. We proceed with the Bernoulli distribution as an example.


\framebox{$\opposbishops$ \textbf{\textit{Example:}}} The conjugate for the Bernoulli distribution is the \emph{Beta} distribution
\begin{equation}\label{eq:beta}
\mathcal{B}_{(\alpha, \beta)}(\theta)
:=
\frac{1}{B(\alpha, \beta)}
\theta^{(\alpha-1)}
(1-\theta)^{(\beta-1)},
\end{equation}
defined in terms of \emph{hyperparameters} (parameters which define the prior distribution)  $\alpha$ and $\beta$. The normalization constant is given by the Beta function, defined as:
\begin{equation}\label{eq:betaIdentity}
B(\alpha, \beta)
:=
\int_0^1
t^{\alpha-1}
(1-t)^{\beta - 1}
dt
=
\frac{\Gamma(\alpha)\Gamma(\beta)}{\Gamma(\alpha + \beta)}.
\end{equation}
The second equality establishes a relationship between the Beta function and the well-known Gamma function $\Gamma(x)$ (see Ref.~\cite{Weisstein}). While these definitions might come across as unnecessary complications, they will greatly simplify some calculations later on. 

Suppose that our prior is given by the Beta distribution with hyperparameters $\alpha_0$ and $\beta_0$, that is: $p(\theta) = \mathcal{B}_{(\alpha_0, \beta_0)}(\theta)$. If we perform $n$ trials in the coin-toss experiment, obtaining $\sum_i X_i$ heads and $n - \sum_i X_i$ tails, the posterior distribution is given by:
\begin{equation}
p(\theta|{\bf X})
\propto
\underbrace{\theta^{\sum_i X_i}(1-\theta)^{n - \sum_i X_i}}_{\Pi_i p(x_i|\theta)}
\underbrace{\theta^{(\alpha_0-1)}(1-\theta)^{(\beta_0-1)}}_{\mathrm{Prior}}
=
\theta^{\alpha_0 + \sum_i X_i - 1}
(1-\theta)^{n + \beta_0 - \sum_i X_i - 1}
.
\end{equation}

However, note that the RHS of the equation above is simply the Beta distribution $\mathcal{B}_{(\alpha', \beta')}(\theta)$ with updated hyperparameters $\alpha' = \alpha_0 + \sum_i X_i$ and $\beta' = \beta_0 + n - \sum_i X_i$, up to the normalization constant. 
Thus, if the normalization constant of the Beta distribution, or any other other property of interest such as the moments, is known beforehand, one can promptly find them for the posterior distribution simply by updating the hyperparameters accordingly. 

\begin{figure}[h!]
	\centering
	\includegraphics[width=0.75\textwidth]{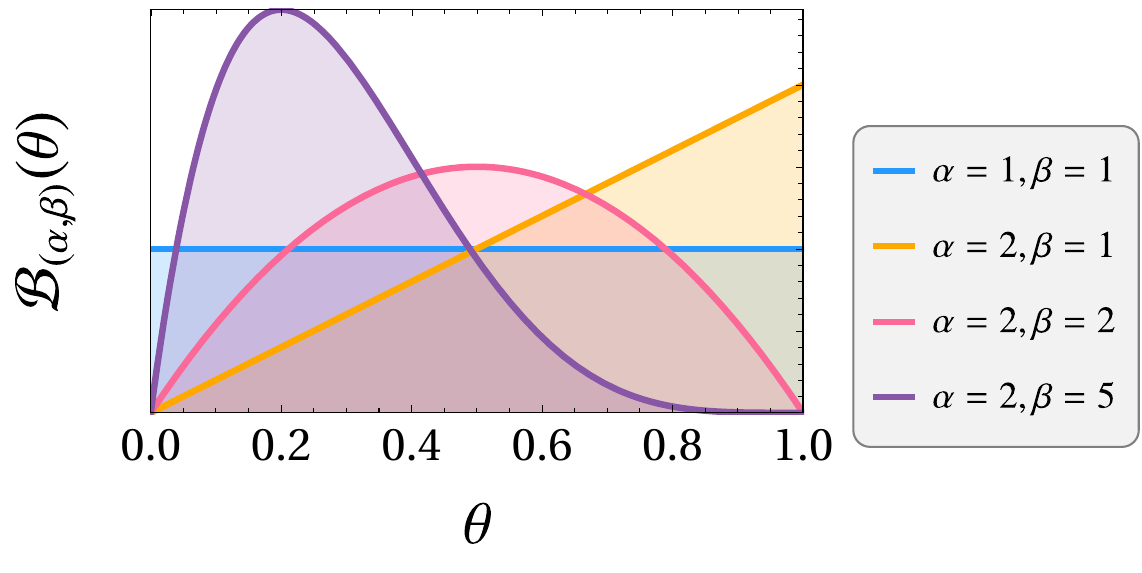}
	\caption[Plot of the Beta distribution for different hyperparameters.]{Plot of the Beta distribution for different hyperparameters.}
	\label{fig:betaPlot}
\end{figure}

As a particular case, we mention that by taking $\alpha_0 = \beta_0 = 1$ we obtain a flat distribution. Thus, the prior we chose in the previous example was actually a conjugate one. In that case, we could write the posterior as $p(\theta|X_2 = 1, X_1 = 0) = \mathcal{B}(2, 2)$. One can also easily find the normalization factor by calculating $B(2, 2) = \Gamma(2)\Gamma(2)/\Gamma(4) = 1! 1!/4! = 1/6$. So we can see that this agrees with our previous brute-force results. For reference, we plot the beta distribution for a few different parameters in Fig.~\ref{fig:betaPlot}. $\bishoppair$

We also discuss an important example for the normal distribution \cite{Kay1993} (p.~332). The following results will be important for us in Sec.~\ref{sec:VTSB}:


\framebox{$\opposbishops$ \textbf{\textit{Example:}}} Suppose we have a detection record given by i.i.d. data 

\begin{equation}
X \sim \mathcal{N}(\mu_0, \sigma)
\end{equation}
generated from a normal distribution with mean $\mu_0 = 0$ and variance $\sigma^2$. 
How can we apply the Bayes rules to infer the mean of the Gaussian? We will suppose the likelihood is modeled by a Gaussian of known variance, given by $P( X | \mu) = \mathcal{N}(\mu ,\sigma)$. For simplicity, we will also suppose that the prior is also a Gaussian, since the normal distribution is its own conjugate distribution \cite{Murphy2007}. We denote the mean and the variance of the prior by $\mu_p$ and $\sigma_p^2$, respectively. By doing so, we can apply Bayes rule to arrive at the following posterior:

\begin{equation}
p(\mu|X) \propto p(X|\mu) p(\mu) = 
\expo{
\frac{(X-\mu)^2}{2\sigma^2}+
\frac{(\mu - \mu_p)}{2\sigma_p^2}
}.
\end{equation}

We can complete the squares by noticing that the term inside the exponential above may be written as

\[
\frac{\sigma_p^2 + \sigma^2}{2 \sigma_p^2 \sigma^2}
\left[
\mu^2 
- 2\mu\frac{\sigma^2 \mu_p + \sigma_p^2 X}{\sigma_p^2 + \sigma^2}
+
\frac{\sigma_p^2 \sigma^2}{\sigma_p^2 + \sigma^2}\left(\frac{X^2}{\sigma^2}\right)
\right],
\]
since the last term does not depend on $\mu$ we can factor this part of the exponential out from the integral, since it will just be incorporated in the normalization factor. We can then complete the square. By doing so we arrive at the following posterior:

\begin{equation}
p(\mu|X) = \frac{1}{\sqrt{2\pi}\tilde{\sigma}} \expo{-\frac{(\mu - \tilde{\mu})^2}{2\tilde{\sigma}^2}}.  
\end{equation}

\begin{wrapfigure}{r}{7cm}
		\centering
		\includegraphics[width = 7cm]{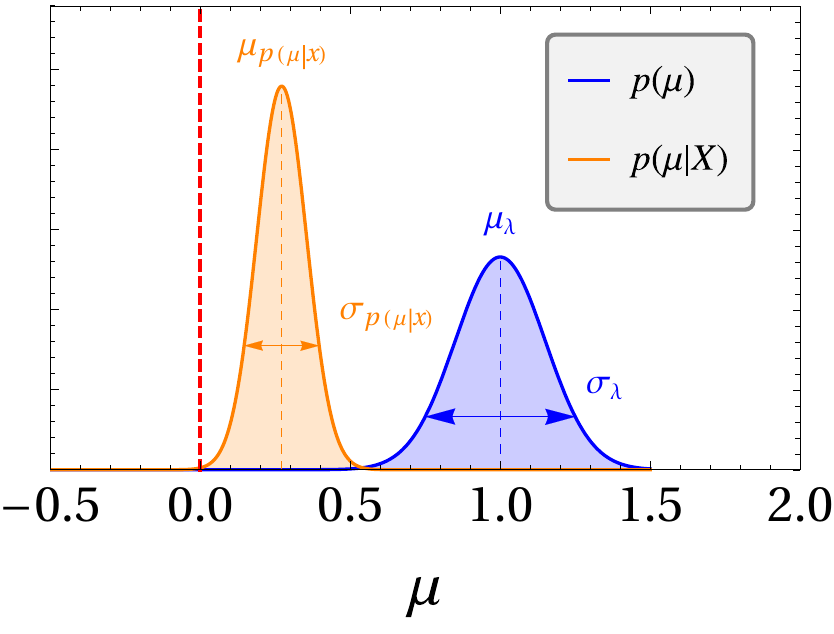}
		\caption[Bayesian update of the normal distribution.]{Bayesian update of the normal distribution.}
		\label{fig:GaussianUpdate}
\end{wrapfigure}
Note how the normal distribution is indeed its own conjugate. Here, we defined the variance of the posterior as:
\begin{equation}
\tilde{\sigma}^2 := \frac{\sigma_p^2 \sigma^2}{\sigma_p^2 + \sigma^2},
\end{equation}
and the mean of the posterior as:
\begin{equation}\label{eq:GaussianPosterior}
\tilde{\mu} := \frac{\sigma^2 \mu_p + \sigma_p^2 X}{\sigma_p^2 + \sigma^2}   .
\end{equation}
This result has an intuitive interpretation: the estimated mean is simply the average between the prior's mean $\mu_p$ and the outcome $X$, weighted by the variances. 
Essentially, the new mean is simply a "center of mass" between $X$ and $\mu_p$, and the more they differ, the more the estimation is shifted towards $X$. It is also interesting to study the limiting case: if $\sigma \rightarrow 0$ then $\mu_p \rightarrow X$. 
The reasoning behind this is that since the variance of the likelihood is small, it is very unlikely that $X$ was sampled from a point far away from the mean, so we have a strong reason to believe that $X$ itself is a good representative of the true average. See Fig.~\ref{fig:GaussianUpdate} for an illustration. $\bishoppair$

To close this section, note the crucial difference from the frequentist paradigm in our example: earlier, we were mainly interested in quantities which are averaged over all possible stochastic realizations, such as the Fisher information in Eq.~\eqref{eq:Fisher_def2} or the MSE from Eq.~\eqref{eq:mse}. Here on the other hand, we can see that the  posterior is conditioned \emph{on the data}, that is: different realizations obviously lead to different posteriors. In the classical case, we were very often conditioning over the \emph{parameter} instead. This contrast in the approaches motivates many of the new definitions and results that we are going to encounter further into this chapter.

\section{Bayesian estimators}\label{sec:bayes_est}

We have seen in the last section how the Bayesian framework yield posterior distributions. These are simply probability densities containing information about the parameter of interest $\theta$. Here we will show that the Bayesian framework also admits point estimation strategies, in the same spirit of the classical estimation theory. In particular, the posterior distributions themselves can be used to construct concrete estimators. By doing so, one can condense the information contained in a distribution into a single random variable: the (Bayesian) estimator. This strategy is, of course, of great practical interest; specially due to the fact that we can take advantage of big part of the formalism constructed in Chapter~\ref{chp:frequentist} for frequentist estimation.

This framework requires two main ingredients. The first one is the use of Bayes theorem for constructing posteriors distributions. The second one, which we are going to introduce now, is the notion of Bayesian figures of error (or cost functions). 
For instance: now that the parameter $\theta$ is \emph{also} a random variable, how can one define the MSE in this situation? We are going to show that typical Bayesian estimators are defined as functions which minimize some Bayesian figure of error with respect to the posterior distribution. 
In particular, the resulting classes of estimators depend on the type of cost function that is chosen.  

More concretely, it is possible do define a cost, or loss, function $C(\theta, \hat{\theta}({\bf X}))$ which decides how to penalize the estimator $\hat{\theta}$. Note that $C$ is a function of both the estimator $\hat{\theta}$ and the parameter $\theta$ (so it is also a random variable). 
We define the Bayesian error, also known as Bayesian loss, or risk, as the cost function integrated over $p(\theta, {\bf x})$:
\begin{equation}\label{eq:defbayesrisk}
\mathcal{R}
:=
\avg{\avg{C(\theta, \hat{\theta}({\bf X}))}{\bf X}}{\theta}
=
\int d\theta
\int C(\theta, \hat{\theta}({\bf x})) p(\theta, {\bf x}) d{\bf x}.
\end{equation}
By writing the joint pdf as $p(\theta, {\bf X}) = p({\bf X}|\theta) p(\theta)$, we also have:
\begin{equation}\label{eq:defbayesrisk1b}
\mathcal{R}
=
\int p(\theta) d\theta
\int C(\theta, \hat{\theta}({\bf x})) p({\bf x}|\theta)  d{\bf x}.
\end{equation}
We use $\avg{\cdot}{\theta}$ and $\avg{\cdot}{\bf X}$ to refer to the mean over the prior distribution and the stochastic average, respectively. By minimizing the error in Eq.~\eqref{eq:defbayesrisk} we obtain the \emph{Bayesian estimator}. Note how this is equivalent to the integration of the frequentist error, such as the MSE from Eq.~\eqref{eq:mse}, over the prior. In other words, we are \emph{averaging} the error over the parameter $\theta$ and weighting it with respect to the prior: we give more importance to the values of $\theta$ which are more likely to occur or to be true, and this is encoded precisely by the prior $\theta$. Deviations for unlikely values of $\theta$ are penalized less, and vice-versa.

\begin{figure}[h!]
		\centering
		\includegraphics[width=0.8\textwidth]{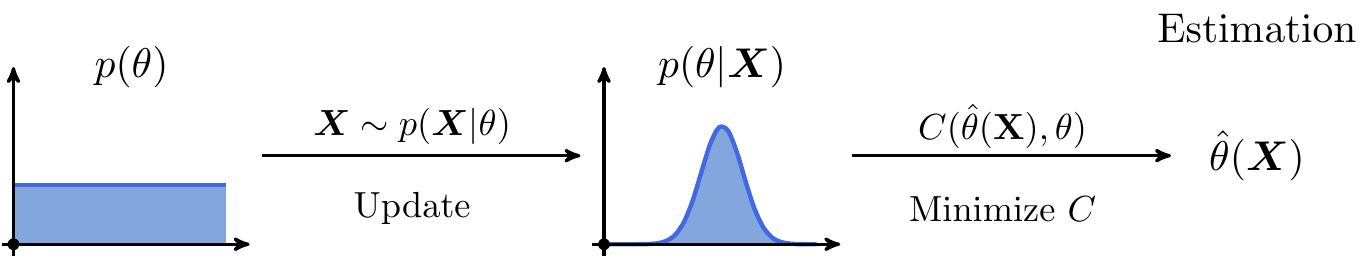}
		\caption[Steps for obtaining Bayesian estimators.]{Steps for obtaining Bayesian estimators.}
		\label{fig:estimator_diagram}
\end{figure}

We illustrate this procedure in Fig.~\ref{fig:estimator_diagram}. As we will show in the following example, the procedure for obtaining a MMSE estimator in the Bayesian sense is much simpler than obtaining its analogue from the frequentist scenario.

\framebox{$\opposbishops$ \textbf{\textit{Example:}}} Let us investigate the MSE in the Bayesian framework. For that, we take the cost function $C(\theta, \hat{\theta}({\bf X})) = (\theta - \hat{\theta}({\bf X}))^2$. The Bayesian mean-squared error (BMSE) reads:
\begin{equation}\label{eq:defbayesrisk2}
\epsilon_B (\hat{\theta})
:=
\int p(\theta) d\theta
\int
(\theta - \hat{\theta}({\bf X}))^2 p({\bf x}|\theta)  d{\bf x}
=
\avg{\epsilon(\hat{\theta}|\theta) }{\theta}
.
\end{equation}
Note how the BMSE is simply the frequentist MSE from Eq.~\eqref{eq:mse} integrated over the prior. We find that the equation above is minimized if we take:
\begin{equation}\label{eq:bayesianMean}
\frac{d \epsilon_B (\hat{\theta})}{d \hat{\theta}}
\bigg|_{\hat{\theta}=\hat{\theta}_{\mathrm{BA}}}
= 0
\implies
\hat{\theta}_{\mathrm{BA}}({\bf X}) = \int \theta p(\theta|{\bf X}) d\theta.
\end{equation}
The estimator above minimizes the MSE in the Bayesian sense. Note how this it is simply the mean value of the posterior distribution. For that reason, from hereafter we refer to $\hat{\theta}_{\mathrm{BA}}({\bf X})$ either as the Bayesian average (BA) or as the posterior mean. 

We also mention a few other common types of Bayesian estimators, showcasing the usefulness of this approach. 
In the case of the absolute error, we take the cost function $C(\theta, \hat{\theta}({\bf X})) = |\theta - \hat{\theta}({\bf X})|$. 
It can be shown that the corresponding Bayesian estimator $\hat{\theta}_{\mathrm{abs}}$ is the \emph{median} of the posterior, which satisfies:
\begin{equation}\label{eq:median}
\int_{-\infty}^{\hat{\theta}_{\mathrm{abs}}({\bf X})}
p(\theta|{\bf X})d\theta
=
\int_{\hat{\theta}_{\mathrm{abs}}({\bf X})}^{-\infty}
p(\theta|{\bf X})d\theta.
\end{equation}
Finally, we mention a last important example. There exists a third estimator, which is also very standard, known as the \emph{maximum a posteriori} (MAP) estimate. Rigorously defining the cost function which is minimized by the MAP is slightly difficult and we will omit the details here, but the proper construction can be found in Ref.~\cite{Bassett2019}. In practice, this estimator is simply the mode of the posterior:
\begin{equation}\label{eq:map}
\hat{\theta}_{\mathrm{map}}({\bf X})
=
\argmax_{\theta}
p(\theta|{\bf X}).
\end{equation}
One of the usefulness of this estimator, besides its simplicity, is that in certain conditions it agrees with the maximum likelihood estimate discussed in Sec.~\ref{sec:MLE}. 
In particular, we are going to argue in Sec.~\ref{sec:VTSB} that the MAP actually converges to the MLE in the asymptotic limit.  $\bishoppair$

The take away message from these examples is that each cost function has an associated estimator which minimizes it, per Fig.~\ref{fig:estimator_diagram}. 
For simplicity, we will focus mainly on the posterior mean; especially when discussing thermometry in later chapters.
While the choice of a cost function may seem quite subjective, the MS estimator is actually also the optimal estimator for a wider class of cost functions. 
One such example happens when $C(\theta, \hat{\theta}({\bf X}))$ is simultaneously convex and symmetric \cite{VanTrees2001} (p.~239). 
Thus, this estimator provides a good balance between generality and simplicity. This is an appropriate moment to employ these results in the practical example from the previous section. We will explicitly obtain the estimators, and their error, for the Bernoulli trials.

\begin{figure}[t!]
	\centering
	\includegraphics[width=0.5\textwidth]{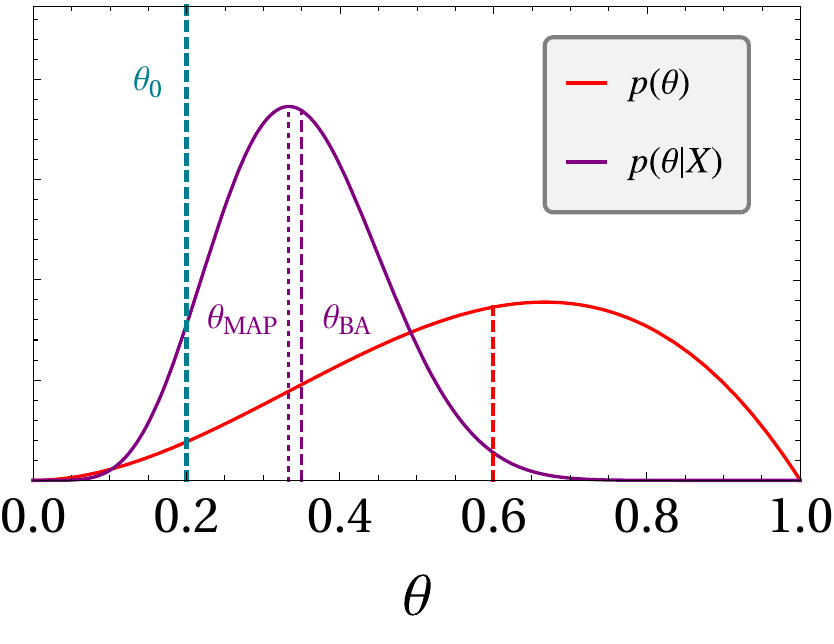}
	\caption[The \emph{maximum a posteriori} and Bayesian average estimators for the Beta distribution.]{The \emph{maximum a posteriori} and Bayesian average estimators for the Beta distribution. In this particular example we choose $\mathcal{B}(3, 2)$ as the prior. The true value of the parameter is $\theta_0 = 0.2$ and we update the prior with $n=15$ outcomes. In this particular realization we had $k=4$, which yield the estimates $\hat{\theta}_{\mathrm{BA}}({\bf X})=0.35$ and $\hat{\theta}_{\mathrm{BA}}({\bf X})=1/3$. The red dashed line shows the BA estimate of the prior.}
	\label{fig:BetaUpdate}
\end{figure}

\framebox{$\opposbishops$ \textbf{\textit{Example:}}} We found that a for a prior distribution $\mathcal{B}_{\alpha_0, \beta_0}(\theta)$, the posterior becomes, after $n$ Bernoulli trials:
\begin{equation}
p(\theta|{\bf X}) = \mathcal{B}_{\alpha_0 + k, \beta_0 + n - k}(\theta).
\end{equation}
Here we wrote $k := \sum_i X_i$. How does the MAP and the BA estimates look like under the posterior distribution written above? By using the identity from Eq.~\eqref{eq:betaIdentity} together with the definition in Eq.~\eqref{eq:beta} we can can show that the mean of the Beta distribution is given by $\alpha/(\alpha + \beta)$. Similarly, the mode is given by $(\alpha - 1)/(\alpha + \beta - 2)$. This means that the BA and the MAP evaluate to
\begin{equation}\label{eq:BAbeta}
\hat{\theta}_{\mathrm{BA}}({\bf X})
=
\frac{\alpha_0 + k}{\alpha_0 + \beta_0 + n}
\end{equation}
and
\begin{equation}
\hat{\theta}_{\mathrm{MAP}}({\bf X})
=
\frac{\alpha_0 + k - 1}{\alpha_0 + \beta_0 + n - 2},
\end{equation}
respectively. We show a visual example in Fig.~\ref{fig:BetaUpdate}. The estimator in Eq.~\eqref{eq:BAbeta} has a particularly nice interpretation. It can be written as:
\begin{equation}
\frac{\alpha_0 + k}{\alpha_0 + \beta_0 + n}
\equiv
\frac{n}{\alpha_0 + \beta_0 + n}
\left(
\frac{k}{n}
\right)
+
\frac{\alpha_0 + \beta_0}{\alpha_0 + \beta_0 + n}
\left(
\frac{\alpha_0}{\alpha_0 + \beta_0}
\right)
,
\end{equation}
which is the weighted average between the MLE and prior estimate \cite{Ruschendorf2014} (pp.~42-43). To conclude this example, we will explicitly calculate the MSE for the posterior mean in Eq.~\eqref{eq:BAbeta}. Using the properties for the mean and the variance of the Bernoulli distribution, we find that the variance and the bias of the BA are given by:
\begin{equation}
\Var{(\hat{\theta}_{\mathrm{BA}})}
=
\frac{1}{(\alpha_0 + \beta_0 + n)^2}
\Var{\left(\sum_i X_i \right)}
=
\frac{n \theta (1-\theta)}{(\alpha_0 + \beta_0 + n)^2}
\end{equation}
and
\begin{equation}
b(\hat{\theta}_{\mathrm{BA}})
=
\frac{\alpha_0 + \avg{\sum_i X_i}{}}{\alpha_0 + \beta_0 + n}
-
\theta
=
\frac{\alpha_0 (1-\theta) - \beta_0 \theta}{\alpha_0 + \beta_0 + n},
\end{equation}
respectively. Here we have used the definitions from Eqs.~\eqref{eq:variance} and \eqref{eq:bias}. Moreover, note how the BA is, in general, a biased estimator, as this example shows.

By using the bias-variance decomposition obtained in Eq.~\eqref{eq:bias_var}, we finally arrive at the (frequentist) MSE for the BA estimator:
\begin{equation}
\epsilon(\hat{\theta}_{\mathrm{BA}}|\theta)
=
\frac{n \theta (1-\theta) + (\alpha_0 (1-\theta) - \beta_0 \theta)^2}{(\alpha_0 + \beta_0 + n)^2}
.
\end{equation}
We can now evaluate the \emph{Bayesian MSE} the integrating the expression above over the prior. In the case of the flat prior, with $\alpha_0 = \beta_0 = 1$, we obtain:
\begin{equation}
\epsilon_B(\hat{\theta}_{\mathrm{BA}})
=
\int_0^1 p(\theta) \epsilon(\hat{\theta}_{\mathrm{BA}}|\theta) d\theta
=
\frac{1}{6(n + 2)}.
\end{equation}
Now, let us compare this result with the MLE estimator from the last chapter, which is just the sample mean. Note that the MLE in this case is simply the MAP for a flat prior. The estimator in this case was unbiased and the MSE was given by Eq.~\eqref{eq:MLEbinomial}:
\begin{equation}
\epsilon(\hat{\theta}_{\mathrm{MLE}}|\theta)
=
\frac{\theta (1-\theta)}{n}.
\end{equation}
This means that the BMSE in this case evaluates to:
\begin{equation}
\epsilon_B(\hat{\theta}_{\mathrm{MLE}})
=
\frac{1}{6n}.
\end{equation}
Note how the BMSE of the BA is slightly lower than the one for the sample mean. 
\begin{figure}[h!]
	\centering
	\includegraphics[width=0.8\textwidth]{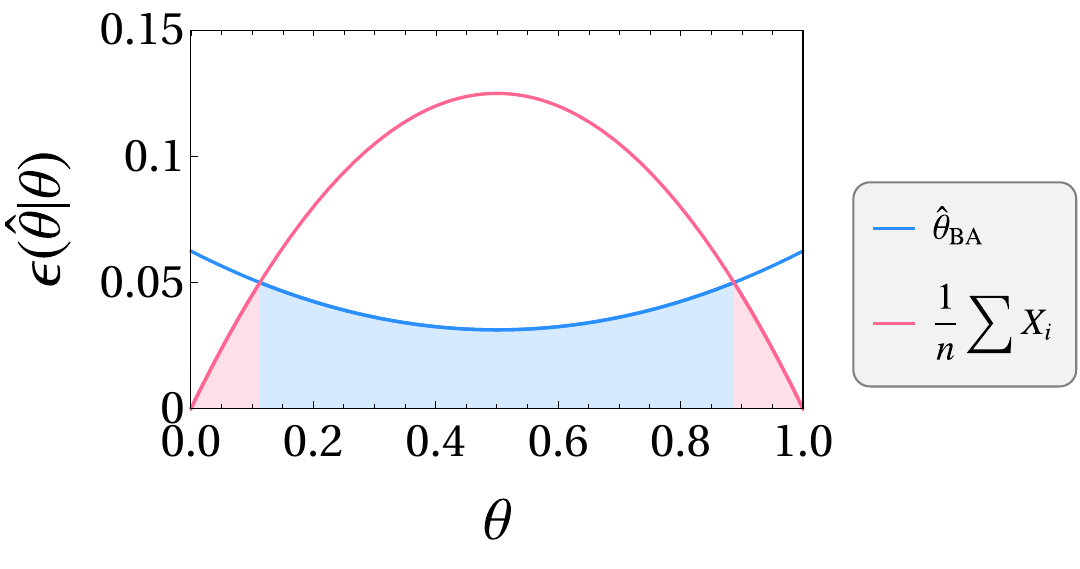}
	\caption[MSE for the Beta distribution with $n=2$.]{MSE for the Beta distribution with $n=2$. We plot the results for the posterior mean (blue) and the sample mean (magenta) estimators. We shade the plot according to the optimal estimator. The blue (magenta)-shaded region corresponds to the interval where the BA (sample mean) estimator is better.}
	\label{fig:MSEBeta}
\end{figure}

We also plot the results for the MSE of the BA and the MLE estimators in Fig.~\ref{fig:MSEBeta}. This picture give us some intuition about the relevance of the different definitions of error that we have introduced throughout the last two chapters. 
First, note how the optimal estimator among these two choices depends on the particular value of $\theta$. 
The usefulness of the Bayesian error in this case is that it encapsulates the optimality of these estimates into a single figure: the BMSE of the BA is lower due to the fact that, \emph{on average}, it performs better that the MLE. In particular, while the MLE is better for values around $\theta=0$ or $\theta=1$, the BA works better for intermediate values of $\theta$.  $\bishoppair$

We just saw how to obtain concrete estimators in this example by means of Bayesian tools. These results are, of course, a further merit of this approach. As we mentioned in the last chapter, MMSE estimators, or even MVU estimators, are very often unrealizable. 
If one consider these figures of error in the Bayesian sense instead, achieving MMSE estimators, or similar, is perfectly possible in general \cite{VanTrees2001} (p.~239). 
Thus, we abandon the task of finding the best estimator for \emph{every} $\theta$ and we focus instead on finding the best estimator in the Bayesian sense, as we defined in Eqs.~\eqref{eq:defbayesrisk} and \eqref{eq:defbayesrisk1b}. By doing so, we obtain estimators which are good on \emph{average}.

This formulation results in a sensible strategy in the scenario where, e.g., one is initially very ignorant about $\theta$, so it might be hard to choose a reasonable estimator which works well for the true value of the parameter. 
Of course, this example also serves to further stress that Bayesian and frequentist paradigms should not be seen as entirely antagonistic approaches, and neither that one of these formulations is necessarily better than the other. 
Their techniques might be actually very analogous or complementary at times: the last example just showed us how one can use Bayesian techniques as the starting point to construct concrete estimators. The BA and MAP estimators are perfectly valid estimators even beyond the Bayesian setting, as we have even calculated their MSE in the frequentist sense. Thus, even if one is not interested in the Bayesian interpretation of the data, it might still be useful to consider the Bayes theorem as a starting point and as a \emph{tool} for obtaining concrete estimators. 

The Bayesian framework and its notion of integrated error is also useful for, e.g., formalizing the notion of ordering of the risk, in consonance with much of our discussions throughout Sec.~\ref{sec:MVU} and \ref{sec:est_properties} regarding the optimality of estimators in a purely frequentist sense. 
A very famous problem, known as Stein's example results in what is known as the James-Stein estimator \cite{Stein1956, James1992}. This is a scenario where a biased estimator is \emph{always} better than the MLE estimator and where we can make use of, e.g., the idea of \emph{admissibility} to compare them \cite{Berger1985}. 

We mention that some authors might not  consider Eq.~\eqref{eq:defbayesrisk} a concept which is \emph{entirely} Bayesian. A purely Bayesian loss is conditioned on the data instead, as we defined in Eq.~\eqref{eq:bayese_error_pure}. Thus, we could rather define a \emph{posterior} loss
\begin{equation}\label{eq:posteriorloss}
\epsilon(\hat{\theta}|{\bf X})
:=
\int
C(\theta, \hat{\theta}({\bf X}))
p(\theta|{\bf X})d\theta
\end{equation}
which disregards other realizations and is conditioned on the measured data ${\bf X}$. Thus, what we call the Bayesian error in this work, per Eq.~\eqref{eq:defbayesrisk}, is sometimes called \emph{preposterior} loss or integrated risk. It is an error which is averaged over \emph{all} possible realizations, while Eq.~\eqref{eq:posteriorloss} takes only the observed data into consideration. 
Therefore, the preposterior loss is the error that we expect to obtain \emph{before} we perform any measurements. 
Hence, one could argue that Eq.~\eqref{eq:defbayesrisk} is useful in numerical or theoretical investigations where the experimenter does not have any data at hand. In this case he or she would like to have a general picture of what type of error should expected for different realizations. 
With the data in hands, it is often more useful to use the posterior loss instead. Fortunately, this difference is not very significant when discussing estimators, since minimizing either the posterior or preposterior losses in Eqs.~\eqref{eq:defbayesrisk} and~\eqref{eq:posteriorloss} is equivalent and yields the same estimators \cite{Robert1994} (p.~62-63). Our current results are valid even in this scenario.

Dwelling into the philosophical and technical implications of the Bayesian and frequentist theories is well beyond the scope of this work \cite{Aitchison1964, Bland1998, Vallverdu2008}. For that, many appropriate references can be found. Refs.~\cite{Jaynes2003} and \cite{Robert1994} discuss many of the implications and distinctions between the two schools of statistics. A special attention to this topic is given in Ref.~\cite{Samaniego2010}.

Furthermore, obtaining appropriate priors and a sensible loss function, which we considered to be some of the main ingredients for inference, is another very important aspect of the Bayesian theory. This is specially true when one is interested in employing a fully Bayesian approach. 
In those scenarios, it is much more important to be careful with such details. We omit these topics since this is not the focus of this dissertation. However, we mention a few techniques which are standard. 
We have, throughout this chapter, used flat priors as a synonyms of uninformative priors. This is not strictly true in general. Many models might require more sophisticated distributions to better convey the idea of ignorance. Among possible techniques to derive them, the idea of Jeffrey's priors and the maximum entropy principle \cite{Jaynes2003} (pp.~181-183, 372-378) configure some of the main techniques. 
Sometimes, it is possible to obtain an appropriate prior by analyzing the symmetries and constraints of problem \cite{Bernardo1994} (p.~366), such as scale estimation and phase estimation problems \cite{Jaynes2003} (pp.~378-394). Thus, it is clear that assigning appropriate priors and error functions is a whole topic of study on itself. A self-contained introduction can be found in Ref.~\cite{Robert1994} (Chapter 3.5).

\begin{figure}[h!]
	\centering
	\includegraphics[width=\textwidth]{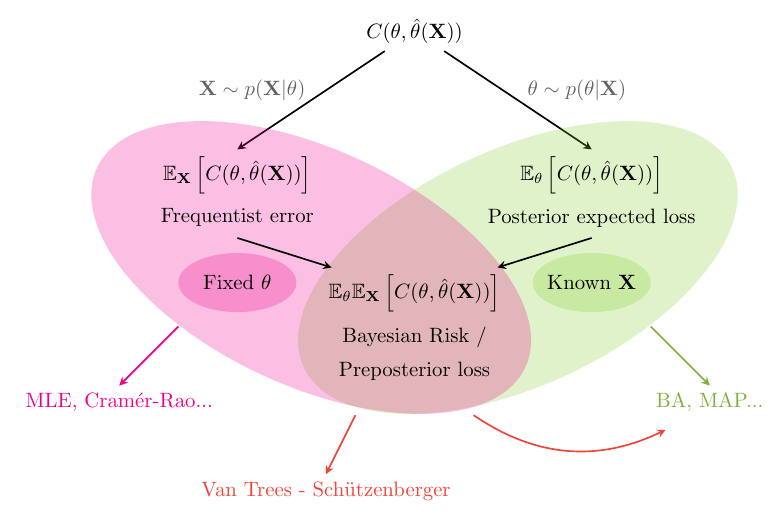}
	\caption[A diagram with the summary of different inference approaches.]{A diagram of the summary of different inference approaches.}
	\label{fig:errortypes}
\end{figure}

While we are not going to directly employ most of these concepts in the development of our original results, we have nevertheless mentioned them in this thesis for completeness. This is motivated by the fact that part of the current literature on quantum thermometry borrows from many of these ideas: a certain familiarity with some of these topics might be useful in keeping up to date with the bibliography we are concerned with. We summarize this discussion, and the different definitions of error, such as Eqs.~\eqref{eq:classical_error},~\eqref{eq:defbayesrisk} and~\eqref{eq:posteriorloss} in Fig.~\ref{fig:errortypes}. We introduce the remaining concept from the figure, the Van Trees-Schützenberger inequality, in the coming section.

\section{The Van Trees-Schützenberger inequality}\label{sec:VTSB}

As we saw, due to the gap between the Bayesianism and the frequentism, transposing some concepts from one paradigm to another requires some care. 
In particular, it is not immediately clear how the Cramér-Rao bound should look like in terms of an integrated loss function, in the sense of Eq.~\eqref{eq:defbayesrisk2}. Since the classical CRB is essentially a frequentist concept, it naturally depends on the parameter of interest in general.
Not only that, but the bound is also unable to capture any of the information contained in the prior distribution.
In this chapter, we discuss an extension to the Bayesian case. This result was independently obtained by Van Trees~\cite{VanTrees2001} in 1968 and by Sch\"utzenberger~\cite{Schutzenberger1957} a few years earlier. We refer to their work as the Van Trees-Schu\"tzenberger bound (VTSB). 
This approach avoids one of the conundrums of the usual CRB; the bound is no longer dependent on the parameter being estimated. Instead, we obtain an inequality which depends only on the estimator and the prior distribution. 

As it turns out, we can further adapt the Cramér-Rao inequality as a Bayesian bound through the use of the Cauchy-Schwarz theorem and a clever choice of joint densities. We can then turn it into an inequality for the MSE while also embedding information contained within the prior. This derivation is of very similar spirit to the one for the CRB in Sec.~\ref{sec:CRB}. The regularity conditions required for this bound \cite{VanTrees2001} (p.~261), together with some generalizations and applications, can be found in Ref.~\cite{Gill1995}.

First, we shall assume that integration and derivation are interchangeable, in the sense that: 
\begin{equation}\label{eq:VanTreesCondition1}
\int \ppx{p(x|\theta)}{\theta} dx 
= 
\ppx{}{\theta} \int p(x|\theta)dx
= 
0.     
\end{equation}
We will also assume that the prior density $p(\theta)$ vanishes at the limits of integration. By doing so, we can use integration by parts to show that the following identity holds:
\begin{equation}\label{eq:BayesianCramer1}
\int  
\theta \ppx{(p(x|\theta) p(\theta))}{\theta}d\theta 
= 
-\int p(x|\theta)d\theta,
\end{equation}
and, by analogy, these conditions also imply that:
\begin{equation}
\int  \ppx{(p(x|\theta) p(\theta))}{\theta}d\theta = 0.
\end{equation}
Since the estimator $\hat{\theta}(X)$ is \emph{not} a function of the random parameter itself, we can multiply the integral above by $\hat{\theta}$, subtract by  Eq.~\eqref{eq:BayesianCramer1} and then integrate with respect to $x$ on both sides to obtain:
\begin{equation}
\int \int 
(\hat{\theta}(x) - \theta)
\ppx{(p(x|\theta) p(\theta))}
{\theta}p(\theta) d\theta dx = \int \int p(x|\theta)
dx d\theta   
= 
1.
\end{equation}
Note however that we can write the derivative above as:
\[
\ppx{(p(x|\theta) p(\theta))}{\theta}
=
\ppx{(p(x|\theta) p(\theta))}{\theta}
{\color{blue} \frac{p(x|\theta) p(\theta)}{p(x|\theta) p(\theta)}}
=
p(x|\theta) p(\theta)
\ppx{\ln{(p(x|\theta) p(\theta))}}{\theta},
\]
so that the previous equation becomes:
\begin{equation}
\int \int (\hat{\theta}(x) - \theta)
\ppx{\ln{(p(x|\theta) p(\theta))}}{\theta}
p(x|\theta) p(\theta)
d\theta dx 
=
1 
\end{equation}
The LHS can be written as an expectation value: 
\begin{equation}
\EX_\theta 
\EX_X \left[
(\hat{\theta}(X) - \theta)
\ppx{\ln{(p(x|\theta) p(\theta))}}{\theta}
\right]
=
1. 
\end{equation}
Using the Cauchy-Schwarz inequality, in analogy to the derivation from Sec.~\ref{sec:CRB}, we can rewrite this as:
\begin{equation}\label{eq:CauchyVanTrees}
\EX_\theta 
\EX_X \left[
(\hat{\theta}(X) - \theta)^2
\right]
\EX_\theta
\EX_X \left[
\left(
\ppx{\ln{(p(x|\theta) p(\theta))}}{\theta}
\right)^2
\right]
\geqslant
1.
\end{equation}
By using the product rule for logarithms, we can rewrite the expectation in Eq.~\eqref{eq:CauchyVanTrees} as:
\begin{equation}\label{eq:BayesianCramer2}
\begin{split}
\EX_\theta
\EX_X \left[
\left(
\ppx{\ln{(p(x|\theta) p(\theta))}}{\theta}
\right)^2
\right]
&=
\EX_\theta
\EX_X \left[
\left(\ppx{\ln{p(x|\theta)}}{\theta}\right)^2
\right]\\
&+
2 \EX_\theta
\EX_X \left[
\left(\ppx{\ln{p(x|\theta)}}{\theta}\right)
\left(\ppx{\ln{p(\theta)}}{\theta}\right)
\right]\\
&+
\EX_\theta
\EX_X \left[
\left(\ppx{\ln{p(\theta)}}{\theta}\right)^2
\right],
\end{split}
\end{equation}
where we can use the condition \eqref{eq:VanTreesCondition1} to verify that the middle term vanishes:
\[
\EX_\theta
\EX_X \left[
\left(\ppx{\ln{p(x|\theta)}}{\theta}\right)
\left(\ppx{\ln{p(\theta)}}{\theta}\right)
\right]
=
0.
\]
Now, we can see that the inner expectation in the first term of Eq.~\eqref{eq:BayesianCramer2} is simply the Fisher Information of the likelihood. Hence, the first expectation in Eq.~\eqref{eq:BayesianCramer2} is simply the Fisher information averaged over the prior:
\begin{equation}
\EX_\theta 
\EX_X \left[
\left(\ppx{\ln{p(x|\theta)}}{\theta}\right)^2
\right]    
=
\EX_\theta [F(\theta)]
\end{equation}
Meanwhile, the expectation of the last term in Eq.~\eqref{eq:BayesianCramer2} with respect to $X$ is trivial, since it only depends on $\theta$. Therefore, it results in an expression for the Fisher information of the prior distribution, which we denote by:
\begin{equation}
F_p
:=
\EX_\theta\left[
\left(\ppx{\ln{p(\theta)}}{\theta}\right)^2
\right].
\end{equation}
Thus, having identified all the tree terms, we can finally use the inequality \eqref{eq:CauchyVanTrees} to arrive at the desired result:

\textbf{The Van Trees-Sch\"utzenberger inequality}

The Bayesian mean square error~\eqref{eq:defbayesrisk2} of any estimator $\hat{\theta} = \hat{\theta} (X)$ of $\theta$, with $X \sim p(X|\theta)$, is bounded by the inequality
\begin{equation}\label{eq:VTSB}
\epsilon_B
=
\EX_\theta 
\EX_X \left[ 
(\hat{\theta} - \theta)^2
\right] 
\geqslant
\frac{1}{\EX_\theta[F(\theta)] + F_p},
\end{equation}
given the prior distribution $p(\theta)$ and the appropriate regularity conditions. In analogy to Eq.~\eqref{eq:efficient_estimatorN}, one has
\begin{equation}\label{eq:VTSBN}
\epsilon_B
\geqslant
\frac{1}{n \EX_\theta[F(\theta)] + F_p}
\end{equation}
for a sequence ${\bf X} = X_1, ..., X_n$ of i.i.d. outcomes.

This bound is quite general in the sense that, differently from the standard CRB in Eq.~\eqref{eq:efficient_estimator} for \emph{unbiased} estimators, no requirement about  unbiasedness was made here. Moreover, we can immediately see that the VTSB does \emph{not} depend on $\theta$. The FI of the likelihood is actually averaged over the prior. This is an upside of this result; we can obtain the limits of precision of an estimator independently $\theta$.  Additionally, when comparing this inequality with the usual CRB we immediately see an extra term which accounts for the prior knowledge about the parameter.

Of course, this does not mean that the VTSB is more useful than the CRB. Ultimately, they both tell us very different things: the CRB is an inequality for the MSE given by Eq.~\eqref{eq:mse} in the frequentist sense. 
This bound is of course much more appropriate when one is interested in the precision of the estimator for a particular value of $\theta$. Thus, it might be reasonable to say that some information about the maximum achievable precision is "lost" when one integrates over the prior to use the VTSB. And the VTSB is valid, of course, only for the \emph{Bayesian} analogue of the MSE in Eq.~\eqref{eq:defbayesrisk2}. For instance, maybe a certain estimator performs very well for a certain range of  the parameter and very bad for others. By using the Bayesian risk and the VTSB we would have information of their Bayesian averages only, and not of specific parametrizations. Of course, it is also possible to argue that this is not so bad because the VTSB can incorporate a localized knowledge about $\theta$ pretty well through $F_p$ and the expectation over the FI anyway. Nevertheless, both approaches are just different perspectives on how to \emph{present} the data and the associated bounds. It is on the hands of the theoretician and the experimentalist which of these two presentations would yield better interpretations for his or her particular problem; in some sense, the two constructions are just showing the same information in different ways. 

More strikingly, and also going back to technical aspects, while the CRB is very often achievable in the asymptotic limit for, e.g., maximum likelihood estimators (as discussed in Sec.~\ref{sec:MLE}), the VTSB is not tight in general. 
This happens due to the asymptotic behavior of the posterior distribution and the structure of the VTSB. 
For independent outcomes, a result known as the Bernstein-von Mises theorem \cite{Cam1986, Vaart1998} assures that the posterior converges, in the limit of large $n$, to a Gaussian whose mean is centered around the true value of the parameter $\theta_0$ with variance $1/nF(\theta_0)$. In symbols,
\begin{equation}\label{eq:vonMises}
P(\theta|{\bf X}) 
=
\sqrt{\frac{n F(\theta_0)}{2\pi}}
e^{-\frac{n F(\theta_0) (\theta - \theta_0)^2}{2}},
\quad
(\text{large n}).
\end{equation}
This is very similar to the asymptotic properties of the MLE which we mentioned in Sec.~\ref{sec:MLE}. A consequence of the result above is that the MSE scales with $1/nF(\theta)$ in the asymptotic limit. This implies that the BMSE scales as
\begin{equation}\label{eq:BMSEAsymptotic}
\epsilon_B(\hat{\theta})
\sim
\avg{\frac{1}{n F(\theta)}}{\theta},
\quad
(\text{large n, MSE}).
\end{equation}
In other words, the asymptotic BMSE scales simply with $1/nF(\theta)$ averaged over the prior. Meanwhile, the VTSB scales with
\begin{equation}
\epsilon_B(\hat{\theta})
\sim
\frac{1}{\avg{n F(\theta)}{\theta}},
\quad
(\text{large n, VTSB}),
\end{equation}
in the limit of many measurements. Thus, as a consequence of Jensen's inequality \cite{Bickel2015} we have that $\avg{\frac{1}{F(\theta)}}{\theta} \geqslant \frac{1}{\avg{F(\theta)}{\theta}}$, meaning that the VTSB is, in general, not tight \cite{VanTrees2001} (p.~265). In summary, this means that estimators which are asymptotically efficient, such as the MLE and the MAP, will converge to the \emph{expected} Cramér-Rao bound, but not to the VTSB. 

An elementary proof of the the result by Berstein and von Mises \cite{Lindley1965, Hartigan1983} can be sketched in terms of the Laplace approximation \cite{Butler2007}, which is often used to approximate posterior distributions \cite{Tierney1986}. 
A short and comprehensive summary of the method can be found in Ref.~\cite{MacKay2002} (p.~341). 
A rigorous proof, with consideration for the appropriate regularity conditions, can be found in more technical texts \cite{Cam1986, Walker1969}.

We also mention a generalization which eliminates the requirement for a prior with bounded supports where the distribution does not vanish at the end points \cite{Ramakrishna2020}. This result might be useful when one is interested in, e.g., truncated Beta distributions. Needless to say, the VTSB is not the unique Bayesian lower bound of interest. Research concerning alternative, and possibly tighter, bounds is also active \cite{Bacharach2019}. A compilation of several works concerned with Bayesian bounds can be found in Ref.~\cite{VanTrees2007}. We conclude this section with an example.

$\opposbishops$ \textbf{\textit{Example:}} In the problem of estimating the mean of a normal distribution, we were able to obtain the mean of the posterior distribution in Eq.~\eqref{eq:GaussianPosterior}. Because the posterior is a Gaussian, the MAP and the BA naturally coincide, and the single-shot estimator is given by:
\begin{equation}
\hat{\mu}_{BA} 
= 
\hat{\mu}_{MAP} 
= 
\frac{\sigma_p^2 X + \sigma^2 \mu_p}{\sigma_p^2 + \sigma^2}.   
\end{equation}
The \emph{squared} error associated with the estimator above will be:
\begin{equation}
(\hat{\mu}_{BA} - \mu)^2 =
\frac{
\sigma_p^4 (x-\mu)^2 
+ 2 \sigma_p^2  \sigma^2 (x-\mu) (\mu_p - \mu)
+ \sigma^4 (\mu_p - \mu)^2}
{(\sigma_p^2 + \sigma^2)^2}.
\end{equation}
Now, per the definition~\eqref{eq:defbayesrisk2}, we are expected to take the stochastic average (with respect to $X$) in order to calculate the BMSE. We first should note that:
\begin{equation}
\EX_X[X - \mu] = \EX_X[X] - \mu = 0    
\end{equation}
This means that the second term in the RHS of the equation above vanishes, and we also have that
\begin{equation}
\EX_X[(X - \mu)^2] = \sigma^2,
\end{equation}
so:
\begin{equation}
\EX_X[(\hat{\mu}_{BA} - \mu)^2]
=
\frac{\sigma_p^4 \sigma^2 + \sigma^4 (\mu_p - \mu)^2}
{(\sigma_p^2 + \sigma^2)^2}.   
\end{equation}
Now what remains is to take the average over the prior. The first term is independent of $\mu$, so we may as well leave it alone. The second term on the other hand depends on the parameter $\mu$ we want to estimate, so we can verify that:
\begin{equation}
\EX_\mu[(\mu_p - \mu)^2]
=
\int
(\mu - \mu_p)^2
p(\mu)
d\mu
=
\int
(\mu_p - \mu)^2
\frac{e^{\frac{(\mu - \mu_p)^2}{2\sigma_p^2}}}{\sqrt{2\pi}\sigma_p}
d\mu
=
\sigma_p^2.
\end{equation}
Here, we explicitly wrote the variance calculation as an integral for clarity. Now, we can finally write the BMSE error for this estimation procedure:
\begin{equation}
\EX_\mu \EX_X[(\hat{\mu}_{BA} - \mu)^2]
=
\frac{\sigma_p^4 \sigma^2 + \sigma_p^2 \sigma^4}
{(\sigma_p^2 + \sigma^2)^2}   
= 
\frac{\sigma_p^2 \sigma^2}{\sigma_p^2 + \sigma^2}.
\end{equation}

Finally, we can compare the result above with the Van Trees-Schu\"tzenberger inequality. The Fisher Information for the prior will be $F_p = \frac{1}{\sigma_p^2}$.
Similarly, we have$F(\mu) = \frac{1}{\sigma^2}$ for the likelihood. Since it does not depend on the parameter $\mu$ we are estimating, it is trivial to check that
\begin{equation}
\EX_\mu(F(\mu)) =  \frac{1}{\sigma^2}.   
\end{equation}
Hence, it is easy to see that:
\begin{equation}
\frac{1}{\EX_p(F(\mu)) + F_p} = \frac{\sigma_p^2 \sigma^2}{\sigma_p^2 + \sigma^2} 
\end{equation}
Therefore, we can finally verify that:
\begin{equation}
\EX_\mu \EX_X[(\hat{\mu}_{MAP} - \mu)^2]
=
\frac{1}{\EX_\mu(F(\mu)) + F(p)}  
= 
\frac{\sigma_p^2 \sigma^2}{\sigma_p^2 + \sigma^2} 
\end{equation}
Thus the MAP and the BA also \emph{saturate} the VTSB it in this case. This happens precisely due to the fact that the Fisher information $F(\mu)$ [Eq.~\eqref{eq:Fisher_def2}] depends only on the known $\sigma$, and not on the unknown parameter $\mu$, as we mentioned. $\bishoppair$

%% file: chapters/metrology.tex
Quantum metrology is the field which is concerned with measurement protocols of high precision \cite{Giovannetti2011, Pezze2018, yuQuantumFisherInformation2022} and with the problem of estimation theory \cite{Braunstein1994} in quantum mechanics. 
In particular, much of the current advances and the literature are either preoccupied with the theoretical and experimental idiosyncrasies of the quantum theory in this context \cite{Kiilerich2015}, or with the use of quantum resources and how they can bring quantum advantages in metrological settings \cite{escherQuantumMetrologyNoisy2011, Marzolino2014, Braun2018}. 

Quantum advantages, in a very broad sense, have since long been discussed. Ubiquitous instances of this are the closely related field of quantum optics and \cite{Ou1996}, and also in the field of quantum algorithms since the 1990s \cite{Cleve1998}, albeit in a very different context. 
An omnipresent topic which, even though we are not going to focus on, serves as an important example, is the idea of achieving the \emph{Heisenberg limit}, where one can obtain a scaling of $1/n^2$ for the variance of the estimation by means of quantum protocols, in contrast with, e.g., the usual bound of $1/n$ for the CRB, arising from the classical estimation theory \cite{Shapiro1989, Braunstein1992, Lee2002, Luis2002}. The latter is often called the \emph{the standard quantum limit} (SQL). 

Such type of improvement might be explored through entanglement \cite{Maccone2013, Wang2018}, quantum criticality \cite{Zanardi2008, Garbe2020, Ying2022} and squeezing \cite{Caves1981, Milburn1994}, to mention a few instances. A typical implementation consists in interferometric setup which makes use of highly entangled states, such as NOON states or GHZ states \cite{leibfriedHeisenbergLimitedSpectroscopyMultiparticle2004}.
See Ref.~\cite{horodeckiQuantumEntanglement2009} (p.~889-891) for the definition and further discussion about such entangled states. While this type of investigation is not exactly our focus in Chapter~\ref{chp:results}, they are naturally a great motivation for the underlying framework discussed in the upcoming sections.

On a historical note, we cite the canonical works by Personick \cite{Personick1971}, Helstrom \cite{Helstrom1969, Helstrom1976} and Holevo \cite{Holevo2011} as early and very important contributions to the field, which were vital in kick-starting the foundations of quantum estimation theory. Furthermore, we also mention Giovanneti, Lloyd and Maccone's work as one of the earliest accounts of a more modern formulation of the field, which was central in shaping the current landscape for quantum metrology \cite{Giovannetti2006}. Further examples and applications of quantum metrology can be found in Ref.~\cite{tothQuantumMetrologyQuantum2014}.

Needless to say, the evident usefulness of quantum metrology makes it a powerful instrument in several different fields, ranging from biology \cite{Taylor2016} to photonics \cite{Polino2020}. Our focus in this dissertation is the application of this framework in the context of quantum thermodynamics, in a subfield known as \emph{quantum thermometry} (see Ref.~\cite{Mehboudi2019} for a review), which is concerned with the particular problem of temperature estimation. We introduce this topic in Sec.~\ref{sec:thermometry}, following with a review of a concrete and recently proposed \cite{Seah2019} implementation in Sec.~\ref{sec:collisional}. 

\section{The quantum Fisher information}

\begin{figure}[h!]
	\centering
	\includegraphics[width=\textwidth]{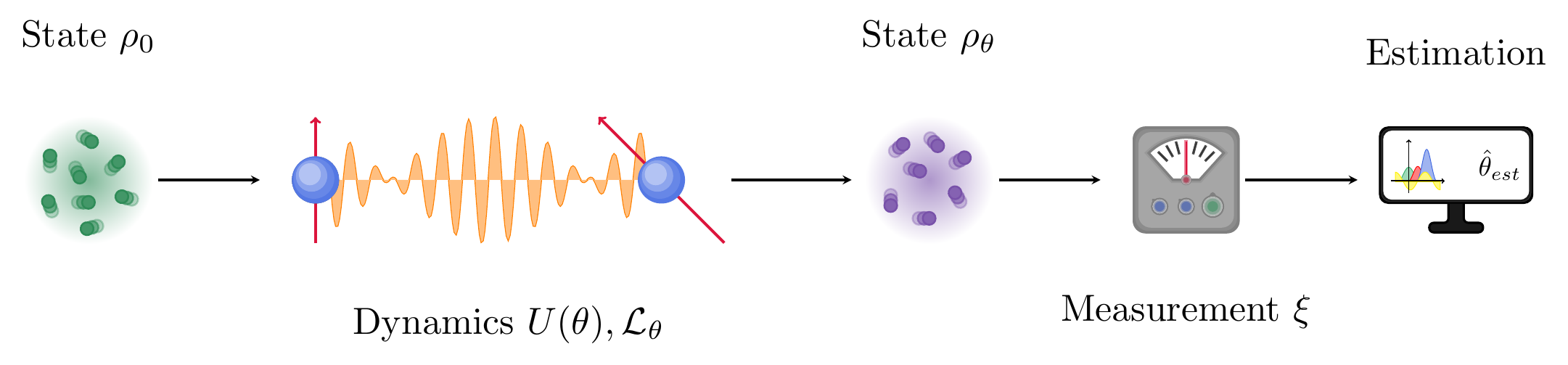}
	\caption[Typical protocol in quantum metrology.]{Typical protocol in quantum metrology.}
	\label{fig:metrology_drawing}
\end{figure}

We start this section by describing the standard steps in a typical metrology experiment, as illustrated by Fig.~\ref{fig:metrology_drawing}. We then introduce the relevant definitions, discussing the idea behind quantum metrology and quantum estimation theory. 
For that, let $\rho$ be the density matrix describing the state of a (possibly mixed) system. 
One usually starts with an input (or probe) state denoted by $\rho_0$. A parameter of interest $\theta$, which can be, e.g., a phase or a frequency, is encoded into an output state $\rho_\theta$ by means of a dynamical evolution. 
This embedding of the parameter is very often done trough a unitary evolution $U(\theta)$, a superoperator $\mathcal{L}_\theta$ describing a dissipative evolution or even through the measurement back-action described by Kraus operators \cite{Clark2019}. 
Here, we will focus on protocols where the final state $\rho_\theta$ is measured after unitary and dissipative dynamics. This measurement is described by a set of positive
operator-valued measures (POVMs) \cite{Nielsen2012}, denoted by $\{M_X\}$. 
These measurements are then associated with outcomes $X$ and, by the Born rule, with a probability distribution $p(X|\theta) = \tr{M_X \rho_\theta }$. 
This pdf essentially plays the role of the likelihood that we have encountered in statistics and estimation theory. 
The empirical data and the theoretical distribution $p(X|\theta)$ can then be used to perform estimations about the parameter $\theta$, in the same spirit of point estimation theory and Bayesian inference. 

Of course, the discussion above shows that the quantum realm introduces an extra complication: we now have the freedom of selecting the measurement basis, or more generally, the POVMs. This means that the chosen POVM $M$ defines the probability distribution associated with the measurements. These distributions might each have different degrees of sensibility upon the parameter, meaning that one should carefully consider how to perform a measurement: some choices might result in a better distinguishability within the neighbourhood of states $\rho_\theta$.  Ideally, this suggests a maximization of the Fisher information over all possible measurements. The resulting quantity is known as the \emph{quantum Fisher information}
\begin{equation}\label{eq:QFIdef1}
\mathcal{F}(\theta)
:=
\max_{M}
F(\theta; M),
\end{equation}
where $F(\theta; M)$ is the FI associated to the distribution $p(X|\theta) = \tr{M_X \rho_\theta }$.

The definition above is very intuitive on its meaning. However, maximizing over \emph{operators} is, in general, not trivial at all. Fortunately, it is possible to arrive in a closed form expression for the optimal POVMs and for the QFI. For that, we closely follow Ref.~\cite{Paris2009}, which provides a rich discussion about quantum metrology, furnishing us with some useful operational results. Moreover, we also include a result from Ref.~\cite{Safranek2018}, which consists in a very efficient trick trick for calculating the QFI matrix based on the vectorization process (see the Appendix~\ref{appendix:vectorization}).

We start by defining an operator $\Lambda_\theta$ known as the \emph{Symmetric Logarithmic Derivative} (SLD), which is a self-adjoin operator satisfying a Lyapunov equation of the form
\footnote{For a more detailed derivation of the upcoming results and a better motivation of the definitions of the SLD, see Refs.~\cite{Personick1971} and \cite{Helstrom1976} (p.~266).}
\begin{equation}\label{eq:SLDdefinition}
\Lambda_\theta \rho_\theta +  \rho_\theta  \Lambda_\theta 
= 
2 \partial_\theta \rho_\theta.
\end{equation}
If we derive the expression for the probabilities for the Born rule with respect to $\theta$
\begin{equation}
\partial_\theta p(X|\theta) = \Tr{(\partial_\theta \rho_\theta) M_x},
\end{equation}
we can simply plug Eq.~\eqref{eq:SLDdefinition} in the equation above to find:
\begin{equation}\label{eq:FisherPOVMFirst}
F(\theta; M)
=
\int dx 
\frac{\Re(\Tr{\rho_\theta M_x \Lambda_\theta})^2}{\Tr{\rho_\theta M_x}}.
\end{equation}
Here we have used the cyclic property of the trace and the identity $\Tr{A_1A_2 ...A_n}=(\Tr{A_n...A_2A_1})^*$ for Hermitian operators $A_1, ..., A_n$ \cite{Helstrom1976} (p.~46). 
\footnote{To prove this, note that we can write 
$\Tr{A_1A_2 ...A_n}=\Tr{(A_n^\dagger ...A_2^\dagger A_1^\dagger)^\dagger} =\Tr{(A_n ...A_2 A_1)^\dagger}=(\Tr{A_n...A_2A_1})^*$, since the operators inside the trace are all hermitian. 
Naturally, $\rho_\theta$, $\Lambda_\theta$ and $M_x$ are all hermitian as well, so this property holds in the derivation of Eq.~\eqref{eq:FisherPOVM}.}
In the case of discrete outcomes, the integral above can be substituted by a summation sign.

Proving the quantum analogue of the CRB is very similar to our derivations of the standard CRB in Chapter~\ref{chp:frequentist} and to the derivation of the VTSB in Chapter~\ref{chp:bayes}, where we used the Cauchy-Schwarz inequality in both cases, based on the standard inner product for integrals. 
Here, we consider instead the Cauchy-Schwarz inequality for the Hilbert-Schmidt inner product (See the Appendix~\ref{appendix:vectorization} once again), which is given by $|\Tr{A^\dagger B}|^2 \leqslant \Tr{A^\dagger A} \Tr{B^\dagger B}$. 
Thus, our objective is to bound the expression for the "measurement-based" FI in Eq.~\eqref{eq:FisherPOVMFirst} by its maximum (over the space of POVMs). By doing so, we will be able to obtain an operational equation for the QFI, which we defined in Eq.~\eqref{eq:QFIdef1}. Algebraically, what we do is:
\begin{equation}\label{eq:QFIproof}
\begin{split}
F(\theta; M)
& \leqslant
\int dx 
\left|
\frac{\Tr{ \rho_\theta M_x \Lambda_\theta}}{\sqrt{\Tr{\rho_\theta M_x}}}
\right|^2\\
& =
\int dx
\left|
\Tr{
\frac{\sqrt{\rho_\theta} \sqrt{M_x}}{\sqrt{\Tr{\rho_\theta M_x}}}
\sqrt{M_x} \Lambda_\theta \sqrt{\rho_\theta}
}
\right|^2\\
& \leqslant
\int dx
\Tr{M_x \Lambda_\theta \rho_\theta \Lambda_\theta}\\
& =
\Tr{\rho_\theta \Lambda_\theta^2}.
\end{split}
\end{equation}
Note that we can bound the FI by the RHS in the first line simply because $\Re(\cdot)^2 \leq |\cdot|^2$. Furthermore, in the second inequality we chose $A^\dagger = \sqrt{\rho_\theta} \sqrt{M_x}/\sqrt{\Tr{\rho_\theta M_x}}$ and $B = \sqrt{M_x} \Lambda_\theta \sqrt{\rho_\theta}$. Note how $\Tr{A^\dagger A} = {\bf I}$, with these choices. Finally, in the last line we have used the "normalization" property of the POVM, that is, $\int M_x dx = {\bf I}$. The last expression in the equation above is precisely the QFI. In other words, we can write that the FI associated with the distribution of the measurements of $\rho_\theta$, with respect to the set of POVMs $\{ M_x \}$ is bounded by the quantum Fisher Information:
\begin{equation}\label{eq:QFIdef2}
F(\theta; M)
\leqslant
\mathcal{F}(\theta)
=
\max_{M}
F(\theta; M)
=
\Tr{\rho_\theta \Lambda_\theta^2}.
\end{equation}
If we use the QFI in conjuction with the CRB, we arrive at the quantum Cramér-Rao bound (QCRB):
\begin{equation}\label{eq:QCRB}
\Var{\hat{\theta}({\bf X}; M)}
\geqslant
\frac{1}{\mathcal{F}(\theta)}.
\end{equation}
In the case of $n$ independent measurements where no correlations are present, we can simply include a factor of $n$ in the denominator of the RHS of the equation above, in analogy with the usual CRB. We included the POVM $M$ in Eqs.~\ref{eq:FisherPOVMFirst} and \ref{eq:QCRB}  to make it explicit that the estimators (and the underlying pdf) \emph{do} depend on the POVMs. 

Finally, we comment on the attainability of the QFI, which translates into the choice of an optimal POVM (or measurement basis). 
The equality in Eq.~\eqref{eq:QFIdef2} is satisfied if we choose the POVMs as the set $\{\ket{\Lambda_\theta}_x\bra{\Lambda_\theta}_x\}$. That is, the optimal POVMs are projectors constructed from the eigenvectors of the SLD \cite{Paris2009}. Refer also to \cite{Liu2020} (p.25). To see this, note that if calculate the trace appearing in the QFI we have:
\begin{equation}
\Tr{\rho \Lambda_\theta^2} = \int \lambda_x^2 \bra{\Lambda_\theta}_x \rho_\theta \ket{\Lambda_\theta}_x dx,
\end{equation}
where we have used the spectral decomposition for the SLD, with $\lambda_x$ representing its eigenvalues. 
Analogously, if we take $M_x = \ket{\Lambda_\theta}_x\bra{\Lambda_\theta}_x$ then Eq.~\eqref{eq:FisherPOVM} also becomes:
\begin{equation}\label{eq:FisherPOVM}
F(\theta; \ket{\Lambda_\theta}_x\bra{\Lambda_\theta}_x)
=
\int dx 
\frac{\Re(\Tr{\rho_\theta \ket{\Lambda_\theta}_x\bra{\Lambda_\theta}_x \Lambda_\theta})^2}{\bra{\Lambda_\theta}_x \rho_\theta \ket{\Lambda_\theta}_x}
=
\int \lambda_x^2 \bra{\Lambda_\theta}_x \rho_\theta \ket{\Lambda_\theta}_x dx.
\end{equation}
where we use that $\lambda_x$ is real (because the SLD is hermitian) and that $\ket{\Lambda_\theta}_x\bra{\Lambda_\theta}_x \Lambda_\theta =  \lambda_x \ket{\Lambda_\theta}_x\bra{\Lambda_\theta}_x$. Showing that this choice of POVM indeed achieves the QFI.

We also provide a short summary here:
\begin{itemize}
	\item In order to calculate the QFI in Eq.~\eqref{eq:QFIdef2} the SLD must be known. It is obtained by solving Eq.~\eqref{eq:SLDdefinition} for $\Lambda_\theta$.
	\item Therefore, the QFI depends \emph{only} on the state $\rho_\theta$ and its parametrization. 
	\item Since the SLD depends on $\theta$, this means that the optimal POVM might also depend on $\theta$ as well, which is \emph{unkown}. We refer back to this point at the end of Sec.~\eqref{sec:trickQFI}. 
\end{itemize}
We treat the operational aspects of the QFI and the SLD in Sec.~\ref{sec:trickQFI}, where we show a simple way of obtaining the QFI and the optimal POVMs based on the technique of of Ref.~\cite{Safranek2018}. Older and conventional approaches are based on, e.g., the diagonalization of the density matrix \cite{Helstrom1976, Paris2009}.

The second point is also intuitive. The QFI essentially depends on the "statical manifold", or distinguishability of the states, around the parameter $\theta$. There is actually a deep connection between quantum estimation theory with geometrical aspects. See, for example, \cite{Braunstein1994} for an earlier work. A very thorough review on this topic can be found in Ref.~\cite{Sidhu2020}.
Moreover,  Ref.~\cite{Paris2009} also briefly discusses the connection of the QFI with the Bures metric. A few recent publications discuss some further aspects of these geometrical links \cite{xingMeasureDensityQuantum2020a, tsangPhysicsinspiredFormsBayesian2020}, including a very interesting relationship between the Berry curvature and the QFI \cite{Guo2016}. 
Note that this also explains why the notion of a Fisher information, or of a quantum version of a CRB, is slightly more complicated than the formulation from standard estimation theory. In the latter, we only consider probability distributions, which are much simpler than quantum states. The mathematical structure of quantum mechanics introduces non-commuting operators and also the idea of measurements, which makes the structure of the problem much richer and more complex. The reader may check any of the references above for a rigorous discussion on this topic.

\section{A trick for calculating the QFI}\label{sec:trickQFI}

In this section we discuss the procedure described in Ref.~\cite{Safranek2018}, which will aid us in obtaining the SLD from  Eq.~\eqref{eq:SLDdefinition} and the QFI from Eq.~\eqref{eq:QFIdef2}. This trick is based on the vectorization technique, which we discuss in Appendix~\ref{appendix:vectorization}. To start, note that we can express the first term on the LHS in Eq.~\eqref{eq:SLDdefinition} as a product of three matrices: ${\bf I} \Lambda_\theta \rho_\theta$, allowing us to rewrite it as $(\rho^T \otimes {\bf I})\choi{\Lambda_\theta}$ through vectorization. We can proceed analogously for the second term. Therefore, this equation simply becomes a linear system in a bigger Hilbert space:
\begin{equation}\label{eq:SLDtrick}
(\rho_\theta^T \otimes {\bf I} + {\bf I} \otimes \rho_\theta)\choi{\Lambda_\theta} 
= 
2 \choi{\partial_\theta \rho_\theta}.    
\end{equation}
The solution is straightforward with orthodox methods: 
\begin{equation}\label{eq:SLDtrick2}
\choi{\Lambda_\theta} 
= 
2 (\rho_\theta^T \otimes {\bf I} + {\bf I} \otimes \rho_\theta)^{-1}\choi{\partial_\theta \rho_\theta}.    
\end{equation}
This avoids the hassle of diagonalizing the density matrix: we convert the problem of obtaining the QFI and the SLD to a simple linear algebra problem based on a linear system. Obtaining the inverse matrix becomes the numerical bottleneck in this case.  Moreover, by using the identity $\tr{A^\dagger B} = \choi{A}^\dagger \choi{B}$, we can rewrite the QFI in terms of the vectorized expressions as
\begin{equation}\label{eq:QFItrick}
\mathcal{F}
=
\choi{\Lambda_\theta}^\dagger
(\rho_\theta^T \otimes {\bf I} + {\bf I} \otimes \rho_\theta)^{-1}
\choi{\Lambda_\theta}.
\end{equation}
We now discuss an example, adapted from the contents of Ref.~\cite{Safranek2018}, showing how the result above can be used in practice.

$\opposbishops$ \textbf{\textit{Example:}} Consider the pure state given by $\ket{\psi} = \frac{1}{\sqrt{2}}\ket{0} + \frac{e^{i \varphi}}{\sqrt{2}}\ket{1}$, where $\varphi$ is \emph{known} a phase. We will consider a scheme describing noise, which is quantified through a parameter $\nu \in (0,1)$ which we want to estiamate. The state of interest is a convex combination of the pure state $\ket{\psi}$ and the maximally mixed state:
\begin{equation}\label{eq:stateExampleQFI}
    \rho_{\nu \varphi}  = (1 - \nu)\ket{\psi}\bra{\psi} + \frac{\nu}{2}{\bf I}
    =
    \frac{1}{2}
    \begin{matrix}
    \begin{pmatrix}
        1 & e^{i\varphi}(1-\nu)\\
        e^{-i\varphi}(1-\nu) & 1
    \end{pmatrix}
    \end{matrix}.
\end{equation}
The mixed state $\rho_{\nu \varphi}$ describes, e.g., some noise deterioration due to dephasing or an experiment where a machine prepares the state $\ket{\psi}$ with a certain degree of imperfection. Our objective in this example is to obtain the QCRB for the parameters $\nu$ and $\varphi$, given the state $\rho_{\nu\varphi}$. Note that in this example we skipped the first two steps in the diagram show in Fig.~\eqref{fig:metrology_drawing}. We will consider the parametrized state above as given, ignoring the underlying input states and the parametrization dynamics. 

We begin by obtaining the QFI and the optimal POVMs for the noise parameter $\nu$. According to Eq.~\eqref{eq:SLDtrick}, we may start by vectorizing the expression
\[
`   \choi{\partial_\nu \rho_{\nu\varphi}} 
    =
    -\frac{1}{2}
    \begin{matrix}
    \begin{pmatrix}
        0 \\
        e^{-i\varphi} \\
        e^{i\varphi}\\
        0
    \end{pmatrix}
    \end{matrix},
\]
and the LHS of Eq.~\eqref{eq:SLDdefinition}
\[
(\rho_{\nu\varphi}^T \otimes I + I \otimes \rho_{\nu\varphi})
=
\begin{matrix}
\begin{pmatrix}
1&\frac{1}{2} e^{i \varphi } (1-\nu )&\frac{1}{2} e^{-i \varphi } (1-\nu )&0\\
\frac{1}{2} e^{-i \varphi } (1-\nu )&1&0&\frac{1}{2} e^{-i \varphi } (1-\nu )\\
\frac{1}{2} e^{i \varphi } (1-\nu )&0&1&\frac{1}{2} e^{i \varphi } (1-\nu )\\
0&\frac{1}{2} e^{i \varphi } (1-\nu )&\frac{1}{2} e^{-i \varphi } (1-\nu )&1
\end{pmatrix}
\end{matrix},
\]
respectively. By inverting the matrix above we obtain the following expression for the QFI:
\begin{equation}
\mathcal{F}_\nu
=
2 
\text{vec}(\partial_\nu \rho_{\nu\varphi})^\dagger
(\rho_{\nu\varphi}^T \otimes I + I \otimes \rho_{\nu\varphi})^{-1}
\text{vec}(\partial_\nu \rho_{\nu\varphi})
=
\frac{1}{\nu(2-\nu)}.
\end{equation}
Furthermore, we can now use Eq.~\eqref{eq:SLDtrick2} to explicitly obtain the SLD
\begin{equation}
\Lambda_\nu 
=
\frac{1}{\nu(2-\nu)}
\begin{matrix}
\begin{pmatrix}
1-\nu&e^{i \varphi}\\
e^{-i \varphi}&1-\nu
\end{pmatrix}
\end{matrix}.
\end{equation}
By calculating its eigenvectors we can finally arrive at the optimal measurement basis:
\begin{equation}
\ket{B_{\nu,1}}
=
\frac{1}{\sqrt{2}}
\begin{matrix}
\begin{pmatrix}
e^{i\varphi} \\
1
\end{pmatrix}
\end{matrix},
\quad
\ket{B_{\nu,2}}
=
\frac{1}{\sqrt{2}}
\begin{matrix}
\begin{pmatrix}
-e^{i\varphi}\\
1
\end{pmatrix}
\end{matrix}.
\end{equation}
\begin{figure}[h!]
	\centering
	\includegraphics[width=0.45\textwidth]{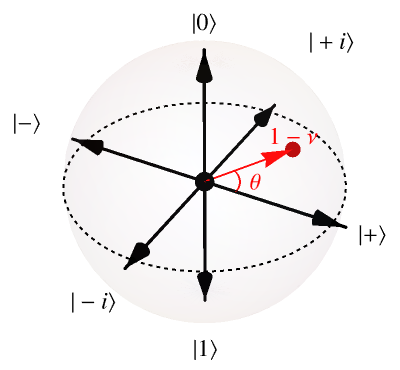}
	\includegraphics[width=0.45\textwidth]{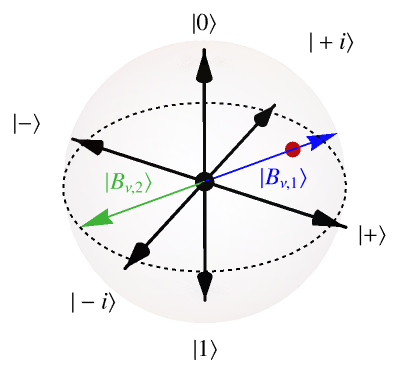}
	\caption[(Left) State $\rho_{\nu\varphi}$ and (Right) the optimal measurement basis on the Bloch sphere.]{(Left) State $\rho_{\nu\varphi}$ and (Right) the optimal measurement basis on the Bloch sphere.}
	\label{fig:BlochQFI}
\end{figure}

The POVMs are then constructed simply by taking the projectors $\ketbra{B_{\nu,1}}{B_{\nu,1}}$ and $\ketbra{B_{\nu,2}}{B_{\nu,2}}$. In Fig.~\ref{fig:BlochQFI} we illustrate the state and the basis on the Bloch sphere. 
Note how $\ket{B_{\nu, 1}}$ is in the same direction as the state. By using this measurement basis, we obtain the distributions
\begin{equation}\label{eq:distributionsexampleQFI}
p(X = B_{\nu, 1}|\nu) = 1-\frac{\nu}{2} \quad \text{and} \quad p(X = B_2|\nu) =\frac{\nu}{2}
\end{equation}
for the outcomes. 
\footnote{Here we commit a slight abuse of notation. 
By writing $p(X = B_i|\nu)$ we refer to the probability of measuring the system in the state $\ket{B_i}$, conditioned on the parameter $\nu$.}
Note that measurements with this POVM are analogous to a coin toss experiment with  $\nu/2$ as the parameter in the Bernoulli distribution.
In the ideal scenario where $\nu = 0$, we would \emph{always} measure the system in the state $\ket{B_{\nu, 1}}$, but \emph{never} in $\ket{B_{\nu,2}}$, since the pure state $\ket{\psi}$ defined at the beginning of this example is orthogonal to $\ket{B_{\nu,2}}$. 
On the other hand, in a configuration where $\nu$ is non-zero, we would sometimes measure the state in $\ket{B_2}$. This happens not because of a superposition in the $\{\ket{B_{\nu, 1}}, \ket{B_{\nu,2}}\}$ basis, but rather, because of the classical probabilities associated with the mixture in $\rho$. 
Choosing a basis other than ${\ket{B_{\nu, 1}}, \ket{B_{\nu, 2}}}$ would make it harder to decide whether the results arise from the superposition in the state $\ket{\psi}$ or because of the impurity of $\rho$. 

Finally, note that the QFI imparts the ultimate limit of precision associated to the estimation of the parameter of concern, while the eigenvectors of the SLD provide the optimal basis which is associated with the saturation of this bound. 
Nevertheless, they give no information about how the measurement data should be used. It is then clear that one should also formulate a concrete post-processing strategy. In this example, we can estimate $\nu$ by counting the number of different outcomes in each state, in a large analogy with the scenario of Bernoulli trials. 
Thus, the obvious and best candidate for an estimator in this case is the sample mean, which we discussed in previous chapters. 
The FI associated with the distributions in Eq.~\eqref{eq:distributionsexampleQFI} is simply the expression that we have obtained in Eq.~\eqref{eq:bernoulliFisher}, but with respect to the parameter $\nu/2$ instead. 
You can see that the FI in this case will \emph{coincide} with the QFI, as it should be, since we are purposely performing measurements in the optimal basis. Finally, note that due to the properties of the sample mean as an estimator for the Bernoulli distribution, we actually saturate the bound even in the non-asymptotic regime. This estimator is \emph{efficient}, as we have discussed before in Sec.~\ref{sec:CRB}. These steps give a complete picture of how the noise parameter could be estimated in this toy model.

There are a few further things worth mentioning in this example. The first one is that the QFI, as it very often happens, depends on the unknown parameter $\nu$. 
Fortunately, that was not the case for the optimal POVM here, which depends only on $\varphi$. That, however, is not guaranteed to happen every time. 
If we decided to probe the phase $\varphi$ instead of the parameter $\nu$, for instance, the optimal POVMs would also depend on the unkown $\varphi$ \cite{Safranek2018}. 
\footnote{See, e.g., Ref.~\cite{Chapeau-Blondeau2015} (p.~3) and the discussion therein.} 
We can repeat the same steps with respect to the parameter $\varphi$ instead to see this in practice. By doing so, we find the SLD and the optimal measurement basis to be
\begin{equation}
\Lambda_\varphi 
=
i(\nu - 1)
\begin{matrix}
\begin{pmatrix}
0 & -e^{i \varphi}\\
e^{-i \varphi}&0 
\end{pmatrix}
\end{matrix}
\end{equation}
and
\begin{equation}
\ket{B_{\varphi,1}}
=
\frac{1}{\sqrt{2}}
\begin{matrix}
\begin{pmatrix}
i e^{-i\varphi} \\
1
\end{pmatrix}
\end{matrix},
\quad
\ket{B_{\varphi,2}}
=
\frac{1}{\sqrt{2}}
\begin{matrix}
\begin{pmatrix}
-i e^{-i\varphi}\\
1
\end{pmatrix}
\end{matrix},
\end{equation}
respectively. Note how the latter explicitly depends on $\varphi$. This example illustrates that it is not uncommon for the optimal basis to depend on the unknown parameter itself, introducing another layer of complexity to the problem. 
This often calls for adaptative strategies whenever possible, where one continuously change the measurement basis, or, e.g., parameters and the general configuration of the interaction \cite{Demkowicz-Dobrzanski2017, Rodriguez-Garcia2021, Wan2022}. $\bishoppair$

\begin{figure}[h!]
	\centering
	\includegraphics[width=\textwidth]{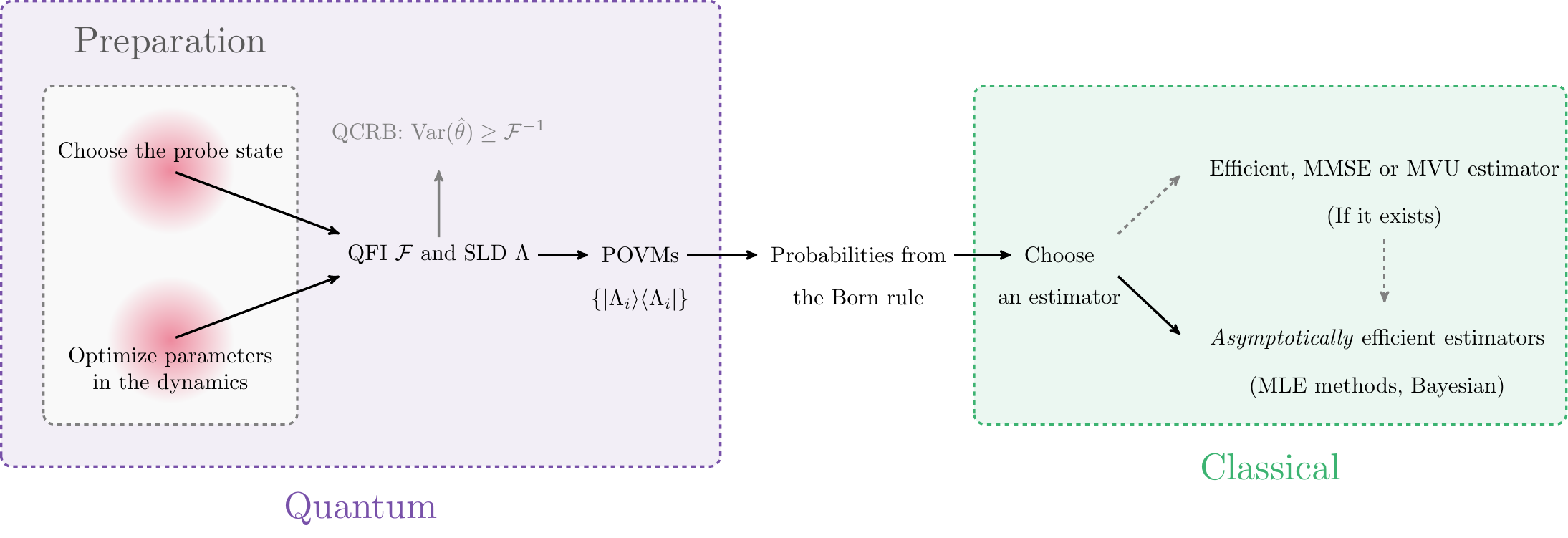}
	\caption[Standard optimization strategies in metrology.]{Standard optimization strategies in metrology.}
	\label{fig:QFIstrategy}
\end{figure}

The previous example briefly showed the typical steps which constitute a metrology experiment in quantum mechanics. One of most striking aspects is the importance of the measurement choice in this scheme. 
We illustrate the main ideas with a diagram in Fig.~\ref{fig:QFIstrategy}. 
A very common point of focus in the literature is the notion of optimization over \emph{the probe states} and quantities appearing in the parametrization process, such as parameters in a unitary evolution. This was not a point of concern in the previous example because we considered a given state. In more concrete scenarios these might be central points of investigation.
The experimenter will typically try to choose a probe state which better encodes the parameters of interest after undergoing a predefined dynamics. Likewise, it is also important to perform this process in the best configuration possible. 
If some parameters of the interaction are known, they might be tuned in a way which results in a state with a larger QFI. 
For that reason, many works are devoted to establishing experimental setups and protocols with a sensible configuration which allow for very precise estimations. This is, in a first instance, quantified through the QFI, further justifying its importance. 

Once the parametrized state has been obtained, it is then possible to calculate the QFI and the SLD. The latter is, in turn, used to obtain the optimal POVMs. 
More specifically, the projectors are constructed from the eigenvalues of the SLD,  providing the best possible measurements, whose precision is bounded by the QCRB. 
These are quantities which are obtained entirely from the state, and no other elements are necessary in principle. An important point in this step, which has actually been omitted in the diagram, is that a few other strategies might result in very sharp improvements. 
One can, for example, exploit superposition and entanglement, performing joint measurements on several copies of the system or ancillas. 
By doing so it is often possible to capture correlations which result in more precise estimations \cite{modiQuantumCorrelationsMixedState2011}. Furthermore, protocols which employ other strategies or (quantum) resources, such as quantum discord, also configure a very active field in quantum metrology \cite{bromleyThereMoreQuantum2017, khalidMeasurementbasedQuantumCorrelation2018}. See Ref.~\cite{Braun2018} for a review.

Regardless, once the POVMs have been decided upon, the estimation task essentially becomes a classical problem. At this point, the state and the POVM are the relevant objects, and we can use the Born rule to obtain the probability distribution for each of the outcomes. 
By doing so, the formalism that we have discussed in the previous chapters can be used as a powerful tool in the post-processing tasks, where use the resulting distributions and the empirical data to estimate the parameters of interest. 
This is the point where it is necessary to pick a sensible estimator following the considerations that we have thoroughly discussed. Additionally, it is also possible to calculate the variance, or any figure of error which might be relevant, associated with the estimator. 

These steps are illustrated on the green region highlighted on the diagram in Fig.~\ref{fig:QFIstrategy}. It is then  possible to look for efficient estimators, or even MMSE and MVU estimators. 
This should provide an optimal post-processing treatment. If these do not exist, it is then common to resort to estimators which are \emph{asymptotically} efficient, such as the ones obtained by MLE methods. 
Still, it is also important to remember that even MLE strategies might contain some limitations. This is a hurdle encountered in, e.g., phase estimation \cite{Rodriguez-Garcia2021}. These considerations also highlight the importance of the FI and the QFI: even if we are either unable to obtain an optimal estimator (or to employ the best measurement strategy), the (quantum) Fisher information still provides an important benchmark.

For a good application of this discussion in practice, the reader might refer to the Section III from Ref.~\cite{Chapeau-Blondeau2016} for example, where the problem of the qubit phase estimation is investigated. This publications provide a very good illustration of each of these steps that we have discussed. There, the author discusses the optimization of the QFI, the measurements and the estimators, in this order. An optimization over the probe states is given in a previous work by the same author, in Ref.~\cite{Chapeau-Blondeau2015}. Note how these contents overlap with the diagram shown in Fig.~\ref{fig:QFIstrategy}.

To complete our discussion, we dwell into question of the attainability of the QCRB.
In the single-parameter case the QCRB can usually be achieved, at least asymptotically, by employing measurements in the appropriate basis per our discussions. Additionaly, adaptative protocols might have to be employed. \footnote{See Ref.~\cite{Barndorff-Nielsen2000} for a discussion in the case of pure states.} 
The situation becomes much more involved when the multi-parameter case is considered: since different parameters will, in general, require a different set of POVMs for optimality, one can seldom saturate the multi-parameter QCRB. 
Surprisingly, the geometrical aspects of the QFI come at hand in this discussion: one can establish a link between the simultaneous indistinguishably of different parameters and the impossibility of attaining the multi-parameter QCRB with the non-zero Berry curvature and the geometrical properties of the quantum statistical model. 
These very elegant conclusions, which cleverly connect geometric and statistical aspects, can be found in Refs.~\cite{Guo2016} and \cite{Li2022} (see also \cite{Yu2022} for an experimental follow-up). On the same spirit, other results provide useful insights about this problem: the QCRB is saturable if, and only if, the state $\rho$ belongs to what is called the \emph{quantum exponential family} \cite{Liu2020} (p.~25). 
We are not going to define this class of states here, but the statement of the theorem can be found in Ref.~\cite{Hayashi2005} (pp.~118-120). 
This result depicts one of the  differences between classical and quantum estimation theories quite well: in the former the CRB is saturated through the exponential family of \emph{distributions}, while in the latter we consider a family of \emph{states}. 
This illustrates how distributions are the fundamental objects in standard statistics, while it makes more sense to consider the manifold of states to play the analogous role in quantum metrology.

Finally, we shall conclude this section with a practical example. We begin by mentioning Ref.~\cite{Teklu2009}, where the authors provide some optimization over single-qubit probes for one-parameter qubit gates. 
Moreover, they also study the \emph{stability} of these states as probes; in the sense that they investigate how small perturbations in the probe states affect the FI of the model. 
By doing so they were able to show that these smalls perturbations might introduce large deviations in the FI, making the protocol unstable. Surprinsingly, one of their main conclusions was that using entangled two-qubit probes in the Bell states actually \emph{protects} the protocol against fluctuations in the probe state. This is a concrete example in the literature which shows how the preparation process, and choosing good probe states, is important in metrology. 
An experimental approach of a very similar problem was done in~\cite{Brivio2010}. Therefore, we introduce an example below which adapts the discussion from Refs.~\cite{Barndorff-Nielsen2000, Teklu2009, Brivio2010} in a simpler scenario.
See also Refs.~\cite{wasakOptimalMeasurementsPhase2016, schmiedQuantumStateTomography2016} for very pedagogical approaches to the problem.

$\opposbishops$ \textbf{\textit{Example:}} Consider the initial state $\ket{\psi_0} = \cos{\frac{\alpha}{2}} \ket{0} + e^{i \varphi}\sin{\frac{\alpha}{2}} \ket{1}$, parametrized in terms of the polar angle $\alpha$ in the Bloch sphere. 
Now, let us suppose that this state undergoes a unitary dynamics of the type $U(\theta) = \exp{-i \frac{\theta}{2} \sigma_z}$, resulting in the state $\ket{\psi_\varphi} = \cos{\frac{\alpha}{2}} \ket{0} + e^{i (\varphi - \theta)}\sin{\frac{\alpha}{2}} \ket{1}$ (up to a phase factor) which encodes a phase-shift angle $\varphi$ that we wish to estimate.  
Note that we are able to generate the pure state in Eq.~\eqref{eq:stateExampleQFI} from the previous example through the same process. Our objectives with this example are two-fold:

\begin{itemize}
	\item We want to see how the angle $\theta$ in the probe states influences the QFI and the precision of the estimation.
	\item  We extend the investigation to measurement bases beyond the optimal protocol. We would like to quantify how sub-optimal measurements fare.
\end{itemize}

We start by calculating the QFI. By repeating the steps of the previous example, we find that the term $(\rho_\theta^T \otimes I + I \otimes \rho_\theta)$ is given by
\[
\begin{split}
\left(
\begin{array}{cccc}
 \cos ^2\left(\frac{\alpha }{2}\right) & \frac{1}{4} \sin (\alpha ) e^{i (\theta -\varphi )} & \frac{1}{4} \sin (\alpha ) e^{-i (\theta -\varphi )} & 0 \\
 \frac{1}{4} \sin (\alpha ) e^{-i (\theta -\varphi )} & \frac{1}{2} & 0 & \frac{1}{4} \sin (\alpha ) e^{-i (\theta -\varphi )} \\
 \frac{1}{4} \sin (\alpha ) e^{i (\theta -\varphi )} & 0 & \frac{1}{2} & \frac{1}{4} \sin (\alpha ) e^{i (\theta -\varphi )} \\
 0 & \frac{1}{4} \sin (\alpha ) e^{i (\theta -\varphi )} & \frac{1}{4} \sin (\alpha ) e^{-i (\theta -\varphi )} & \sin ^2\left(\frac{\alpha }{2}\right) \\
\end{array}
\right),
\end{split}
\]
for $\rho_\theta = \ketbra{\psi_\theta}{\psi_\theta}$. This matrix is singular, so it cannot be used to calculate the QFI in this case. We avoid this conundrum by employing one of the strategies from Ref.~\cite{safranekDiscontinuitiesQuantumFisher2017} (See Theorem 4, p.~5).  Consider the state $\tilde{\rho}_\nu := (1 - \nu)\tilde{\rho} + \frac{\nu}{2}I$, where $0 < \nu < 1$ is a real parameter. Note that this is the same state from the previous exercise. This time, however, we are not interested on its physical features, but rather, on its mathematical convenience. We can invert the previous term by taking the limit:

\begin{equation}
(\rho_\theta^T \otimes I + I \otimes \rho_\theta)
=
\lim_{\nu \rightarrow 0}
(\tilde{\rho}_\nu^T \otimes I + I \otimes \tilde{\rho}_\nu).
\end{equation}

That is, we calculate the QFI for a mixed state and then obtain the limiting case of pure states by taking $\nu \rightarrow 0$ \cite{Safranek2018}. 
This convenient technique avoids the problem of singular matrices.\footnote{In more generality, the result from Ref.~\cite{safranekDiscontinuitiesQuantumFisher2017} states that by defining the state $\tilde{\rho}_\nu := (1 - \nu)\tilde{\rho} + \frac{\nu}{\dim \mathcal{H}} I$, where $\mathcal{H}$ is the finite-dimensional underlying Hilbert space, one can write the QFI of the state $\tilde{\rho}$ as $\mathcal{F}(\rho) = \lim_{\nu \rightarrow 0} \mathcal{F}(\tilde{\rho}_\nu)$. This strategies is also used to deal with discontinuities.} 
Hence, by employing very similar steps to the previous exercise, we obtain that the QFI of the state $\rho_\theta$ is given by:
\begin{equation}\label{eq:QFIqubit}
	\mathcal{F}(\rho_\theta) = \sin^2(\alpha).
\end{equation}

This is a well-known result for qubits \cite{Barndorff-Nielsen2000} (p.~4486) \footnote{Since we are dealing with pure states, simple formulas which do \emph{not} depend on vectorization can be found in any standard reference \cite{Paris2009, Liu2020}. We insist in these formulas for conciseness, because these are the only ones we are going to use in future sections.}, with some intuition behind it. 
Notice how Eq.~\eqref{eq:QFIqubit} vanishes for $\alpha = 0$ or $\alpha = \pi$, which corresponds to the states $\ket{0}$ and $\ket{1}$, respectively, and how it achieves it maximum for $\alpha = \pi/2$, corresponding to states in the equator of the Bloch sphere, such as the $\ket{+}$ and $\ket{-}$ states, defined by $(\ket{0} \pm \ket{-})/\sqrt{2}$, respectively. 

\begin{wrapfigure}{r}{0.45\textwidth}
	\includegraphics[width=0.45\textwidth]{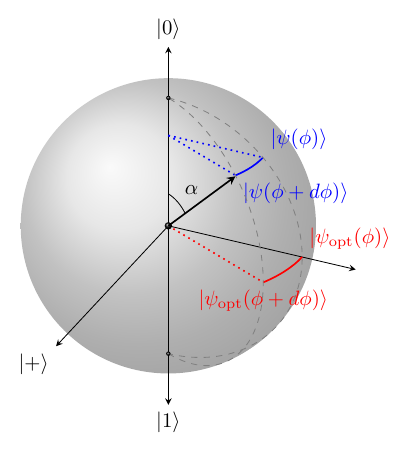}
	\caption[Illustration of qubit states in the Bloch sphere after an infinitesimal phase shift.]{Illustration of qubit states in the Bloch sphere after an infinitesimal phase shift. Here $\ket{\psi_{\mathrm{opt}}}$ illustrates an optimal probe state lying in the equator.}
	\label{fig:minimumvariance}
\end{wrapfigure} 
The reason for that is very simple: it is very hard to distinguish the states around the pole by changing the phase-shift $\theta$. 
In the limiting case where the state is situated in one of the poles of the Bloch sphere, applying the unitary $U(\theta)$ does not affect the state of the qubit at all, making it impossible to infer the parameter. 
Conversely, we find that distinguishing states on the equator of the Bloch sphere after applying the phase shift is the easiest: infinitesimal displacements of the azymuthal angle will be the most detectable in this region. One might notice how the QFI in Eq.~\eqref{eq:QFIqubit} coincides with one of elements of the metric tensor of the unit sphere.

Let us proceed by considering the measurements. This time, we will go beyond the optimal measurements. We shall, instead, consider a wider class of POVMs (which are not necessarily constructed from the SLD):
\begin{equation}
M_0 = \ketbra{\Psi}{\Psi}, \quad 
M_1 = I - M_0, \quad 
\text{with} \quad
\ket{\Psi}
=
\cos{\frac{\beta}{2}} \ket{0} + e^{i \phi}\sin{\frac{\beta}{2}} \ket{1}.
\end{equation} 
Here $\beta$ and $\phi$ are angles which parametrize the measurement basis. Our objective is to see how the angles $\alpha$ and $\varphi$ in the probe states and the angles $\beta$ and $\phi$ parametrizing the POVMs affect the protocol. Namely, we will be able to quantify how well different probe states and POVMs perform. This is quantified through the FI. By using the projectors above, we obtain the probabilities
\begin{equation}\label{eq:probPhaseShift}
p_{0,1}(\theta) = \frac{1}{2} \pm \frac{f(\theta)}{2}, \quad \text{with} \quad
f(\theta):=\cos (\alpha ) \cos (\beta ) - \sin (\alpha ) \sin (\beta ) \cos (\theta -\varphi +\phi ),
\end{equation} 
associated with $M_0$ and $M_1$, respectively. If we obtain $k$ outcomes in $0$ and $n-k$ outcomes in $1$, we can write the distribution for the outcomes of this experiments as as $p(X|\theta) = p_0(\theta)^X (1-p_0(\theta))^{1-X}$, allowing us to explicitly calculate its FI. 
We should be careful with this distribution however. While it closely resembles the Bernoulli distribution, the "Bernoulli parameter" actually depends on $\theta$ in a complicated manner, per Eq.~\eqref{eq:probPhaseShift}. 
The simplest way to calculate the FI in this case, since we have already done that for the Bernoulli distribution, is to use the reparametrization rule for the Fisher Information \cite{Ly2017}.
\footnote{Suppose that a parameter $\eta$ depends on another parameter of interest $\theta$. The rule states that $F(\eta) = \left(\frac{d\eta}{d\theta}\right)^2 F(\theta)$.} By doing so we can calculate the FI to be:
\begin{equation}\label{eq:FIqubitfull}
F(\theta|\alpha, \beta)
=
\left(\frac{d p_0(\theta)}{d\theta}\right)^2
\frac{1}{p_0(\theta)(1-p_0(\theta))}
=
\frac{\sin ^2(\alpha ) \sin ^2(\beta ) \sin ^2(\theta -\varphi +\phi )}{
(1-f(\theta)^2) 
}.
\end{equation}

\begin{figure}[t!]
	\centering
	\includegraphics[width=\textwidth]{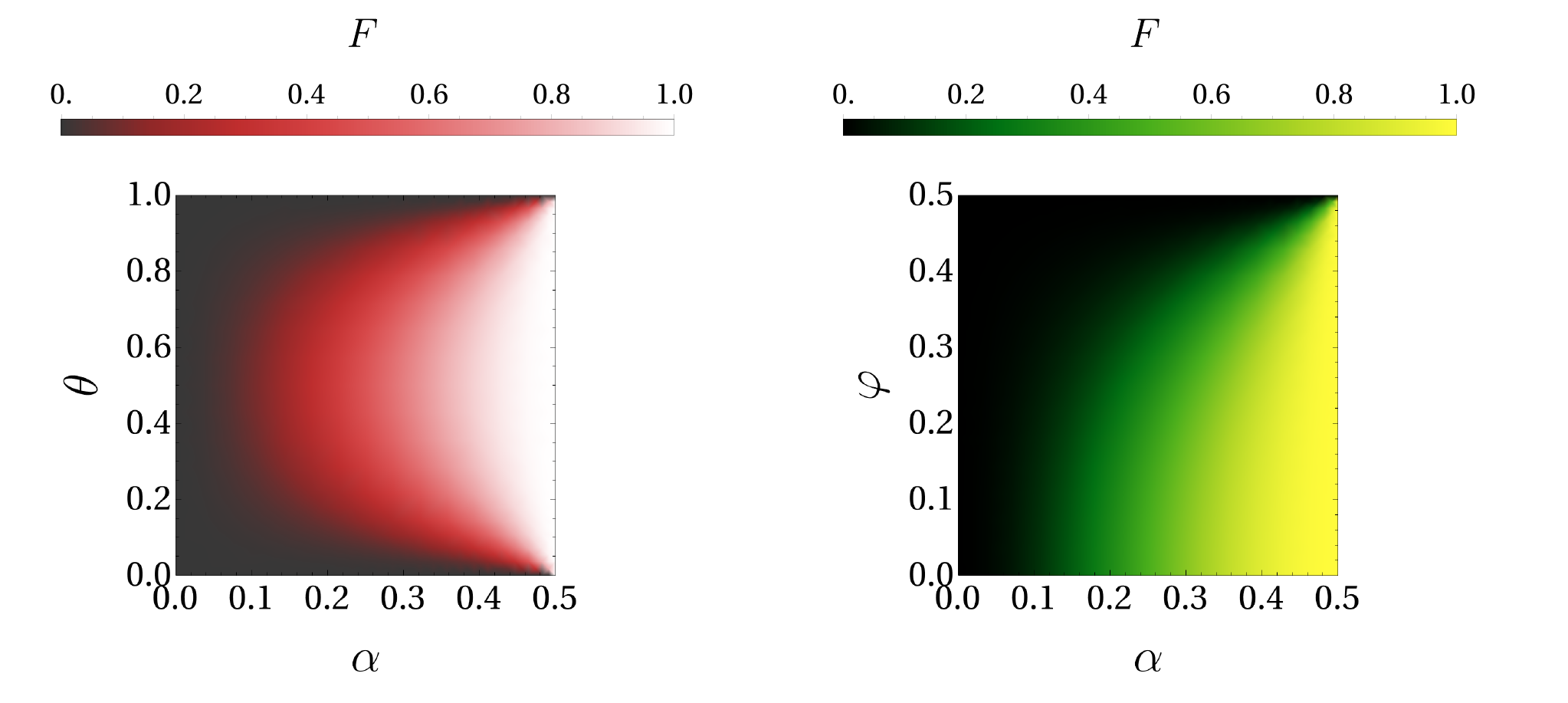}
	\caption[FI for the qubit in different scenarios.]{FI for the qubit in different scenarios. On the left we plot the FI as a function of the polar angle $\alpha$ for the qubit state on the Bloch sphere and the interferometric parameter $\theta$. 
On the right we plot the QFI as a function of $\alpha$ and $\varphi$, describing all the possible probe states. We can see that the optimality is achieved when $\alpha = \pi/2$, regardless of the azymuthal angle $\varphi$ in the Bloch sphere.}
	\label{fig:FIQubitStability}
\end{figure}

We can optimize the expression above by taking $\alpha = \beta = \pi/2$. 
This means that one should take a probe state and a measurement basis which lie on the equator of the Bloch sphere. By doing so, we obtain $F(\theta) = 1 $ for all $\theta$, independent of $\phi$ and $\varphi$. 
The FI in this case coincides with the maximum achievable QFI, with $\alpha = \pi/2$.
\footnote{However, we briefly mention that mixed states are also considered, the optimal configuration is much different \cite{wasakOptimalMeasurementsPhase2016}.} We plot these results in Fig.~\ref{fig:FIQubitStability}.

These calculations by themselves are interesting enough as an example. However, in Ref.~\cite{Teklu2009} they bring a very interesting point into attention, which concerns the \emph{stability} of the probe states: that is, how much do small fluctuations in the parameters of the probe state and of the POVMs influence the optimality of the protocol? 
We can follow Ref.~\cite{Teklu2009} and Taylor expand Eq.~\eqref{eq:FIqubitfull} around $\alpha=\pi/2$ to obtain (we consider that $\varphi = \phi = 0$, for simplicity):
\begin{equation}
F(\theta) 
\approx 
1-\delta \alpha^2 \frac{1}{\sin^2 \theta},
\end{equation}
with $\delta \alpha = \pi/2 - \alpha$. Surprisingly, for small $\theta$ we can see that this becomes $F(\theta) \approx \delta \alpha^2/\theta^2$. In other words, small deviations in the preparation of the probe \emph{greatly} affect the achievable precision of the protocol, specially when small angles are considered. 

This example showcases why the \emph{preparation} step is very subtle when devising experiments in quantum metrology, and why one should also consider other aspects beyond only maximizing the QFI. 
The striking conclusion from the authors in Ref.~\cite{Teklu2009} was to show that entangled probes are able to improve \emph{the stability} of the protocol, making it optimal for a wider range of configurations and less sensitive to small perturbations in the preparation of the probe. This is an interesting example to show how aspects such as quantum improvements might be useful even beyond the scenario of surpassing the standard quantum limit. $\bishoppair$

We finish this section with a brief disclaimer about some caveats which arise in quantum metrology. The previous discussion, for instance, very interestingly  showed how entanglement could be used to improve the \emph{stability} of a protocol. 
And while we ommited such an example here, an even larger body of the literature is concerned with the challenge of surpassing the SQL \cite{micadeiCoherentMeasurementsQuantum2015}; as we mentioned in the beggining of this chapter. 
However, employing quantum adtanvantages is in practice, a difficult task. For instance, the choice of the estimator in phase estimation problems requires some care 
\cite{belliardoAchievingHeisenbergScaling2020a}. Additionaly, it is very well known at this point
that schemes based on the Heisenberg scaling are typically very sensitive to noise \cite{pezzePhaseDetectionQuantum2007, albarelliRestoringHeisenbergScaling2018, leibfriedHeisenbergLimitedSpectroscopyMultiparticle2004}.
In Ref.~\cite{escherGeneralFrameworkEstimating2011} the authors provide some further bounds and concretely shows how estimations in the presence of noise might gradually transit from the Heisenberg scaling to the usual scaling in the SQL.
Maybe even more importantly, the question of what counts as a "resource" for quantum metrology and the QCRB is certainly a fundamental question which we should dwell into. For instance, Ref.~\cite{zwierzGeneralOptimalityHeisenberg2010} provides a unifying discussion of these aspects, establishing a connection between several differents experiments and models. 
In summary, these observations, allied to the aforementioned references, allows us to better understand the current literature along with its challenges, and show us that many subtleties, such as technological feasibility, robustness and post-processing aspects should be taken into account 
when devising a sensible experiment in quantum metrology. 

%
%
%
%
%

\section{Quantum thermometry}\label{sec:thermometry}

Measuring and estimating temperature with high precision in nanoscale and quantum systems is an effort of increasingly importance in modern technologies \cite{brudererProbingBECPhase2006, mckayCoolingStronglyCorrelated2011, sabinImpuritiesQuantumThermometer2015, fujiwaraDiamondQuantumThermometry2021}. 
We can find uses in, e.g., ultra-cold gases \cite{marzolinoPrecisionMeasurementsTemperature2013a, mehboudiUsingPolaronsSubnK2019, boutonSingleAtomQuantumProbes2020,  khanSubnanokelvinThermometryInteracting2022}, biology \cite{kucskoNanometrescaleThermometryLiving2013} and even relativity \cite{ranganijahromiRelativisticQuantumThermometry2023}, with proposals for the detection of the Unruh temperature \cite{mannQuantumThermometry2014} and the thermometry of gravitational waves \cite{pereiradesanetoTemperatureEstimationGravitationalWave2022}. 
Lying in a very rich intersection between quantum metrology and quantum thermodynamics, the field concerned with such problem is known as quantum thermometry, and it has seen a  fast development in the past decade
\cite{depasqualeEstimatingTemperatureSequential2017, hovhannisyanMeasuringTemperatureCold2018, mukherjeeEnhancedPrecisionBound2019, mitchisonSituThermometryCold2020, salado-mejiaSpectroscopyCriticalQuantum2021, planellaBathInducedCorrelationsEnhance2022}.
For a recent collection of important developments and works in the field, consult Refs.~\cite{binderThermodynamicsQuantumRegime2018} (Part~III, Sec.~3) and~\cite{mehboudiThermometryQuantumRegime2019}.
\footnote{For a review in quantum thermodynamics instead, see Ref.~\cite{deffnerQuantumThermodynamicsIntroduction2019}.
}
Rather than \emph{solely} transposing the tools from quantum estimation theory, quantum thermometry emerges as a very careful framework which incorporates the physical and mathematical traits from thermodynamics into the underlying framework of quantum metrology. Due to its own peculiarities, quantum thermometry comes across theoretical and experimental challenges which are particular of it as its own field - some challenges which are not generally present in quantum metrology and other subfields. 
With that in mind, we use the discussions of the preceding chapters as the foundation of the upcoming results.  
\begin{figure}[b!]
	\centering
	\includegraphics[width=\textwidth]{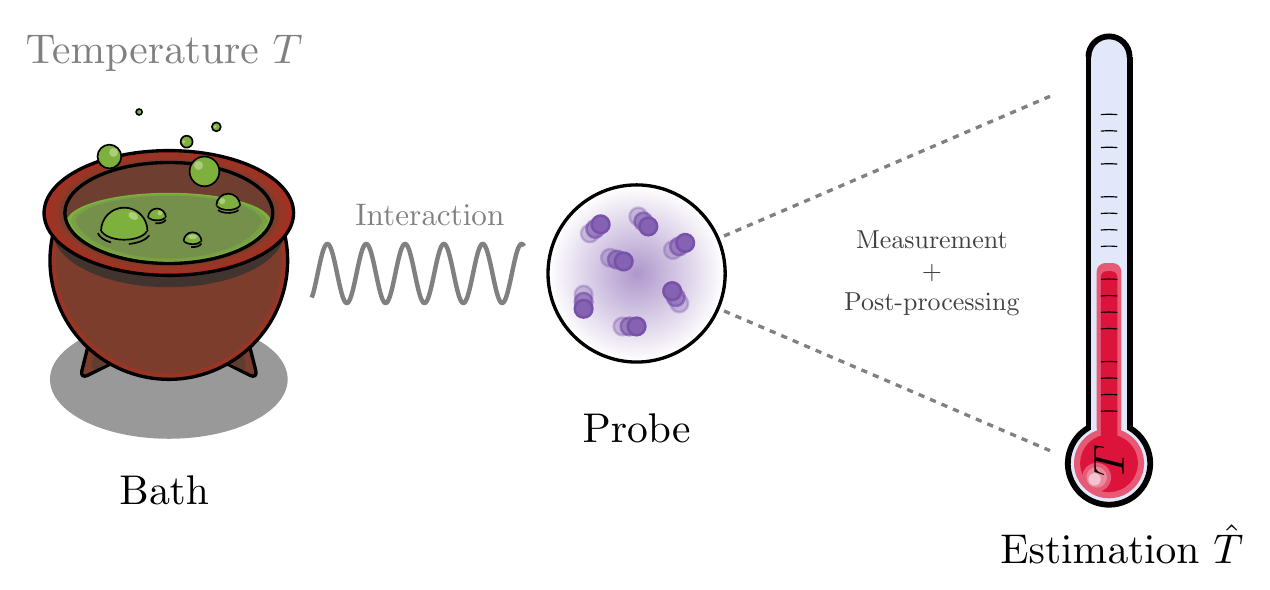}
	\caption[Diagram illustrating a thermometry experiment.]{Diagram illustrating a thermometry experiment.}
	\label{fig:thermometry}
\end{figure}

One of the main building blocks in the field is the notion of a \emph{thermal} quantum Fisher information, which can be used as the main benchmark for thermometric protocols -  specially for near-equilibrium configurations \cite{jevticSinglequbitThermometry2015}.
The standard setup is illustrated in Fig.~\ref{fig:thermometry}. It consists on a probe
which is left to interact with a thermal bath at a temperature $T$. The probe eventually thermalizes to the Gibbs state
\begin{equation}\label{eq:ThermalStateDef}
\rho_{th} := \frac{e^{-\beta H}}{Z},
\end{equation}
associated with the Hamiltonian $H$, where we define $\beta = 1/T$ as the reciprocal of the temperature (from hereafter we take the Boltzmann constant to be $k_b = 1$). 
Hence, by performing measurements and post-processing strategies (per Fig.~\ref{fig:QFIstrategy}, for example) on the thermalized probe, we can infer the temperature $T$ of the bath. 
All things considered, this is very similar to previous setups we have analyzed, with the complication of an extra step where the probe acts as a proxy for measuring the bath temperature. Modeling this interaction might be a big challenge on itself. We omit such a discussion here, but an example can be found in Ref.~\cite{correaEnhancementLowtemperatureThermometry2017a}.

More concretely, the thermal Fisher information (TFI) is defined as the quantum fisher information of the thermal state, and it is given by
\begin{equation}\label{eq:TFI}
\mathcal{F}_{\mathrm{th}} = \frac{C}{T^2},
\end{equation}
where 
\begin{equation}\label{eq:heatcapacity}
C = \frac{\expec{H^2} - \expec{H}^2}{T^2}
\end{equation}
is the heat capacity of the system. To arrive at this result we can write the Lyapunov equation for the thermal state
\begin{equation}\label{thermalstatelyapunov}
\Lambda_T \rho_{th} + \rho_{th} \Lambda_T = 2 \partial_{T} \rho_{th}, 
\end{equation}
and solve it for the "thermal" SLD $\Lambda_T$. We briefly review some results from statistical physics in order to proceed. In particular, we will need to use some common results in order to obtain the derivative of the density matrix in Eq.~\eqref{eq:ThermalStateDef}. 
First, notice that it can be written in a more convenient way through the use of the chain rule:
\[
\partial_T \rho_{th}= -\beta^2 \ppx{\rho_{\mathrm{th}}}{\beta} = -\beta^2 \left[ - H \frac{e^{-\beta H}}{Z} - \frac{e^{-\beta H}}{Z} \left( \frac{1}{Z}\ppx{Z}{\beta} \right) \right],
\]
or equivalently,
\begin{equation}\label{thermalstatederivative}
\partial_T \rho_{\mathrm{th}}=  \beta^2 \left[  H - \frac{1}{Z}\ppx{Z}{\beta} \right] \rho_{\mathrm{th}},
\end{equation}
since $\partial_T \beta = - 1/T^2$. Moreover, knowing that
\[
\frac{\partial Z}{\partial \beta} = - \sum_n E_n e^{-\beta E_n} = - Z \expec{H},  
\]
we have:
\begin{equation}
\frac{1}{Z} \ppx{Z}{\beta} = \expec{H}.
\end{equation}
Thus we can see that \eqref{thermalstatederivative} becomes:
\begin{equation}
\partial_T \rho_{\mathrm{th}} = \beta^2(H - \expec{H})\rho_{\mathrm{th}},     
\end{equation}
and, in turn, Eq.~\eqref{thermalstatelyapunov} yields:
\begin{equation}
\Lambda_T \rho_{\mathrm{th}} + \rho_{\mathrm{th}} \Lambda_T = 2  \beta^2(H - \expec{H})\rho_{\mathrm{th}}    . 
\end{equation}
Given that the thermal state is simply the exponential of the Hamiltonian, we know that it commutes with the Hamiltonian itself, i.e. $[H, \rho_{\mathrm{th}}] = 0$. This fact which makes the equation above much easier to solve, and the solution is straightforward if we assume that $\Lambda_T \rho_{\mathrm{th}} = \rho_{\mathrm{th}} \Lambda_T$, leading us to:
\begin{equation}
\Lambda_T = \beta^2 (H - \expec{H}).    
\end{equation}
We can immediately verify that this solution for SLD is valid. Now, since $\trp{H\expec{H}} = \expec{H}^2$ it is also easy to see that the QFI $\trp{\rho_\mathrm{th} \Lambda_T^2}$ will be given by the expression:
\begin{equation}
\mathcal{F}_{\mathrm{th}} = \frac{\expec{H^2} - \expec{H}^2}{T^4},
\end{equation}
thus proving the results in Eqs.~\eqref{eq:TFI} and~\eqref{eq:heatcapacity}. Finally, we mention that the measurement optimal basis in this scenario is simply a projection onto the energy eigenbasis, since the thermal state~\eqref{eq:ThermalStateDef} is diagonal in this basis.

$\opposbishops$ \textbf{\textit{Example:}} As an example, we shall apply the results above to a qubit. We start by considering the very simple case where the qubit is described by the Hamiltonian $H = \frac{\Omega}{2} \sigma_z$. Here, $\sigma_z$ refers to the usual Pauli matrix $\sigma_z = \ketbra{0}{0} - \ketbra{1}{1}$. We find that the partition function in this case is given by $Z = 1 + e^{\Omega/T}$, and that
\begin{equation}\label{eq:GibbsStateQubit}
\rho_{\mathrm{th}}
=
\frac{1}{1 + e^{\Omega/T}}
\begin{matrix}
\begin{pmatrix}
1 & 0\\
0 & e^{\Omega/T}
\end{pmatrix}
\end{matrix}.
\end{equation}
Since $\sigma_z$ is idempotent, the second moment very easily evaluates to $\expec{H^2} = \Omega^2/4$. Meanwhile, we find that $\expec{H} = \Tr\{H\rho_{\mathrm{th}}\} = -\frac{\Omega}{2}\tanh\left(\frac{\Omega}{2T}\right)$. Therefore, by substituting these results into \eqref{eq:TFI} and by using the identity $\sech^2 x = 1 - \tanh^2 x$ we obtain
\begin{equation}\label{eq:qubitTFI}
\mathcal{F}_{\mathrm{th}}
=
\left(\frac{\Omega}{2T^2}\right)^2 {\rm sech}^2\left(\frac{\Omega}{2T}\right)
\end{equation}
as the thermal fisher information of a (fully thermalized) qubit. 

We plot this result in Fig.~\ref{fig:thermometry}. The first observation we can make is that the TFI depends on the temperature itself, reaching a maximum for a certain value of $T$, depending on the size $\Omega$ of the energy gap. 
We perform a maximization over the temperature, plotting the maximum achievable TFI as a function of the gap in the inset of Fig.~\ref{fig:thermometry}. 
We can see smaller gaps can result in much higher values for the TFI. We should be careful however in how we interpret these results. This plot tells us that smaller gaps yield higher precision for \emph{certain} values of the temperature, but not for all of them. We can see from the plots that while smaller values of $\Omega$ are more precise for lower temperatures, they quickly decay for larger values of $T$. Hence, it might be preferable to opt out for a larger gap as $T$ increases. 
This scenario is a further illustration on the challenges of optimization in metrology that we have encountered before, but this time in the context of thermometry. We can see how, e.g., optimization might also emerge as a very natural strategy in thermometry, for example. $\bishoppair$

\begin{figure}[t!]
	\centering
	\includegraphics[width=0.8\textwidth]{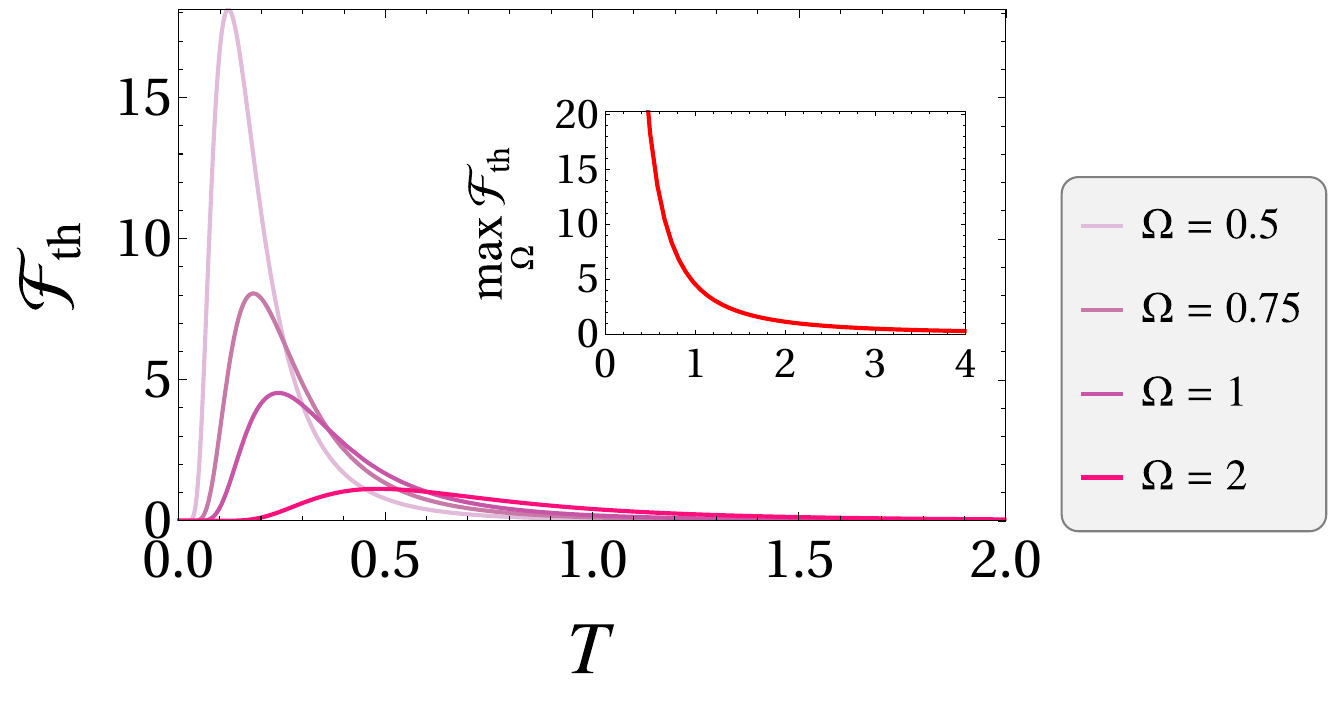}
	\caption[Thermal Fisher information of a qubit.]{Thermal Fisher information of a qubit. We plot the TFI as a function of the temperature $T$ for diffent values of the energy gap $\Omega$. In the inset we plot the maximum TFI (maximized over $T$) as a function of $\Omega$. We can see that smaller gaps result in a higher achievable precision for the optimal temperature.}
	\label{fig:thermometry}
\end{figure}

Another very important observation about the plots in Fig.~\ref{fig:thermometry} and Eq.~\eqref{eq:qubitTFI} which is worth mentioning is that the TFI (and the heat capacity) both converge to zero as $T \rightarrow 0$. 
Thus, standard thermometry protocols become very imprecise in this regime. Many works in the existing literature are concerned with this challenge \cite{correaEnhancementLowtemperatureThermometry2017, mukherjeeEnhancedPrecisionBound2019, jorgensenTightBoundFiniteresolution2020}. 
See for example Sec.~4.3 in Ref.~\cite{Mehboudi2019}. Even worse, the heat capacity usually decays \emph{exponentially} as $T \rightarrow 0$. Many schemes mitigate this problem by employing configurations and techniques which results in a sub-exponential, and sometimes, polynomial decay \cite{hovhannisyanMeasuringTemperatureCold2018, pottsFundamentalLimitsLowtemperature2019}.

A natural generalization would be to investigate how different probes, such as a quantum harmonic oscillator substituting the qubit in the previously described setup, would affect the protocol. See Sec.~2.1.1 from Ref.~\cite{deffnerQuantumThermodynamicsIntroduction2019}, for instance. There are already many thorough investigations in this sense \cite{brunelliQubitassistedThermometryQuantum2012}. 
Moreover, there also exist some surprisingly general results concerning other aspects of thermometry. To name a few, we mention the results concerning the dimensionality and the structure of the eigenspectrum of the system from the discussion in Ref.~\cite{correaIndividualQuantumProbes2015}. There, the authors show that the optimal configuration in the (equilibrium) probe-based thermometry is to employ highly degenerate Hamiltonians. Similarly, a discussion on coarse-grained measurements in quantum thermometry can be found in Ref.~ \cite{hovhannisyanOptimalQuantumThermometry2021}.

\section{Collisional quantum thermometry}\label{sec:collisional}

Throughout this chapter we have discussed the general aspects of quantum metrology and quantum thermometry, with a few concrete applications serving as examples. Our objective in this section is to introduce a concrete platform for thermometry, which will be the foundation for our investigations in Chapter~\ref{chp:results}. 
During this section, in particular, we shall discuss mostly model-specific results. 
Hence, we close this chapter by introducing the idea of \emph{collisional quantum thermometry}, which employs collisional models in a thermometric setup. 
Collisional models have encountered a growing interest in the recent years in the field of open quantum systems \cite{giovannettiMasterEquationsCorrelated2012, seahNonequilibriumDynamicsFinitetime2019, campbellCollisionModelsOpen2021, liDissipationInducedInformationScrambling2022}. 
Among  important uses, we mention applications in quantum optics \cite{ciccarelloCollisionModelsQuantum2017a}, quantum thermodynamics \cite{dechiaraReconciliationQuantumLocal2018, rodriguesThermodynamicsWeaklyCoherent2019} and, more recently, quantum batteries \cite{landiBatteryChargingCollision2021, seahQuantumSpeedUpCollisional2021,  salviaQuantumAdvantageCharging2022}. 
The reader might refer to Ref.~\cite{cusumanoQuantumCollisionModels2022} for a short and pedagogical introduction, and to Ref.~\cite{ciccarelloQuantumCollisionModels2022} for a comprehensive review.
\footnote{The terms \emph{collision models} or \emph{repeated interaction models} ofter appear in the literature as well. Moreover, note how some adjacent formulations - which might not be strictly regarded as collisional models - also exist in the literature, such as the recent work from Ref.~\cite{grossQubitModelsWeak2018}.}

\begin{figure}[h!]
	\centering
	\includegraphics[width=0.7\textwidth]{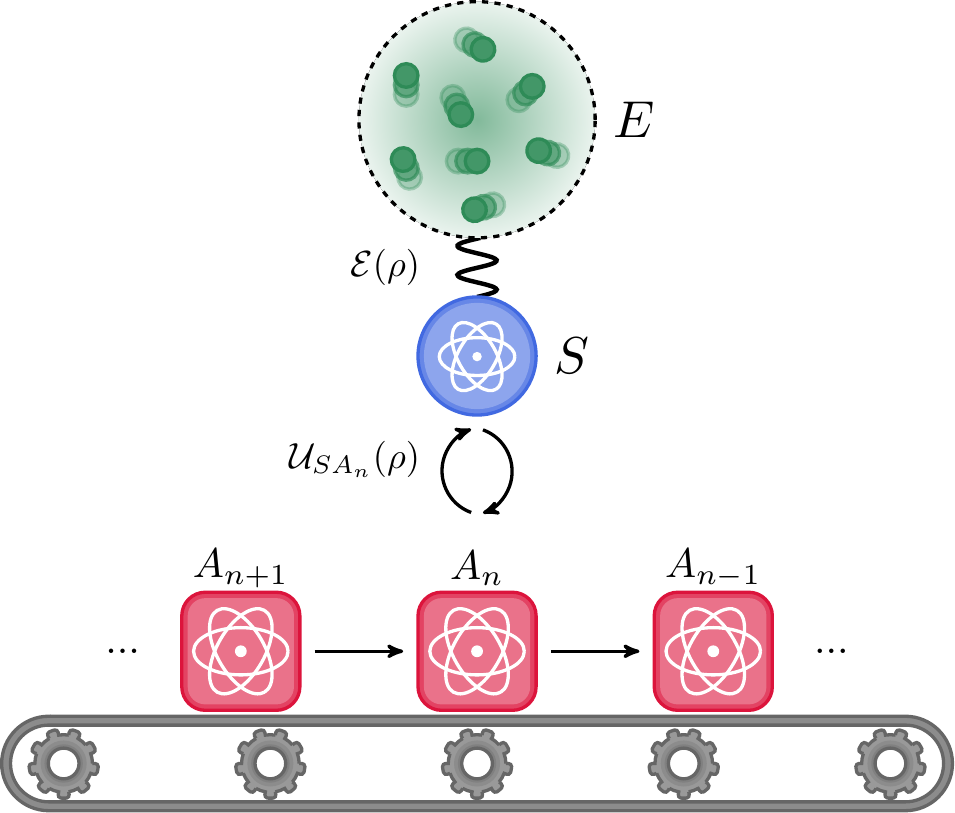}
	\caption[An illustration of the setup used in collisional quantum thermometry.]{An illustration of the setup used in collisional quantum thermometry.}
	\label{fig:collisional_model}
\end{figure}

We show now how collisional models can be tailored as quantum thermometry, closely following the original proposal for collisional quantum thermometry~\cite{Seah2019}. 
The basic idea behind collisional models is that an object of interest repeatedly interact (in other words, "collides") with other constituents of the model. 
In our case, an intermediate system, which we denote by $S$, is placed between two other constituents: the environment $E$, which usually refers to a thermal bath (at a temperature $T$), and a trail of ancillae $A_n$ which are eventually measured, as depicted in Fig.~\ref{fig:collisional_model}. 
One might notice that this setup is in contrast with the standard configuration for probe-based thermometry, where the ancillae (or probes), directly interact with the system of interest.
This is precisely the objective of such a construction: the model from Ref.~\cite{Seah2019} is intrinsically out-of-equilibrium, since the resulting ancillae are, in general, not fully thermalized. This departure from the standard strategy gives rise to two advantages. The first one is that the resulting states of the ancillae might better encode information about the temperature due to their dynamical relaxation rate, surpassing the TFI which we introduced in Eq.~\eqref{eq:qubitTFI}. This will be our main interest in the upcoming discussions. 
The second one is that we are also able to introduce correlations between the ancillae, which allow us to beat the SQL when collective measurements are used.

We shall consider a model where both the system $S$ and the ancillae $A_n$ are resonant qubits, described by $H_S = \Omega \sigma_Z^S/2$ and $H_{A_n} = \Omega \sigma_Z^{A_n}/2$, respectively. Additionaly, the interaction between the $S$ and the environment $E$ is described through the GKLS \cite{breuerTheoryOpenQuantum2007} master equation:
\begin{equation}\label{eq:masterequation}
\frac{d\rho_S}{dt} = \mathcal{L}(\rho_S) = \gamma(\bar{n} + 1)\mathcal{D}[\sigma_-^S] + \gamma \bar{n} \mathcal{D}[\sigma_+^S],
\end{equation}
where $\mathcal{D}[L] = L\rho L^\dagger - \frac{1}{2}\{L^\dagger L, \rho \}$ is the dissipator, $\gamma$ is the coupling strength and $\bar{n} = 1/(e^{\Omega/T} - 1)$ is the Bose-Einstein occupation. 
Therefore, we can write the $SE$ interaction as described by the map $\mathcal{E}(\rho_S) = e^{\tau_{SE} \mathcal{L}}(\rho_S)$. The system-ancilla interaction, on the other hand, is chosen to be a partial-swap \cite{scaraniThermalizingQuantumMachines2002}: \footnote{This choices not unique. For instance, Refs.~\cite{shuSurpassingThermalCramerRao2020} and 
~\cite{oconnorStochasticCollisionalQuantum2021} employ a different interaction, where very analogous results can be obtained.
} 

\begin{equation}
    U_{SA_n} = \exp\Big\{ - i \tau_{SA} g(\sigma_+^S \sigma_-^{A_n} + \sigma_-^S \sigma_+^{A_n})\Big\}.
\end{equation}
By choice, all ancillae are initialized in the ground-state $\rho_A^0 = |0\rangle\langle 0|$. Additionally, the coupling strengths $\gamma \tau_{SE}$ and $g \tau_{SA}$ the free parameters of our model: they can be tuned in order to improve the QFI. 

Moving forward, we formalize the repeated interaction that we have described before. The system undegoes what is called a \emph{stroboscopic} evolution, which consists in an  alternating application of the two maps above,
\begin{equation}\label{eq:stroboscopic_map}
\rho_S^n =  \text{tr}_{A_n}\{\mathcal{U}_{SA_n} \circ \mathcal{E}(\rho_S^{n-1} \otimes \rho_A^0)\} := \Phi(\rho_S^{n-1}),
\end{equation}
where $n = 1,2,3,\ldots$ labels the collisions. 
Here $\mathcal{U}_{SA_n}(\bullet) = U_{SA_n} \bullet U_{SA_n}^\dagger$ and $\circ$ denotes map composition.
We will only consider states in the steady-state regime, in the sense of the strobocopic dynamics from Eq.~\eqref{eq:stroboscopic_map}. In other words, several ancillae are let to interact with the system, which eventually reaches a fixed point $\rho_S^* = \Phi(\rho_S^*)$. 
This regime eliminates any transient effects and introduces a translational invariance with respect to the collisions, greatly simplifying the analysis.

Now that the dynamics is set, we are able to discuss the limits of thermometric precision for the protocol. By following the procedures above, we calculate the steady-state of the \emph{system} to be:
\begin{equation}\label{eq:SSsystem}
\rho^*_S
=
\begin{matrix}
\begin{pmatrix}
\frac{1}
{(1 + e^{\Omega/T})
\left(
1 +
\frac{\sin^2 (g \tau_{SA})}{e^{\Gamma}-1}
\right)}
& 0 \\
0 &
1 - \frac{1}
{(1 + e^{\Omega/T})
\left(
1 +
\frac{\sin^2 (g \tau_{SA})}{e^{\Gamma}-1}
\right)}
\end{pmatrix}
\end{matrix},
\end{equation}
where $\Gamma := \gamma (2 \bar{n} + 1) \tau_{SE}$ is what we define as the thermal (or dynamic) relaxation rate. Note however that this is not our main object of interest. We perform measurements on the ancillae, not on the system. So our objective is to calculate the QFI (and the FI) associated with the ancillae after they interact with the system in the steady state above. That is, we are interested in the state 
\begin{equation}\label{eq:dynamicsSSancilla}
\rho_{A_n}^* 
=  
\text{tr}_{S}\{\mathcal{U}_{SA_n} \circ \mathcal{E}(\rho_S^* \otimes \rho_A^0)\} 
\end{equation}
which, from Eq.~\eqref{eq:SSsystem}, explicitly evaluates to
\begin{equation}\label{eq:SSancilla}
\rho_{A_n}^*
=
\begin{matrix}
\begin{pmatrix}
\frac{1}
{(1 + e^{\Omega/T})
\left(
\csc^2 (g \tau_{SA}) +
\frac{1}{e^{\Gamma}-1}
\right)}
& 0 \\
0 &
1 - 
\frac{1}
{(1 + e^{\Omega/T})
\left(
\csc^2 (g \tau_{SA}) +
\frac{1}{e^{\Gamma}-1}
\right)}
\end{pmatrix}
\end{matrix}.
\end{equation}
\begin{figure}[t!]
     \centering
     \includegraphics[width=\columnwidth]{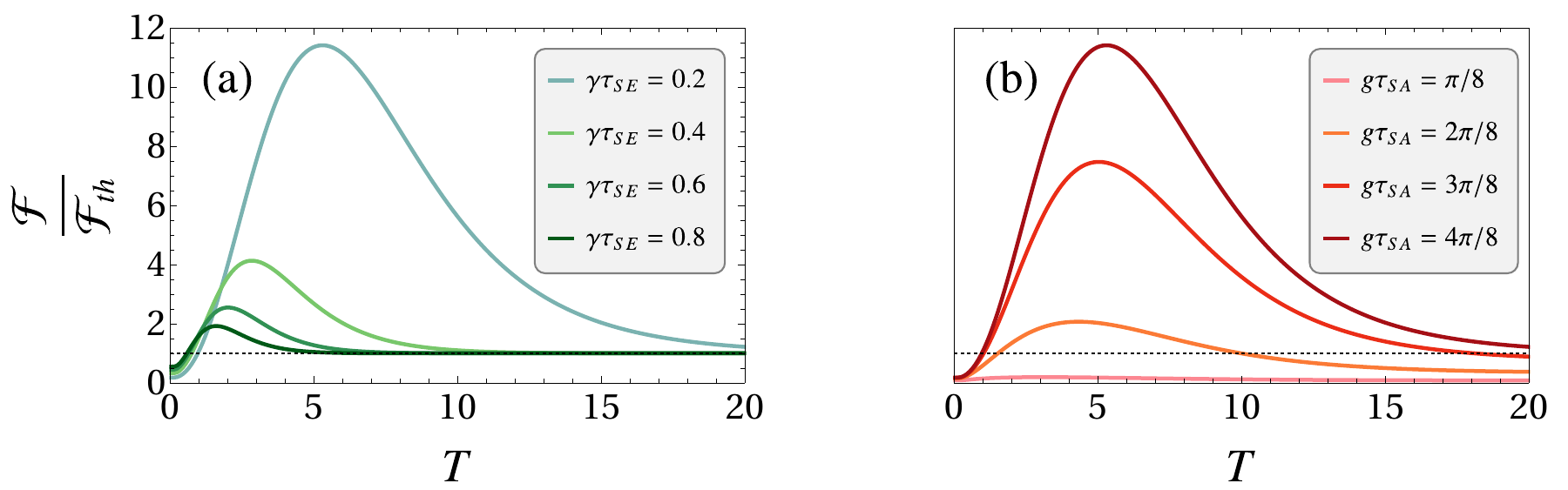}
     \caption[QFI in the collisional thermometry scheme.]{QFI in the collisional thermometry scheme. We plot the ratio between the quantum Fisher information and the thermal Fisher information for (a) full swaps and for (b) $\gamma \tau_{SE} = 0.2$. The dotted line corresponds to $\mathcal{F} = \mathcal{F}_{\mathrm{th}}$.}
     \label{fig:FisherComparison}
\end{figure}
Of special interest is the case of the full-swap, with $g\tau_{SA} = \pi/2$, where the ancilla and the system completely exchange states. In this case the ancillae state reduces to:
\begin{equation}\label{eq:SSfullswap}
\rho_{A_n}^*
\xrightarrow[\text{Swap}]{\text{Full}}
\begin{matrix}
\begin{pmatrix}
\frac{1-e^{-\Gamma}}{1 + e^{\frac{\Omega}{T}}} & 0 \\
0 & 1 -\frac{1-e^{-\Gamma}}{1 + e^{\frac{\Omega}{T}}}
\end{pmatrix}
\end{matrix},
\end{equation}
which will be our case of interest from hereafter. One of the main results from   Ref.~\cite{Seah2019} was to show that the QFI associated with the state~\eqref{eq:SSancilla} actually \emph{surpasses} the TFI which we calculated in Eq.~\eqref{eq:TFI}. More explicitly, they obtained the ratio
\begin{equation}\label{eq:ratioTFI}
\frac{\mathcal{F}}{\mathcal{F}_{th}} 
=
\frac{(\bar{n}+1) \left(e^{\Gamma}+2\bar{n} \Gamma -1\right)^2}{e^{2 \Gamma } (\bar{n}+1)-e^{\Gamma }-\bar{n}},
\end{equation}
in the case of full-swaps. This can be done if we apply the technique from Eq.~\eqref{eq:QFItrick} to the state~\eqref{eq:SSfullswap}.  We plot these results, including the numerical solutions for configurations besides the full-swap, in Fig.~\ref{fig:FisherComparison}.

The most striking aspect in this figure is that the non-equilibrium protocol may offer a very significant improvement over the default experiment. By comparing Figs.~\ref{fig:thermometry} and~\ref{fig:FisherComparison}, the improvement is largest for intermediate temperatures (around $2 \lesssim T \lesssim 10$), which is precisely the region where the TFI quickly falls off. 
This improvement can be attributed to the terms containing the parameter $\Gamma$, appearing in Eqs.~\eqref{eq:SSsystem}~-~\eqref{eq:SSfullswap}. Notice that these terms vanish for fully termalized ancillae. 
In other words, we can say that the thermal relaxation parameter $\Gamma$ helps to better encode the temperature in those states, furnishing us with information beyond what the populations of the thermal state typically offers \cite{Seah2019}. Another important aspect, mentioned in Ref.~\cite{Seah2019}, is that for large $\Gamma$, the ratio in Eq.~\eqref{eq:ratioTFI} behaves as $\frac{\mathcal{F}}{\mathcal{F}_{th}} 
\approx 1 + \Gamma \bar{n} e^{-\Gamma} > 1$. Thus, even though the ratio inevitably tends to unity for large temperatures, it is \emph{always} advantageous to opt for this scheme, as long as an advantage over the thermal sensitivity is concerned. 

Furthermore, the particular equations that we have obtained for the steady-state, and the corresponding QFI, are conditioned on the choice of ancillae which are initially on the ground-state. Other choices, such as ancillae prepared in the excited state, would yield different results. 
Notwithstanding, the qualitative conclusions and general considerations about the model are the same regardless of most input states that we choose for the probes. A careful consideration on this point has been given in Ref.~\cite{shuSurpassingThermalCramerRao2020}, which further expands a few aspects from Ref.~\cite{Seah2019}. 
In particular, the authors obtain the optimal ancillae state with respect to the parameters of the model (including the temperature $T$ itself). 
One of their conclusions is that ancillae initialized in the ground state are either optimal, or at least nearly-optimal, for a wide-range of regimes. For that reason, in the upcoming chapter we will only consider ground-state ancillae and the analytical results above, since other choices offer no substantial differences.


%% file: chapters/results.tex
Throughout this thesis we have become well aware of the conceptual and practical relevance of the quantum Fisher information as a central concept in quantum metrology and quantum estimation theory. 
Our concerns with optimality in chapter~\ref{chp:metrology} revolved entirely around the calculation of the QFI for different models. In particular, we were interested in the suitability of a collisional model as a platform for quantum thermometry \cite{Seah2019} and how its non-equilibrium configuration could be used to provide an advantage over the standard probe-based schemes, which was mainly quantified through the QFI. 
This type of examination is useful useful due to its universality: they provide important benchmarks and a global view of the problem, independent of the particular choice of estimators.

Nonetheless, the post-processing aspects are complementary and equally important. Those are vital when discussing concrete thermometry protocols and, as discussed earlier in chapter~\ref{chp:frequentist}, maximum likelihood methods constitute the default choice as far as concrete estimators are concerned \cite{fiurasekMaximumlikelihoodEstimationQuantum2001, lyTutorialFisherInformation2017, suzukiAmplitudeEstimationPhase2020}. Moreover, further layers of complexity are introduced whenever physical constraints or symmetries of the problem are considered \cite{MacKay2002, vontoussaintBayesianInferencePhysics2011}. 
\footnote{As a practical example, consult the Sec.~6 of Ref.~\cite{escherQuantumMetrologyNoisy2011}, where they discuss different problems in quantum metrology, such as the phase-estimation in a quantum harmonic oscillator. This example also shows how insisting on unbiased estimators might result in unphysical scenarios.}
In the grounds of quantum thermometry, concrete implementations and practical choices of estimators have been a concern in recent works and, as we discussed in Chapter~\ref{chp:bayes}, the Bayesian theory provides a very powerful framework in this sense \cite{Teklu2009, kiilerichBayesianParameterEstimation2016, hanamuraEstimationGaussianRandom2021, morelliBayesianParameterEstimation2021}. 
\footnote{The reader may consult Ref.~\cite{liFrequentistBayesianQuantum2018} for a very through discussion on the differences and the application of frequentist and Bayesian statistics in quantum metrology, albeit with some particular considerations for the problem of phase estimation.
See also Ref.~\cite{vontoussaintBayesianInferencePhysics2011} for a more general review on Bayesian statistics and its applications in physics.
}
To the best of our knowledge, the first use of Bayesian estimation tailored for quantum thermometry has been provided in Ref.~\cite{rubioGlobalQuantumThermometry2021}
\footnote{In the case of \emph{classical} thermometry, Prosper was able to employ Bayesian estimation in his seminal work, putting forth temperature estimation as a scale estimation problem in an early study \cite{prosperTemperatureFluctuationsHeat1993a}.}
, with many important and adjacent investigations following shortly after~
\cite{boeyensUninformedBayesianQuantum2021, mokOptimalProbesGlobal2021, jorgensenBayesianQuantumThermometry2022, mehboudiFundamentalLimitsBayesian2022, glatthardOptimalColdAtom2022, rubioQuantumScaleEstimation2023}. 

Our objective in this chapter is to build a bridge between collisional quantum thermometry and bayesian inference. Bayesian estimation (BE) provides a very natural framework for a sequential updating schemes, where one is able to continuously update the state of knowledge about the temperature distribution in a experimentally friendly manner. From these, it is very simple to implement concrete estimators with the aid of the Bayesian tool set \cite{rubioGlobalQuantumThermometry2021, boeyensUninformedBayesianQuantum2021}. This chapter concludes our investigations about estimation theory, quantum metrology and thermometry, encompassing original results of ours published in~\cite{alvesBayesianEstimationCollisional2022}. 

\section{Ancilla-ancilla correlations}

In Sec.~\ref{sec:collisional} we summarized many of the physical features for collisional thermometry provided in Refs.~\cite{Seah2019} and~\cite{shuSurpassingThermalCramerRao2020}.
However, an important point which was left off from the discussion in Sec.~\ref{sec:collisional} pertains the non-local aspects of the model and the correlations between the ancillae. 
The interaction of the ancillae $A_n$ with the intermediate system $S$ will typically introduce correlations between them in varying degrees. To examine this point in more detail, let us consider a block $A_i...A_{i+n}$ of $n$ ancillae which interact with $S$ after it reaches its steady state $\rho_S^*$. That is, in analogy with Eq.~\eqref{eq:dynamicsSSancilla} for a single ancilla, consider instead the collective state:
\begin{equation}\label{eq:jointAncillastate}
    \rho_{A_i \ldots A_{i+n}} = {\rm tr}_S \Big\{ \mathcal{U}_{SA_{i+n}} \circ \mathcal{E} \circ \ldots \circ \mathcal{U}_{SA_i} \circ \mathcal{E} \big(\rho_S^* \otimes \rho_A^0 \otimes \ldots \otimes \rho_A^0 \big)\Big\}.
\end{equation}
Remember that $\rho_{A_i \ldots A_{i+n}}$ is independent of $i$ due to the translational invariance of steady-state regime. Here, we have the freedom of choosing joint measurements of several ancillae, or even the whole block, at once. However, besides the experimental challenges, such type of measurement inevitably make the theoretical analysis more involved. For that reason, we focus on single-ancilla measurements instead. If we denote the set of POVMs by ${M_X}$, with possible outcomes $X=0,1$, we can associate the pdf
\begin{equation}\label{eq:likelihood_geral}
    P(X_n,\ldots,X_1|T) = {\rm tr} \Big\{ M_{X_n} \ldots M_{X_1} \rho_{A_1 \ldots A_{n}}\Big\}.
\end{equation}
with the state~\eqref{eq:jointAncillastate}. For that reason, we use the mutual information, defined as:
\begin{equation}
I(A_i{:}A_{i+n}) = S(\rho_{A_i}) + S(\rho_{A_{i+n}}) - S(\rho_{A_i A_{i+n}}),
\end{equation}
\begin{figure}[t!]
     \centering
     \includegraphics[width=0.75\textwidth]{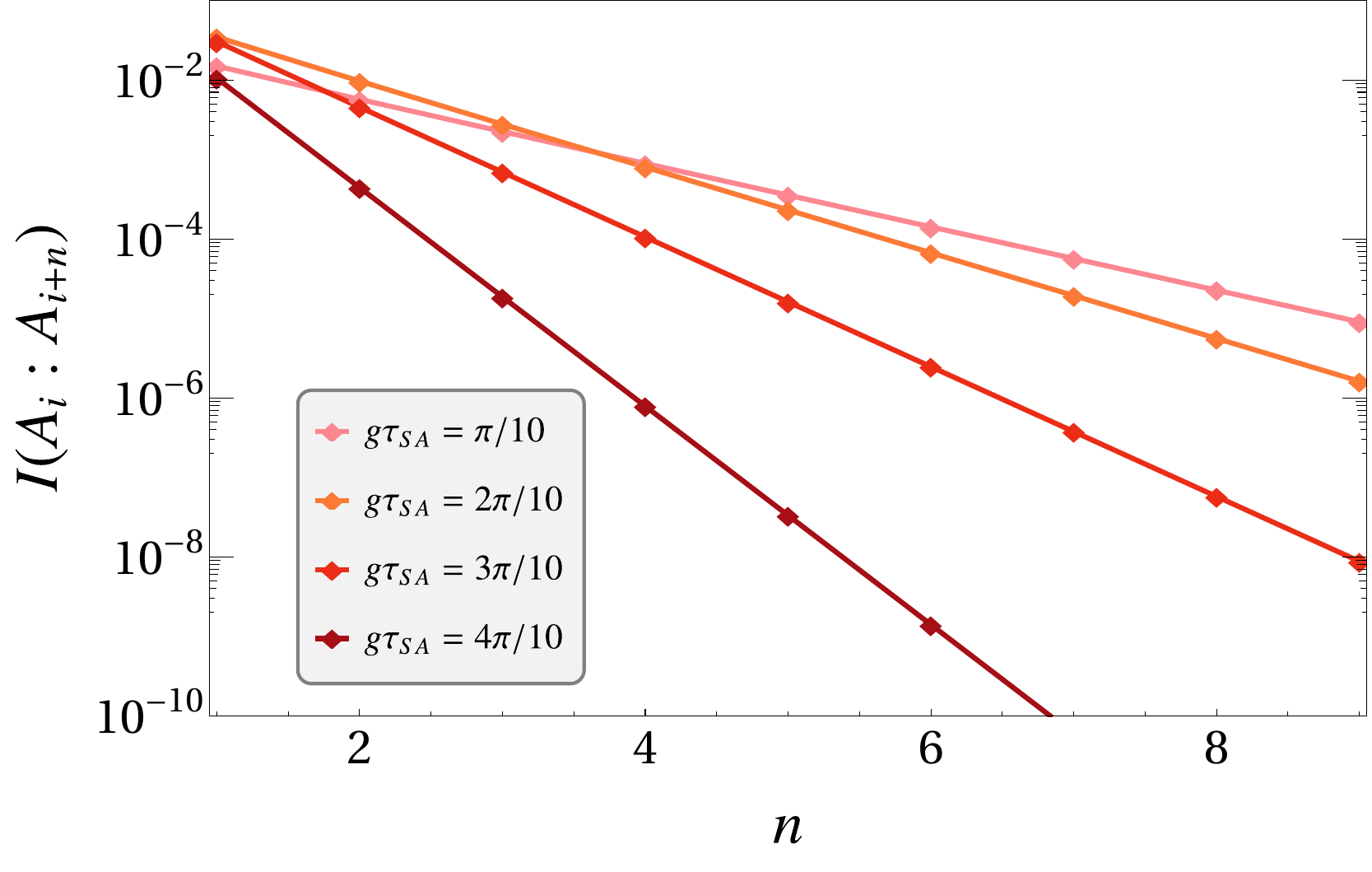}
     \caption[A log-linear plot for the correlation between an ancilla and its $n$-th neighbor for different $SA$ coupling strengths.]{A log-linear plot for the correlation between an ancilla and its $n$-th neighbor for different $SA$ coupling strengths. We take $T/\Omega = 2$ and $\gamma \tau_{SE} = 0.2$.
     }
     \label{fig:LogMutualInformationDecay}
\end{figure}
where $S(\rho) = - {\rm tr}(\rho \ln \rho)$ is the von Neumann entropy, to quantify the correlations which can arise between different ancillae.
We show the result in Fig.~\ref{fig:LogMutualInformationDecay}, plotting $I(A_i{:}A_{i+n})$ as a function of $n$. We can observe that the correlations decay exponentially with each neighbor. Additionally, the rate at which they decay depends on the $SA$ decoupling, decaying faster for ancillae near the full-swap regime (where the correlations eventually vanish). 
Therefore, the pdf associated with the measurement of the ancillae can be written, to a good approximation, as
\begin{equation}\label{eq:DistributionApproximation}
P(X_n, ..., X_1|T) \approx P(X_n|T)...P(X_1|T),
\end{equation}
where 
\begin{equation}\label{eq:CollisionalLikelihood}
P(X_i|T) = \text{tr}(M_{X_i} \rho_{A_{i}}).
\end{equation}
We stress that such an approximation is not an requirement from a Bayesian point of view, but merely a simplification of the problem. For a brief discussion on how the correlations should be computed, consult Appendix~\ref{appendix:correlations}.



\section{Numerical analysis}\label{sec:bayesAnalysis}

We can now introduce numerical apparatuses which will allow us to employ the Bayesian updating scheme for the collisional thermometry. 
The procedure here is very analogous to the examples discussed in chapter~\ref{chp:bayes}. 
Given the outcomes ${\bf X} = (X_1, ..., X_n)$, which are obtained by measuring the ancillae $A_1, ..., A_n$, we can update the state-of-knowledge about the temperature though the Bayes theorem:
\begin{equation}\label{eq:BayesTemp}
P(T|\boldsymbol{X}) = \frac{P(\boldsymbol{X}|T)P(T)}{P(\boldsymbol{X})}.
\end{equation}

The prior is usually chosen either by mathematical convenience or by experimental-motivations. An instance of the latter case is when one might be aware of typical experimental intervals, such as in, e.g., Bose-Einstein condensates~\cite{olfThermometryCoolingBose2015}, or even previous iterations of similar experiments. 
Since we are interested in a purely theoretical analysis, we shall focus on nearly-flat priors, which we shall define later on. In terms of figures of merit, we shall use the MSE and the BMSE, defined in Eqs.~\eqref{eq:mse} and Eq.~\eqref{eq:defbayesrisk2}, respectively. For convenience, we rewrite them here, in terms of the temperature, as
\begin{equation}\label{eq:FrequentistRisk1Temp} 
\epsilon(\hat{T}(\boldsymbol{X})|T)
=
\int(T - \hat{T})^2P(\boldsymbol{X}|T)d\boldsymbol{X},
\end{equation}
and
\begin{equation}\label{eq:FrequentistRisk2Temp} 
\epsilon_{B}(\hat{T}(\boldsymbol{X}))
=
\int 
\epsilon(\hat{T}(\boldsymbol{X})|T)
P(T)
dT.
\end{equation}
Likewise, we will also pick the Bayesian average from Eq.~\eqref{eq:bayesianMean} as our estimator of choice, which minimizes the BMSE~\eqref{eq:defbayesrisk2}, in the mean-square sense:
\begin{equation}\label{eq:estimatorTemp}
\hat{T}(\boldsymbol{X}) = \int T P(T|\boldsymbol{X}) dT 
\end{equation}
We further stress that the posterior loss~\eqref{eq:posteriorloss} is sometimes regarded as the fully Bayesian error, due to the fact that it is conditioned on the actual realization of the experiment, so this is a quantity which might be of great interest for the experimenter whenever concrete data is available. 
We also mention the MAP estimator~\eqref{eq:map} as an equally simple alternative, or
even the median estimator, defined in Eq.~\eqref{eq:map}, as a further option. However, the problem with the latter is that the numerical computation for the median is less straightforward than the other two choices. 
It is not our objective here to dwell into such technical discussions regarding the estimators themselves. A very thorough investigation regarding how different Bayesian estimators - and also different priors - can be used in the collisional thermometry scheme has been done in~\cite{boeyensUninformedBayesianQuantum2021}. The authors also give a very special care to \emph{uninformative} priors in this context. 

While more efficient methods might be available, we describe our procedures for numerically simulating Eqs.~\eqref{eq:BayesTemp}~-~\eqref{eq:FrequentistRisk2Temp}.
Our objective is to sequentially compute the posterior~\eqref{eq:BayesTemp} from a set of $n$  random outcomes ${\bf X}$, with the possibility of easily updating the distribution as new data arrives~\cite{gammelmarkBayesianParameterInference2013, kiilerichBayesianParameterEstimation2016}. 
Additionally, we also need to be able to compute any other corresponding quantities, such as the moments of the distribution. 
However, two complications pop up in regard to a naive numerical implemementation: we have to deal both with a large number of multiplying probabilities in $P(X_1|T)\ldots P(X_n|T)$ and with the computation of the normalization factor $P(\bm{X}) = \int  P(\bm{X}|T) P(T) dT$ from Eq.~\eqref{eq:BayesTemp}.

We start by discretizing the interval of interest $[T_{\rm min}, T_{\rm max}]$ into $N_T$ points $T_k$. Here we denote the lower and upper cutoffs by $T_{\rm min}$ and $T_{\rm max}$, respectively. By doing so, the prior distribution is discretized into a distribution $P_k$ into an analogous manner. Similarly, we can discretize the log-likelihood as
\begin{equation}\label{loglike_discrete}
    L_{kn} = \sum\limits_{i=1}^n \ln P(X_i|T_k).
\end{equation}
Using the log-likelihood here instead of the standard likelihood provides a stable countermeasure against the increasingly smaller probabilities. For practical purposes, we can also visualize $L_{kn}$ as a matrix (or a \emph{grid}) of dimensions $N_T \times n$.
For instance, $L_{k,3}$ describes the log-likelihood of the string $X_1,X_2,X_3$. 
Also notice how $L_{kn} = L_{k,n-1} + \log P(X_n|T_k)$, which can be incremented sequentially with each new piece of data. In this discretized form, the Bayesian updating from Eq.~\eqref{eq:BayesTemp} can now be written as 
\begin{equation}\label{num_step1}
    P_{k|n} = \frac{e^{L_{kn}} P_k}{\sum_q e^{L_{qn}}P_q},
\end{equation}
where $P_{k|n}$ is a shorthand for $P(T_k|X_1\ldots X_n)$.

Furthermore, by defining and using the max of the log-likelihood at each $n$, as $L_{n}^{\rm max} = \max_k L_{kn}$, we circumvent the numerical instabilities.
Namely, we rewrite Eq.~\eqref{num_step1} as 
\begin{equation}\label{num_step2}
    P_{k|n} = \frac{e^{L_{kn}-L_n^{\rm max}} P_k}{\sum_q e^{L_{qn}-L_n^{\rm max}}P_q}.
\end{equation}
This modification ensures that the most likely events will have the best numerical precision. 
By doing so, Eq.~\eqref{num_step2} assumes the form 
\begin{equation}
    P_{k|n} = \frac{P_{kn}}{\sum\limits_q P_{qn}},
\end{equation}
where $P_{kn} = e^{L_{kn}-L_n^{\rm max}} P_k$ can be interpreted as a grid of dimensions $N_T \times n$, which is readily constructed from the log-likelihood matrix $L_{kn}$ and the vector $P_k$ which describes the discretized prior. 

This procedure allows us to readily compute any quantity of interest from the numerical posterior. In particular, the BA~\eqref{eq:estimatorTemp} is given by:
\begin{equation}\label{num_BA}
    \hat{T}_n 
    = \sum\limits_k T_k P_{k|n}.
\end{equation}
We can obtain the MSE [Eq.~\eqref{eq:FrequentistRisk1Temp}] by sample averaging the square error by $(\hat{T}_n - T_0)^2$ over multiple realizations, where we draw the vector ${\bf X}$ according the likelihood $P({\bf X}|T)$, that is: ${\bf X} \sim P({\bf X}|T) = P(X_n|T)...P(X_1|T)$. 
Meanwhile, we obtain the BMSE~\eqref{eq:FrequentistRisk2Temp} in a similar manner, but with an extra step where we generate the data ${\bf X}$ from the likelihood $P({\bf X}|T)$ by randomly sampling the temperature $T$ from the prior $P(T)$. 

\section{Bayesian estimation and collisional thermometry}\label{sec:bayesCollisional}

\begin{figure}[t!]
     \centering
     \includegraphics[width=\textwidth]{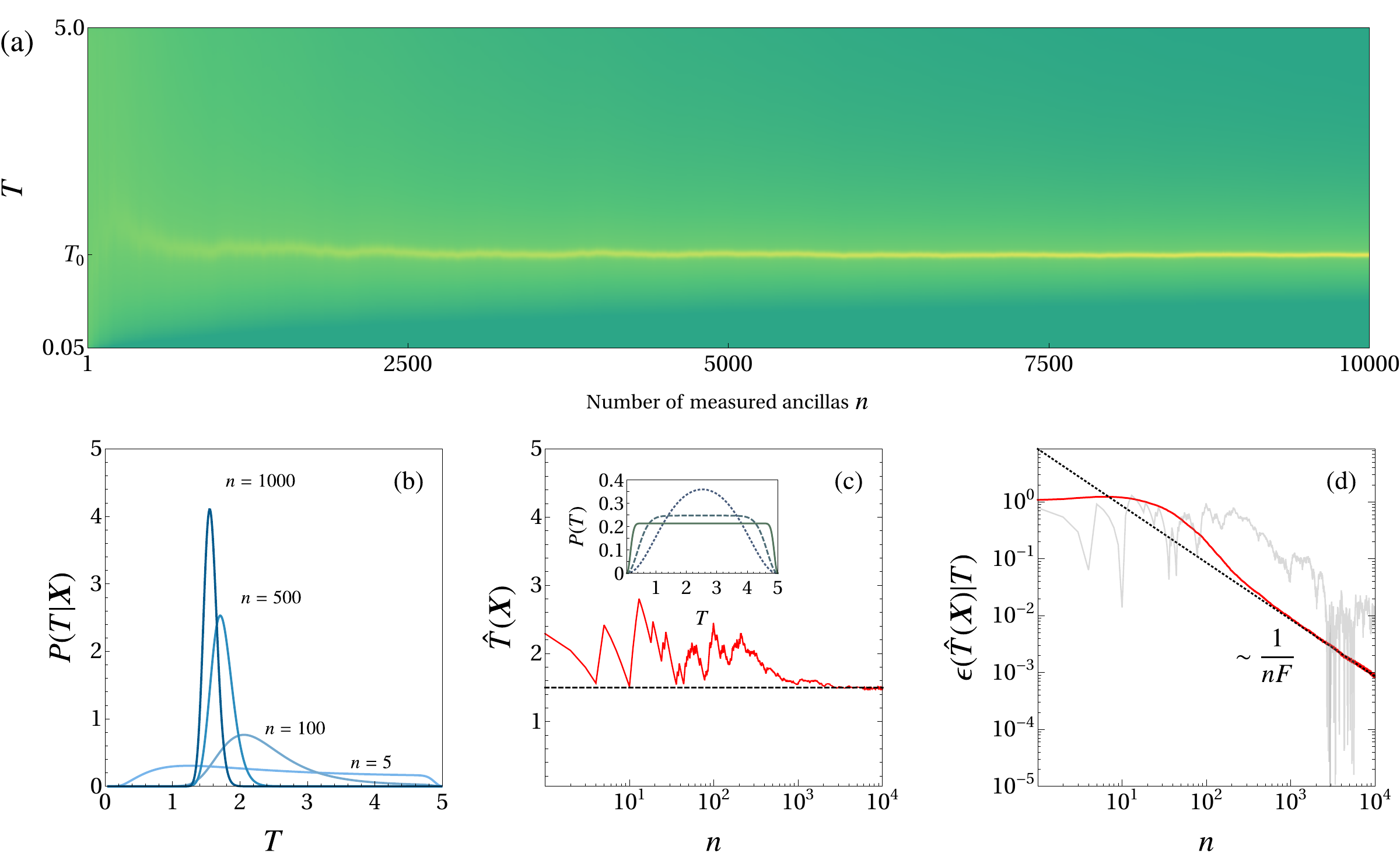}
     \caption[Bayesian estimation for collisional thermometry.]{Bayesian estimation for collisional thermometry.
     The true temperature was chosen as  $T_0/\Omega = 1.5$.
     (a) Density plot of the posterior $P(T|X_1\ldots X_n)$ [Eq.~\eqref{num_step2}]. 
     (b) Same, but for select values of $n$. In both figures, the distribution is clearly seen to converge towards $T_0$ as $n$ increases.
     (c) Random realization of the BA~\eqref{eq:estimatorTemp}, as a function of $n$. 
     Inset: prior distribution~\eqref{eq:prior}, for  $\alpha = -1$ (dotted),  $\alpha = -10$ (dashed) and $\alpha = -100$ (solid).
     (d) MSE [Eq.~\eqref{eq:FrequentistRisk1Temp}] for a single stochastic realization (gray), and averaged over multiple realizations (red). 
     For large $n$ it converges to $1/nF(T_0)$ (dotted), which saturates the CRB.
     All curves were plotted using $\gamma \tau_{SE} = 0.4$, $g\tau_{SE} = \pi/2$ and $\alpha = -100$. The temperature was discretized in steps of $N_T = 500$, from $T_{\rm min} = 0.05$ to $T_{\rm max} = 5$. Caption from~\cite{alvesBayesianEstimationCollisional2022}.
     }
     \label{fig:BayesUpdate}
\end{figure}

We are finally equipped with all the tools necessary for applying the Bayesian framework into collisional thermometry. As we have mentioned, choosing an appropriate prior is a very extensive topic in Bayesian inference \cite{Jaynes2003}, specially when physical considerations or further constraints are taken into account \cite{prosperTemperatureFluctuationsHeat1993a, vontoussaintBayesianInferencePhysics2011, rubioGlobalQuantumThermometry2021, boeyensUninformedBayesianQuantum2021}. 
Since we are interested in generic non-equilibrium scenarios and in numerical simplicity, we choose a near-flat prior over the interval $[T_{\rm min}, T_{\rm max}]$. More specifically, we employ \cite{liFrequentistBayesianQuantum2018}
\begin{equation}\label{eq:prior} 
P(T)
=
\frac{1}{(T_\mathrm{max}-T_\mathrm{min})} \lambda_{\alpha} \left(\frac{T-T_\mathrm{min}}{T_\mathrm{max}-T_\mathrm{min}}\right),
\end{equation} 
where 
\begin{equation} 
\lambda_{\alpha}(\theta) 
= 
\frac{e^{\alpha  \sin ^2(\pi  \theta  )}-1}{e^{\alpha /2} I_0\left(\frac{\alpha }{2}\right)-1},  
\end{equation}
and $I_0$ is the modified Bessel function of the first kind. A major reason for choosing this distribution lies in a technicality concerning the VTSB: the bound, in its standard form, does not hold for priors which do not vanish at the endpoints, such as the flat prior itself. 
Choosing the distribution~\eqref{eq:prior}  avoids this problem \cite{Gill1995, Ramakrishna2020}. 
Moreover, this prior very conveniently allows us to smoothly interpolate between flat and sharp distributions, without altering the endpoints. This can be done by changing $\alpha$, as shown on the inset of Fig.~\ref{fig:BayesUpdate}~(c). The distribution is flat-like for $\alpha$ large and \emph{negative}. Conversely, taking $\alpha > 0$ results in increasingly sharper distributions.

We summarize the basic results in Fig.~\ref{fig:BayesUpdate}. We start by fixing the value of the true temperature $T_0$, which is then used to generate a string of random outcomes $X_i$ from the likelihood $P(X_i|T = T_0)$. We consider the system to be in the full-swap regime $g\tau_{SA} = \pi/2$, so we obtain  the steady from state~\eqref{eq:SSfullswap}. In this case, the likelihood reads
\begin{equation}\label{eq:collisional_likelihood}
P(X_i=1|T) = 
\frac{1-e^{-\Gamma}}{1 + e^{\frac{\Omega}{T}}}
\end{equation}
when we perform measurements in the computational basis, which conveniently happen to be optimal for this class of states, as we discussed in Sec.~\ref{sec:collisional}. This also implies that the FI $F(T)$ associated with the likelihood above coincides with the QFI given in~\eqref{eq:ratioTFI}.

In Fig.~\ref{fig:BayesUpdate}~(a) we plot the posterior distribution $P(T|X_1\ldots X_n)$ [Eq.~\eqref{num_step2}]. The temperature is represented along the vertical axis, while the horizontal axis shows the number of measured ancillae. 
A similar visualization scheme for this sequential updating strategy has been given in Refs.~\cite{gammelmarkBayesianParameterInference2013} and \cite{kiilerichBayesianParameterEstimation2016}, for problems in quantum metrology and open quantum systems. 
Meanwhile, Fig.~\ref{fig:BayesUpdate}~(b) plots the same results as a function of $T$ and for a few selected values of $n$. We can explicitly see how it gradually peaks around the true value of the parameter. 
Additionaly, we can also verify that it indeed starts to resemble a Gaussian for large $n$, with variance $1/nF(T)$, as predicted by the Bernstein–von Mises theorem, discussed in Sec.~\ref{sec:VTSB}~[Eq.~\eqref{eq:vonMises}]. 
Similarly, in Fig.~\ref{fig:BayesUpdate}~(c) we explictly plot a stochastic realization of the estimator. The error associated with this particular realization of $\hat{T}(\bm{X})$ in  is plotted as a gray curve in Fig.~\ref{fig:BayesUpdate}~(d). 
From these panels we can see how the estimator converges to the true vale of the temperature as the number of measured ancillae increases (notice the log scale). Its average over multiple realizations $\bm{X}$ results in the MSE~\eqref{eq:FrequentistRisk1Temp}, plotted as the solid red line in Fig.~\ref{fig:BayesUpdate}~(d). 
\begin{wrapfigure}{r}{0.5\textwidth}
		\includegraphics[width=0.5\textwidth]{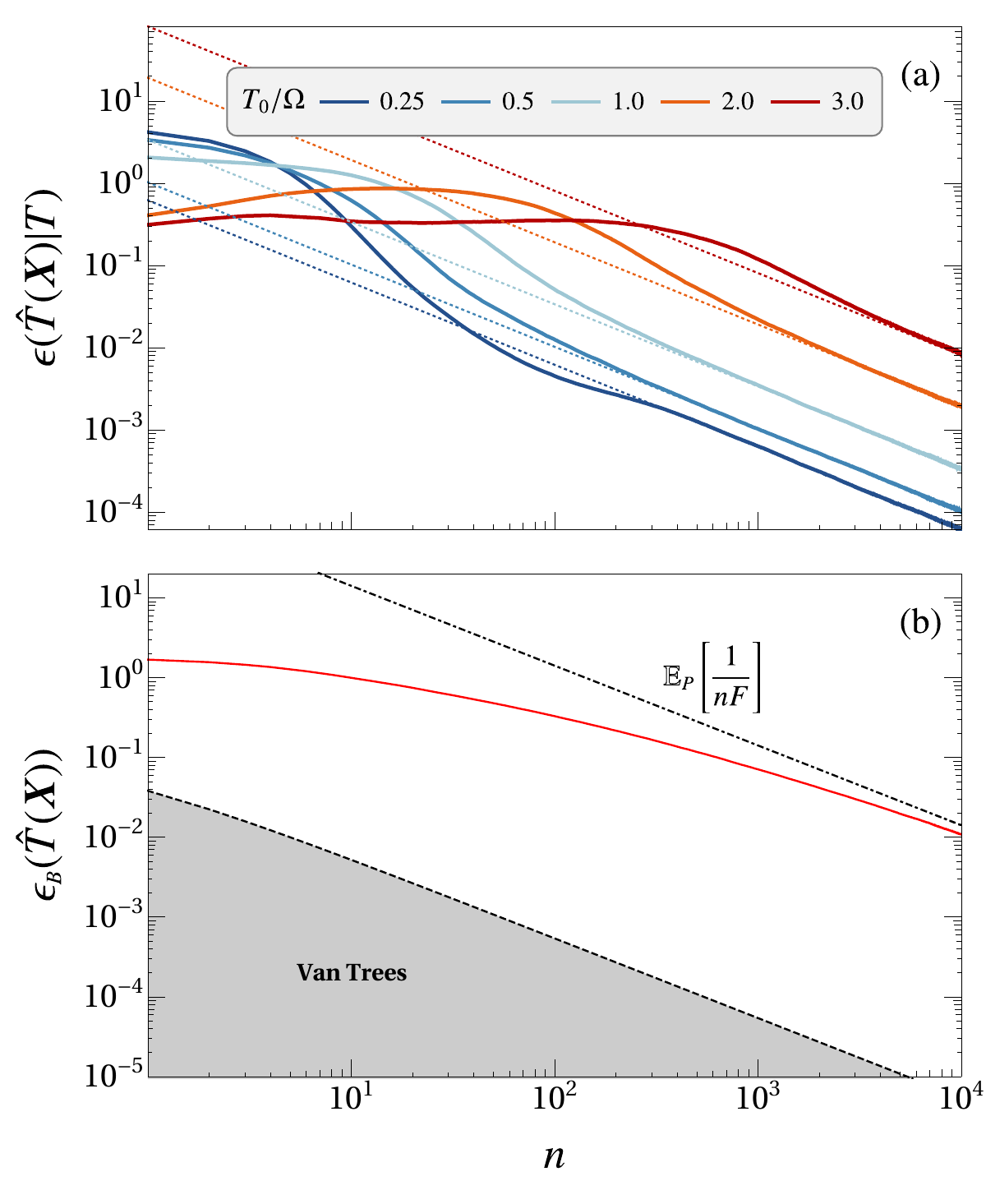}
		\caption[CRB and VTSB in the collisional model.]{(a) MSE [Eq.~\eqref{eq:FrequentistRisk1Temp}] for different values of $T$, averaged numerically over 3000 different trajectories. The dotted lines correspond to the CRB for the different temperatures shown in the legend. (b)  We repeat the procedure in (a), averaging the MSE over 500 trajectories and also over the prior distribution \eqref{eq:prior}, obtaining the BMSE [Eq.~\eqref{eq:FrequentistRisk2Temp}]. The integral over the prior is also performed numerically through a temperature discretization with $N_T = 150$. We also show the asymptotic limit defined in \eqref{eq:BMSEAsymptotic} (dot-dashed).  Other parameters are the same as in Fig.~\ref{fig:BayesUpdate}. Caption from~\cite{alvesBayesianEstimationCollisional2022}.
     }
     \label{fig:BayesianMSE}
\end{wrapfigure}
We can explicitly observe that the estimator is \emph{asymptotically} efficient, as the MSE eventually converges to $1/nF(T_0)$, shown by the dotted line, saturating the CRB. 
	 
Another very important aspect is how well a protocol fares for different temperatures. Thus, in Fig.~\ref{fig:BayesianMSE}~(a) we plot the MSE for different values of $T_0$. The dashed lines highlight the asymptotic limit $1/nF(T_0)$. We can see that the estimation is more accurate as the temperature decreases, which is attributed to the larger sensitivity on $T$ in the likelihood Eq.~\eqref{eq:CollisionalLikelihood}.

As we have thoroughly discussed in chapters~\ref{chp:frequentist} and ~\ref{chp:bayes}, the MSE and the FI often depend on the very parameter being estimated. In the same manner, the MSE in Fig.~\ref{fig:BayesianMSE}~(a) depends on the \emph{unkown} value of the temperature $T_0$. 
This is one of the situations where the Bayesian framework shines as a convenient tool. 
The BMSE $\epsilon_B$ from Eq.~\eqref{eq:FrequentistRisk2Temp} gives us a more concise and global view of the problem, averaging the accuracy of the protocol over the range $[T_{\rm min},T_{\rm max}]$ of temperature, which is independent of particular value of the parameter being estimated.
This is shown in Fig.~\ref{fig:BayesianMSE}~(b). 

We also compare this curve for the BMSE with the VTSB~\eqref{eq:VTSBN}, depicted by the gray region in Fig.~\ref{fig:BayesianMSE}~(b). Notice, for reasons pointed in Sec.~\ref{sec:VTSB}, that the bound is not saturated, even in the asymptotic limit. Instead, the BMSE converges to the average (over the prior) of the reciprocal FI, per Eq.~\eqref{eq:BMSEAsymptotic}. 
In other words, the dashed line in Fig.~\ref{fig:BayesianMSE}~(b) to which the BMSE converges is simply an average of the lines depicting the CRB in Fig.~\ref{fig:BayesianMSE}~(a), but averaged over the prior. 
We could say that the asymptotic value of the \emph{Bayesian} MSE is, in a sense, described as an "expected" Cramér-Rao bound.
Therefore, computing the asymptotic BSME will gives a succinct and sufficiently global view of the aspects of this scheme henceforth. More specifically, we can investigate the behavior of $\epsilon_B$ for large $n$ as we change the system parameters.

\begin{figure}[t!]
     \centering
     \includegraphics[width=\textwidth]{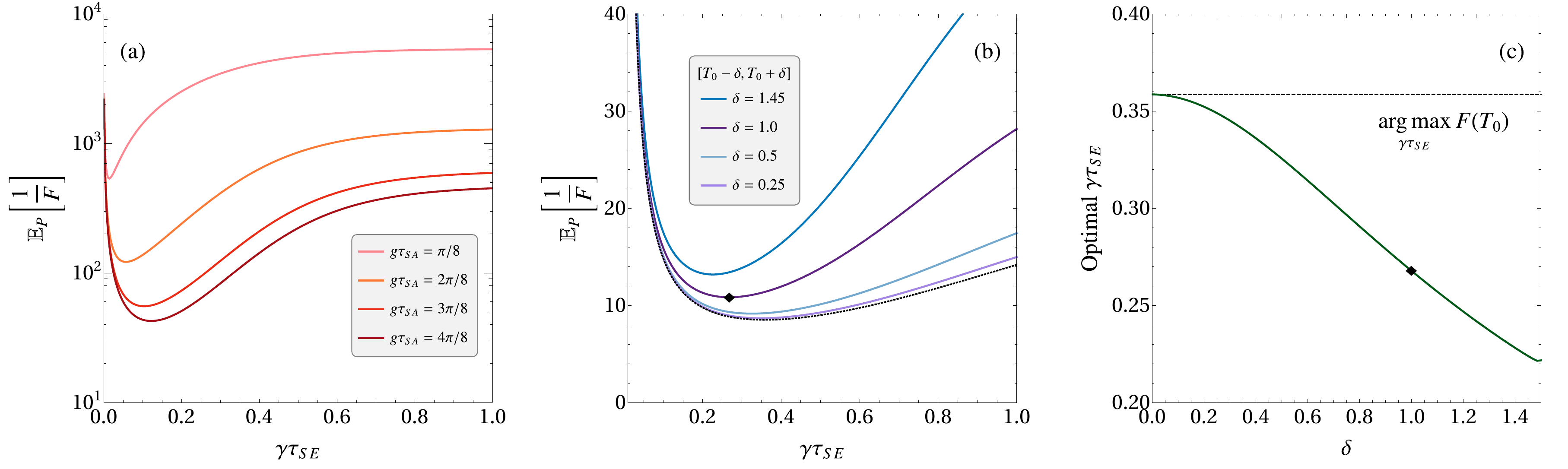}
     \caption[Influence of the temperature interval and cutoff in the asymptotic accuracy of the model, with an optimization over the $SE$ coupling strength.]{
    (a) We plot the expectation in the RHS of Eq.~\eqref{eq:BMSEAsymptotic} as a function of $\gamma \tau_{SE}$ considering the same temperature range $[T_\mathrm{min}, T_\mathrm{max}]$ from Fig.~\ref{fig:BayesUpdate}.
    (b) We fix the SA coupling in the full swap regime $g\tau_{SA} = \pi/2$ and plot Eq.~\eqref{eq:BMSEAsymptotic} for different temperature intervals. Note how the minima shifts to the left as $\delta$ increases. (c) We plot the optimal value of $\gamma \tau_{SE}$ as we increase the size of the interval $[T_0 - \delta, T_0 + \delta]$. We also show the optimal value (dashed) for $F(T_0)$ at the particular temperature $T_0/\Omega = 1.5$. The diamond symbol hightlights the minimum for $\delta = 1$. The prior used here is also given by Eq.~\eqref{eq:prior}, with $\alpha=-100$, but the endpoints are changed as described above. Caption from~\cite{alvesBayesianEstimationCollisional2022}.
     }
     \label{fig:ECRBPlotIntervals}
\end{figure}

With that in mind, we plot the asymptotic value of the BMSE from Eq.~\eqref{eq:BMSEAsymptotic} in Figs.~\ref{fig:ECRBPlotIntervals}~(a) and~(b). In the panel~(a) we describe the asymptotic accuracy of the protocol as a function of $\gamma \tau_{SE}$ for different values of the $SA$ coupling. 
Naturally, since we are employing a Bayesian figure of error here, this plot is \emph{independent} of the temperature, and what we have here is an \emph{average} behavior (with respect to the prior). What is relevant in this case is the temperature interval, that is, the cutoff in the temperatures $T_\mathrm{min}$ and $T_{\mathrm{max}}$, and also the prior.
A smaller value of $\mathbb{E}_P[1/nF]$ results in a more accurate estimation, as seen in the asymptotic behavior of the BMSE, described in Eq.~\eqref{eq:BMSEAsymptotic}. 
Very interestingly, this plot shows us that (i) increasing $g\tau_{SA}$ always yields better results and also that (ii) the accuracy ultimately depends on the value of $\gamma \tau_{SE}$. 
The latter has also been briefly discussed in Ref.~\cite{boeyensUninformedBayesianQuantum2021}.

As we mention above, the achievable precision - in the Bayesian sense - depends on the prior and the chosen interval. Here we try to establish an relationship between the parameters of the model and temperature cutoff. 
In Fig.~\ref{fig:ECRBPlotIntervals}~(b) we plot $\EX_P[1/F]$ as a function of 
the $SE$ coupling, considering a symmetric interval from $T_{\rm min} = T_0 - \delta$ to $T_{\rm max} = T_0 + \delta$, centered at $T_0/\Omega = 1.5$ for different values of $\delta$.  That is, $2\delta$ can be interpreted as the "interval width".

We can see that in either panel, the optimal value of $\gamma \tau_{SE}$ depends on the other parameters of the model, such as $g \tau_{SA}$ and the interval in consideration.
In particular, we can notice from this plot that larger intervals are more sensitive to $\gamma \tau_{SE}$: the optimal regime is narrower, and the error quickly increases with sub-optimal choices. 
For larger intervals, we should tune $\gamma \tau_{SE}$ to smaller values in order to achieve the optimal regime. 
Conversely, we can notice that as the interval narrows, we recover the results found for a deterministic temperature $T_0$, in the same spirit of Fig.~\ref{fig:FisherComparison}~(a) from the previous chapter.
For instance, notice from Fig.~\ref{fig:ECRBPlotIntervals}~(c) how the optimal parameters continuously decrease as $\delta$ increases. 
We also mention that it is known that the optimal value of $\gamma \tau_{SE}$, in terms of the QFI and the FI, actually depends on the temperature itself~\cite{boeyensUninformedBayesianQuantum2021}, so this is one of the scenarios where an adaptative strategy could shine: by continuously performing measurements and updating the state-of-knowledge about the temperature $T$, the coupling parameter $\gamma \tau_{SE}$ could also be changed accordingly. 
In that sense, the work done in Ref.~\cite{oconnorStochasticCollisionalQuantum2021} provides a very clever implementation. There, the authors employ an \emph{stochastic} interaction time $\tau_{SE}$ between the system and the environment. 
\footnote{Curiously, it has been shown recently that it is possible to link such a scenario of stochastic waiting-times to collisional models where ancilla-ancilla correlations are present~\cite{ciccarelloStochasticPeriodicQuantum2022a}. This type of implementation is often associated with non-Markovian behaviour \cite{camascaMemoryKernelDivisibility2021, comarCorrelationsBreakingHomogenization2021}.}
By doing so, one can obtain a better regime of operation for a broader range of parameters. This might also avoid the necessity of being certain of the particular value of the temperature.
Thus, while there might be some loss with respect to a deterministic QFI, by employing random collision times we average out, in a sense, the sub-optimal effect of being unaware of the exact temperature of the bath and the best choice of parameters.

The current discussion showcases how the BMSE might be helpful in the optimization of thermometry protocols. 
As we have made clear in Chapters~\ref{chp:frequentist} and~\ref{chp:bayes}, and also with Fig.~\ref{fig:BayesianMSE} in the current problem, the Fisher information and the CRB both depend on the unknown temperature of the bath. 
Moreover, as shown in Fig.~\ref{fig:ECRBPlotIntervals}, this temperature dependence is also true for the optimal values of $\gamma\tau_{SE}$ and $g\tau_{SA}$. 
This lead us to a conundrum, where the optimal configuration of the protocol depends on the very parameter we are trying to estimate. 
This is a hurdle which was similarly made clear in Ref.~\cite{shuSurpassingThermalCramerRao2020} in regard to the optimization of the probe states.
The Bayesian frameworks avoids this problem with an alternative formulation, focusing on an entire range of temperatures, whose likelihood is quantified by the prior $P(T)$. 
In this sense, we can also see the prior as quantifying a certain "degree of importance" to different temperatures: when calculating the BMSE and the optimal parameters, greater consideration is given to temperatures in which the system is more likely to be found.
If the system is unlikely to be found at a certain interval of temperatures, there is no reason to take it into much consideration when performing any type of optimization.
Additionally, this scenario also makes clear why the BMSE might also be called a \emph{preposterior} error: it is a quantity of special importance \emph{before} any experimental run is performed (hence, when the posterior distribution and any information which might come with it, is unavailable). 
When the parameter is unknown and no data is available, it makes sense to use the BMSE: it is a figure of merit which averages out over all the possible experiments, by integrating over all the possible values of $T$ (weighted by the prior) and also over all the possible stochastic realizations (weighted by their likelihood).
This is a scenario where we are yet to acquire any experimental data and where no knowledge about T, besides what is contained within the prior, is present.
Thus, this type of analysis might be very valuable from an experiment design point of view and for optimization purposes (in the Bayesian sense), as also discussed in Refs.~\cite{rubioGlobalQuantumThermometry2021} and~\cite{boeyensUninformedBayesianQuantum2021}.
Hence, by focusing on the asymptotic BMSE ($n\to \infty$), as compared to the asymptotic MSE $1/nF(T)$, we were able to show in Fig.~\ref{fig:ECRBPlotIntervals} how the BMSE in Fig.~\ref{fig:BayesianMSE}~(b) can be optimized over $\gamma\tau_{SE}$ and $g\tau_{SA}$, to yield a strategy which is good for the entire temperature range. 
Therefore, even though the temperature is in principle unknown, the Bayesian approach provides us with a strategy which, while possibly sub-optimal for a given temperature, will work well \emph{on average}, avoiding more situations where the estimation is too imprecise. 

\section{Effect of noisy probes}

We conclude our investigations by discussing the effect of noisy ancillae in the protocol.
Initially we assumed an ideal scenario where one has perfect control over the probe state. Now we further generalize our approach to the context  where the observer cannot always initialize the ancilla in the desired state. 
For concreteness, we first begin with a scenario where we assume that the ancillae are initialized in a thermal state, investigating how the temperature of the probes affect the asymptotic precision of the estimation, in the same spirit of Fig.~\ref{fig:ECRBPlotIntervals}. In this more general scenario, the likelihood now assumes the form
\begin{equation}\label{eq:collisional_likelihood_probe}
P_{T_P}(X_i=1|T) = 
\frac{e^{-\Gamma}}{1+e^{\frac{\Omega}{T_p}}}+
\frac{1-e^{-\Gamma}}{1 + e^{\frac{\Omega}{T}}}
\end{equation}
instead, replacing Eq.~\eqref{eq:collisional_likelihood}.
As a consequence of the linearity of the stroboscopic channel~[Eq.~\eqref{eq:stroboscopic_map}], the resulting likelihood for the thermalized probe is a convex combination of the likelihood associated with ancillae initialized in the states $|0\rangle\langle 0|$ and $|1\rangle\langle 1|$, weighted by their Gibbs probabilities~[Eq.~\eqref{eq:GibbsStateQubit}]. 
Furthermore, the convexity of the FI \cite{cohenFisherInformationConvexity1968} implies that the resulting precision will be smaller for noisy probes, when compared to ideal probes initialized in the ground state.

\begin{figure}[b!]
     \centering
     \includegraphics[width=\textwidth]{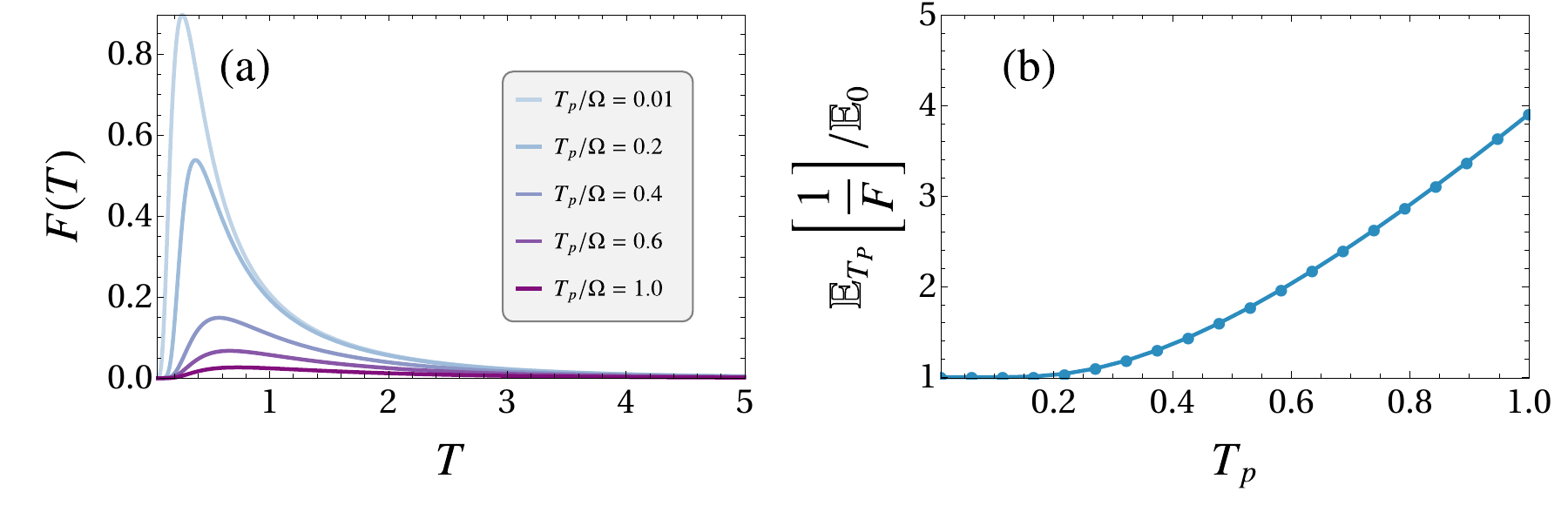}
     \caption[Effect of non-ideal probles in the model.]{(a) We plot the Fisher information of the likelihood in Eq.~\eqref{eq:collisional_likelihood_probe} for different values of $T_p$. (b) We calculate the ratio between the asymptotic Bayesian risk~\eqref{eq:BMSEAsymptotic} obtained by integrating the FI in (a), from $T_\mathrm{min} = 0.1$ to $T_\mathrm{max} = 5$ and the asymptotic Bayesian risk for $T_p = 0$. All the other parameters are the same as in Fig.~\ref{fig:BayesUpdate}. Caption from~\cite{alvesBayesianEstimationCollisional2022}.
     }
     \label{fig:TProbeTest}
\end{figure}

A quantitative description can be given in terms of the asymptotic value of the Bayesian error given by Eq.~\eqref{eq:BMSEAsymptotic}, in the same vein of Fig.~\ref{fig:ECRBPlotIntervals}. 
We first calculate the asymptotic error for ground-state ancillas, which we denote by $\mathbb{E}_0$. 
This quantity can be obtained from the ideal FI given by Eq.~\eqref{eq:ratioTFI}. 
In Fig.~\ref{fig:TProbeTest}~(a) we show how the Fisher information $F(T)$ depends on the temperature $T_p$ of the probe. Meanwhile, in panel~(b) we show how much precision is lost for non-ideal probes, i.e., we calculate the ratio between 
the asymptotic factors $\mathbb{E}_{T_p}[1/F]$ and $\mathbb{E}_0$ as a function of the probe temperatures.

We finish by analyzing a second, but similar, scenario: we are interested in what happens when there is an uncertainty, or systematic error, in the preparation of the ancillae. 
Namely, the ancillae prepared either in the states $|0\rangle\langle 0|$ and $|1\rangle\langle 1|$ with classical probabilities $q$ and $1-q$, respectively, but the experimenter is unaware of their value. Mathematically, this scenario is described by the likelihood
\begin{equation}\label{eq:collisional_likelihood_bias}
P(X_i|T) = 
q P(X_i|T, \rho_{A, 0})
+
(1-q)  P(X_i|T, \rho_{A, 1}),
\end{equation}
where $\rho_{A, k} = | k \rangle \langle k |$, with $k=0,1$. 
\begin{figure}[t!]
     \center
     \includegraphics[width=0.75\textwidth]{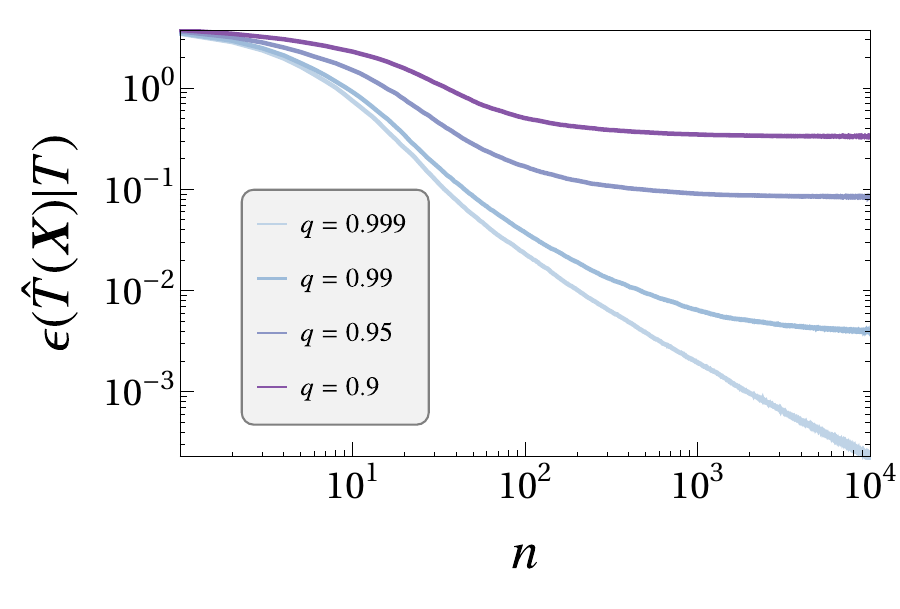}
     \caption[Persistent error in the model due to error in the probe preparation.]{Bayesian MSE~\eqref{eq:FrequentistRisk2Temp} calculated for outcomes generated from the likelihood in Eq.~\eqref{eq:collisional_likelihood_bias} for different values of $q$.  This type of bias introduces a systematic error; the Bayesian risk initially decreases with a $1/n$ scaling but eventually saturates, since the estimation converges to a wrong value of temperature. All the other parameters in the simulation are the same as in Fig.~\ref{fig:BayesUpdate}. Caption from~\cite{alvesBayesianEstimationCollisional2022}.
     }
     \label{fig:ProbeError}
\end{figure}
Nevertheless, the estimation is still performed with respect to the ideal model from Eq.~\eqref{eq:collisional_likelihood}. In other words, what we are describing here is a scenario where we perform inference using an \emph{incorrect}, or imprecise, model. 
This introduces a persistent error into the estimation, as we show in Fig.~\ref{fig:ProbeError}.
Since the experimenter is not using the proper model for the likelihood, the resulting estimation diverges from the correct parameter. 
The error eventually saturates at a certain value, which will correspond to the difference between the true value $T_0$ of the temperature and the temperature one would get from the ideal likelihood~\eqref{eq:collisional_likelihood} for a given realization 
${\bf X} = (X_1, \ldots, X_n)$.

%% file: chapters/geometric.tex
During the mid-twentieth century,  Ehrenberg and Siday \cite{Ehrenberg1949, PeterA.Sturrock}, and later on Aharanov and Bohm \cite{Aharonov1959, Aharonov1961}, were responsible for laying the groundwork \cite{Cohen2019} for some of the most influential results on non-local aspects of quantum theory during the earlier days of quantum mechanics. These pioneering investigations were fundamental to Berry's seminal work \cite{Berry1984}, where he showed the physical consequences of adiabatic geometrical phases (GP) and how they arise in quantum mechanical systems. \footnote{An earlier description of the phenomenon was given by Pancharatnam \cite{Pancharatnam1956} for polarized light in terms of the Poincaré sphere \cite{Gamel2012}. T. Kato also made significant observations about the adiabatic theorem and its geometrical aspects during the early 50's \cite{Kato1950}. However, the physical relevance of the geometrical phases as we know today still went unnoticed by him at the time. A through discussion and the historical considerations of his work can be found in \cite{Simon2019}.} Further advancements were made by Wilczek and Zee, in a generalization to the degenerate case, showing how a non-Abelian structure emerges. A few years later, important contributions by Aharonov and Anandan \cite{Aharonov1987, Anandan1988} also came out, further expanding upon the non-Abelian geometric phase from Wilczek and Zee \cite{Wilczek1984} to the non-adiabatic case. In this section, we briefly review the results by Berry, Aharonov and Anandan. Later on, their results will be important for us in order to motivate the construction of geometric quantum gates. For a review on GPs in quantum information, see \cite{Sjoqvist2015, Sjoqvist2016b}. A rigorous formulation of GPs in terms of differential geometry is given in \cite{Chruscinski2004}.

\section{The Berry phase}

\begin{figure}[t!]
    \centering
    \includegraphics[width=\textwidth]{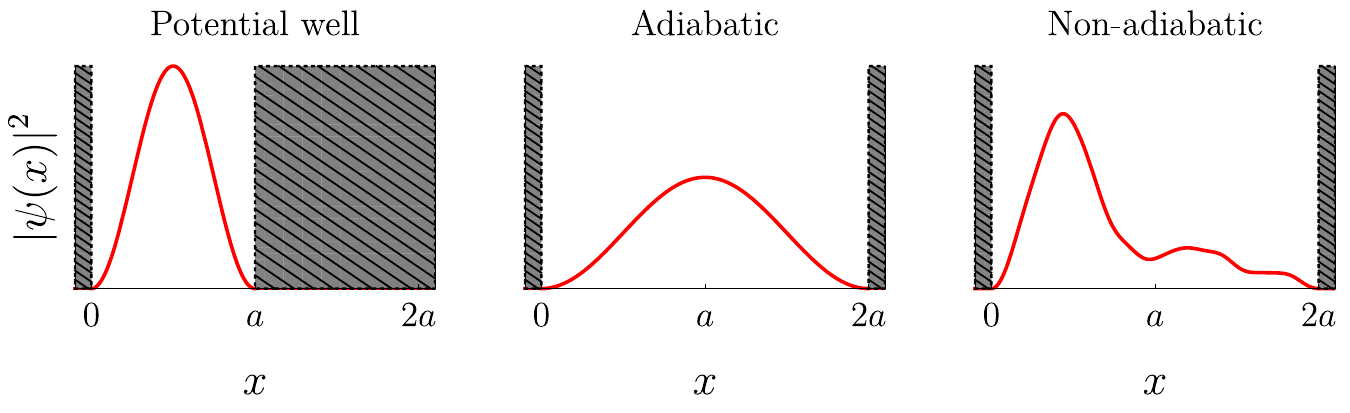}
    \caption[Adiabatic and non-adiabatic dynamics of an infinite potential well.]{Adiabatic and non-adiabatic dynamics of an infinite potential well. In this example we illustrate how the wave function evolves under two different scenarios for an infinite potential well by doubling its width. Consider that the wave function is initially in the ground-state. When the width $a$ of the well is increased \emph{adiabatically}, the w.f. follows the instantaneous ground-state of the well, preserving its trigonometric form. In the non-adiabatic scenario where the width suddenly changes, we get a much more complicated solution, which is a combination of several eigenstates.}
    \label{fig:well_berry}
\end{figure}

The Berry phase is a geometrical phase which can
emerge when the system evolves adiabatically, i.e., when the Hamiltonian changes very slowly. By doing so, the evolution in the Hilbert space follows the
instantaneous eigenstates of the Hamiltonian. We discuss an example of this regime in Fig.~\ref{fig:well_berry}.

More concretely, let us consider a Hamiltonian $H({\bf R}(t))$ which depends on a vector of parameters ${\bf R}(t) = (R_1, R_2, ...)$. Initially, we also consider a non-degenerate case. The eigenstates $\ket{n({\bf R}(t))}$ of this Hamiltonian will naturally depend on the parameters:
\begin{equation}\label{eq:berrySE}
    H({\bf R}(t))\ket{n({\bf R}(t))} = \epsilon_n({\bf R}(t))\ket{n({\bf R}(t))}.
\end{equation}
According to the adiabatic theorem, a state which is initially the $n$-th eigenstate of $H$,
\begin{equation}
    \ket{\psi_n(0)} = \ket{n({\bf R}(0))},
\end{equation}
will evolve into the instantaneous eigenstate of $H$ at a later time $t$, that is
\begin{equation}\label{eq:eigen_evolution}
\ket{\psi_n(t)} = c_n(t) \ket{n({\bf R}(t))},
\end{equation}
given that the evolution is sufficiently slow. In this sense, the time scale is defined by the inverse of the energy gap between the energy of the $n$-th state and the neighboring states; a valid adiabatic regime requires an increasingly slower process the smaller the gap. In this regime, we can establish an useful relationship between the path in the parameter space and a path in the Hilbert space of the wave functions. The coefficient in Eq.~\eqref{eq:eigen_evolution} can be written as
\begin{equation}
c_n(t) = e^{i \gamma_n (t)} 
\exp{\left[ -i \int_0^t dt' \epsilon_n(t') \right]},
\end{equation}
which is the main result of the adiabatic theorem~\cite{Sakurai2020}. The second factor in the equation above is called \emph{the dynamical phase} \cite{Sakurai2020}. Our object of interest, however, is precisely $\gamma_n$, which is known as the geometric phase. We will show why it takes this name very shortly. By plugging the equation above and Eq.~\eqref{eq:eigen_evolution} into the Schrödinger equation 
 $i \hbar \frac{d}{dt}\ket{\psi_n} = H ({\bf R}(t)) \ket{\psi_n}$, we get, after taking the inner product with $\bra{\psi(t)}$ \cite{Girvin2019}:
\begin{equation}\label{eq:geophase1}
\gamma_n(t) = i \int_0^t dt' 
\bra{n({\bf R}(t'))}
\frac{d}{d t'}
\ket{n({\bf R}(t'))}.
\end{equation}

\begin{wrapfigure}{r}{5.5cm}
    \includegraphics[width=5.5cm]{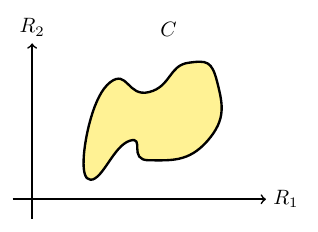}
    \caption[A closed path $C$ in the parameter space.]{A closed path $C$ in the parameter space.}
    \label{fig:parameter_space}
\end{wrapfigure}

The geometrical aspect of the above equation becomes evident if we make one further modification. By using the chain rule to write
\begin{equation}
\frac{d}{d t}
\ket{n({\bf R}(t))}
=
\nabla_{{\bf R}} \ket{n({\bf R}(t))}
\cdot
\frac{d{\bf R}}{dt},
\end{equation}
we may recast Eq.\eqref{eq:geophase1} as
\begin{equation}\label{eq:geophase2}
\gamma_n(t) = 
\int_C {\bf \mathcal{A}}^n({\bf R}) \cdot d{\bf R},
\end{equation}
where
\begin{equation}\label{eq:berryconnection}
{\bf \mathcal{A}}^n({\bf R})
=
i \bra{n({\bf R})} \nabla_{{\bf R}} \ket{n({\bf R})}
\end{equation}
is called the \emph{Berry connection}. This result is remarkable because it reveals a novel interpretation of $\gamma_n$. This phase is not an explicit function of the time, but rather, it depends only on the path C of ${\bf R}$ in the parameter space, irrespective of the dynamical details of the evolution (see Fig.~\ref{fig:parameter_space}). For instance, it is irrelevant, from the point of view of the GP, the rate at which the path C in the parameter space is traversed, as long as the adiabatic regime is valid. Hence, this is the origin of the term \emph{geometric phase}.

One important detail is that this phase is, in general, not gauge invariant. If we change the eigenstate by a phase $\delta({\bf R})$, a change in the Berry connection~\eqref{eq:berryconnection} also incurs:
\begin{equation}
\ket{n({\bf R})} 
\rightarrow e^{\delta\bf(R)} 
\ket{n({\bf R})} 
\implies
\bf{\mathcal{A}}^n({\bf R})
\rightarrow
\bf{\mathcal{A}}^n({\bf R})
-
\nabla_{{\bf R}} \delta({\bf R}).
\end{equation}
This results in a phase change which depends on the endpoints of this gauge, in other words:
\begin{equation}
\gamma_n \rightarrow 
\gamma_n + \delta({\bf R}(0)) - \delta({\bf R}(t)). 
\end{equation}
A important observation which can be made here \cite{Berry1984} is that \emph{cyclic} adiabatic processes in the \emph{parameter} space, i.e., evaluations for which ${\bf R}(t_{\rm end}) = {\bf R}(0)$, dispel the gauge-dependence in the geometrical phase, unambiguously defining a Berry connection for this cyclic evolution:
\begin{equation}\label{eq:geophase3}
\gamma_n(t) = 
\oint \bf{\mathcal{A}}^n({\bf R}) \cdot d{\bf R}.
\end{equation}
The geometrical interpretations of this result are even more far reaching. For instance, one may notice that the Berry connection, due to its gauge properties, is strongly analogous to the vector potential of a magnetic field. It is also possible to define a Berry curvature, which is a tensor constructed from the Berry connection. This allows us to use Stokes' theorem to express the geometrical phase~\eqref{eq:geophase3} in  terms of the area enclosed by the loop in the parameter space \cite{Girvin2019}.

\begin{figure}[h!]
    \centering
    \includegraphics[width=.8\textwidth]{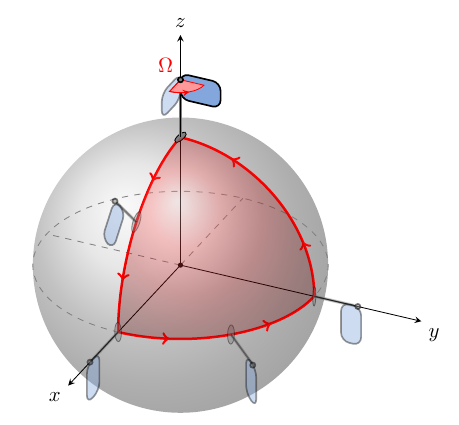}
    \caption{Illustration of the geometrical consequences of the transport of a vector on curved surfaces.}
    \label{fig:sphere_transport}
\end{figure}
In Fig.~\ref{fig:sphere_transport} we show a purely geometrical analogy of this phenomenon. For that, we can picture the following scenario: a flag is put up very close to the north pole of a sphere, pointing south. Afterwards, we carry this flag along a closed path on the sphere, always making sure that it points south. More rigorously, we say that the object is being parallel transported, i.e., we perform no local rotations \cite{Zhang2021}.  Upon returning to the initial point in the north pole, the flag will acquire an azymuthal angle   which is proportional to the solid angle $\Omega$ enclosed by the path. Since no local rotations were performed throughout the process, the resulting rotation is an effect which is completely due to the curvature of the surface. 

Geometrical phases in quantum mechanics are typically discussed in terms of more abstract and higher dimensional spaces, so this description should be seen as a motivation for the problem 
 and as a pedagogical analogy with a clear geometric visualisation of the effect, and not as a complete or rigorous description \cite{Bohm2003}. An analogous formulation for classical physics, which is more akin to this illustration, was given in terms of action-angle variables by J. H. Hannay \cite{Hannay1985}, as an extension to Berry's result. There, as an example, he discusses a spinning symmetric top,  in order to explain the distinction between the two contributions to the rotation: one of them being due to the angular velocity of the top, and the other one being due to the geometrical contribution of the curved surface of the earth. Nevertheless, in the next session we will see a result in quantum mechanics for a two-level system in terms of the Bloch sphere, which, in the same spirit as Fig.~\ref{fig:sphere_transport}, can also be easily visualized.

\section{Non-adiabatic geometrical phases}

A natural question which arises from Berry's framework is whether a similar result holds for a non-adiabatic evolution. Built upon Berry's work and also on the Wilczek-Zee phase, which is a previous generalization of the BP for the degenerate case, it was shown that it is possible to obtain a  non-adiabatic non-Abelian geometrical phase \cite{Aharonov1987, Anandan1988} under certain conditions. This result will be vital for the construction of quantum gates later on, as we shall see.

\begin{figure}[h!]
    \centering
    \includegraphics[width=.8\textwidth]{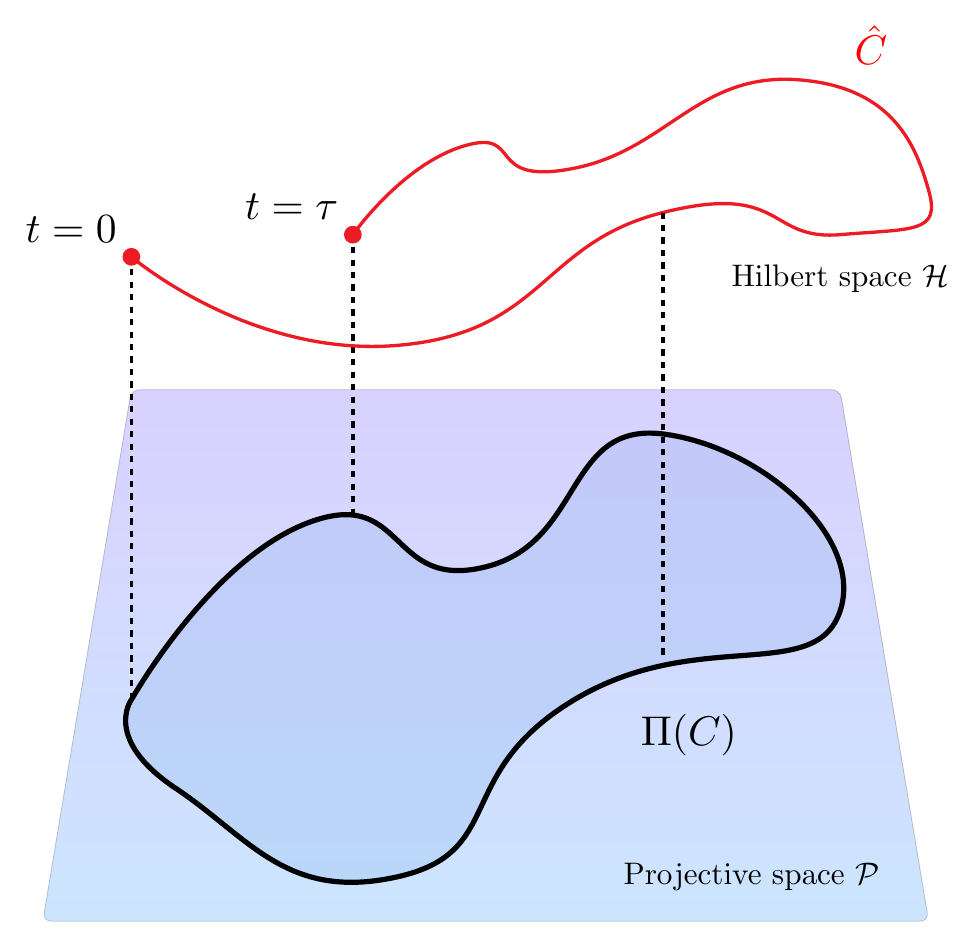}
    \caption[Schematic diagram of the Hilbert space and the ray space $\mathcal{P}$.]{Schematic diagram of the Hilbert space and the ray space $\mathcal{P}$. The open path $\hat{C}$ in the Hilbert space and the closed loop $C = \Pi(\hat{C})$ in the projective space are both shown in the figure.}
    \label{fig:hilbert_space}
\end{figure}

Our objective in this section is to follow the approaches in \cite{Aharonov1987}  and \cite{Anandan1988} to obtain a closed form expression for the desired non-adiabatic non-Abelian geometrical phases.  We begin by addressing the non-adiabatic (but Abelian) phase studied in \cite{Aharonov1987}. This approach is slightly more general in the sense that we consider the Hilbert space $\mathcal{H}$ (with $\dim \mathcal{H} = n$) and its projective space of rays, i.e. the set of rays $\mathcal{P}$ (with $\dim \mathcal{P} = n-1$) defined by the projection map $\Pi:\mathcal{H} \rightarrow \mathcal{P}$ with $\Pi(\ket{\psi}) = \{ \ket{\psi'}: \ket{\psi'} = c \ket{\psi},\quad \text{s.t. } c \in \mathbb{C} \}$, as we illustrate in Fig.~\ref{fig:hilbert_space}.  In short, what Aharonov and Anandan did was to show that it is possible to compute a geometrical phase for \emph{all} cyclic evolutions, not only the adiabatic ones \cite{Aharonov1987, Anandan1988}. Therefore, this framework encompasses other well known special cases, such as the Berry phase, periodic dynamics (such as the precession of a particle in a constant magnetic field) and so on. 

Trying to analyze Berry's phase and the trajectory of the state in the parameter space for a system with a degenerate subspace is one of the examples which illustrate why this new approach can be seen as more general. Curves in the parameter space, presented in the previous section, can show a certain redundancy in this case, since for a degenerate part of the eigenspace the curve in the projective Hilbert space will simply be a point in the parameter space \cite{Anandan1988}. This dynamics nevertheless still describes a proper loop in the space of rays $\mathcal{P}$, which is associated to a non-trivial geometrical phase.

More concretely, let us start by considering a normalized state $\ket{\psi(t)} \in \mathcal{H}$ which evolves according to the Schrödinger equation 
\begin{equation}
    H(t)\ket{\psi(t)} = i \hbar \frac{d}{dt}\ket{\psi(t)}    
\end{equation}
up to a time $\tau$, which satisfies $\ket{\psi(\tau)}=e^{i\phi}\ket{\psi(0)}$. This evolution defines an arbitrary curve $\hat{C}$ in the Hilbert space $\mathcal{H}$. In the ray space $\mathcal{P}$ however this represents a closed loop $C=\Pi(\hat{C})$.

Now, define a state $\ket{\bar{\psi}(t)} = e^{-if(t)}\ket{\psi(t)}$ with $f(\tau) - f(0) = \phi$. The evolution will be cyclic in the ray space since $\ket{\bar{\psi}(\tau)} = \ket{\bar{\psi}(0)}$, and the SE yields:
\begin{equation}\label{eq:SEPhase}
    \frac{df}{dt} = 
    \bra{\bar{\psi}(t)}i\frac{d}{dt}\ket{\bar{\psi}(t)}
    - \frac{1}{\hbar} \bra{\psi(t)} H \ket{\psi(t)}.
\end{equation}
By defining
\begin{equation}\label{eq:geometricalAharonov}
    \beta \equiv \phi + \frac{1}{\hbar}\int_0^\tau \frac{1}{\hbar} \bra{\psi(t)} H \ket{\psi(t)} dt    
\end{equation}
we can integrate Eq.~\eqref{eq:SEPhase} to obtain:
\begin{equation}\label{eq:geometricalAharonov2}
    \beta = \int_0^\tau    \bra{\bar{\psi}(t)}i\frac{d}{dt}\ket{\bar{\psi}(t)} dt. 
\end{equation}
This identity shows that the phase $\beta$ is purely geometrical and depends only on the curve $C$ in $\mathcal{P}$: $\beta$ is independent of both $\phi$ and $H$. Moreover, $H(t)$ can even be chosen in an alternative way such that the dynamical term in \eqref{eq:geometricalAharonov} vanishes. Note how no approximations were made here, and the expression for the phase $\beta$ is exact: the evolution needs not to be slow neither the state $\ket{\psi(t)}$ needs to be an eigenstate of $H(t)$. As a side note, we can also check that by taking $\ket{\bar{\psi}(t)} \approx \ket{n(t)}$ we recover Berry's result \eqref{eq:geophase1}, where $\ket{n(t)}$ corresponds to an instantaneous eigenstate of $H(t)$.

If one is interested in measuring the geometrical phase~\eqref{eq:geometricalAharonov}, a few strategies are possible. One may tune $H$ in such a way that the second term in Eq.~\eqref{eq:geometricalAharonov} is zero, i.e. effectively eliminating the dynamical contribution from the phase, so that $\phi$ becomes purely geometrical. Another possible strategy is to evolve two different states such that their dynamical part is the same, so the geometrical phase (difference) can be measured through $\phi_1 - \phi_2 = \beta_1 - \beta_2$.
\begin{figure}[b!]
    \centering
    \includegraphics[width=.8\textwidth]{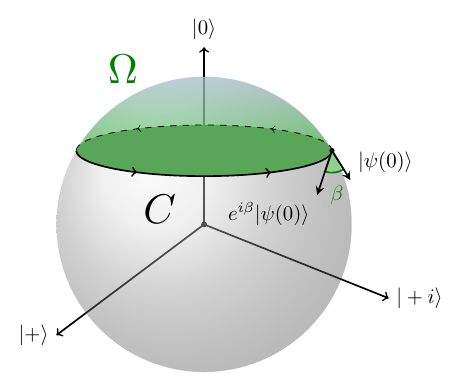}
    \caption[Loop $C$ on the Bloch sphere for the periodic evolution of a qubit under a constant magnetic field.]{Loop $C$ on the Bloch sphere for the periodic evolution of a qubit under a constant magnetic field. The geometrical phase acquired by the qubit corresponds to half the value of the solid angle $\Omega$ associated with the loop.}
    \label{fig:geometric_phase}
\end{figure}

An interesting example is the periodic evolution of a qubit subject to a constant magnetic field in the z-direction, given by $H_B = -\Omega \sigma_z$, where $\Omega = \mu B$, with $\mu$ as the magnetic moment and $B$ as the strength of the magnetic field. Consider an initial state of the form 
$\ket{\psi(t)} =\cos{(\theta/2)} \ket{0} + \sin{(\theta/2)}\ket{1}$. At a later time $t$ the wave function will be $\ket{\psi(0)} = e^{i \Omega t/\hbar}\cos{(\theta/2)} \ket{0} + e^{-i \Omega t/\hbar} \sin{(\theta/2)}\ket{1}$. The period of the evolution is $\tau=\pi \hbar/\Omega$. By direct evaluation we can see that $\ket{\psi(\tau)} = - \ket{\psi(0)}$ and $\phi = \pi$. Thus, by solving the integral in Eq.~\eqref{eq:geometricalAharonov} we get:
\begin{equation}
    \beta = \pi(1 - \cos{\theta}).
\end{equation}
A noteworthy property of this example is that the geometrical phase corresponds to half the angle enclosed by the loop on the Bloch sphere \cite{Berry1984, Aharonov1987}. This example is thus clarifying because this description in terms of the Bloch state helps us visualize what is happening geometrically. However, as we have stressed before, what is really important is the loop in the ray space $\mathcal{P}$, and not the path on the Bloch sphere itself, as this intuitive interpretation in terms of the Bloch sphere is a very particular property of our model and does not hold for all cyclic evolutions (just as the loop in the parameter space for the Berry phase is also just a particular case of a more general result).  

\section{Non-adiabatic non-abelian geometrical phases}\label{sec:NANAphase}

We can now extend this result to the non-Abelian case, following the steps from \cite{Anandan1988}. Consider a $n$-dimensional subspace $V_n(t)$ of $\mathcal{H}$. We shall consider that evolution is cyclic so this subspace is the same for the endpoints of the dynamics, i.e., $V_n(0) = V_n(\tau)$. Now, consider a decomposition $H = V_n(t) \oplus V_m(t)$ of the Hilbert space into two subspaces of dimension $n$ and $m$, respectively. As we will see later, this type of division can be really natural. For example, in several applications, such as one we will choose in subsequent sections, the computational space of the qubits is just a subspace of a larger Hilbert space, which may include auxiliary states into the implementation of a quantum gate. Another common application occurs when $V_n$ is a degenerate eigenspace of $H$, such as it happens in the Wilczek-Zee phase \cite{Wilczek1984}.

 Now, consider two orthonormal bases. First a basis $\{ \ket{\bar{\psi}_a(t)} , a=1,...,n\}$ of $V_n$ which satisfies the cyclic condition $\ket{\bar{\psi}_a(\tau)} = \ket{\bar{\psi}_a(0)}$. And second, a basis $\{ \ket{\psi_a(t)} , a=1,...,n\}$ which follows the SE:
\begin{equation}\label{eq:SEsubspace}
    i\hbar \frac{d}{dt}\ket{{\psi}_a(t)}  
    =
    H  \ket{{\psi}_a(t)}.
\end{equation}
Both bases initially coincide: $\ket{\bar{\psi}_a(0)}= \ket{\psi_a(0)}$. These two bases will be related by a unitary matrix:
\begin{equation}
     \ket{\psi_a(t)}
     =
     \sum_{b=1}^n U_{ba}(t) \ket{\bar{\psi}_b(t)}.
\end{equation}
By inserting the previous equation into Eq.~\eqref{eq:SEsubspace} we obtain an explicit form for the unitary, given by:
\begin{equation}\label{eq:geometric_unitary}
    \bm{U}(t)
    =
    \mathcal{T}
    \exp{\left( \int_0^t i (\bm{A}-\bm{K})dt \right)}.
\end{equation}
Here $\mathcal{T}$ is the time-ordering operator, $K_{ab} = (1/\hbar) \bra{\bar{\psi_a}}H\ket{\bar{\psi_b}}$ corresponds to the dynamical part of evolution and $A_{ab} = i \bra{\bar{\psi_a}} d/dt \ket{\bar{\psi}_b}$ corresponds to the geometrical part of the evolution. To see this, note that $A_{ab}$ does \emph{not} depend on $H$, but rather, only on the structure of the Hilbert space, i.e., it can be computed entirely from the basis $\ket{\bar{\psi}_a(t)}$.

Investigating how $A_{ab}$ transforms further clarifies its nature. By choosing a different basis $\ket{\bar{\psi}'} = \bm{\Omega} \ket{\bar{\psi}}$, where $\bm{\Omega}$ is a unitary, we can see that the two matrices transform as:
\begin{equation}
    \bm{A} \rightarrow i \bm{\Omega}^\dagger \dot{\bm{\Omega}}    
    + \bm{\Omega}^\dagger \bm{A} \bm{\Omega}, 
    \quad
    \bm{K} \rightarrow \bm{\Omega}^\dagger \bm{K} \bm{\Omega},
\end{equation}
showing once again that $ \bm{A}$ transforms as a vector potential and $\mathcal{A} = i \bra{\bar{\psi_a}} d \ket{\bar{\psi}_b}$ is a matrix-valued connection one-form~\cite{Chruscinski2004}. 
In this sense, we can say that $\bm{A}$ is a holonomy matrix for non-adiabatic evolutions \cite{Sjoqvist2012}. 

This observation is important because it dictates the type of model and evolution we will be interested in. Namely, we are interested in loops for which the dynamical part $\bm{K}$ vanishes. This guarantees that evolution is purely geometric for the reasons we have presented before. In this case, from Eq.~\eqref{eq:geometric_unitary} we can write the unitary implemented by a loop $C$ as
\begin{equation}\label{eq:geometric_unitary2}
    \bm{U}(C) = \mathcal{P} \exp{\left( i \oint_C \mathcal{A} \right)},
\end{equation}
where $ \mathcal{P} $ is the path-ordering operator. And finally, we are interested in the non-Abelian property of these phases, in other words, we should be able to obtain two different loops $C$ and $C'$ for which the corresponding unitaries~\eqref{eq:geometric_unitary2} do not commute. One motivation for this is in universal quantum computing, which obviously requires non-commuting gates. As we will see in the next section, the $\Lambda$-type systems are a promising platform which satify both of these requirements.
\newpage

%% file: chapters/lambda.tex
The use of geometrical phases in quantum computing, while robust against certain types of noise and error \cite{Solinas2004, Solinas2012, Viotti2021}, still suffers from decoherence and other open quantum system effects \cite{Shen2021}. This calls for strategies that are able to circumvent this type of problem \cite{Zhang2016}. Non-adiabatic holonomic quantum computing (NHQC) \cite{Sjoqvist2012} has shown promise in performing this task \cite{Johansson2012}. The considerable speed-up in the operation time makes the system much more robust against decoherence. This configuration has been used before in a couple of different experimental systems, such as nitrogen-vacancy centers in
diamond \cite{Zu2014, Arroyo-Camejo2014} and superconducting qubits \cite{AbdumalikovJr2013}. Several extensions or alternative formulations of the original proposal have already been investigated, such as implementations which shorten the original protocol by employing single-loop holonomies \cite{Sjoqvist2016, Herterich2016, Zhou2017, Xu2018} and a generalization to discrete holonomies \cite{Mommers2021}. In this section we review the original implementation of NHQC using $\Lambda$-type systems \cite{Sjoqvist2012} and its robustness against noise. 

\section{The $\Lambda$-type system}

\begin{figure}[h!]
    \centering
    \includegraphics[width=0.5\textwidth]{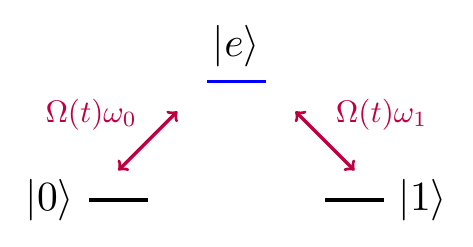}
    \caption[Basic setup for the $\Lambda$-type system.]{Basic setup for the $\Lambda$-type system.}
    \label{fig:setup}
\end{figure}

Here we follow the original setup for NHQC in the $\Lambda$-type system from Ref.~\cite{Sjoqvist2012}.  In this system, we couple two states $\ket{0}$ and $\ket{1}$ to an auxiliary excited state $\ket{e}$, while $\ket{0}$ and $\ket{1}$, which span the qubit space, are uncoupled between themselves. Thus, the system acquires a $\Lambda$-like structure, depicted in Fig.~\eqref{fig:setup}. 

The starting point to describe the model is the Hamiltonian:
\begin{equation}
H(t) = H_0 + \bm{\mu} \cdot \bm{E}(t),
\end{equation}
where $H_0 = -f_{e0} \ket{0}\bra{0} - f_{e1} \ket{1}\bra{1}$ is the bare Hamiltonian and
\begin{equation}
\bm{E}(t) = g_0(t)\cos(\nu_0 t) \bm{\epsilon}_0+ g_1(t) \cos(\nu_1 t) \bm{\epsilon}_1,    
\end{equation}
is the applied oscillating electric pulse. Here, $g_j(t)$ and $\nu_j$ (with $j=0,1$) are the pulse envelope and the oscillation frequency, respectively. Additionally, $\bm \mu$ is the magnetic dipole moment operator and $\bm{\epsilon}$ is the polarization. From hereafter, we also take $\hbar = 1$. By moving to the interaction picture Hamiltonian $H_I(t) = e^{-i H_0 t}H(t)e^{i H_0 t}$ we obtain: 
\begin{equation}\label{eq:Lambda_hamiltonian_nonres}
\begin{split}
H_I(t) 
&=
\Omega_0(t) (e^{-i(f_{e0}+\nu_0)t} + e^{-i(f_{e0} - \nu_0)t})  \ket{e}\bra{0} \\
&+ 
\Omega_1(t) (e^{-i(f_{e1}+\nu_1)t} + e^{-i(f_{e1} - \nu_1)t}) \ket{e}\bra{1}  
+ \text{h.c} .
\end{split}
\end{equation}
where $\Omega_j = \bra{e} \bm{\mu}  \cdot \bm{\epsilon} \ket{j} g_j(t)/2$ are transition frequencies which depend only on the parameters of the applied field. In a final step, we tune the frequencies $\nu_j$ so they are resonant with the bare transition frequencies $f_{ej}$, i.e. $\nu_j = f_{ej}$. By doing so one finds
\begin{equation}\label{eq:Lambda_hamiltonian}
H_I(t) 
=
\Omega_0(t) (1 + e^{-2if_{e0} t})  \ket{e}\bra{0} + \Omega_1(t) (1 + e^{-2if_{e1} t})  \ket{e}\bra{1}  + \text{h.c},
\end{equation}

Note how the bare Hamiltonian introduces counter-rotating terms of the type $(1+e^{-2if_{ej} t})$. These terms are a source of nonideality and the focus of our research, therefore, in the next chapter we will be concerned with how to handle these terms and how they affect the quantum gates. 
\begin{figure}
    \centering
    \includegraphics[width=\textwidth]{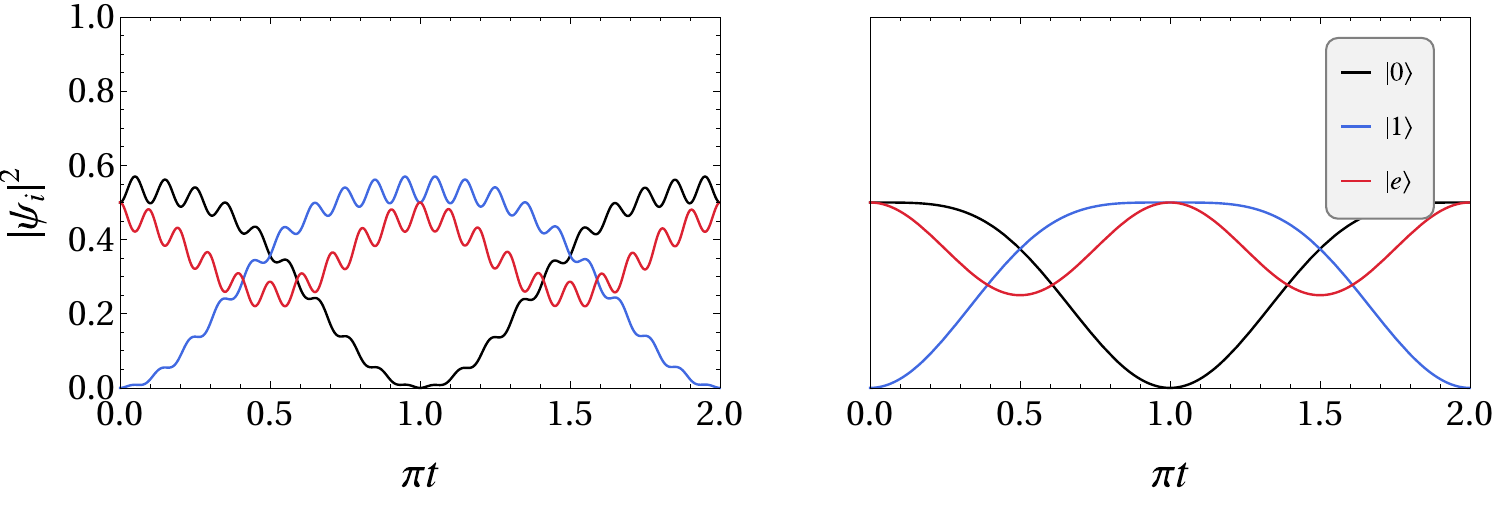}
    \caption[Effects of the counter-rotating terms in the dynamics of the $\Lambda$-system.]{Effects of the counter-rotating terms in the dynamics of the $\Lambda$-system. We compare the solutions of Eq.~\eqref{eq:Lambda_hamiltonian}. We plot the populations (left) in the presence of counter-rotating terms, where $f_{e0} = f_{e1} = 10$ and $\Omega_0(t) = \Omega_1(t) = 1/\sqrt{2}$, and (right) in the RWA regime. In both cases we take the initial state to be $\ket{\psi_0} = (\ket{0} + \ket{e})/\sqrt{2}$. As the frequencies $f_{ei}$ increase, the smaller the ripples on the non-RWA solution get in comparison to overall period of the dynamics.}
    \label{fig:dynamics1}
\end{figure}
However, for now we will be interested in the regime of the rotating-wave approximation (RWA). The argument here is that whenever $f_{ei}$ is much bigger  the other typical frequencies of the system, $e^{\pm 2 i f_{ej} t}$ become rapidly oscillating terms which average out to zero. 
\footnote{One way to justify this claim is by employing the Magnus expansion \cite{Blanes2009,Zeuch2020} for the unitary evolution. We get leading correction terms of order $O(\Omega/f_{ej})$, which are negligible whenever $\Omega/f_{ej} \ll 1$.}
In this scenario the counter-rotating terms and their effects are negligible. This allows us to describe the system and the implementation of NHQC in an ideal setting.
This can be seen in more detail in Fig.~\eqref{fig:dynamics1}. The counter-rotating terms introduces "ripples" in the solution. Meanwhile, when we average out these terms, we get the RWA Hamiltonian
\begin{equation}\label{eq:Lambda_hamiltonian_rwa}
H^{\rm RWA}_{I}(t) 
=
\Omega_0(t) \ket{e}\bra{0} + 
\Omega_1(t) \ket{e}\bra{1}  + \text{h.c},
\end{equation}
which leads to a smooth solution, seen on Fig.~\eqref{fig:dynamics1}~(b). Thus, it is possible to smooth out the dynamics of the system by increasing the energy gap of the bare states, approaching the RWA regime. 

To further comprehend the dynamics of the system, we can perform a very illustrative analysis on the Hamiltonian in Eq.~\eqref{eq:Lambda_hamiltonian_rwa} in terms of its eigenstates. Here we introduce a frequency envelope $\Omega(t)$ and the relative amplitudes $\omega_0$ and $\omega_1$, which allows us rewrite the transition frequencies as $\Omega_0(t) = \omega_0 \Omega(t) $ and $\Omega_1(t) = \omega_1 \Omega(t)$, where we assume that $|\omega_0|^2 + |\omega_1|^2 = 1$. We then define the dark state as $\ket{d} = -\omega_1\ket{0}+\omega_0\ket{1}$ and the bright state as $\ket{b} = \omega_0^*\ket{0}+\omega_1^*\ket{1}$. By doing so, we can rewrite Eq.~\eqref{eq:Lambda_hamiltonian_rwa} as:
\begin{equation}\label{eq:Lambda_hamiltonian_rwa_bd}
H^{\rm RWA}_{I}(t) 
=
\Omega(t)
(\ket{e}\bra{b} + \ket{b}\bra{e}).
\end{equation}

\begin{figure}
    \centering
    \includegraphics[width=0.5\textwidth]{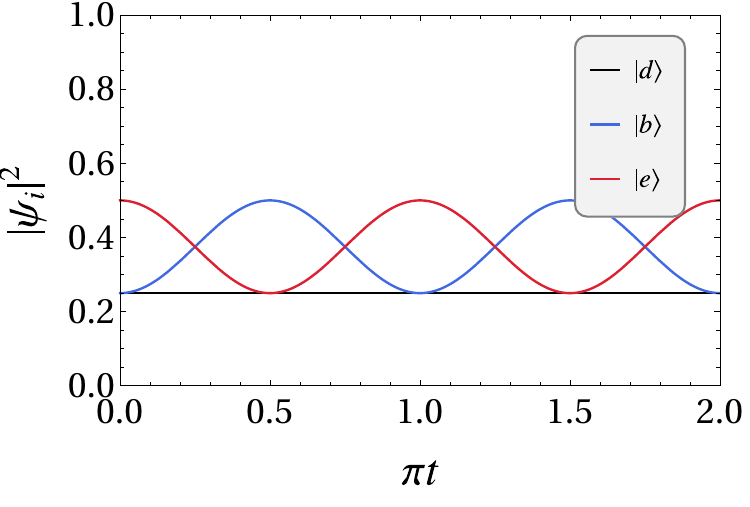}
    \centering
    \begin{tikzpicture}[scale=.5]
        \draw [ultra thick] (0,0) -- (1,0);
        \draw [ultra thick] (4,0) -- (5,0);
        \draw [ultra thick, blue] (2,2) -- (3,2);
    
        \node [left, scale = 2] at (0, 0) {$\ket{b}$};
        \node [right, scale = 2] at (5, 0) {$\ket{d}$};
        \node [above, scale = 2] at (2.5, +2) {$\ket{e}$};
    
        \draw [ultra thick, <->, purple] (0.5,0.5) -- (1.5,1.5);
        \node [above left, scale = 1.5, purple] at (1, 1) {$\Omega(t)$};
        
        \draw [white] (-5, 0) -- (0, -4);
        
    \end{tikzpicture}
    \caption[Dynamics of the $\Lambda$-system in the bright-dark basis.]{Evolution of the amplitudes of the state $\psi_0$ defined in Fig.~\eqref{fig:dynamics1} in terms of the bright and dark states (left). We also depict the $\Lambda$-type-system in this case (right).}
    \label{fig:dynamics2}
\end{figure}
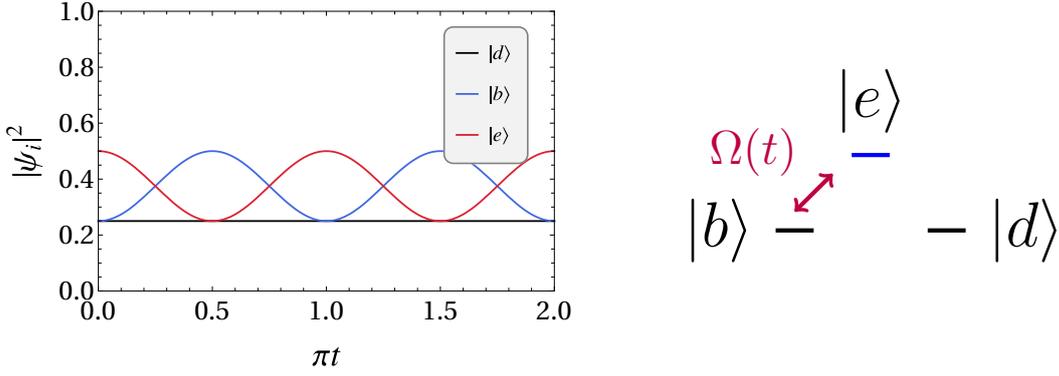
Thus, we can see the convenience of this change of basis; the state $\ket{d}$ is decoupled from the evolution, and we effectively get a two-level system with oscillations between the bright state $\ket{b}$ and the auxiliary state $\ket{e}$. In Fig.~\eqref{fig:dynamics2} we depict the Rabi oscillations between these two states. Note how the amplitude of $\ket{d}$ remains unchanged throughout the evolution.

\section{Implementing single-qubit gates}\label{sec:singlequbit_gates}

Now we can finally make use of the formalism from the previous chapter to show how the geometrical properties of the $\Lambda$-type system can be used in order to implement universal single-qubit gates. In our implementation we are interested in the qubit subspace $M(0) = \text{span }\{\ket{0}, \ket{1}\}$, while the state $\ket{e}$ plays the important auxiliary role. This space evolves into $M(t)$, which is spanned by:
\begin{equation}\label{eq:se_solution}
\ket{\psi_k(t)} = \exp\left( - i \int_0^t H_I^{\rm RWA}(t') dt' \right) \ket{k}
=
\mathcal{U}(t, 0) \ket{k},
\end{equation}
where $k=0, 1$ and $\mathcal{U}(t, 0)$ is the time-evolution operator. Notice that in the case of the RWA Hamiltonian~\eqref{eq:Lambda_hamiltonian_rwa} it is not necessary to include a time-ordering operator. It is also possible to get some insight into the dynamics by writing these expressions in terms of the bright-dark basis. The unitary matrix becomes
\begin{equation}
\mathcal{U}_{bd}(t, 0)
=
\ket{d}\bra{d}+
\cos(\Phi)(\ket{b}\bra{b} + \ket{e}\bra{e}) - i \sin(\Phi)(\ket{e}\bra{b} + \ket{b}\bra{e})
\end{equation}
where 
\begin{equation}\label{eq:pulse_area_int}
\Phi
=
\int_0^t \Omega(t')dt'
\end{equation}
is the area enclosed by the pulse. Consequently, the bright and dark states evolve as
\cite{Herterich2016}:
\begin{equation}\label{eq:evo_states}
\begin{split}
\ket{\psi_d(\Phi)} &= \ket{d},\\
\ket{\psi_b(\Phi)} &= \cos(\Phi)\ket{b} - i\sin(\Phi)\ket{e}.
\end{split}
\end{equation}
In particular, we are interested in pulses which satisfy
\begin{equation}\label{eq:pulse_area}
\Phi = \Phi_C := \pi.
\end{equation}
In this case, we can see that the states evolve as
\begin{equation}\label{eq:bd_basis_holonomic}
\begin{split}
\ket{\psi_d(\Phi_C)} &=  \ket{d},\\
\ket{\psi_b(\Phi_C)} &= -\ket{b},\\
\ket{\psi_e(\Phi_C)} &= -\ket{e}.
\end{split}
\end{equation}
The effect of this geometrical evolution whenever $\Phi = \pi$ is to implement a holonomy matrix $Z_{bd}$ which acts by flipping the sign of $\ket{b}$ and $\ket{e}$ in the bright-dark basis. In other words:
\begin{equation}
\mathcal{U}_{bd}(C)
=
\begin{matrix}
\begin{pmatrix}
1 & 0 &0\\
0 & -1 &0\\
0 & 0 &-1
\end{pmatrix}
\end{matrix}
.
\end{equation}
An explicit calculation shows that this matrix, in the computational basis, becomes, after properly projecting it back onto the qubit space:

\begin{equation}\label{eq:single_loop_uni}
U(C) 
=
\mathcal{U}_{bd}(C) \mathbb{P}
= 
\begin{matrix}
\begin{pmatrix}
\cos \theta & e^{-i \phi } \sin \theta\\
 e^{i \phi } \sin \theta\ & -\cos \theta\\
\end{pmatrix}
\end{matrix}
=
\bm{n} \cdot \bm{\sigma}
\end{equation}
where $\bm{n} = (\sin \theta \cos \phi, \sin \theta \sin \phi, \cos \theta)$ and $\mathbb{P} = \ket{0}\bra{0} + \ket{1} \bra{1}$. 
\footnote{
This is a slight abuse of notation, since $U(C)$ is still a $3$ x $3$ matrix and proper projection operators should be rectangular matrices for our intended purpose. Nevertheless, $U(C)\ket{e} = 0$ so, unless explicitly stated, we write all the unitaries and relevant operators in the subspace of the qubit states for simplicity.}
Here, we have  parametrized the frequencies $\omega_0$ and $\omega_1$ as $\omega_0=\sin(\theta/2)e^{i\phi}$ and $\omega_1=-\cos(\theta/2)$. Besides the convenient representation for the unitary, this also guarantees that $|\omega_0|^2 + |\omega_1|^2 = 1$. Thus, this process implements a $\pi$ rotation around $\bm{n}$ on the Bloch sphere .
\footnote{
To see this, note that Eq.~\eqref{eq:single_loop_uni} can be alternatively written as $U(C) = ie^{-i \frac{1}{2} \pi (\bm{n} \cdot \bm{\sigma})}$
\cite{Sjoqvist2012, Herterich2016}. 
}
This unitary, however, is not universal; Eq.~\eqref{eq:single_loop_uni} only implements traceless operations. By employing a second loop $C_{\bm{m}}$ we can implement the universal gate:
\begin{equation}\label{eq:ideal_holonomy}
U(C) = U(C_{\bm{m}})U(C_{\bm{n}}) 
=
\bm{n} \cdot \bm{m}
- i \bm{\sigma} \cdot (\bm{n} \times \bm{m}).
\end{equation}

This transformation has a clear geometrical meaning as well. The universal gate $U(C)$ above corresponds to a rotation in the plane spanned by $\bf{n}$ and $\bf{m}$ by an angle of $2\cos^{-1}(\bf{n} \cdot \bf{m})$. Therefore, any single-qubit gate can be obtained by properly choosing the appropriate pulses, which will determine $\bf{n}$ and $\bf{m}$.

\begin{figure}
    \centering
    \includegraphics[width=\textwidth]{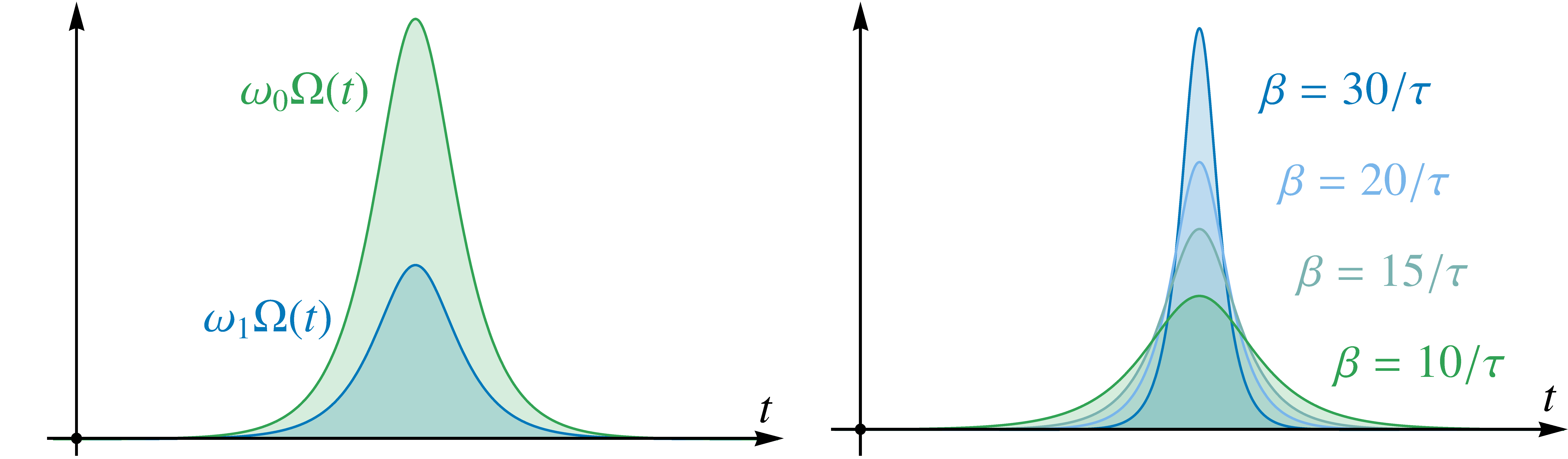}
    \caption[Illustration of $\sech$ pulses.]{(a) A representation of the hyperbolic secant pulse from Eq.~\eqref{eq:sech_pulse}. (b) Representation of the pulse for different values of $\beta$. Note how $1/\beta$ can be seen as the length of the pulse. Here $\tau$ is an arbitrary time scale.}
    \label{fig:pulses}
\end{figure}

Under these considerations, we are going to choose a very convenient type of pulse, namely, pulses which have the shape of a hyperbolic secant:
\begin{equation}\label{eq:sech_pulse}
\Omega(t) =  \beta \sinh(\beta t).
\end{equation}
This parametrization is convenient because it guarantees that condition~\eqref{eq:pulse_area} is satisfied for whichever $\beta$ is chosen. This choice gives us freedom of using different pulse lengths just by changing the parameter $\beta$. As shown in Fig.~\eqref{fig:pulses}, $1/\beta$ can be interpreted as the length of the pulse: pulses are sharper for large values of $\beta$ and wider when $\beta$ is small. Moreover, by appropriately choosing the angles $\theta$ and $\phi$ we can determine the vector $\bf{n}$ (and by choosing a different set of angles for the second pulse we determine $\bf{m}$). Pulses of this type are experimentally friendly and exhibit desirable properties in quantum optics and quantum control \cite{Rosen1932, Roos2004, Economou2006, Economou2007}. Of course, other pulses can also be used, such as Gaussian and square pulses \cite{Spiegelberg2013, Sjoqvist2016}.

As a practical example, we consider the Hadamard gate $H$ and the $S$ gate, which are defined as
\begin{equation}
H
=
\frac{1}{\sqrt{2}}
\begin{matrix}
\begin{pmatrix}
1 & 1\\
1 & -1
\end{pmatrix}
\end{matrix}
\quad
\text{and} 
\quad
S
=
\begin{matrix}
\begin{pmatrix}
1 & 0\\
0 & i
\end{pmatrix}
\end{matrix},
\end{equation}
respectively. For the Hadamard gate we need only a single pulse, choosing $\theta = \pi/4$ and $\phi = 0$. For the $S$-gate the protocol is slightly more complicated, since we need the two pulses. In this case, we can take $\theta_0 = \pi/2$ and $\phi_0 = \pi/2$ for the first pulse and $\theta_1 = 0$ and $\phi_1 = \pi/4$ for the second pulse.

The drawback of gates which require two loops, such as the $S$-gate, is that the exposure time to decoherence effects is longer. We will see this explicitly in the next chapter. Protocols which implement non-adiabatic quantum gates for single-loops which can mitigate this effect can be found in \cite{Sjoqvist2016} and \cite{Herterich2016}.

On a closing note, we comment on the geometrical aspect of the model, checking for the condition discussed in Sec~\eqref{sec:NANAphase} for a fully geometrical evolution. We can see, from Eqs.~\eqref{eq:Lambda_hamiltonian_rwa_bd} and~\eqref{eq:evo_states}, that the dynamical contribution is $\bra{\psi_k(t)} H \ket{\psi_l(t)} = 0$. for $k, l = b, d$. Since $\ket{d}$ decouples from the Hamiltonian in Eq.~\eqref{eq:Lambda_hamiltonian_rwa_bd}, and $\ket{\psi_d(t)} = \ket{d}$, it is easy to see that $\bra{\psi_k(t)} H \ket{\psi_l(t)}$ vanishes for any inner product involving the dark state. Finally, we can show that $\bra{\psi_b(t)} H \ket{\psi_b(t)} = 0$ by direct calculation, since
\[
\bra{\psi_b(t)} H \ket{\psi_b(t)} = \Omega(t)(\cos \Phi \bra{e} + i \sin \Phi \bra{b})(\ketbra{e}{b} + \ketbra{b}{e})(\cos \Phi \ket{e} - i \sin \Phi \ket{b}) = 0.
\]
This proves the geometric nature of the gate: the dynamical contributions in $K_{kl} = \bra{\psi_k(t)} H \ket{\psi_l(t)}$ in Eq.~\eqref{eq:geometric_unitary} vanish, and we are left with the geometrical contributions in Eq.~\eqref{eq:geometric_unitary2} only \cite{Sjoqvist2012}.

\section{Implementing two-qubit gates}\label{sec:twoqubitgates}

In this section we once more follow the original proposal in \cite{Sjoqvist2012}, which also includes a protocol based on the  S\o rensen–M\o lmer scheme \cite{Sorensen1999} for implementing two-qubit gates (an adiabatic implementation is presented in \cite{Duan2001}. See also Ref.~\cite{ZhaoXu2019} for a multi-qubit generalization of the NHQC scheme). The two-qubit gate can be constructed with two ions in the same three-level $\Lambda$ configuration. 
The transition $0 \rightarrow e$ ($1 \rightarrow e$) is driven by a laser with detuning $\nu \pm \delta$ $[\mp (\nu \pm \delta)]$, where $\nu$ is the phonon frequency of a vibrational mode \cite{Duan2001, Kirchmair2009} and $\delta$ is an additional detuning [Fig.~\eqref{fig:iontwoqubit}]. Moreover, two extra conditions which the setup should satisfy are: the Lamb-Dicke criterion $\eta \ll 1$, where $\eta$ is the Lamb-Dicke parameter, and $|\Omega_i(t)| < \nu$ in order to suppress the off-resonant couplings \cite{Sorensen1999}. The parameter $\eta$ is related to the zero-point spread of the ion \cite{Wineland1998} and the coupling strength between its motional states and internal degrees of freedom \cite{Li2002}. The effective Hamiltonian describing this interaction assumes the form (see Ref.~\cite{Kim2008} for a derivation):
\begin{figure}[b!]
    \centering
    \includegraphics[width=0.5\textwidth]{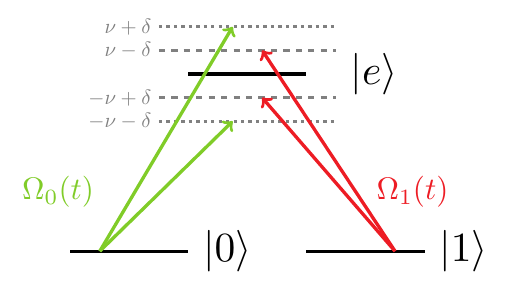}
    \caption[Setup for the ions in the two-qubit gate.]{Setup for the ions in the two-qubit gate. The green (red) arrows and the dotted (dashed) lines correspond to the $\nu \pm \delta$ ($\nu \mp \delta$) detuning of the $0 \rightarrow e$ ($1 \rightarrow e$) transition. Both ions have the same configuration.}
    \label{fig:iontwoqubit}
\end{figure}
\begin{equation}
    H^{(2)} = \frac{\eta^2}{\delta}\left(
    |\Omega_0(t)|^2 \sigma_0(\phi, t) \otimes \sigma_0(\phi, t) - |\Omega_1(t)|^2 \sigma_1(-\phi, t) \otimes \sigma_1(-\phi, t)
    \right),
\end{equation}
where
\begin{equation}
\begin{split}
    \sigma_0(\phi, t) = e^{i \phi/4}(1 + e^{-2i f_{e0} t})\ket{e}\bra{0} + {\rm h.c.},\\
    \sigma_1(-\phi, t) = e^{-i \phi/4}(1 + e^{-2i f_{e1} t})\ket{e}\bra{1} + {\rm h.c}.
\end{split}
\end{equation}
After eliminating off-resonant couplings of the singly excited states $\ket{0e}$ , $\ket{e0}$, 
$\ket{1e}$ and $\ket{e1}$ and performing the RWA the Hamiltonian reads:
\begin{equation}\label{eq:twoqubit_total}
    H^{(2), {\rm RWA}}(t) 
    =
    \frac{\eta}{\delta^2}
    \sqrt{|\Omega_0(t)|^4 + |\Omega_1(t)|^4}
    \left(H^{(2), {\rm RWA}}_0(t)
    +
    H^{(2), {\rm RWA}}_1(t)\right), 
\end{equation}
with
\begin{equation}\label{eq:twoqubit1}
    H^{(2), {\rm RWA}}_0(t) 
    =
    \sin{\frac{\theta}{2}} e^{i \phi/2} \ket{ee}\bra{00} 
    - \cos{\frac{\theta}{2}} e^{-i\phi/2} \ket{ee}\bra{11} +\rm h.c.
\end{equation}
and
\begin{equation}\label{eq:twoqubit2}
    H^{(2), {\rm RWA}}_1(t) 
    =
    \sin{\frac{\theta}{2}} \ket{e0}\bra{0e} - \cos{\frac{\theta}{2}}  \ket{e1}\bra{1e} + \rm h.c.
\end{equation}

The single and two-qubit gates share similar criteria: the phase $\phi$ should be kept constant throughout the evolution, while the frequencies satisfy $|\Omega_0(t)|^2/|\Omega_1(t)|^2 = \tan(\theta/2)$. Additionaly, we should once again respect the criterion~\eqref{eq:pulse_area} for the pulse area:
\begin{equation}
    \frac{\eta^2}{\delta}
    \int_0^\tau
    \sqrt{|\Omega_0(t)|^4 + |\Omega_1(t)|^4}dt
    =
    \pi.
\end{equation}
Moreover, one extra observation is in order; since $H^{(2)}_0(t)$ and $H^{(2)}_1(t)$ commute, it is possible to decompose the evolution of the total Hamiltonian~\eqref{eq:twoqubit_total} as
\[
    \exp{\left(-i \int_0^\tau H^{(2), {\rm RWA}} (t) dt\right)}
    =
    \exp{\left(-i \pi H_0^{(2), {\rm RWA}} \right)}
    \exp{\left(-i \pi H_1^{(2), {\rm RWA}} \right)}.
\]
The Hamiltonian $H^{(2)}_1(t)$ however acts trivially on the relevant computational subspace $\{\ket{00}, \ket{01}, \ket{10}, \ket{11}\}$ \cite{Sjoqvist2012}, making $H^{(2)}_0(t)$ the relevant term in the evolution. Due to this fact, by redefining some parameters in Eq.~\eqref{eq:twoqubit_total}, $H^{(2)}(t)$ can effetively be seen as a $\Lambda$-type-like Hamiltonian \cite{ZhaoXu2019} 
\begin{equation}
H_{\mathrm eff}^{(2), {\rm RWA}}
= 
\Omega^{(2)}(t) 
(\omega_0^{(2)}\ketbra{00}{ee} + \omega_1^{(2)} \ketbra{11}{ee})
+
\rm h.c.
\end{equation}
with $\Omega^{(2)}(t) = \sqrt{|\Omega_0(t)|^4 + |\Omega_1(t)|^4}$,  $\omega_{1}^{(2)} = e^{i\frac{\phi}{2}}\sin{(\theta/2)}$ and $\omega_{0}^{(2)} = -e^{-i\frac{\phi}{2}}\cos{(\theta/2)}$. Thus, by analogy to the single qubit gate, as done in Eqs.~\eqref{eq:se_solution}-\eqref{eq:single_loop_uni}, we get the corresponding unitary:
\begin{equation}
\begin{split}
    U^{(2)}(C_n)
    &=
    \cos{\theta}\ket{00}\bra{00}+
    e^{-i\phi}\sin{\theta}\ket{00}\bra{11}+
    e^{i\phi}\sin{\theta}\ket{11}\bra{00}\\
    &-\cos{\theta}\ket{11}\bra{11}
    +\ket{01}\bra{10}
    +\ket{10}\bra{10}.
\end{split}
\end{equation}
This unitary acts just like a single qubit gate in the space $\{\ket{00},\ket{11}\}$ and it leaves the components in the space  $\{\ket{01}, \ket{10}\}$ invariant. 

By choosing $\theta = 0$ we construct a CZ gate
\begin{equation}
\begin{split}
    U^{(2)}_{CZ}
    &=
    \ket{00}\bra{00}
    +\ket{01}\bra{10}
    +\ket{10}\bra{10}
    -\ket{11}\bra{11},
\end{split}
\end{equation}
which is an entangling gate and can form a universal set together with other single-qubit gates \cite{Bremner2002}.

Finally, if we take the counter-rotating terms into account, Eq.~\eqref{eq:twoqubit1} and Eq.~\eqref{eq:twoqubit2} become
\begin{equation}\label{eq:twoqubit1nonRWA}
    H^{(2)}_0(t) 
    =
    (1 +  e^{-2 i f_{e0} t})^2\sin{\frac{\theta}{2}} e^{i \phi/2} \ket{ee}\bra{00} 
    - (1 +  e^{-2 i f_{e1} t})^2 \cos{\frac{\theta}{2}} e^{-i\phi/2} \ket{ee}\bra{11} +\rm h.c.
\end{equation}
and
\begin{equation}\label{eq:twoqubit2nonRWA}
    H^{(2)}_1(t) 
    =
    4 \cos^2{(f_{e0} t)} \sin{\frac{\theta}{2}} \ket{e0}\bra{0e} - 4 \cos^2{(f_{e1} t)} \cos{\frac{\theta}{2}}  \ket{e1}\bra{1e} + \rm h.c.,
\end{equation}
respectively.

\section{Gate robustness}\label{sec:robustness}

Throughout this work, we have mentioned the robustness of the geometric quantum gates  several times. Our objective in this section is to make this claim more precise, explicitly showing, by numerical means, how the NHQC implementation in the $\Lambda$-system is sufficiently robust against open quantum systems effects. We discuss the impact of   amplitude damping in the system, as it was originally done in Ref.~\cite{Sjoqvist2012}. 
The interested reader can find an investigation on the robustness of the model against a few other error sources in Ref.~\cite{Johansson2012}. For instance, it was shown by the authors that contributions due to errors in the pulse envelope and relative pulse amplitude are all of \emph{second} order and, more strikingly, independent of the pulse duration. We omit these results here, since we will not consider this type of error in our analysis in the next chapter.

\begin{wrapfigure}{r}{.5\textwidth}
    \includegraphics[width=.5\textwidth]{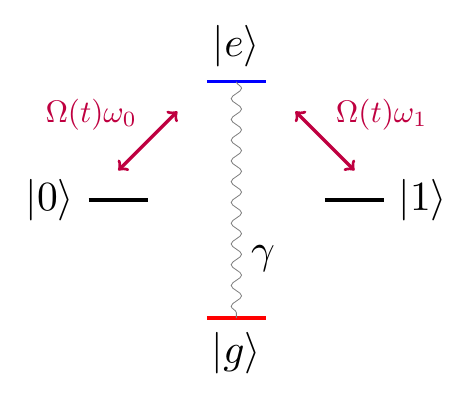}
    \caption[Basic setup for the $\Lambda$-type system when  an open quantum system approach is considered.]{Basic setup for the $\Lambda$-type system when  an open quantum system approach is considered. The excited state decays with a rate $\gamma$. There is no coupling or decay with the computational subspace itself. This means that dissipation effects only occur while the pulse is being applies and the excited state is populated.}
    \label{fig:setup_dissipative}
\end{wrapfigure}
We consider the following model: in the $\Lambda$-type system, the excited state $\ket{e}$ is  an unstable state which undergoes dissipation, while the computational states $\ket{0}$ and $\ket{1}$ can be regarded as stable ground states. Thus, a physical description of the model will look like Fig.~\eqref{fig:setup_dissipative}. In our formulation we assume that the excited state decays to an auxiliary ground state $\ket{g}$ which is \emph{not} coupled to any other states through any type of unitary dynamics. 
This means that whenever the ground state is populated, these excitations are "lost" from the point of view of the computational subspace. This will result in a non-ideal mixed state at the end of the computation, which will decrease the overall fidelity of the gate. Our purpose is precisely to investigate how to mitigate this effect.\footnote{Of course, other formulations are also possible. 
For instance, in transmon qubits the state $\ket{0}$ is already encoded into the ground state and there is no further low-lying level $\ket{g}$. 
Thus, in \cite{AbdumalikovJr2013} decoherence is modeled by considering the dephasing dissipators $D_\rho(\ketbra{1}{1}-\ketbra{0}{0})$ and $D_\rho(\ketbra{e}{e}-\ketbra{0}{0})$.}

We will model the decay with an amplitude-damping jump operator given by $L=\ket{g}\bra{e}$. Under a purely dissipative evolution this term will push the excitations in $\ket{e}$ towards the ground state $\ket{g}$ (and, of course, this process will also damp the coherences of the system, together with the populations). Initially considering the RWA case, the dynamics of the system will then be given by
\begin{equation}\label{eq:RWAdissipative}
 \frac{d\rho}{dt}
    =
    i[\rho, H^{\rm RWA}_I(t) ]
    +
    \gamma D_\rho(\ket{g}\bra{e}), 
\end{equation}
where $H^{\rm RWA}_I(t)$ is the RWA Hamiltonian in  Eq.~\eqref{eq:Lambda_hamiltonian_rwa}. Robustness against this type of noise has been shown in \cite{Sjoqvist2012}. Meanwhile, the effect of dephasing is qualitatively similar and has also been investigated in \cite{Johansson2012}. The takeaway message in this case was that the inverse pulse length should be much larger than the typical coupling strength $\gamma$, i.e. we should have $\beta \gg \gamma$. By increasing $\beta$ the fidelity also monotonically increases, approaching unity. This is precisely one of the advantages of this non-adiabatic scheme: by removing constraints on the operation time one can use shorter pulses, diminishing the effects of decoherence.

As a sanity check, we can, by turning off the unitary part in Eq.~\eqref{eq:RWAdissipative}, also understand how the dissipative part acts on the system. For a pure initial state of the type $\ket{\psi} = c_0\ket{0} + c_1\ket{1} + c_e\ket{e}$, the density matrix evolves as:
\begin{equation}
    \rho(t)
    =
    \begin{matrix}
    \begin{pmatrix}
    |c_0|^2 & c_0 c_1^* & e^{-\gamma t/2} c_0 c_e^* & 0\\
    c_1 c_0^* & |c_1|^2 & e^{-\gamma t/2} c_1 c_e^* & 0\\
    e^{-\gamma t/2} c_e c_0^* &  e^{-\gamma t/2}  c_e c_1^* & e^{-\gamma t} |c_e|^2 & 0\\
    0 & 0 & 0 & |c_e|^2(1-e^{-\gamma t})\\
    \end{pmatrix}
    \end{matrix}.
\end{equation}

We can see that the process does not interfere with the computational subspace, but it dampens the population of the excited state and destroys the coherences. In the steady state limit $\gamma \tau \rightarrow \infty$, all the population from the excited state is transferred to the ground state. Hence, what happens when we turn off the unitary interaction is that the dissipative process occurs only during the pulse application, since there is no coupling between the computational subspace and the ground state. Thus, by shortening the pulse we also shorten the time during which the excited state is occupied and we minimize the errors due to dissipation. This is of course consistent with what we expected when choosing the jump operators to model the dissipation in this setup, so this result should not be seen as surprising.

We can now turn on the unitary part of interaction and try to quantify how well this protocol performs. For that, we will take the initial state $\ket{\psi_0} = \ket{0}$. Our gate of choice will be the Hadamard gate, whose implementation was discussed in Sec.~\eqref{sec:singlequbit_gates}. Ideally, the application of the Hadamard gate upon $\ket{0}$ should yield $\ket{+} = (\ket{0} + \ket{1})/\sqrt{2}$. 

The figure of merit for gate performance will be the fidelity, defined for two arbitrary mixed states $\rho$ and $\sigma$ as \cite{Uhlmann1976, Jozsa1994}
\begin{equation}\label{eq:fidelity_full}
   F(\rho, \sigma)
    = 
    \left(\tr{\sqrt{\sqrt{\rho} \sigma \sqrt{\rho}}}\right)^2.
\end{equation}
The fidelity essentially quantifies how similar two states are, being $0$ for orthogonal states and $1$ when both states are equal. Since we want to compare our output state with an expected result $\ket{\psi}$ which is always pure, the fidelity assumes a much simpler form. In this case, it will be
\begin{equation}\label{eq:fidelity}
    \mathcal{F}
    =
    F(\rho, \ket{\psi}\bra{\psi})
    = 
    \bra{\psi} \rho \ket{\psi}, 
\end{equation}
where $\ket{\psi} = U(C) \ket{\psi_0}$ is the ideal output upon the application of the ideal quantum gate $U(C)$ and $\rho$ is the density matrix obtained by evolving the system under Eq.~\eqref{eq:RWAdissipative} for the whole duration of the pulse. We may also use the infidelity $1-\mathcal{F}$ whenever it is convenient.

\begin{figure}
    \centering
    \includegraphics[width=\textwidth]{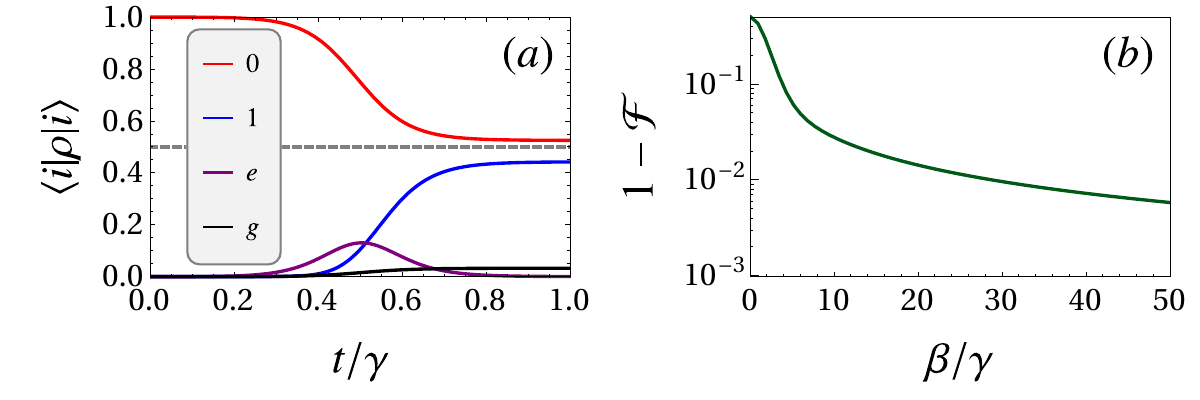}
    \caption[Open quantum system effects in the dynamics of the $\Lambda$-system.]{We plot (a) the populations of the density matrix for the implementation of the Hadamard gate and (b) the infidelity $1-\mathcal{F}$ of the gate as a function of the pulse duration $\beta$. In both cases the input state is $\ket{0}$. The gray dashed line highlights the point $\bra{i}\rho\ket{i}=0.5$.}
    \label{fig:dissipative}
\end{figure}

Basic results are shown in Fig.~\eqref{fig:dissipative}. On panel (a) we can see that the application of the pulse temporarily populates the excited state $\ket{e}$, transferring excitations from $\ket{0}$ to $\ket{1}$. Meanwhile, the ground state is also slightly populated due to these dissipative effects, and some of these excitations are "lost", decaying to the ground state $\ket{g}$. And while we did not plot the coherences here, they are also affected by this dissipative process. At the end of the protocol we get a mixed state which deviates from the pure target state. We can quantify this effect by investigating how the end result changes depending on the (inverse) duration $\beta$ of the pulse. In  Fig.~\eqref{fig:dissipative}(b) we plot these results, showing the infidelity $1-\mathcal{F}$ as a function of the inverse pulse length relative to the dissipation strength. We can see that it monotonically decreases as we increase $\beta/\gamma$. Thus, as long as the RWA is valid, a decrease in the pulse length results in a monotonically increasing fidelity, approaching unity asymptotically.


\newpage

%% file: chapters/results_qc.tex
In the previous section we considered an ideal implementation of the $\Lambda$-type system
where the counter-rotating frequencies from Eq.~\eqref{eq:Lambda_hamiltonian} are rapidly oscillating. This yields the RWA Hamiltonian from Eq.~\eqref{eq:Lambda_hamiltonian_rwa} whose robustness under amplitude noise we analyzed in Sec.~\ref{sec:robustness}. In this ideal scenario the fidelity can be arbitrarily improved simply by shortening the pulse duration as much as necessary. However, we are interested in what happens in a regime where the RWA is not valid. 

\begin{figure}[t]
    \includegraphics[width=\textwidth]{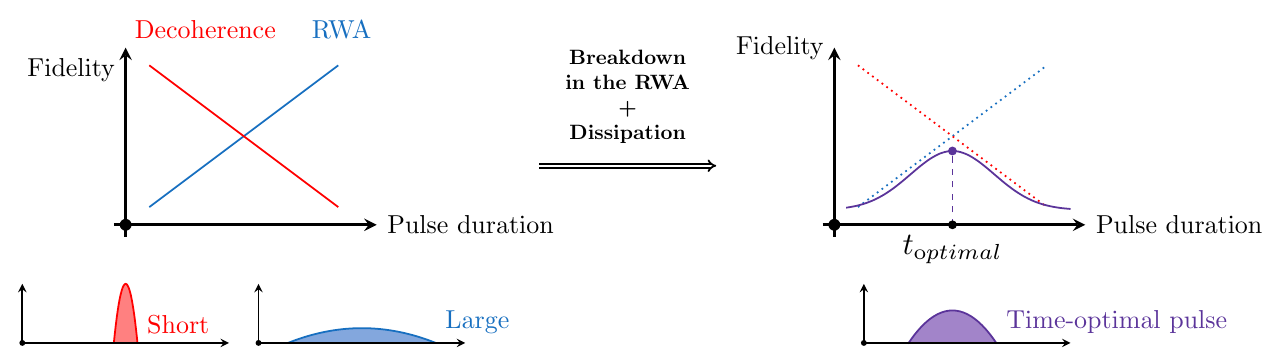}
    \caption[A diagram illustrating the competition between decoherence and the counter-rotating contributions.]{A diagram illustrating the competition between decoherence and the counter-rotating contributions. By \emph{increasing} the pulse length we approach the RWA regime, mitigating the effects of the counter-rotating terms. On the other hand, by \emph{decreasing} the pulse duration the dissipative losses are diminished. When both effects are considered at once the optimal pulse duration lies in an interval in-between these two scenarios.}
    \label{fig:time_optimal_drawing}
\end{figure}

An extension of the original proposal in Ref.~\cite{Sjoqvist2012} for the non-RWA case has been made in the past \cite{Spiegelberg2013}, albeit for the dissipationless model. There, the authors show that the RWA starts to break down at very small operation times due to the comparable time scale between the pulse and counter-rotating frequencies. Unfortunately, this constrains how fast these gates can operate in a scenario where one is trying to mitigate decoherence, making it impossible to employ increasingly shorter pulses beyond a certain limit. 
Instead, it is also important to account for the validity of the RWA; while it is necessary to use sufficiently short pulses in order to avoid dissipative losses, the pulses should also be long enough so as not to introduce significant dynamical corrections due to the counter-rotating terms. Our contribution here is to consider both effects at once, investigating the interplay between the counter-rotating oscillations and the dissipative phenomena in the system. 
We show how the combination of the breakdown in the RWA and the open quantum system effects introduce a trade-off in the pulse length, limiting the maximum achievable fidelity of the gate. This is illustrated in Fig.~\ref{fig:time_optimal_drawing}.

In this chapter we perform a series of numerical investigations in order to calculate the optimal pulse duration and the corresponding fidelity in a few cases of interest. Moreover, we also analyze the Hamiltonian in the bright-dark basis when counter-rotating corrections are considered. Surprisingly, these terms introduce a coupling between the dark and the excited state which, differently from what is seen in Eq.~\eqref{eq:Lambda_hamiltonian_rwa_bd}, is not present in the RWA regime, further hampering the protocol. Finally, we also investigate the effect of heterogeneous counter-rotating frequencies in the system. We show that they may, very slightly, improve the fidelity of two-loop gates. This chapter constitutes on our original results, discussed in Ref.~\cite{Alves2022b}. Note that some similar investigations in a different context also exist. In such cases, the trade-off occurs between decoherence and the (non-ideal) finite run-time of adiabatic implementations instead \cite{Florio2006, Lupo2007}.

\section{Counter-rotating effects in the bright-dark basis}

\begin{figure}
    \centering
    \includegraphics[width=\textwidth]{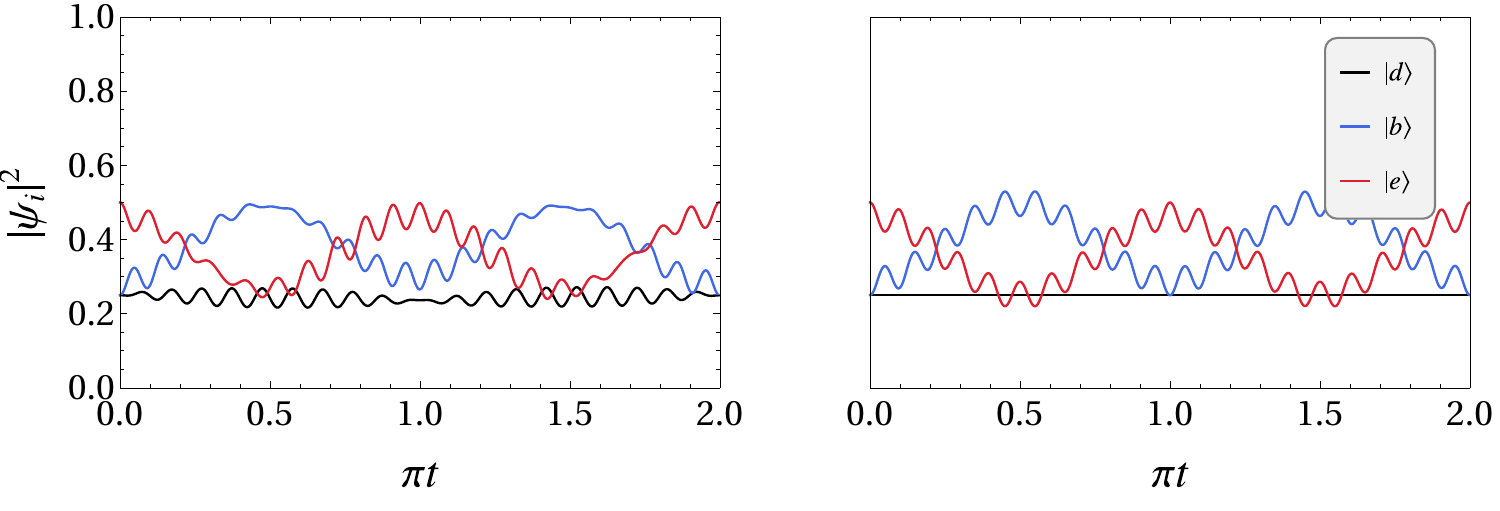}
    \caption[Dynamics of the amplitudes of the bright and dark states in the $\Lambda$-system]{Dynamics of the amplitudes of the bright and dark states in the $\Lambda$-system, according to Eq.~\eqref{eq:Hbd_rotating}, with $f_{0e} = 10$ and $f_{1e} = 11$ (left) and Eq.~\eqref{eq:Hbd_rotating2} with $f_{0e}=f_{1e}=f=10$ (right). The initial state is $\ket{\psi_0} = (\ket{0} + \ket{e})/\sqrt{2}$ and the pulse is specified by the parameters $\Omega(t)=1$ and $\omega_0 = \omega_1 = 1/\sqrt{2}$.}
    \label{fig:dynamicsBD1}
\end{figure}

\begin{figure}[h!]
    \centering
    \includegraphics[width=\textwidth]{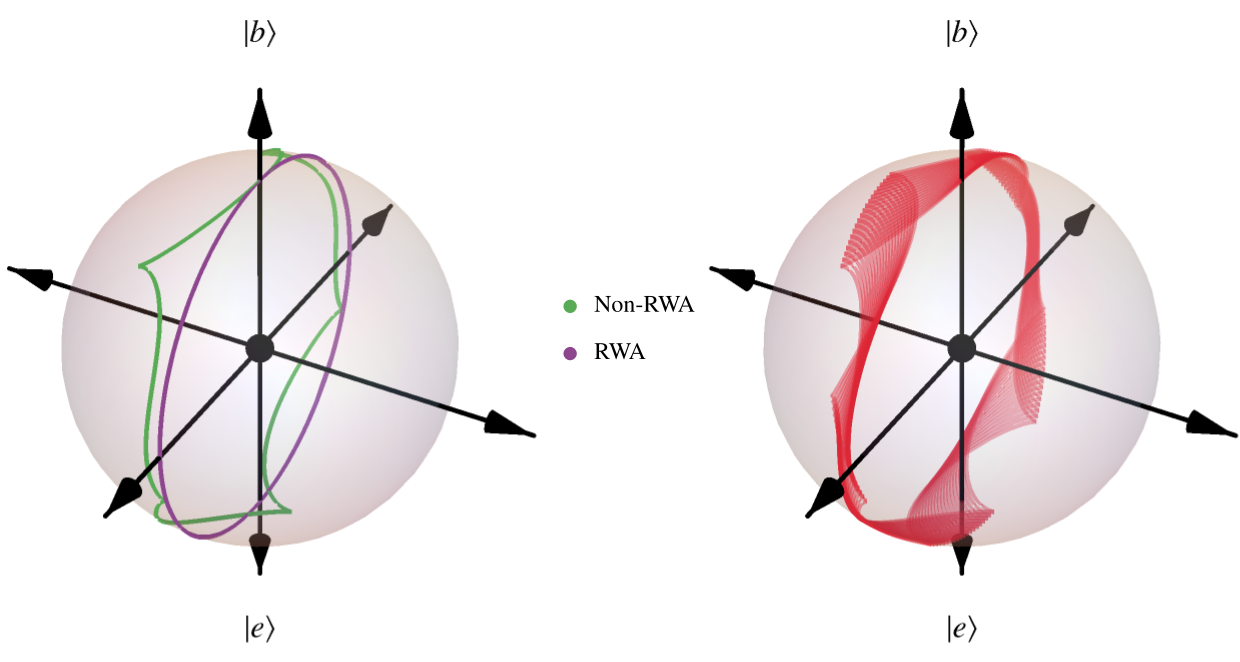}
    \caption[$\Lambda$-system in the Bloch sphere.]{(Left) Trajectory of RWA and non-RWA evolution on the Bloch sphere for the state $\ket{\psi_0} = (\ket{0}+\ket{1})/\sqrt{2}$. (Right) Trajectory on the Bloch sphere over a long time. Other parameters are the same as in Fig.~\ref{fig:dynamicsBD1}.}
    \label{fig:BlochDynamics1}
\end{figure}

During this section we will, once again, briefly disregard decoherent effects and go back to the Hamiltonian in Eq.~\eqref{eq:Lambda_hamiltonian}. It is meaningful to ask ourselves what happens when we rewrite this full Hamiltonian in terms of the dark and the bright states. When we do that, we get
\begin{equation}\label{eq:Hbd_rotating}
\begin{split}
H_{bd}(t) & =
\Omega(t)
(1+|\omega_0|^2e^{-2i f_{e0} t} + |\omega_1|^2e^{-2i f_{e1} t} )
\ket{e}\bra{b}\\
& \quad + \Omega (t)\omega_0 \omega_1(e^{-2i f_{e1} t} - e^{-2i f_{e0} t})
\ket{e}\bra{d}
+\text{h.c}
.
\end{split}
\end{equation}
Notice from the second line that the counter-rotating terms, differently from RWA case, couple the dark state with the rest of the system. Thus, we lose this interesting property. As we will see later on, this will play a large role for the single-qubit gates. However, the expression above assumes a particularly simple form when $f_{e1}=f_{e0}=f$, which is:
\begin{equation}\label{eq:Hbd_rotating2}
H_{bd}(t)  =
\Omega(t)
(1+e^{-2i f t} )
\ket{e}\bra{b}
+\text{h.c.}
\end{equation}
When the two counter-rotating frequencies are the same this extra term cancels out and the dark state coupling vanishes once more. We show this effect concretely through numerical simulations in Fig.~\ref{fig:dynamicsBD1}. 
It may also be elucidating to examine what happens when we plot the trajectory of the state on the Bloch sphere when the Hamiltonian~\eqref{eq:Hbd_rotating2} is considered. If we initialize the system in the state $\ket{\psi_0} = \ket{b}$ the amplitude of the dark state remains zero at all times, so it is possible to depict the evolution of the wave function on the Bloch sphere with poles $\ket{b}$ and $\ket{e}$. Results are shown in Fig.~\ref{fig:BlochDynamics1} for $\omega_0 = \omega_1 = 1/\sqrt{2}$.

We can make a few observations about this picture. As one would expect, the trajectory for the RWA is just a great circle around one of the axis, since the Hamiltonian just implements a rotation in the Bloch sphere. Meanwhile, the counter-rotating terms introduce a wobbling movement around the ideal path. These corrections introduce a slight deviation from the starting point after the end of the evolution. In Fig.~\ref{fig:BlochDynamics1} we can see that the wave function does not return to the initial state. Instead, what we observe is a new trajectory with a slight offset in comparison to the previous "cycle". If we run the simulation for several cycles it is possible to see, in the right panel of Fig.~\ref{fig:BlochDynamics1}, the cumulative effect of these slight deviations. Thus, even when the dark state coupling is eliminated by choosing equal counter-rotating frequencies, we do not, in general, return to the initial subspace. 

\section{Single-qubit gates}\label{sec:results_single}

Now that we have gained some further intuition on the effect of the counter-rotating contributions to the system, we are in a position to discuss the time-optimal regime of the $\Lambda$-Hamiltonian.  In this section we investigate the performance of single-qubit gates when both counter-rotating and open quantum system effects are present. For that, we compute the fidelity
\begin{equation}\label{eq:fidelity}
    \mathcal{F} 
    = 
    \bra{\psi_0}
    U(C)^\dagger \rho U(C) \ket{\psi_0},
\end{equation}
where $\rho$ evolves according to the master equation
\begin{equation}\label{eq:master_equation_gate}
    \frac{d\rho}{dt}
    =
    i[\rho, H_I ]
    +
    \gamma D(\ket{g}\bra{e}),
\end{equation}
defined in Eq.~\eqref{eq:masterequation}. Here, $\ket{\psi_0}$ refers to the input state. Meanwhile, the unitary $U(C)$ is the ideal gate unitary, given by Eq.~\eqref{eq:ideal_holonomy}. Therefore, $U(C)\ket{\psi_0}$ represents the ideal output. Hence, in Eq.~\eqref{eq:fidelity} we calculate the fidelity between the ideal outcome and the density matrix $\rho$ of the non-ideal process, which is obtained by solving  the master equation in Eq.~\eqref{eq:master_equation_gate}. The Hamiltonian $H_I$ is, as we saw, given by Eq.~\eqref{eq:Lambda_hamiltonian} and contains the counter-rotating terms. Thus, we can see that Eq.~\eqref{eq:master_equation_gate} accounts for both the dissipative and counter-rotating contributions. 

Note however that these steps yield the fidelity for a particular input state $\ket{\psi_0}$. A final step in our simulations is to compute the \emph{average} fidelity for several different states which are uniformly distributed in the Bloch sphere, in order to get a more representative picture. We describe the algorithm for uniformly choosing states in the Bloch sphere in Appendix~\ref{appendix:fibonacci}. Moreover, for better visualization we shall focus on the infidelity $1-\mathcal{F}$  when plotting the figures. 

\begin{figure}
	\centering
	\includegraphics[width=\textwidth]{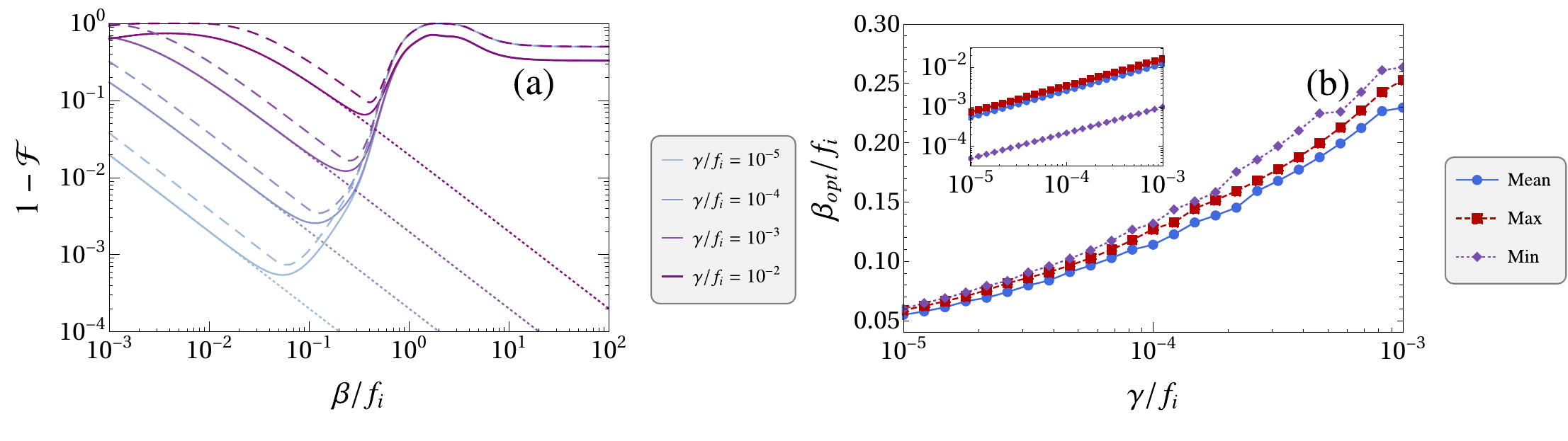}
	\caption[Time-optimal configuration of the S gate.]{Time-optimal configuration of the S gate. (a)  We plot the mean (maximum) infidelity $1-\mathcal{F}$ with the solid (dashed) lines for different values of $\gamma/f_i$. Results are shown as a function of the ratio $\beta/f_i$ between the inverse pulse length and the counter-rotating frequency. The dotted lines show the RWA solution from Eq.~\eqref{eq:RWAdissipative}, i.e., we assume that $\beta/f_i \rightarrow 0$. We avoid any significant overlap between the pulses by choosing a spacing of $\Delta t = 10/\beta$ between them. (b) Plot of the optimal (inverse) pulse-length $\beta_{\rm opt}$ as a function of $\gamma/f_i$. The corresponding infidelity is shown in the inset. The grid in this case is a sample space of $1000$ points uniformly distributed in the interval $[0.03, 0.3]$. In both panels the average infidelity was calculated for $100$ input states uniformly distributed over the Bloch sphere, as described in Appendix~\ref{appendix:fibonacci}.}
	\label{fig:timeoptimalS}
\end{figure}

We start by performing simulations for the $S$ gate. More generally, the phase-shift gate $\ket{k} \rightarrow e^{2 i k \pi (\phi' - \phi)}\ket{k}$ can be implemented through two unit vectors:  ${\bf n} = (\cos \phi,\sin \phi,0)$ and  ${\bf m} = (\cos \phi', \sin \phi', 0)$, which corresponds to the choice $\theta = \theta' = \pi/2$,  as defined in Eq.~\eqref{eq:ideal_holonomy}. Thus, as we mention in Sec.~\ref{sec:singlequbit_gates}, the $S$ gate can be constructed by choosing $\phi = \pi/4$ and $\phi' = \pi/2$. 
First results are shown in Fig.~\ref{fig:timeoptimalS}(a), relating the infidelity $1-\mathcal{F}$ to the inverse pulse length $\beta$ (in units of $f_i$). 
In  Fig.~\ref{fig:timeoptimalS}(b) we also plot the optimal inverse pulse length $\beta_{\rm opt}$, which minimizes the infidelity from panel (a). For now, we consider the counter-rotating frequencies to be homogeneous, i.e., $f_{0e} = f_{1e} = f_i$, as described by the Hamiltonian in Eq.~\eqref{eq:Hbd_rotating2}. Besides the average infidelity, we also include the maximum and minimum infidelities (relative to the input states). Their qualitative behavior in panel (a), and also the optimal value of $\beta$ in Fig.~\ref{fig:timeoptimalS}(b), are largely unchanged. Therefore, from hereafter we shall focus solely on the mean infidelity in our analysis, as it captures the optimal configuration and the general behaviour of the gates well enough.

We can observe the optimal regime of the system for intermediate values of $\beta/f_i$. If possible, one should tune the parameters of the system such that the relation $f_i > \beta \gg \gamma$ holds. That is, the inverse pulse length $\beta$ should be much larger than the dissipation rate $\gamma$, and simultaneously, the counter-rotating frequencies $f_i$ should be sufficiently larger than $\beta$. By doing so, one can achieve the minimum infidelity shown on the inset of panel (b) for curves such as the ones seen on panel (a). 

From Fig.~\ref{fig:timeoptimalS}(b), for example, we can observe that the ratio between the inverse pulse length and the counter-rotating frequencies should be of the order of $\beta/f_i \approx 0.1$, for $\gamma/f_i \approx 10^{-4}$, which is of the same order as parameters found in the literature \cite{AbdumalikovJr2013}. The precise value of the optimal $\beta/f_i$, however, will depend on how strong the decoherence is. Note, in Fig.~\ref{fig:timeoptimalS}(a), how the minimum shifts depending on the value of $\gamma/f_i$. In particular, as $\gamma/f_i$ decreases, we achieve the minimum infidelity by using larger pulses (smaller $\beta/f_i$). This result, of course, agrees with intuition: if dissipation plays a lesser role, then it makes sense to use larger pulses in order to improve the validity of the RWA. That is why, following this idea, in Fig.~\ref{fig:timeoptimalS}(b) we plotted the optimal (inverse) pulse length $\beta_{\rm opt}$ as a function of $\gamma/f_i$. The results seem to agree with what we just described. Another interesting observation is that, from the inset in panel (b), we can see that the minimum achievable infidelity scales with a power of $\gamma/f_i$. 

We can also discuss two other relevant regimes by analyzing Fig.~\ref{fig:timeoptimalS}(a).  We begin with the regime of small $\beta/f_i$, describing longer pulses. In particular, we first focus on the limiting case of $\beta/f_i \rightarrow 0$. In this situation, all the curves should converge to the same fidelity, regardless of the value of $\gamma/f_i$. The reason is that the open quantum system effects become dominant in this scenario, hampering any population transfer between the qubit states and the excited state $\ket{e}$. Thus, what happens in this case is that the qubit state remains largely unchanged, and the output state is approximately the same as the input state. This is not visible in the figure due to the fact that this result would require unfeasibly long simulations. Instead, we will provide an analytical calculation to support this claim near the end of the current section. We will also include a better visual picture of this result when discussing two-qubit gates in the next section.

We used dotted lines to plot the numerical solution for the RWA Hamiltonian in the purely dissipative scenario (see the caption in Fig.~\ref{fig:timeoptimalS}). As one would expect, under this approximation the infidelity decreases monotonically as we shorten the duration of the pulses. Moreover, the solution of Eq.~\eqref{eq:master_equation_gate}, plotted with solid lines, agrees very well with the RWA solution, evidencing the fact that dissipation is much more relevant than the counter-rotating dynamics in this regime. We can also clearly see the breakdown of the RWA~\cite{Johansson2012} in this figure. As we increase $\beta/f_i$, the non-RWA solution eventualy starts to diverge from the RWA results around the minimum of the plot. This change of behavior marks the point where the RWA starts to fail.

In the same spirit, we eventually reach the limiting case $\beta/f_i \gg 1$ of very short pulses. In this other regime, the counter-rotating corrections in Eq.~\eqref{eq:Lambda_hamiltonian} become $1 + e^{-2 i f_i t} \approx 2$. Thus, the physical consequence is that the system undergoes a cyclical evolution twice. In other words, the bright state evolves as $\ket{b} \rightarrow -\ket{b} \rightarrow \ket{b}$, per Eq.~\eqref{eq:bd_basis_holonomic}. Therefore, the net effect of the protocol is simply to implement the identity gate, which, just like the strongly dissipative regime, leaves the input state (approximately) unchanged.

It is possible to quantify the infidelity in both of these cases, and the calculation happens to be the same, as we have just explained. The fidelity of a single input state is given by the overlap $|\bra{\psi_0}U(C)\ket{\psi_0}|^2$ between the target state $U(C)\ket{\psi_0}$ and the input state $\ket{\psi_0}$. If we use the usual parametrization in the Bloch sphere
\begin{equation}
\ket{\psi_0} = 
\cos{\left(\frac{\alpha}{2}\right)} \ket{0}
+
e^{i\varphi} \sin{\left(\frac{\alpha}{2}\right)} \ket{1}
\end{equation}
for the input states, where $\alpha$ and $\varphi$ descibe the polar and the azymuthal angle in the sphere, respectively, the mean infidelity can be obtained through an integration over the Bloch sphere for all input states:
\begin{equation}
1 - \mathcal{F}_{\infty} 
=
\frac{1}{4\pi}
\int_0^{2\pi} d \varphi 
\int_0^\pi d \alpha
|\bra{\psi_0}U(C)\ket{\psi_0}|^2.
\end{equation}
For the S gate, a direct calculation yields $1 - \mathcal{F}_{\infty}  = 2/3$. As we mentioned, this is not visible for $\beta/f_i \ll 1$. However, we can visually verify it for $\beta/f_i \gg 1$ in Fig.~\ref{fig:timeoptimalS}(a).

\section{Heterogeneous frequencies}

Now we move to the case where the two counter-rotating frequencies appearing in Eq.~\eqref{eq:Hbd_rotating} are different. As we have mentioned before, whenever $f_{e0} \neq f_{1e}$ there is a coupling between the dark and the excited state, as seen in the last term of Eq.~\eqref{eq:Hbd_rotating}. We will be interested in what role this difference might play in comparison to the homogeneous case. 

\begin{figure}[b!]
	\centering
	\includegraphics[width=\textwidth]{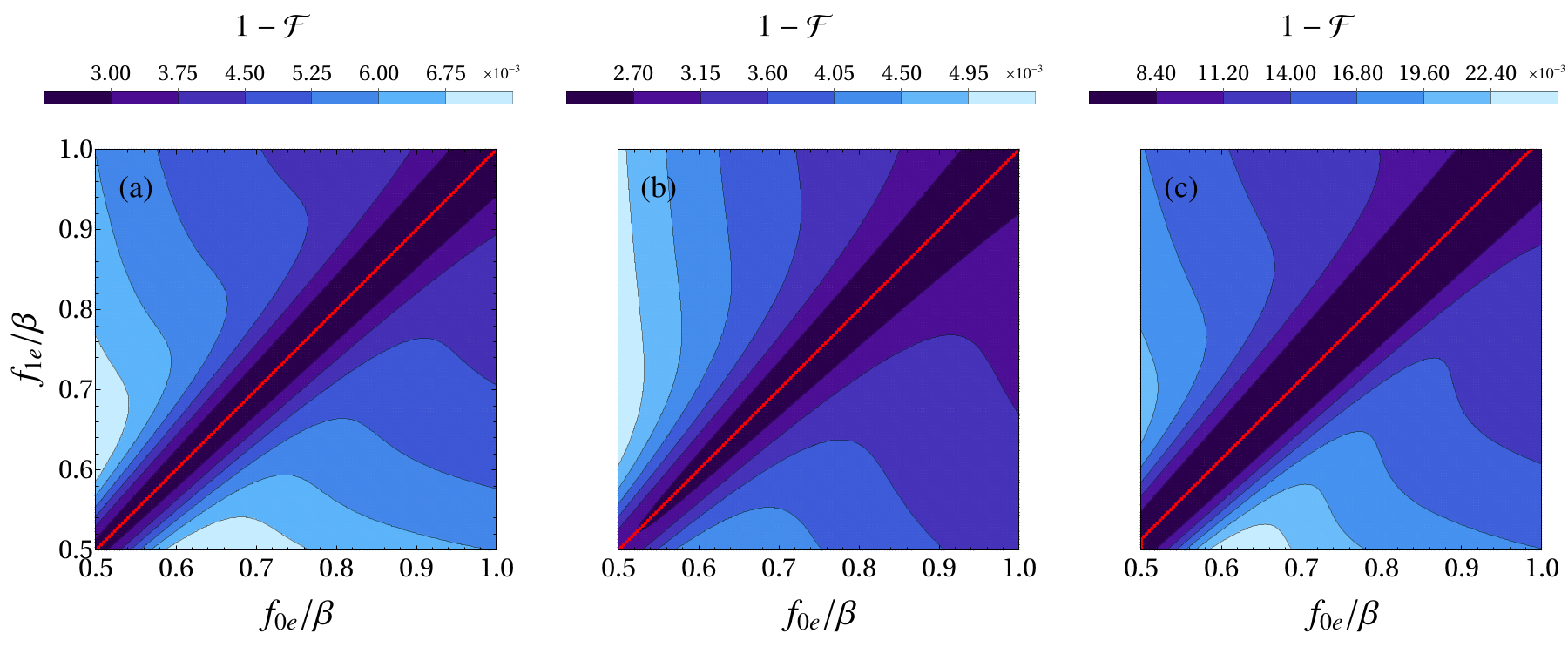}
	\caption[Contour plot of the infidelity for heterogeneous frequencies.]{Contour plot of the infidelity for heterogeneous frequencies. Average infidelity $1-\mathcal{F}$ as a function of the counter-rotating frequencies $f_{0e}$ and $f_{1e}$ for $\gamma/\beta = 0.02$. Results are shown for (a) the $X$ gate, (b) the $H$ gate, and (c) the $S$ gate. Other details and parameters are chosen to be the same as those in Fig.~\ref{fig:timeoptimalS}. The simulations were done for a grid of $150$ × $150$ frequencies.}
	\label{fig:FidelityContour}
\end{figure}

In Fig.~\ref{fig:FidelityContour} we show a countour plot for the infidelity $1-\mathcal{F}$ as a function of $f_{0e}/\beta$ and $f_{1e}/\beta$. We show the results for the $X$, H and $S$ gates, respectively. This plot can give us a few important insights. First, the $S$  gate is more susceptible to decoherence due to the fact that it requires two pulses, as we discussed in the paragraph above Eq.~\eqref{eq:ideal_holonomy}. Additionally, it is also possible to observe that Fig.~\ref{fig:FidelityContour}~(a) is
symmetric, while Fig.~\ref{fig:FidelityContour}~(b) and ~(c) are not. This is due to the fact that we have $\omega_0 = \omega_1$ for the $X$. Meanwhile, we have different amplitudes for the H and $S$ gates.

Another question we can try to answer is how $f_{0e}$ and $f_{1e}$ should be chosen in order to maximize the fidelity. We used a red line to plot the optimal $f_{1e}$ for a given $f_{0e}$. The results shown in panels (a) and (b) show the infidelity is minimized by taking $f_{0e} = f_{1e}$. These results indicate that single-pulse gates are optimal  for homogeneous frequencies. This possibly happens due to the decoupling between the dark and the excited states which occurs in this scenario. A further evidence of this is that we can also observe, in Figs.~\ref{fig:timeoptimalS}~(a) and (b), that  increasing one of the frequencies beyond the point where $f_{0e} = f_{1e}$ actually \emph{decreases} the fidelity of the gate. Therefore, the coupling between the $\ket{e}$ and $\ket{d}$ seems to play an important role in the performance of the protocol.

Surprisingly, in Fig.~\ref{fig:FidelityContour}~(c) we obtain a slightly different behavior for the $S$ gate. The optimal regime occurs when the frequencies $f_{0e}$ and $f_{1e}$ are \emph{different}. 
The red line shows that ideally one should increase the frequencies at a constant ratio, which is, however, different from unity. We believe that the relevant mechanism here is the non-abelian feature of the gates. Since the two unitaries in Eq.~\eqref{eq:ideal_holonomy}, which are necessary for implementing two-pulse gates, are non-commuting, they might be responsible for this behavior.
This is supported by the fact that that the fidelity of noncommuting gates is lower than the product of their fidelities. This has been verified in Ref.~\cite{Spiegelberg2013}.

\begin{figure}[t!]
	\centering
	\includegraphics[width=\textwidth]{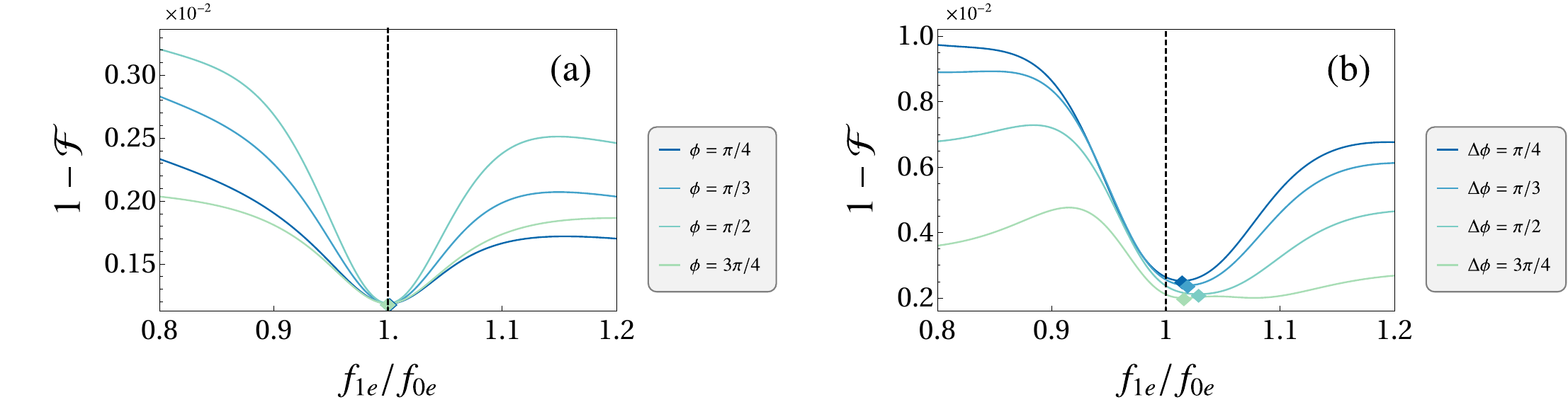}
	\caption[Infidelity as a function of the ratio $f_{1e}/f_{0e}$ for single-qubit
gates.]{Infidelity as a function of the ratio $f_{1e}/f_{0e}$ for single-qubit
gates. We show the results for (a) a single-pulse gate with $\theta = \pi/4$ and
for different values of $\phi$. (b) we plot the results for the
phase-shift gate, with $\theta = \theta' = \frac{\pi}{2}$ and $\Delta \phi = 2(\phi' - \phi)$.  The coupling strength is given by $\gamma/\beta = 10^{-3}$ , the spacing between the pulses is $\Delta t = 20/\beta$ and the frequency $f_{0e}$ is fixed and given by $f_{0e}/\beta = 10$. We use a sampling space of 250 frequencies
and 100 input states, as described in Fig.~\ref{fig:timeoptimalS}.}
	\label{fig:heterogeneous_single}
\end{figure}

Another point of interest is whether other gates display the same behavior as the ones we have just seen. We start by further comparing how different frequencies affect the single and two-loop gates, plotting the infidelity as a function of the ratio $f_{1e}/f_{0e}$ for a fixed $f_{0e}$. Results are shown in Fig.~\ref{fig:heterogeneous_single}. We implement single and a two-loop gates in panels (a) and (b), respectively. In Fig.~\ref{fig:heterogeneous_single}~(a) we consider a class of traceless single-qubit gates given by Eq.~\eqref{eq:single_loop_uni}. In our simulations we fix $\theta = \pi/4$, repeating the simulation for different values of $\phi$. 
For the second gate, shown in panel (b), we consider the phase-shift gate described in the previous section, with $\Delta \phi = 2(\phi' - \phi)$. 

\begin{figure}[b!]
	\centering
	\includegraphics[width=0.5\textwidth]{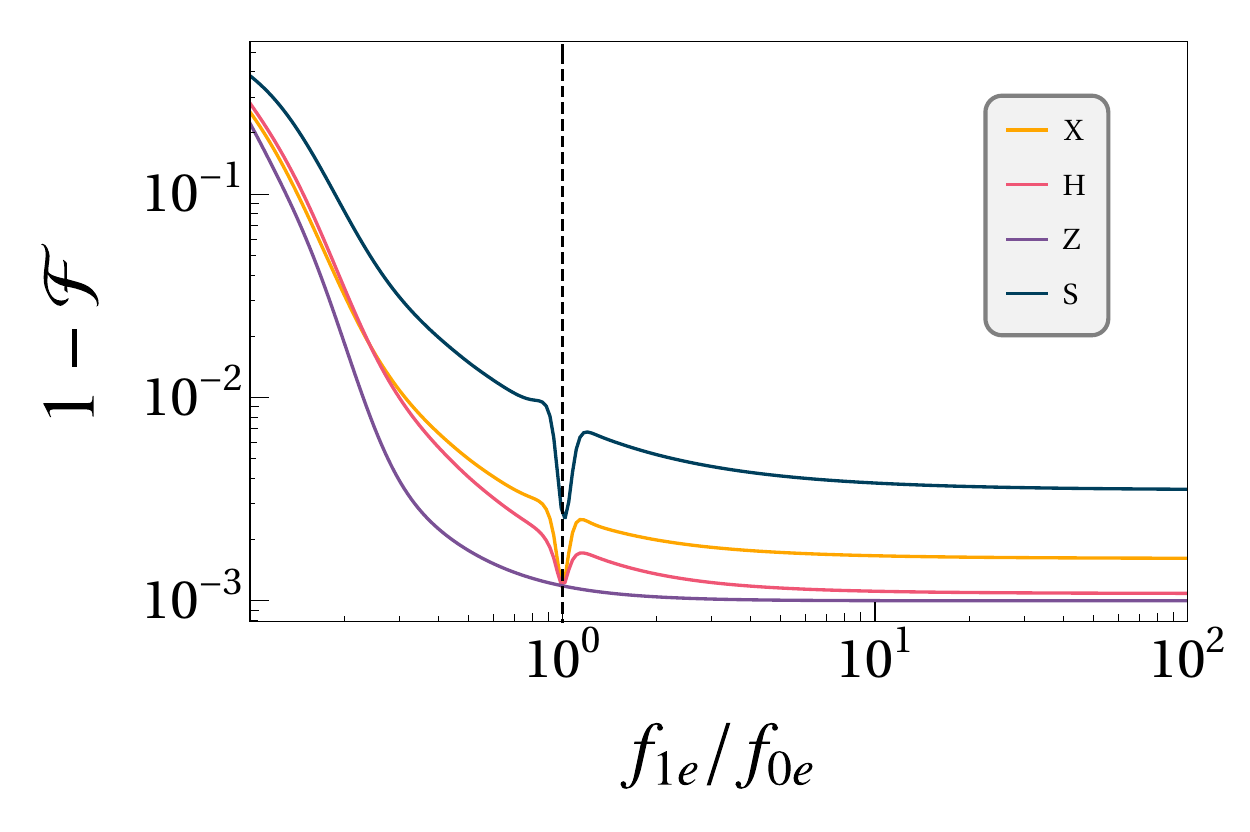}
	\caption[Relevance of the $\ket{e} \leftrightarrow \ket{d}$ coupling in the $\Lambda$-system.]{Relevance of the $\ket{e} \leftrightarrow \ket{d}$ coupling in the $\Lambda$-system. We plot the infidelity as a function of the ratio $f_{1e}/f_{0e}$ for the NOT
gate (yellow), the Hadamard gate (magenta), the Z gate (purple), and
the S gate (dark blue). The decay rate, the sampling space, and $f_{0e}$ are
the same as those in Fig.~\ref{fig:heterogeneous_single}. The dashed line highlights the point where $f_{1e}/f_{0e}=1$. We also highlight the minimum, in each plot with a diamond mark.}
	\label{fig:infidelity_log}
\end{figure}

We are interested in investigating the influence of gate parameters in the optimal ratio between the two frequencies. We do this by varying $\phi$ and $\phi'$. 
We find that single-pulse gates are optimal when the two frequencies are the same. Meanwhile, panel (b) tell us that the optimal configuration occurs when the two frequencies are \emph{slightly} different. We can see, however, that this gain is very marginal. This indicates that a small coupling between the dark state and the excited state is actually the most important mechanism in reducing the infidelity, and that small differences between the two frequencies are not so important. The discussion in the next paragraph makes this point  clearer, where we plot the infidelity as a function of the ratio $f_{1e}/f_{0e}$ for different gates in Fig.~\ref{fig:infidelity_log}. 

Simulations for the $X$, H, and $S$ gates show a sharp decrease in the infidelity when $f_{0e} \approx f_{1e}$. For the $X$ and $S$ gates we have a global minimum at $f_{e0} \approx f_{1e}$, while for the Hadamard gate this point is only a local minimum. For the $Z$ gate on the other hand, we observe no critical point at all; the infidelity is monically decreasing in this case. 
So, whether $f_{0e} = f_{1e}$ is the optimal gate configuration will depend on the gate parameters. Nevertheless, for all the gates shown in Fig.~\ref{fig:infidelity_log} we can see that homogeneous frequencies result in infidelities which are very close to the minimum. Thus, taking $f_{0e} \approx f_{1e}$ seems like a good rule of thumb, and even for the H and $Z$ gates, going to regime where $f_{e1} \gg f_{0e}$ does not seem to yield much improvement. 

The behavior of these curves is very intuitive if we look into the transition amplitudes for each gate. Among the single-loop gates, the  $X$ gates displays the most drastic drop. This is due to the fact that both transition frequencies have the same modulus, so this maximizes the last term in Eq.~\eqref{eq:Hbd_rotating}, which introduces the $\ket{e} \leftrightarrow \ket{d}$ coupling. For that reason, the $X$ gate (or any other gate which differs by a complex phase in the transition amplitudes), is the most sensitive to this effect. Meanwhile, in the $Z$ gate we have $\omega_0 = 0$ and $\omega_1 = 1$. 
Hence, $f_{0e}$ plays no role here and we do not observe any coupling between the dark and the excited states. For that reason we get a monotonically decreasing curve in Fig.~\ref{fig:infidelity_log}. Finally, we should observe an intermediate behavior for other gates, such as the Hadamard gate. Whether taking $f_{0e} = f_{1e}$ or taking $f_{1e} \gg f_{0e}$ is truly optimal will depend on the ratio between the two transition amplitudes $\omega_0$ and $\omega_1$.

\section{Two-qubit gates}

\begin{figure}[h!]
	\centering
	\includegraphics[width=0.75\textwidth]{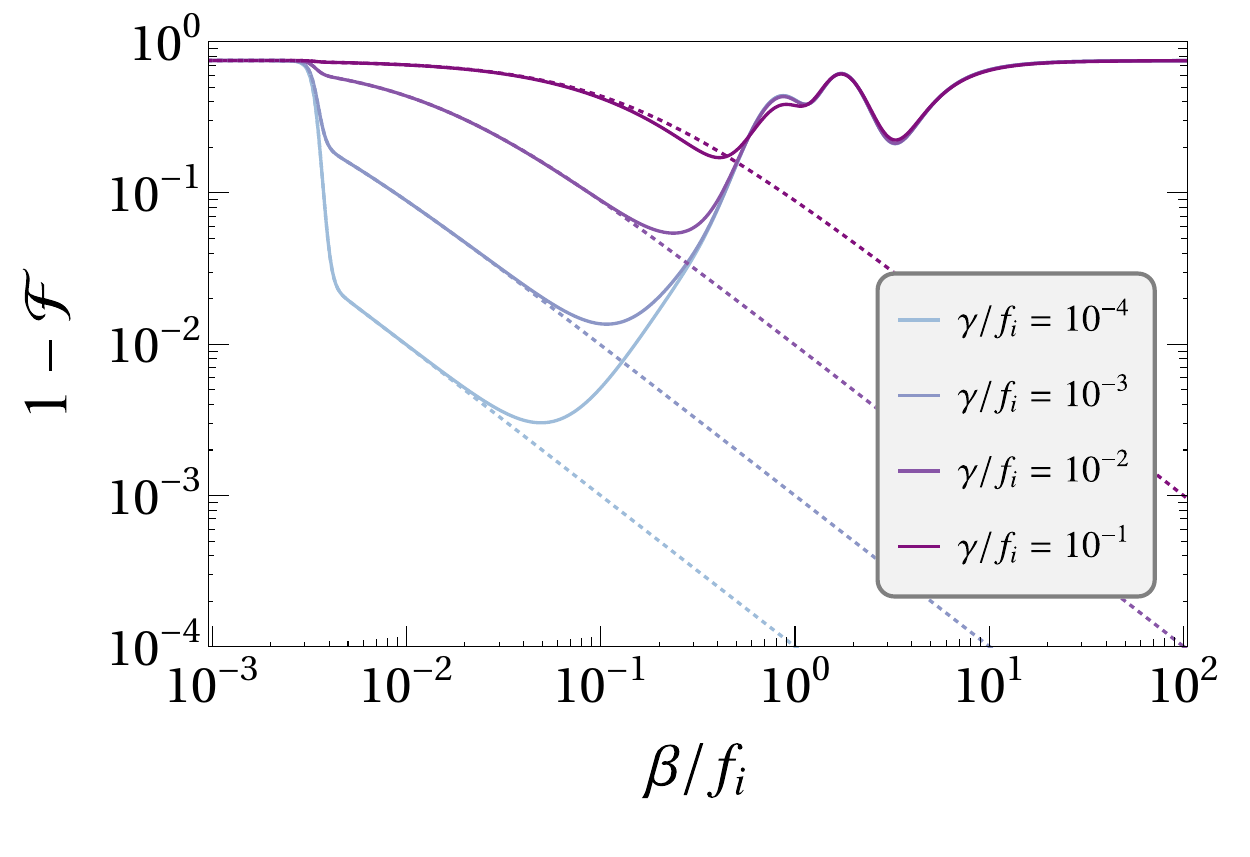}
	\caption[Infidelity of the CZ gate as a function of $\beta/f_i$.]{Infidelity of the CZ gate as a function of $\beta/f_i$ . The dotted lines represent the RWA solution given by the Hamiltonian in Eq.~\eqref{eq:twoqubit1}. The infidelity is averaged over the input states $\ket{+}\ket{+}$, $\ket{+}\ket{-}$, $\ket{-}\ket{+}$, and $\ket{-}\ket{-}$. Other details are the same as those in Fig.~\ref{fig:timeoptimalS}.}
	\label{fig:CZFidelity2}
\end{figure}

We conclude our analysis by investigating a two-qubit gate. Our findings are quite analogous to the single-qubit results. Here we consider the Hamiltonians from Eq.~\eqref{eq:twoqubit1nonRWA}. Just like in the RWA case, here we also have $[H^{(2)}_0(t), H^{(2)}_1(t)] = 0$, so $ H^{(2)}_1(t)$ only introduces a trivial phase in the relevant subspace ${\rm span}\{\ket{00}, \ket{01}, \ket{10}, \ket{11}\}$. Moreover, we only consider local noise (we suppose that there is no collective decay). The master equation in this case reads
\begin{equation}\label{eq:master_equation_gate2}
    \frac{d\rho^{(2)}}{dt}
    =
    i[\rho^{(2)}, H^{(2)}_0(t)]
    +
    \gamma_1 D(\ket{g}\bra{e} \otimes I)
    +
    \gamma_2 D(I \otimes \ket{g}\bra{e}),
\end{equation}
where $\rho^{(2)}$ is the density matrix for the two-qubit state. For simplicity we also consider that $\gamma_1 = \gamma_2 = \gamma$. We perform simulations for the CZ gate, defined as 
\begin{equation}
CZ
=
\ketbra{0}{0} \otimes I + \ketbra{1}{1} \otimes \sigma_z.
\end{equation} 
It can be constructed by choosing $\theta = 0$, and we show the results in Fig.~\ref{fig:CZFidelity2}. In this case, we opt for a much simpler space of input states. A uniform sampling of states for two-qubits is much harder than in the single-qubit scenario \cite{Team2021}. Therefore, we focus on a few cases of interest. In particular, we consider four different input states: $\ket{+}\ket{+}$, $\ket{+}\ket{-}$, $\ket{-}\ket{+}$, and $\ket{-}\ket{-}$. These states are useful because they yield maximally entangled states upon the application of the CZ gate. For instance, $CZ\ket{+}\ket{+} = (\ket{0}\ket{+} + \ket{1}\ket{-})/\sqrt{2}$. Note how the application of the Hadamard gate in the second qubit results in the Bell state $(\ket{00}+\ket{11})/\sqrt{2}$.

Qualitatively, the results are very similar to what we have obtained for the single-qubit gates, demonstrating that two-qubit gates are also reasonably robust against this type of noise. Note however that their performance is much more similar to the $S$ gate in Fig.~\eqref{fig:timeoptimalS}, which is a two-loop gate. So they are not as robust as the single-loop gates, such as the $X$ or H gates. A useful characteristic of this gate, however, is that the optimal inverse pulse length is very similar to the single-qubit scenario. This might be a useful experimental feature, since the optimal pulse duration is of similar magnitude for both the single and two-qubit gates. 
Finally, note how the behavior in the limiting cases of strong dissipation or of very small counter-rotating frequencies is similar to what we described at the end of Sec.~\ref{sec:results_single} for the single-qubit gates. Moreover, note how we can visually observe, in this plot, how all the curves converge to the same value. As we have thoroughly discussed in Sec.~\ref{sec:results_single}; the strong decoherence spoils the protocol, and the final state is essentially unchanged. Thus, we can see that the infidelity converges to a value which is independent of $\beta/f_i$.

%% file: chapters/conclusion.tex
In the first part of this dissertation we incorporated, through the Bayesian lens,  concrete aspects of estimation theory into the collisional thermometry scheme. 
Our results showcased how the Bayesian framework provides an alternative point of view to thermometry,  yielding further insights as a thermometric tool and also some practical advantages \cite{rubioGlobalQuantumThermometry2021, rubioQuantumScaleEstimation2023}. 
The first aspect of this approach was the possibility of sequentially updating the temperature distribution based on the measurement outcomes. 
The estimators where then derived from those distributions, whose performance was then assessed through the Bayesian mean-square error. 
The Cram\'er-Rao bound and its Bayesian counterpart, the van Trees-Sch\"utzenberger inequality, were then used as some of the main benchmarks for the estimator performance. 
Thus, by investigating the asymptotic value of the Bayesian MSE, we performed a global analysis of protocol independently of particular values of the temperature, obtaining some of the optimal configurations for the model among the regimes which we have considered. 
Further generalizations, such as the use of collective measurements, are also possible. 
In that sense, we could also employ adaptive strategies \cite{escherQuantumMetrologyNoisy2011, boeyensUninformedBayesianQuantum2021, oconnorStochasticCollisionalQuantum2021, mehboudiFundamentalLimitsBayesian2022}; allowing us to optimize the coupling parameter $\gamma \tau_{SE}$ between the system and the bath, the probe state \cite{shuSurpassingThermalCramerRao2020} or the basis (in the scenario of collective measurements). An extension of the model to continuous variables systems~\cite{cenniThermometryGaussianQuantum2022}, and also the use of machine learning tools~\cite{luizMachineClassificationProbebased2022}, are also promising research venues.

Naturally, our discussion only scratches the surface of what Bayesian thermometry can offer. A very natural direction to follow is the investigation of different estimators and prior distributions. Much progress has been made in Refs.~\cite{rubioGlobalQuantumThermometry2021, boeyensUninformedBayesianQuantum2021, rubioQuantumScaleEstimation2023} on this topic. 
Additionally, Bayesian estimation very often helps dealing with difficulties which are particular or specific to the problem at hand, such as constraints, symmetries and even experimental challenges, by means of heuristic strategies~\cite{Teklu2009, vontoussaintBayesianInferencePhysics2011, Chapeau-Blondeau2015, Kiilerich2015, schmiedQuantumStateTomography2016, martinez-vargasQuantumEstimationUnknown2017, rubioGlobalQuantumThermometry2021}.

Likewise, investigations which dwell into further and more general aspects of quantum estimation theory in a broader sense than presented here, such as alternative bounds to the CRB and the Van-Trees Sch\"utzenberger bound, are equally important \cite{Personick1971}. See Refs.~\cite{zhangQuantumMetrologicalBounds2014} and~\cite{luQuantumWeissWeinsteinBounds2016} and the discussion therein, which serve as quantum formulations of the Ziv-Zakai and Weiss-Weinstein bounds from classical estimation theory \cite{zivLowerBoundsSignal1969,weissLowerBoundMeansquare1985}. 
The recent results in Refs.~\cite{martinez-vargasQuantumEstimationUnknown2017} and \cite{demkowicz-dobrzanskiMultiparameterEstimationQuantum2020} also provide some illuminating ideas. A bigger focus on situations where data might be scarce is an additional scenario where Bayesian estimation might shine. Some important contributions have been made in, e.g., Refs.~\cite{rubioNonasymptoticAnalysisQuantum2018, rubioQuantumMetrologyPresence2019, rubioBayesianMultiparameterQuantum2020}.
Finally, what should be clear at this point is that the global treatment is not limited either to the Bayesian and frequentist techniques, nor are the two frameworks mutually exclusive~\cite{MacKay2002}. Some classical works \cite{Personick1971, Helstrom1976, Holevo2011} provide important instances of a fully Bayesian treatment. Similarly, the frequentist approach is just as useful when considering practical applications. 
At the end of the day, the advantages and the obstacles in each approach shifts in a case-by-case scenario, and it is left to the experimenter to evaluate their clarity and usefulness in each different context.
It is also important to stress, as we have seen many times, that these tools are not  restricted to thermometry. Rather, Bayesian estimation has found widespread success in the quantum metrology community for decades, in a plethora of different problems, albeit it has been far less employed in quantum thermometry, in part due to the very recent history of the field.

In the second part of this work we studied a time-optimal approach to non-adiabatic holonomic quantum computation based on $\Lambda$-type three-level systems. More specifically, we have further investigated the validity of the rotating-wave approximation in this model and its conjunction with open quantum system effects. 
We showed that there is a trade-off in the gate operation time; shorter pulses can be used to protect the system against decoherence, but they are more fragile against imperfections brought from the RWA. Additionally, we were able to obtain the optimal parameters for typical experimental configurations of the $\Lambda$-system, allowing for an optimization of the gate performance in terms of the pulse length.

Furthermore, we also explored the role of asymmetric frequencies in the performance of single-qubit gates. We found that different counter-rotating frequencies introduce a non-ideal coupling between the dark and the excited states of the $\Lambda$-Hamiltonian. For that reason, similar frequencies suppress the effect of this unwanted coupling, resulting in a sudden improvement of the gate performance in comparison to the asymmetrical configuration. For two-loop single-qubit gates on the other hand, we found that different counter-rotating frequencies result in a very slight improvement. 
We believe that this might be a result of the non-Abelian character of the geometrical phases used in the construction of these gates. Nevertheless, in all scenarios, approximate (but not necessarily identical) frequencies resulted in a significant improvement. 
These aspects are a discussion of interest as far as real experiments are considered, since, besides their own idiosyncrasies and limitations, the platform of choice might deeply constrain how much these frequencies can be fine-tuned.

A few other important aspects of this model have been left out, and might be useful subjects of future investigations. One of these directions is how well other pulse shapes might fare, such as square and Gaussian pulse. The latter, for instance, has been previously used in real experiments for NHQC \cite{AbdumalikovJr2013}. 
It might be interesting to ask what pulse shape yields the maximum achievable fidelity when both dissipation and the breakdown of the RWA are considered. In particular, as far as we are aware, it is an open question whether certain pulse shapes are more robust against any of these effects. Some similar investigations already exist \cite{Zhang2019, Ji2022}, and they might be an useful aid in this direction of study. 

The use of time-dependent perturbation theory using, e.g., the Magnus expansion, might also provide some useful insight into the model for the dissipationless case. Unfortunately, we could not find much success in the use of perturbation theory in the scope of this dissertation, in a large part due to the complexity that Sech pulses introduce into these methods, resulting in unwieldy expressions. It is, however, possible to apply these techniques at lower orders for square and trigonometric pulses. This might provide some further insight into a few parts of this work, such as the optimality of asymmetric frequencies for two-loop single-qubit gates. Nevertheless, applying an analytical approach when both counter-rotating contributions and open quantum system effects are considered is still a challenge.

Finally, among generalizations and future investigations, one can include extensions of our study into different implementations of the NQHC protocol for $\Lambda$-systems, such as the off-resonant scheme \cite{Xu2015, Sjoqvist2016}, the single-loop scheme \cite{Herterich2016}, environment-assisted implementations \cite{Ramberg2019}, path shortening \cite{Xu2018}, and time-optimal-control implementations \cite{Han2020, Chen2020}. It would also be possible to include other type of open quantum system effects, such as dephasing and collective noise, or even other type of errors, such as imperfections on the pulse amplitude itself \cite{Johansson2012}. Extensions to the multi-qubit gates \cite{Zhao2019, Xu2021} and strategies which possibly do not rely on RWA \cite{Gontijo2020} also provide interesting research venues.

%% file: chapters/vectorization.tex

In this appendix we discuss the vectorization procedure. See Ref.~\cite{Dhrymes2000} for an operational approach. In the field of quantum information, the concept is known as the Choi–Jamiołkowski isomorphism \cite{Jamiokowski1972, Choi1975, Zyczkowski2004, Jiang2013}. Under this isomorphism we can very conveniently define an operation known as the Hilbert-Schmidt inner product:
\begin{equation}\label{eq:hilbertschmidt}
\tr{A^\dagger B} = \choi{A}^\dagger \choi{B},
\end{equation}
which, as written in this form, is based on a process known as \emph{vectorization}, where one "stacks" the rows of a matrix into a single column:
\begin{equation}
\choi
{
\begin{matrix}
a & b\\
c & d
\end{matrix}
}
:=
\begin{matrix}
\begin{pmatrix}
a \\
b \\
c \\ 
d
\end{pmatrix}
\end{matrix}.
\end{equation}

In order to prove Eq.~\eqref{eq:hilbertschmidt}, we can write the trace of the product $A^\dagger B$ as:
\begin{equation}
\tr{A^\dagger B} = \sum_i (A^\dagger B)_{ii} = \sum_{ij} A^\dagger_{ij} B_{ji} = \sum_{ij} (A^*_{ji})^\dagger B_{ji}  .   
\end{equation}
Note however that if we can write the matrix $A$ as $A = \begin{matrix}(\vec{A}_1 & \vec{A}_2 & ... &  \vec{A}_n)\end{matrix}$, where we define $\vec{A}_i$ as the $i$-th column of the matrix $A$, we can write the element $A_{ji}$ as $A_{ji} = (\vec{A}_j)_i$. Thus, the previous equality becomes:
\begin{equation}
\tr{A^\dagger B} = \sum_j (\vec{A}^*_i)_j (\vec{B}_i)_j = \choi{A^*}\cdot\choi{B} = \choi{A}^\dagger \choi{B},   
\end{equation}
as we wanted to show. 

Another very important identity is the following:
\begin{equation}
\choi{ABC} = (C^T \otimes A)\choi{B}.
\end{equation}
As a particular case, we can take either $A = I$ or $C = I$ to show that the product between two matrices can be expressed as:
\begin{equation}
AB = (I \otimes A)\choi{B} = (B^T \otimes I)\choi{A}.
\end{equation}
To see this, note that the $k$-th row of this product can be written as:
\begin{equation}
(\overrightarrow{ABC})_k = AB\vec{C}_k = A \sum_i \vec{B}_i C_{ik}.
\end{equation}
However we can split this sum in a more explicit way:
\[
(\overrightarrow{ABC})_k 
= \sum_i (A C_{ik}) \vec{B}_i 
=
\begin{matrix}
\begin{pmatrix}
C_{1k} A & C_{2k} A & \dots
\end{pmatrix}
\begin{pmatrix}
B_1\\
B_2\\
\vdots
\end{pmatrix}
\end{matrix}.
\]
The row vector however is simply $(\vec{C}_k^T \otimes A)$, and the second term can readily be identified as $\vec{B}$. Therefore, by stacking all the $(\overrightarrow{ABC})_k $ together we arrive at:
\begin{equation}
\vec{ABC}
=
\begin{matrix}
\begin{pmatrix}
\vec{C}_1^T \otimes A\\
\vec{C}_2^T \otimes A\\
\vdots
\end{pmatrix}
\end{matrix}
\vec{B}
=
(C^T \otimes A)\choi{B},
\end{equation}
which is the identity we wanted to prove.

%% file: chapters/correlations.tex
In this appendix we show how Eq.~\eqref{eq:DistributionApproximation} can be decomposed in a more general manner. From~\eqref{eq:likelihood_geral}, we can write:
\begin{equation}
\begin{split}
    P(X_1,\ldots, X_n|T) 
    = & ~P(X_n|T,X_1,\ldots,X_{n-1}) 
    \\& \times~P(X_{n-1} | T, X_1, \ldots, X_{n-2})
    \\& \times~\ldots
    \\& \times~P(X_2|T,X_1)
    \\& \times~P(X_1|T).
\end{split}
\end{equation}
Each of these transition probabilities can all be calculated directly from the model. 
For instance, we can write $P(X_2|X_1) = \tr \left( M_2 M_1 \rho_{A_1 A_2} \right)/\tr \left( M_1 \rho_{A_1 A_2} \right)$.

Focusing on the case where $T$ is discretized into steps $T_k$, we can now generalize Eq.~\eqref{loglike_discrete} to 
\begin{equation}\label{loglike_hierarchy}
    L_{kn} = \sum\limits_{i=1}^n \ln p(X_i|T_k,X_1,\ldots,X_{i-1}).
\end{equation}
By using Eq.~\eqref{loglike_hierarchy} in the place of Eq.~\eqref{loglike_discrete}, the construction from chapter~\ref{chp:results} remains valid, even when considering dependent outcomes. In other words, the formalism itself remains the same, and we should just be careful to simply incorporate the correlations into the likelihood.

Since the results of Fig.~\ref{fig:LogMutualInformationDecay} show that the mutual information decays exponentially with the distance between the ancillas, one does not need, in practice, to retain the full hierarchy of distributions in Eq.~\eqref{loglike_hierarchy}. It is possible, instead, to perform a truncation at a finite Markov order.
For the sake of the example, let us consider the case where only nearest-neighbor correlations are important. We may then approximate:
\begin{equation}\label{loglike_mark}
    L_{kn} \simeq \sum\limits_{i=1}^n \ln P(X_i|T_k,X_{i-1}),
\end{equation}
where $P(X_i|T_k,X_{i-1})$ forms essentially a Markov chain. 
This feature is very useful, since it is hard to handle the theoretical model due to the exponentially-increasing dimensions of the Hilbert space as extra ancillae are considered. Meanwhile, dealing with a small Markov order is analytically/numerically tractable.

%% file: chapters/fibonacci.tex
In this section we describe an algorithm to approximately distribute points on a sphere in an uniform manner.  A common approach to this problem consists in randomly sampling
these points with respect to the Haar measure \cite{Johansson2012}. However, since computing the fidelity for a large number of points is computationally expensive,
we turn instead to a simple algorithm called Fibonacci lattice (or nodes), which is
a reasonable approximation to deterministically distribute states over the Bloch sphere.  This approach is used to map points from a Fibonacci lattice in a
square onto the surface of a sphere through a cylindrical equal area projection \cite{Hardin2016}. This is advantageous due to the fact that it is possible to get a uniform distribution even for a small number of points, while random methods usually require a larger sample size.

\begin{figure}[h!]
	\centering
	\includegraphics[width=0.65\textwidth]{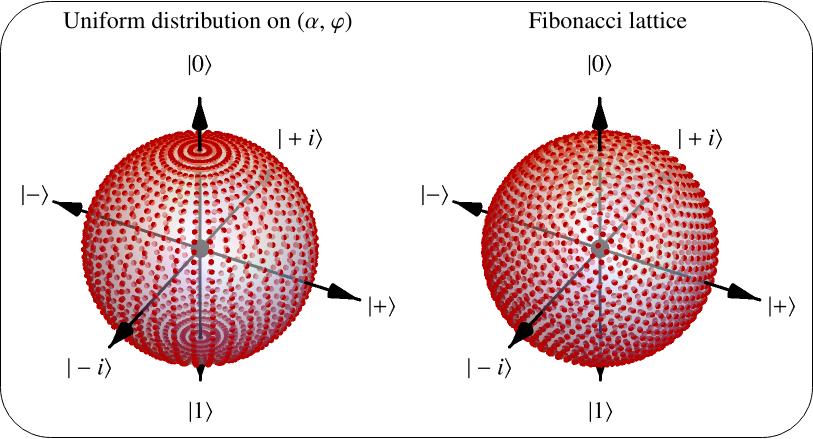}
	\caption[Uniform disitrubion of points using the Fibonnaci nodes algorithm.]{Uniform disitrubion of points using the Fibonnaci notes algorithm. We denote the polar and azymuthal angles by $\alpha$ and $\varphi$, respectively. We plot (left) the points sampled from a uniform distribution on $(\alpha, \varphi)$ and (right) the points sampled by using the algoritm.}
	\label{fig:FibonacciSpiral}
\end{figure}

This type of procedure is motivated by the fact that in the naive approach, where one uniformly samples the spherical angles, the density of points is strongly skewed towards the poles. This can been seen in Fig.~\eqref{fig:FibonacciSpiral}, where we compare the uniform sampling with the the algorithm. The procedure for a set of $N$ points works as follows: 

\begin{enumerate}
  \item Uniformly sample the $z$-coordinates $z_n$ from the interval $[-1, 1]$ 
  \item Distribute the azimuthal angle according to $\varphi_n = 2 \pi \Phi n$, where $\Phi = (1 + \sqrt{5})/2$ is the golden ratio
  \item Take $x_n = \sqrt{1 - z_n^2} \cos{\varphi_n}$ and $y_n = \sqrt{1 - z_n^2} \sin{\varphi_n}$
\end{enumerate}

The desired set of points will be given by the set of triples $(x_n, y_n, z_n)$ for $n=1, ..., N$. See an implementation below, in the Table~\ref{table:FibonacciCode}:

\begin{table}[h!]
	\centering
	\begin{boxedCode}
	\begin{mmaCell}[addtoindex=0,moredefined={FibonacciNodes, x, y, z, r},morepattern={np_, np},morefunctionlocal={n}]{Input}
  FibonacciNodes[np_]:=Module[\{x, y, z, r, \mmaDef{\(\pmb{\alpha}\)}\},
  z = Table[1-2 \mmaFrac{n-1}{np-1},\{n,np\}]; 
  r=\mmaSqrt{1-\mmaSup{z}{2}};                                                                   
  \mmaDef{\(\pmb{\alpha}\)}=(2\mmaDef{\(\pmb{\pi}\)} N@GoldenRatio)(Range@np);
  
  Transpose@\{r Cos[\mmaLoc{\(\pmb{\alpha}\)}], r Sin[\mmaLoc{\(\pmb{\alpha}\)}],z\}
  ]
	\end{mmaCell}
	\end{boxedCode}
	\caption[Mathematica code for the Fibonacci nodes algorithm.]{Mathematica code for the Fibonacci nodes algorithm. \label{table:FibonacciCode}} 
\end{table}

\begin{table}[h!]
	\centering
	\begin{boxedCode}
	\begin{mmaCell}[addtoindex=20,moredefined={FibonacciStates, FibonacciNodes},morepattern={np_, np}]{Input}
  FibonacciStates[np_]:=Module[\{\mmaLoc{\(\pmb{\alpha\varphi}\)List}, \mmaLoc{\(\pmb{\psi}\)}\},
  \mmaLoc{\(\pmb{\psi}\)}[\mmaPat{\(\pmb{\alpha}\)_},\mmaPat{\(\pmb{\varphi}\)_}]:=Cos[\mmaFrac{\mmaPat{\(\pmb{\alpha}\)}}{2}]\{1,0\}+\mmaSup{\mmaDef{e}}{\mmaDef{i}
\mmaPat{\(\pmb{\varphi}\)}}Sin[\mmaFrac{\mmaPat{\(\pmb{\alpha}\)}}{2}]\{0,1\};
  
  (*(x, y, z) -> (Polar, Azymuthal)*)
  \mmaLoc{\(\pmb{\alpha\varphi}\)List}=(ToSphericalCoordinates@FibonacciNodes[np])[[All,\{2,3\}]];
  
  (*Removes singularities*)
  (*Psi_1, Psi_2, ..., Psi_n*)
  \mmaLoc{\(\pmb{\psi}\)}@@@\mmaLoc{\(\pmb{\alpha\varphi}\)List}/.\{Indeterminate\(\pmb{\to}\) 0\}
  ]
	\end{mmaCell}
	\end{boxedCode}
	\caption[Generates a set of points in the Bloch sphere from the Fibonacci nodes algorithm.]{Generates a set of points in the Bloch sphere from the Fibonacci nodes algorithm. \label{table:FibonacciBlochCode}}
\end{table}

For our purposes, this procedure can be used to uniformly generate points on the Bloch sphere. The Mathematica code which generates a list containing uniformly distributed states over the Bloch sphere is shown in Table~\ref{table:FibonacciBlochCode}.

One of the advantages of this method is that it is really simple to implement,
and since the points are not distributed randomly, we can get a reasonably uniform distribution even for a small number of points. Of course, if one is interested in
methods which randomly distribute the points more canonical approaches are also
available \cite{Muller1959}. For a pedagogical discussion on the Haar measure and how to apply it to quantum information theory in the random sampling of unitaries and quantum states, see Ref.~\cite{Team2021}.